\documentclass[aps,floats,prx,showpacs,preprint]{revtex4-1}

\usepackage{amsfonts,amsmath}
\usepackage{bm}
\usepackage{dcolumn}
\usepackage{amssymb}
\usepackage{epsfig} \usepackage{latexsym}
\usepackage{graphicx}
\usepackage{xcolor}
\usepackage{multirow}
\usepackage{subfigure}
\usepackage{hyperref}

\newcommand {\bR} {{\mathbf R}}
\newcommand {\br} {{\mathbf r}}
\newcommand {\ud} {{\textrm d}}

\usepackage{cancel}
\usepackage[normalem]{ulem}
\begin{document}

\title{Lectures on quantum supreme matter.}

\author{Jan Zaanen}

\affiliation{The Institute Lorentz for Theoretical Physics, Leiden University, Leiden, The Netherlands}

\date{\today}

\begin{abstract}

These notes are based on lectures serving the advanced graduate education of the Delta Institute of Theoretical Physics in the Netherlands in autumn 2021. The goal is to explain in a language that can be understood by non-specialists very recent advances in quantum information and especially string theory suggesting the existence of entirely new forms of matter. These are metallic states characterized by an extremely dense many body entanglement, requiring the supremacy of the quantum computer to be completely enumerated. The holographic duality discovered in string theory appears to be a mathematical machinery capable of computing observable properties of such matter, suggesting the presence of universal general principles governing its phenomenology. The  case is developing that these principles may well apply to the highly mysterious physical properties observed in  the high temperature superconductors and other strongly interacting electron systems of condensed matter physics.

\end{abstract}

\maketitle

\tableofcontents

\section{ Introduction and overview.}
\label{section:introduction}

The exceptional powers of physics in the scientific endeavour of mankind originates in the capacity of mathematics to facilitate us to look  beyond the capacity of our ape brain to discern the nature of reality. The history of the subject has all along revolved around advances in mathematics and advances in quantitative experimentation that meet each other once in a while, invariably  leading to scientific revolutions. In the first half of the twentieth century this was exceedingly successful, starting with Einstein's miracle year and climaxing in the landing of the standard model of high energy physics in the 1970's. But since I entered the professional floor in the early 1980's this became a sluggish affair. 

But it is hanging in the air that this train may start moving again. At stake is our understanding of the general nature of matter, and given the source of the new mathematical machinery this may eventually also shed light on the nature of space-time: quantum gravity is luring around the corner.  Matter is of course at the centre of the physicist's view on reality. But the way it has been understood rests on implicit assumptions that were taken as self-evident -- this 'paradigm' emerged gradually and has been greatly successful in explaining the behaviour of common stuff, varying from the solids, liquids and gasses of daily life, the "quantum fluids"  (the metallic state and superfluid states)  all the way up to the Higgs field at the high energy frontier. 

It is shimmering through that very different states of matter may exist, which are not all that rare but went unrecognized for the reason that the "eyeglass" in the form of mathematical equations were not available. It is about empirical mystery meeting novel mathematics, where the latter shines a bright light on the former. The meeting place is most unexpected departing from established wisdom. Humble experimentation on messy pieces of rusted copper that started 30+ years ago (condensed matter physics) appear to communicate with mathematical  contraptions developed over a similar long period by string theorists that were originally intended to address the nature of black hole singularities and so forth. In addition, the mathematics of information that has been evolving inspired by the quantum computer pursuit plays a crucial role as catalyzer of this reaction. 

Arguably, the only serious "new math meets new physics" event that evolved during my career span has been the discovery of topological order. I learned an interesting lesson from a person who played a key role in discovering the math that marries with the experiments in this context: Frank Wilczek. By shear experimental serendipity the fractional quantum Hall effects were discovered in the early 1980's. Laughlin's divine guess work in the form of his wavefunction surely gave away the crucial key. In the same period  a  development  propelled by mathematics was unfolding in Frank's theoretical high energy community, disconnected from the tampering in the condensed matter labs. The role of the mathematical subject of topology had been picked up in the 1970's in the form of the massive Thirring model. Frank had become aware that this had a condensed matter incarnation in the form of the Su-Schrieffer-Heeger model, merely by coffee time conservations with Bob Schrieffer at the ITP Santa Barbara were both of them had a job at the time. Inspired by this, Frank wrote down topological field theory in 2+1D, subsequently discovering that the mathematicians Chern and Simons had already accomplished this feat. The remainder is a famous history, exerting a large influence on contemporary physics. 

The advice that I got from Frank is to be acutely aware that the big leaps forward tend to happen at the border between subjects that are seemingly unrelated. He argued that there is a steady progress in these communities with facts and insights steadily accumulating but it then can happen that  they suddenly match -- CS theory being a case in point. The developments I will describe in these lecture notes may be of a similar kind.

It is still work in progress -- nothing is decided and it may still turn out to be a fluke. But the circumferential evidences are accumulating that we are on the right track.  Moreover, it is in any case a fertile ground to train young physicists for the simple reason that the separate subfields cover a large area of contemporary fundamental physics.  

What is this new form of matter? Only some basic notions of quantum information are required to appreciate it.  One lesson of the quantum computer pursuit is that the view on nature based on mathematical information theory is clarifying. Matter is about stacking microscopic buildings together, forming a wholeness that is governed by {\em emergence}: the wholeness shows behaviours that are  completely different from those of the parts. Macroscopic reality is constructed from quantum parts and upon stacking these together one runs automatically in {\em exponential complexity}. The effort for a classical information processing machine increases {\em exponentially} in the number of parts. This is in turn rooted in the exquisite quantum-physical  phenomenon of entanglement, and a machine that exploits the information processing capacity of entanglement may get this done in a polynomial time. This is captured by the quote "quantum supremacy". 

However, matter as we know it from the textbooks of physics is computable -- the triumph of twentieth physics. But how this can be having just established that a-priori thermodynamically large systems should fall prey to the incomputable exponential complexity? The reason is that the charted forms of matter are of a special kind. The macroscopic ground state ("vacuum") is actually {\em devoid of many body entanglement}. Its wavefunction is called the "short ranged entangled tensor product" in the quantum information language. 

The claim that I want to expose here is that forms of matter may exist, resting on a ground state that is {\em even in the macroscopic realms characterized by the exponential complexity associated with dense many body entanglement}. Dealing with new phenomena we need new names and I made up myself the "quantum supreme matter" quote in the title in reference to such stuff. 

 A natural place to look for such stuff is in the context of the "non-stoquastic" (again, quantum info language: see underneath) quantum systems. The prototypical examples are  systems of  strongly interacting fermions at a finite density.  One does not have to go to exotic  corners of the universe to find them. The electrons populating mundane pieces of rusted copper already fully qualify. These became famous because of the discovery of superconductivity at a high temperature in the 1980's. In the intervening period everything available in the arsenal of physics was thrown at this stuff, while a zoo of other solids were discovered exhibiting similar phenomena. The capacity to have a closer look in these electron worlds using the instruments of experimental condensed matter physics improved spectacularly since the 1980's. The result has been that the more we learned from these observations,  the more we got confused. A remarkable regularity is discernible in the data begging for a mathematical description. But the available equations did not match at all.  
  
Is this eerie regularity telling us a story about physical reality in the grip of quantum supremacy? To get anywhere one needs answers in the form of equations and these were not available. But help arrived from a most unexpected direction. During the last 40 years a monumental mathematical machinery was created by the string theorists, aimed at cracking the quantum gravity problem. Discovered in the late 1990's, the AdS/CFT correspondence became the most prominent part in this toolbox. Some ten years ago the string theorists started to experiment with the conditions met in condensed matter systems, finding that these equations describe a physics that is suggestively similar to what is observed in the "high Tc" type electron systems. 

When one encounters for the first time the way it works it may appear as absurd:  fanciful black holes under the reign of general relativity take the role of quantum computer to describe the observable properties of quantum stuff living in a space-time with a dimension missing! That this is acting as a quantum computer is a recent insight which has not completely settled down yet. In this regard, I may have some news in the offering in these notes. But in my perception these are no-brainers, it is pretty obvious once realized how to look at these matters. 

The stuff that is dealt with in AdS/CFT is entirely different from the humble rusted copper electrons, but the big deal is that it is very quantum supreme. As such, it reveals {\em general emergence principle} tied to the quantum supremacy.   As the usual "classical" matter is governed by emergence principles such as spontaneous symmetry breaking and even the very notion of "particles" as excitations, the correspondence reveals  analogous  meta-principles governing the properties of quantum supreme matter. These are entirely different from nearly anything found in the textbooks -- in fact, only the condensed matter main stream  "stoquastic quantum criticality" comes close. In a very recent development evidences are accumulating in the condensed matter laboratories that these general principles may be operational in the copper rust and related systems. 

The reader may have inferred that this sounds like a peculiar congruence of presently fashionable subjects. Black holes -- the gravitational wave detection of mergers, the event horizon telescope. The race to build a quantum computer, and the more subtle magic of high Tc electrons acknowledged at e.g. the editorial offices of Science and Nature. In my perception it is just a beneficial coincidence that these may be just parts of a greater wholeness and the familiarity with the various parts may help to disseminate the message more quickly.  It is just the dynamics of physics itself that is in the driver seat.  The crucial work was done in the various sub-communities over a long stretch of time. It is precisely Frank's thing  -- the leap forward takes place at the borders between sub-communities. 

But this places at the same time a bit of a burden on the communication. To fully appreciate what is going on one has to be quite well at home in the various portfolios. I see my own humble role in this context. I may have in this regard a bit of an advantage, for the reason that my brains are of the generalist type. I am in physics because of hedonism -- my brains are better entertained by physics than by anything else -- and as such I am also exceedingly promiscuous. I am working a lot with experimentalists, but also with computational specialists. But perhaps I  enjoy most working with the mathematically inclined. I stumbled somewhat by accident in the string theory pursuit, and my physics hungry brains could not get enough of it!

I decided to attempt to capture it on paper, in a maximally accessible elementary language. Much of it is quite conceptual, and the aspects that really matter are captured by simple and elegant equations although years of intense study may be required to become a specialist in  any of the various fields. 

Much of these notes is devoted to the context: up to Section (\ref{AdSCFTgen}) it is all preliminaries, while the news is found mostly in Sections (\ref{holoSM},\ref{holotransport}) and  the rather tentative final Section (\ref{highTcxep}). String theorists can skip Section (\ref{AdSCFTgen}) and scan (\ref{holoSM}-\ref{Intertwined}) to look for my tweaks of the standard AdS/CFT canon. But they should carefully study Section (\ref{fermionsigns}), I just know from close experience that this community has a bit of a blind spot for the sign troubles.   For the computational reader Sections (\ref{qucritical} - \ref{fermionsigns}) may be quite familiar. For condensed matter physicists much of Sections (\ref{SREproducts},\ref{qucritical}) may be staple food. 

To help you further getting a grip on this affair, let me first present an executive summary of the narrative, followed by packing meat on these bare bones  in the 170 pages or so that follow.

\subsection{Semi-classics versus quantum supremacy (Section  \ref{SREproducts}).} 

In the course of the twentieth century a paradigm emerged offering a highly successful  mathematical description for the nature of matter. When I entered physics in the late 1970's there was a sense that it captured the fundamentals of the physics of all matter. It revolves around strong emergence: depart from primitive microscopic building blocks, stack them together using the rules of fundamental physics and a wholeness arises showing behaviors being entirely different from the sum of the parts. 

Historically, this started in the statistical physics tradition, explaining the different phases of mundane matter like fluids and solids in terms of spontaneous symmetry breaking, explaining the transitions between the stable phases as well.  In the era of the standard model revolution in the 1970's it became increasingly manifest that the same kind of physical logic is also at work in the high energy realms. Symmetry is at the heart of this emergence agenda while one accommodates the local conservation of elementary charges employing gauge theory. This can yet be captured by the strategy of employing primitive microscopic degrees of freedom that may even be unrelated to physical reality, eventually renormalizing in a collective physics of the right kind. The astonishing success of "lattice" QCD is case in point, in computing the difference between the proton and neutron mass departing from an in essence artificial spin system living on a lattice infused by the appropriate symmetries. 

One step beyond and one encounters the high energy frontier. Also in these realms a general consensus developed that the standard model is no more than an effective field theory, of the same kind as the phenomenological theories encountered in condensed matter physics such as the Ginzburg-Landau theory that impeccably describes the macroscopic behaviours of superconductors.  The trouble is that the emergence is so absolute that the information regarding what is lying behind the standard model and classical gravity is completely erased from the whole. This captures the essence of the quantum gravity problem. 

A priori, the complicating factor is in the fact that the microscopic building blocks are governed by {\em quantum physics}. A recent development that at least in my head has played a crucial role is inspired by the engineering pursuit that is presently unfolding: the construction of the quantum computer. The origin of this is in the realization that {\em information} as part of physical  reality matters: mathematical complexity theory, devoted to a precise mathematical classification of the effort it takes to compute particular problems. This in turn revolves around the question whether this effort scales in a polynomial- ("computable") or exponential ("incomputable") fashion with the number of constituent bits. It was realized that the exquisite quantum property of {\em entanglement} is a computational "resource", making it possible in principle for the quantum computer that is designed to exploit the entanglement to compute exponential hard problems that are beyond the capacity of any classical computer.  This is captured by the "quantum supremacy" quote that recently made headlines. 

This is in turn backreacting on the understanding of physical reality. Dealing with many body entanglement the mathematical information theory is still in its infancy. To the degree that we need it here, it is actually not going beyond {\em rephrasing} wisdoms from quantum physics that were known all along. But the benefit of using the quantum information language is that it supplies a crisp complementary conceptual framework. It is a source of {\em words} that capture the essence of the quantum physics in this information language, like "SRE products", "(non)-stoquastic" and even my "quantum supreme stuff". This is all quite elementary but  it is helpful in tidying up our comprehension of what really matters.          

Matter is a-priori the stuff formed from an infinite number of microscopic "qubits".  In full generality, it {\em should} run into the exponential complexity associated with the {\em many body entanglement} in the exponentially large many body Hilbert space -- the quantum supremacy affair.  But this is incomputable by conventional means. Why was it then possible for our forefathers to write text books regarding the way that matter works, getting seemingly everything right? Although it still has to penetrate the high energy and condensed matter textbooks, the information language is in this context simple and razor sharp. 

The characterization of matter departs from the nature of the zero temperature ground state ("vacuum"). The ground states of the matter discussed in textbooks is actually invariably of a special kind called "short ranged entangled tensor product states", abbreviated as "SRE-products".  I will review this in detail in Section  (\ref{SREproducts}): the vacuum is in essence "anchored"  in a classical bit string. At the microscopic scale one is dealing with entanglement but in the renormalization towards the macroscopic scale this entanglement fades away. The macroscopic state itself is devoid of any entanglement and it represents a classical information processing machine. It can therefore be described in terms of an effective field theory in the  {\em classical} limit. 

One may then object referring to the traditional "quantum liquids" such as superfluids/superconductors and Fermi liquids, or even the confining state of QCD. Aren't these reflecting quantum properties on the macroscopic scale? The answer is in the microscopic {\em representation} used to construct the "anchor" tensor product. In essence, this conventional "quantum matter" corresponds with "classical bit strings" where the individual bits are picked from the wave side of the quantum mechanical particle-wave duality as discussed at some length in Section  (\ref{SREproducts}). Perhaps not generally realized, even the macroscopic Fermi liquid has to be regarded as a "classical" state of matter in the quantum-information sense of the word. It is devoid of many-body entanglement. 

This macroscopic "classicalness" is actually the key to the success of the textbooks in making it possible to compute reliably the properties of much of the matter that experimentalists have encountered.   This is particularly explicit in condensed matter physics where the microscopic point of departure is known. The classical bit string anchor state underlies the standard mean-field theory relating to order parameters, like the BCS theory. One should subsequently "dress" it  perturbatively -- these are the "diagrams" wiring in the short ranged entanglement, an effort of polynomial complexity. 
 
In high energy physics the deep UV is not known. However, this ESR product affair is at the core of the prevailing paradigm of quantum field theory. It is just the mean-field affair in "reverse gear": {\em semiclassics}. One departs from the {\em classical} field theory, in the form of an action that reproduces the classical physics at the saddlepoint. This is then re-quantized by promoting the free fields modes to harmonic oscillators, associating the quantized classical modes with particles. The non-linearities of the classical theory turn into the interactions between the particles, handled by diagrammatics wiring in the short ranged entanglement.

This program has been very successful in charting the nature of matter in the physical universe. But does it capture {\em everything} that exists? Do states of matter exist that are incomputable with the available machinery because these are controlled by the exponential complexity inherent to the quantum physics of many things? Paraphrasing the quantum computer language, I like to call these "quantum supreme matter".   

\subsection{The intellectual crisis in condensed matter physics: high Tc superconductivity (Section \ref{highTcxep}).}

Mankind needs mathematics to deal with the physical world and without equations we are struck blind. Could it be that such quantum supreme matter is lying in front of us but we do not recognize it lacking the "mathematical eyeglass"?  A place to look for it is condensed matter physics. This is dealing with states of matter formed from constituents that have in principle no secrets: the electron systems in solids. Different from the high energy realms, experiments are relatively easy and it is a rich basin of empirical information. It is also the realm where semiclassics triumphed with the highly successful theories that are at the root of the electronics revolution (band structure) but also explaining the state of normal metals (Fermi-liquids) and superconductivity (BCS theory). 

Once again, in the early 1980's the belief was widespread that the fundamentals were completely known. In this constellation, in 1986 superconductivity at a high temperature (up to a $T_c \sim$ 150 K) was discovered  in the unlikely chemical territory of oxidized copper materials.  This triggered a gold rush -- when $T_c$ could be pushed up  to room temperature engineering applications would be bountiful. But it was also an era when the physicists who were young during the BCS revolution occupied the executive floor of condensed matter physics. This triggered an unprecedented hype, having the beneficial side effect that these electron system were intensely looked at in the laboratories. The prevalent mindset at the time  took the existing paradigm for granted, and the focus was entirely on variations on the established themes. The normal metallic state was assumed to be formed from electron-like quasiparticles that in the guise of the BCS theory are subjected to a strong attractive interaction that should  lead to the formation of electron pairs at a high temperature -- the "superglue" notion. 

However, already early on experiment revealed behaviours that were greatly surprising: in an obvious way, these were at odds with anything that followed from the available equations. Halfway the 1990's the hype came to an end when it became clear that the engineering wishful dreaming would not substantiate. However, the subject stayed alive, in fact driven by advances in the instrumentation in condensed matter laboratories. Although not in the public eye, this progress has been spectacular. Some of the existing "telescopes of the electron world" improved by many orders of magnitude: a case in point is photoemission where the resolution increased by a factor $10^4$. Other techniques sprang into existence such as the powerful Scanning Tunneling Spectroscopy machines. These innovations were often first unleashed on the cuprates. But the more we learned, the more it became unavoidable to accept that the available equations were not delivering. This field of enquiry turned increasingly into a highly empirical pursuit. When it all started the theorists were very vocal occupying a majority of the plenary slots at the big meetings. Presently, these plenary programs are dominated by the experimentalists. 

Much is going on in these electron systems. This is further complicated by the lack of a mathematical framework  that can be used to filter and organize the myriad of experimental results. That is, my claim is that the recent progress in theory which is the subject of these lecture notes may now provide  such a framework. It is far from perfect, let alone that it can be claimed to be decisive but it is in a stage that I find it undeniable that the gross conceptions appear to shed a clear light on the mysteries. But you have to first get acquainted   to this very different way of thinking about these matters. While this story unfolds underneath, I will here and there include reference to cuprate experiments, to zoom in on the big signals in the very last Section (\ref{highTcxep}). 

\begin{figure}[t]
\includegraphics[width=0.9\columnwidth]{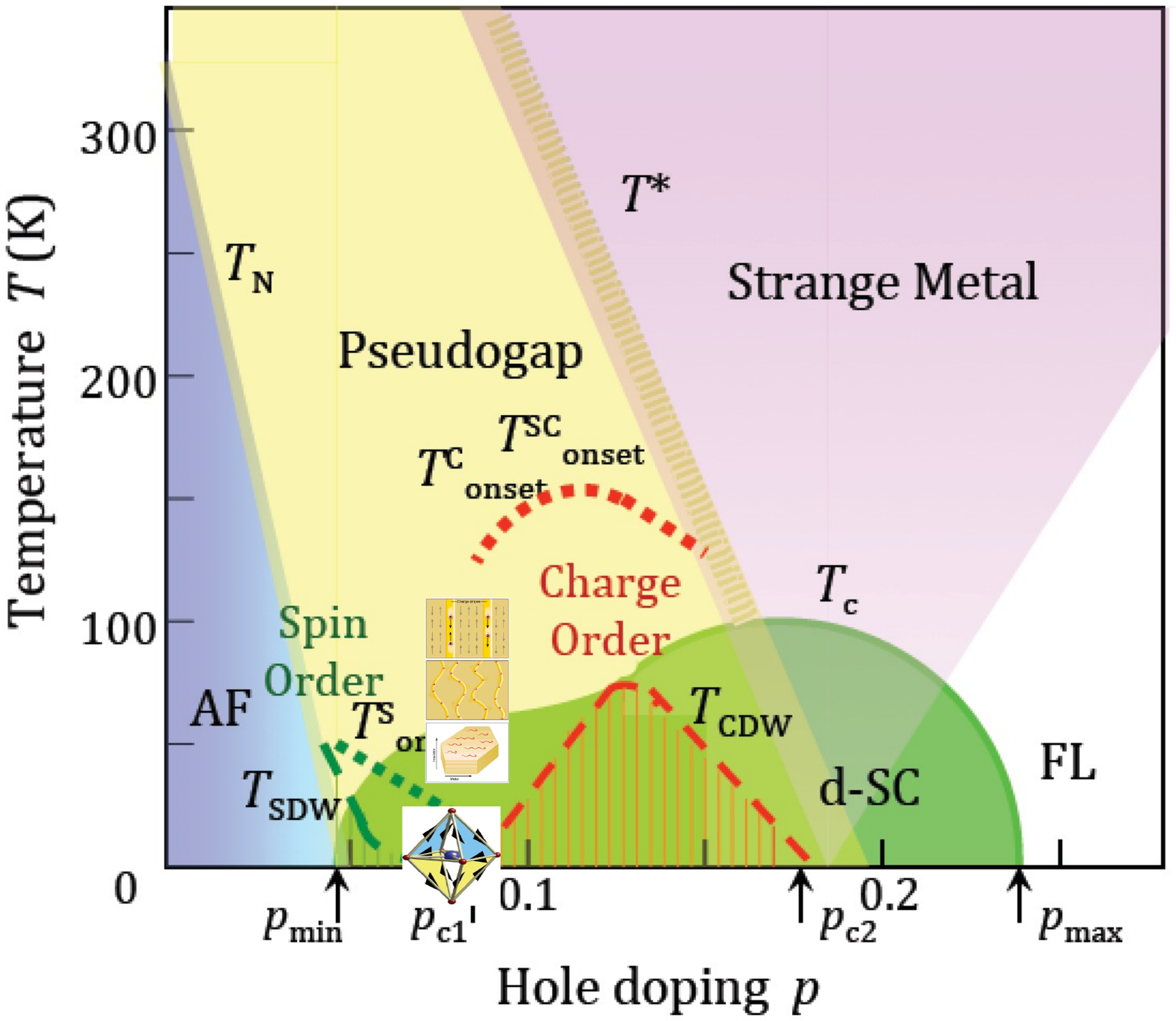}
\caption{ The "phase diagram" of hole doped cuprate high Tc superconductivity, according to a 2015 community consensus \cite{Naturecons15}. One departs from a Mott insulator that is doped (x-axis) and a variety of phenomena is found as function of temperature (y-axis), see the main text.}
\label{fig:cupratephasedia}
\end{figure}

Let me here present a short roadmap of this physics landscape, for orientational purposes \cite{Naturecons15}. It departs from the chemistry: these cuprates are formed from simple "perovskite" $CuO_2$ layers -- you may think about them as simple square lattices. These are kept apart by electronically inert spacer layers formed from highly ionic insulators. All the electronic action is in the copper oxide layers. The story start with stoichiometric "parent" compounds characterized by effectively one valence electron per $CuO_2$ unit cell. These are so-called Mott insulators: different from normal (band) insulators the electrons are localized because of very strong local repulsive interactions. In essence, it is just a traffic jam of electrons  (Section \ref{Mottness}). These are subsequently doped as in semiconductors by dopants in the spaces layers. When the doping level $p$ exceeds 5\% or so metallic behaviour sets in and the cuprates start to superconduct, see Fig. (\ref{fig:cupratephasedia}). At doping levels less than $p_c \simeq 0.19$ one is dealing with some kind of electronic "stop-and-go" traffic and at low temperatures one finds a novel "intertwined" cocktail of exotic ordering phenomena. Ironically, I found it convenient to discuss these in a bit of a detail in the most extreme "black hole" story altogether, Section (\ref{Intertwined}). 

This seems to set in at the "pseudogap" temperature indicated by $T^*$ which is decreasing in a more or less linear fashion with doping while the superconducting $T_c$ is rising. The latter reaches a maximum at the "optimally doped" doping level $p_{\mathrm{opt}}$, which is close to- but not coincident with the "critical doping" $p_c \simeq 0.19$ suggested by extrapolating the $T^*$ to zero temperature. Upon increasing doping further one enters the "overdoped" regime where $T_c$ is decreasing. At a doping $p_{\mathrm{max}}$ superconductivity eventually disappears. 

Above $T^*$ (underdoped) or $T_c$ (overdoped) one enters the metallic state of the cuprates: this is the king of the hill in this landscape when it comes to the big mystery story. This  was already established in the late 1980's when this state was called the "strange metal". This will be the main target in the empirical  theatre of the theoretical developments that I will present in these lecture notes. 

\begin{figure}[t]
\includegraphics[width=0.9\columnwidth]{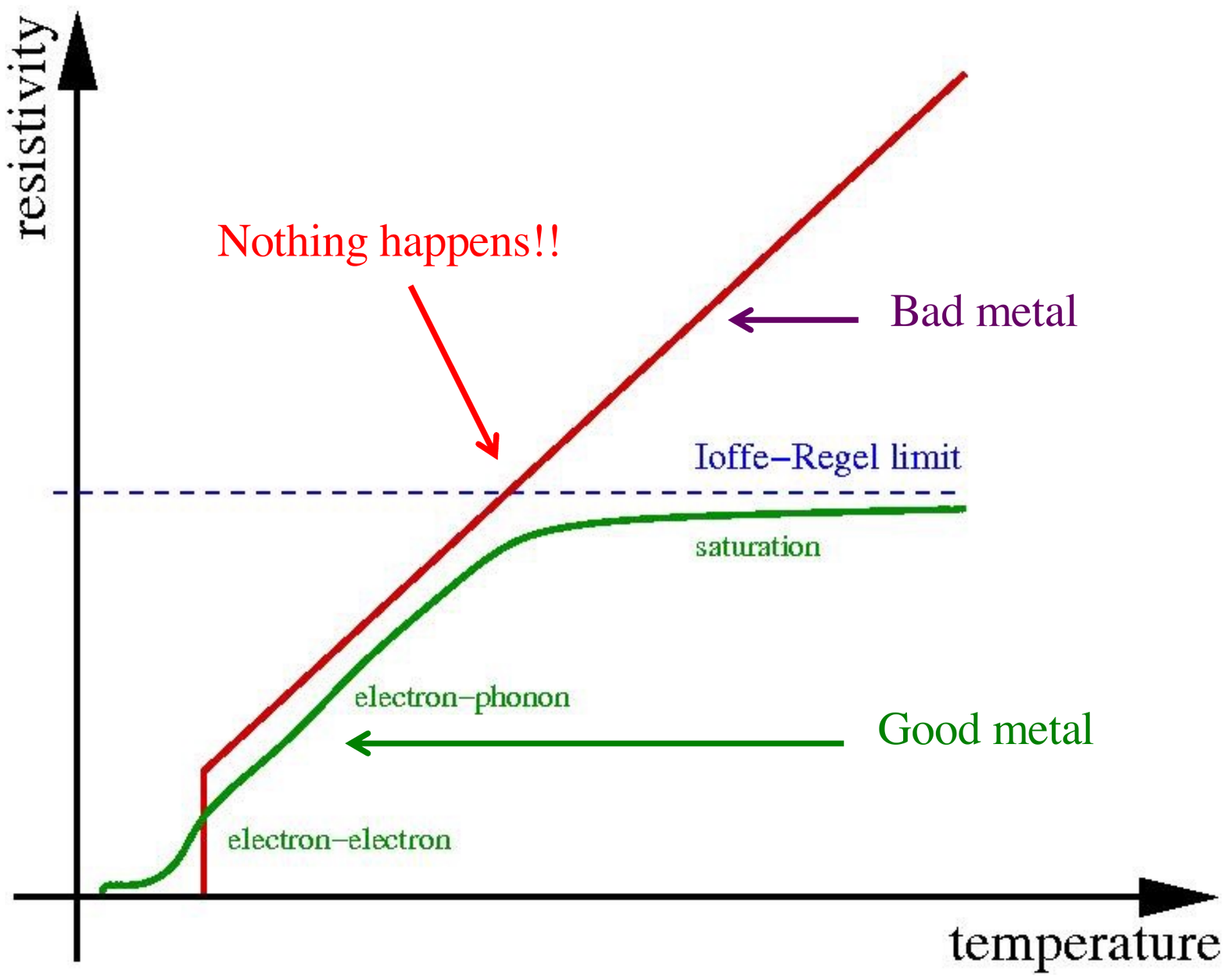}
\caption{ The famous linear-in-temperature electrical resistivity of the cuprate strange metal near optimal doping (red line), compared to the typical outcome for a conventional Fermi-liquid metal (green line).}
\label{fig:linresistivity}
\end{figure}

The strangeness appeals to the physicist's soul (Section \ref{linresemp}). The issue is that the physical properties exhibited by the strange metals behave actually in {\em very simple} ways. But departing from established theory this simplicity should not occur! 

This wisdom pertains to any property that has been measured but one does not have to dig deep: the DC electrical resistivity has been all along a highlight of the strangeness, see Fig. (\ref{fig:linresistivity}). Let us first consider the resistivity of a conventional Fermi-liquid metal. At a finite temperature the Fermi-liquid cannot be distinguished from a classical kinetic gas at the macroscopic scale (Section \ref{Fermiliquid}). Generically one needs a breaking of the translational invariance to obtain a finite resistivity and this is associated with the presence of the background ionic lattice. But in a Fermi-liquid the thermal quasiparticles loose their momentum individually (see Section \ref{holotransport}). This is basically like a pin-ball machine where the scattering mechanism is by default strongly temperature dependent. At low temperature it is dominated by the "Umklapp" electron-electron scattering $\sim T^2$. Subsequently the inelastic scattering involving the phonons takes over that may cause  a variety of temperature dependences. Eventually the mean-free path becomes of order of the lattice constant and the resistivity saturates  (Mott-Ioffe-Regel limit). But nothing of the kind happens in the strange metals: one finds a perfectly straight line all the way from the superconducting $T_c$ up to the melting point of the crystal at a 1000 degrees or so!

This "unreasonable simplicity" reaches even farther. Optical conductivity measurements \cite{vdMarel2003,Heumen21} show that it is a so-called Drude transport, where the resistivity is set by a Drude weight ("carrier density") and a momentum relaxation rate (Section \ref{holotransport}). The former is temperature independent while the relaxation time is $\tau_K \simeq \tau_{\hbar} = \hbar /(k_B T)$! This time has a fundamental status, it is just dimensional analysis: Planck's constant has the dimension of action, energy times time while $k_B T$ has the dimension of energy. I coined myself the name "Planckian dissipation" for this quantity \cite{Planckiandiss}. You will meet this motive a number of times in the notes: the claim is that this is the shortest possible time associated with the production of heat, allowed by the fundamental principles of quantum physics. Moreover, this bound can only be reached in a {\em compressible, densely many body entangled system}. Although there are many more signals in the data pointing in this direction, this "Planckian" resistivity is iconic.  

It was first identified in the context of the "stoquastic" quantum critical states as discussed at length in Section (\ref{QPTgen}). This departs from the idea that a continuous zero temperature "quantum" phase transition occurs where some form of order disappears (see next item). Part of this agenda is that one expects a "quantum critical wedge" anchored at the zero temperature transition. At the time of writing of the Nature review \cite{Naturecons15} it was generally believed that the strange metal behaviour is driven by such a "quantum critical point" involving the disappearance of the "pseudogap order": the purple wedge enclosing the "strange metal" in Fig. (\ref{fig:cupratephasedia}). 
  
But in the mean time this has changed drastically. Recent developments show that an underdoped- and overdoped {\em strange metal phase} exist, meeting at a first order like transition at $p_c$ as discussed in Section (\ref{qucritphase}). The updated phase diagram looks like Fig. (\ref{fig:husseyphasedia}). The central challenge addressed by these notes is to explain such "quantum critical metallic phases" in terms of a mathematical theory. 

\subsection{Stoquastic systems and the quantum critical point (Section \ref{qucritical}).}

The first big step in the right direction is embodied by the  understanding of "stoquastic" quantum phase transitions, that developed in the 1990's, as reviewed in Section (\ref{qucritical}). The quote "stoquastic" is yet again coming from the quantum information side. This refers to a special subclass of quantum systems that map on an equivalent {\em statistical physics} problem. This rests on path integrals. One maps the quantum system onto the path integral "sum of worldhistories" in space time, followed by the analytic continuation to imaginary time. Only under quite special conditions the ensuing path integral may then become coincident with a Boltzmann partition sum. This can be handled with the formidable powers of the stochastic mathematics exploited in the statistical physics tradition: it is equivalent to addressing the occurrence of phases and phase transitions in classical systems. This territory is quite well charted. 

Besides illustrating the stable phases as ESR products, this also reveals that  one may find the "strongly interacting quantum critical state" at the zero temperature quantum phase transition. This is just the quantum equivalent of the strongly interacting critical state realized at thermal phase transitions, a highlight of 1970's statistical physics. The key is that such states exhibit universality (Section \ref{thermaluniversality}): scale invariance emerges, and the physical observables are captured by simple scaling functions. These require data  such as the anomalous scaling dimensions (power law exponents) that are completely determined merely by symmetry and dimensionality.  

The "unreasonable simplicity" that begs for an explanation in e.g. the cuprates is already shimmering through: physical responses are characterized by the quantum incarnation of the scaling functions of the thermal phase transitions. All one needs to do is to "Wick rotate" back from the Euclidean (imaginary time) to the Lorentzian signature (real time) such that the scaling behaviours acquire an entertaining twist. The propagators of the quantum system turn into branch cuts at zero temperature, while these exhibit "energy-temperature" scaling at a finite temperature. Simple scaling arguments reveal the generic occurrence of Planckian dissipation in the finite temperature quantum critical states at macroscopic times. This can be  understood by involving an Eigenstate Thermalization logic \cite{dAlesio} ruled by dense many body entanglement. This is the legitimate motive for the long standing belief that the strange metal behaviour originates in such a quantum phase transition. As I just pointed out, this is not quite true in the cuprates  but there is a host of other systems where the experimental support for strange metal like behaviour associated with a quantum phase transition is very convincing.
 
When this critical state was at the centre of attention in the 1970's, mathematical complexity theory had not yet disseminated in the physics community.  But nowadays it appears to be a common perception among the experts that the difficulties to simulate this state are rooted in the fact that right at the strongly critical point this state is characterized by exponential complexity. 

I will argue in Section  (\ref{thermaluniversality}) that the simplicity embodied by the scaling relations is seemingly paradoxically {\em caused} by the complexity of the states. Although conjectural, this "scaling simplicity of observables as consequence of the quantum supremacy" may be the overarching principle turning to the uncharted territory of "non-stoquastic" physics -- it appears to be the central message signalled by the black holes of holography.

This quantum supremacy claim seems to imply an ultimate complexity, an exponential large number of bits that seemingly has the potential to describe an extremely {\em complicated} world. However, the issue is that this complexity refers to the {\em unitary} quantum evolution that one has to tackle when one wants to compute this stuff in full. But the quantities that can be observed by experimentalists are the "vacuum expectation values" ( in the jargon of high energy physics, VEV's): the {\em expectation values} of operators that remain after the {\em collapse of the wavefunction}. This is the same deal being at the heart of the working of the quantum computer. The unitary evolution implemented by a sequence of one- and two qubit gates processes the exponential complexity but actually information is not processed. The information processing takes place when the state collapses yielding the "read out". The observables in the physics laboratory are equivalent to the read out information. 

The other crucial ingredient is the absence of scale. At the quantum phase transition such a scale invariance is emerging, while from the quantum-information side one learns that the presence of scales (gaps) is detrimental for the near ground state "quantum supremacy". For (effectively) Lorentz invariant systems this get lifted to conformal invariance and this powerful symmetry governs the physical observables, mathematically captured by the powerful conformal field theories (CFT's).

The take home message is that this stoquastic quantum critical state is a relatively well understood form of "quantum supreme matter", revealing a phenomenology where one discerns a number of general principles. These reveal a strong contrast with the "particle physics paradigm" of semi-classics.   The "unreasonable simplicity" implied by universality is a case in point. The main message of holography in this context appears to be that the same type of meta-principle is ruling in the  uncharted realms of finite density systems infested by the sign problem.   

\subsection{Quantum field theory 2.0: what lies behind the sign problem brick wall?  (Section \ref{fermionsigns})}

For the mathematically inclined it is natural to stare away from problems where equations are not available. A case in point is the (fermion) {\em sign problem}. That there is big trouble with generic many body quantum problems was realized a long time ago. But it was effectively shuffled under the rug. In fact, all of the quantum field theory tradition in high energy physics is rooted in the stoquastic machinery that I just discussed, justifiably so since the standard model agenda is characterized by conditions (zero density, time reversal symmetry) that eliminates the signs.  

Computers keep people honest and the sign problem has been on the main stage of the computational community working on condensed matter inspired problems. I will shortly review where they are after their 40 years of struggle with this problem in Section (\ref{Compmethods}). For a long time it was put away by the mainstream as a "technical  problem for software engineers". I started to appreciate it myself already in the early 1990's as a {\em foundational} problem, perhaps even as  the most pressing of all "known-unknowns" in fundamental physics. It has been with me since then -- I tried repeatedly a hack but as everybody else to no avail. 

But in a recent era perceptions are changing.  Yet again the quantum computer has exerted beneficial influences. It is fully acknowledged in this community -- it is a high priority item on their bench marking list.  It cannot be emphasized enough that it is not a programming problem: the sign problem is generic in physical systems and it acts as an impenetrable brick wall for our understanding  of what is  going on. There used to be no equations whatsoever! But at the same time such physical systems should exhibit reasonable physical behaviours. Is the "strangeness" observed experimentally in the cuprates and elsewhere telling a story about the very different nature of "sign-full physics"? 
 
In fact, the Section (\ref{fermionsigns}) devoted to the fermion signs is the last pre-requisite before I turn to the holographic agenda. When the first results for finite density holography started to appear  around 2008 I was immediately intrigued, suspecting that it had dealings with such physics. In the mean time this has settled in, see the second half of these notes. The claim is that {\em holography is the first, and presently only controlled mathematical framework that is addressing  a form of "sign full matter".}  
I like to call it  "quantum field theory 2.0." It is a new release dealing with the collective behaviour of signful matter ("2"), but it is the first quite primitive version ("0"). I repeat, what is called quantum field theory is entirely resting on the stoquastic machinery which is cooked to perfection, something like "QFT 1.61".

What is the sign problem? I just explained that there is the special subcategory of "stoquastic" quantum problems that can be handled by the powerful stochastic methods of statistical physics yielding a quite thorough understanding of how it works. However, this invariably involves special conditions: generic quantum problems are "non-stoquastic" as it is called in the quantum information language. The stochastic approach is failing because of the occurrence of "negative probabilities" which of course do not make sense. 

Given that there is a long tradition to ignore the sign problem I have included an elementary tutorial (Section \ref{fermionsigns}). Much of this is just meant to supply some basic intuition, in the form of elementary exercises such as the first quantized path integral for fermions. But there is not much to tell: once again, I have been myself banging my head against  it for the last quarter of a century not getting anywhere. 

The crucial part is Section (\ref{TroyerWiese}): the mathematical theorem by Troyer and Wiese \cite{Troyerwiese} that boils down to the demonstration that the problem of strongly interacting fermions at a finite density is {\em generically} characterized by a quantum supreme ground state! As a leading character in the computational effort, Matthias Troyer  had a track record in shooting down software-engineering style attempts to "solve" the sign problem. Their seminal contribution was  inspired by the desire to  establish once and for all that it is a foundational affair.  

Since the discovery of cuprate superconductivity much of this computational effort was focussed on the problem of doped fermion Mott insulators. One of the few facts that we know for sure is that the cuprates are of this kind. The essence may be captured in terms of the simple Hubbard and related $t-J$ models, see Section (\ref{Mottness}). In this context one is encountering a particularly bad sign problem that can be diagnosed in an entertaining way.  The  computational community has been battling this particular problem for many years, running invariably into the exponential complexity problem (Section \ref{Compmethods}): even with the present day computational resources and shrewd algorithms really nothing is known for sure.

\subsection{Holographic duality: looking behind the quantum supremacy brick wall (Section \ref{AdSCFTgen}).}  

Not so long ago it used to be that we did not have even a single equation telling us anything regarding the "reasonable physics" of systems forced by the sign problem to be  of the quantum supreme kind. This started to change in 2007 when a big effort erupted in the string theory community revolving around "high Tc" style  condensed matter physics. 

One should value the specific merit of this community. They are the contemporary representatives of the "Einstein method": seek the propulsion to decode the nature of reality in the form of {\em mathematics}. In the 40 years or so that string theory has been around, the mathematics that they produced is breathtaking. The trouble is that serious math cannot be explained in a tweet: one has to study it intensely, do exercises and so forth to wrap the mind around the meanings conveyed by the equations. The tragedy is that these profound equations refused to connect to empirical reality in the domain where it was intended for: quantum gravity and related phenomena like the origin of the standard model. When a mini black hole would have banged during the first run of the LHC -- a highlight prediction of string theory -- it would have looked different. Unfortunately, the  LHC did not produce any surprise beyond the standard model lore. 

I got myself into it in the era before 2007. I could not stand it that string theory talks appeared  in my reference frame as hocus pocus and I decided to just learn it a bit. A Stanford string theory graduate came to Leiden with the desire to cross-over to condensed matter and he volunteered to organize an informal course. This great intellectual adventurer (Darius Sadri, from Persian descent) unfortunately deceased at a very young age. After a year of intense sessions that caused frequently headaches in my ageing brain I started to feel at home and soon thereafter the "Anti-de-Sitter/Condensed Matter Theory" (AdS/CMT) news broke. It was shear ocincidence that I got into it at the right time. All along the main bottleneck for further dissemination has been that it takes a major investment to appreciate how this math works. 

The period 2007-2013 was an amazing roller coaster ride with astonishing leaps forward happening every half year or so. But around 2013 it became clear that about all low lying fruits were picked -- the program was not at all done but further progress required the patience and resilience characteristic of empirical scientific pursuits. But patience is not the strength of the string theorists and in 2013 the mainstream turned suddenly elsewhere. This is best called "its form qubits" -- yet again inspired by quantum information but with the focus on the quantum gravity side. This has produced some spin-off of potential relevance to condensed matter but this is mainly revolving around quantum non-equilibrium \cite{SonnerLiuRev}, a subject beyond the scope of these lectures.        

There is still much work to be done to bring the half-finished products to the condensed matter market place. But the trouble is that a work force is needed that  is primarily inspired by the condensed matter context  -- the string theorists are excused, their heart lies with quantum gravity. Who has the time and energy in the condensed matter community to delve  in this unfamiliar territory that is a threat for funding success? The bottom line is that presently only a small group of physicists is working on the subject. These lecture notes are also intended be part of a recruitment effort. I can assure the reader that this "holography" is very enjoyable once you have figured out how it works. 

Explaining how it works in detail is beyond the scope of these notes. There are excellent books available \cite{Erdmengerbook,holodualbook,lucasbook}  and one may sign up for  specialized lecture courses organized by the string theorists. But the good news is that in the mean time matters have cleared up to a degree that a story can be told which is entirely focussed on the ramifications for the physics of matter. The mathematical machinery is not unlike the hardware in the experimental laboratories. As a theorist one does not need all the gory details of the op-amps and other pieces of hardware that are often critical for the performance of the machines. But it is possible with much less effort to familiarize with the gross workings of the machines, the kind of information they deliver and their shortcomings.

The machine at stake has a number of names: the "AdS/CFT correspondence", "gauge-gravity duality", or "holographic duality", while the application to condensed matter is called "AdS/CMT". It is an outgrowth of the intense pursuit in the 1980's and 1990's that climaxed in the "second string revolution" in 1995. Soon thereafter Maldacena discovered the correspondence as a spin-off. It is a no-nonsense mathematical device that makes possible to compute matters in a completely controlled way. The big deal is that it reveals a profound and surprising mathematical connection between gravitational physics -- general relativity (GR) -- and non-gravitational quantum physics. In the latter the quantum supreme part is on the main stage. The profundity is in the notion that this connection is "holographic", metaphorically referring to the usual holograms: two dimensional photographic plates that reconstruct three dimensional images when pierced by coherent light. According to the correspondence,  quantum physics in D space time dimensions maps on gravitational physics in D+1 dimensions. 

In Section (\ref{Maldacenaholo}) I will attempt to give an impression of how the original Maldacena AdS/CFT works, avoiding as much as possible mathematical intricacies. The take home message is that it addresses precisely the zero density stoquastic quantum criticality of Section (\ref{qucritical}), reconstructing the scaling phenomenology as I highlighted. This  pertains especially also to the finite temperature regime where the bulk dual turns out to be governed by a Schwarzschild black hole. This is an amazing story with as highlight the "fluid gravity duality". The macroscopic behaviour of the fluid realized at a finite temperature should be governed by the Navier-Stokes theory of hydrodynamics. It has been demonstrated that there is a precise mapping between the near-horizon dynamical geometry of this black hole in the gravitational "bulk" and hydrodynamics in the "boundary". The influence of the zero temperature strongly interacting quantum critical state shines through in the latter in the form of the "minimal viscosity" which is the way that Planckian dissipation manifests itself in this context. 

But this "holographic oracle" is still littered with mystery. It is only brought under mathematical control in a special limit of the boundary quantum physics, which has to be in the large $N$ limit of a matrix field theory at large 't Hooft coupling as I will explain in  Section (\ref{Maldacenaholo}). Only in this limit the bulk theory turns into classical GR  in the bulk -- the correspondence is general but for finite $N$ one has to tackle stringy {\em quantum} gravity in the bulk. Although the initial promise was that AdS/CFT would shine light on this, it is still largely in the dark. This large $N$ limit has of course nothing to do with electrons in solids,  but this does not appear to matter for the  quantum criticality phenomenology that I announced in the above. 

This unreasonable success of classical gravity to capture the phenomenological description of stoquastic densely entangled matter seems to be rooted in a deep mathematical relation. Dealing with the curved space-times of Riemannian geometry  one meets the notion of "isometry", the sense of "symmetry" in this context. Consider for instance  the two dimensional surface of a perfect ball. The (scalar) curvature is everywhere the same and the ball is therefore characterized by a maximal isometry. Mathematically there is a relation between the isometries of a curved manifold in  D+1 dimensions and the symmetry in D dimensional flat spacetime, of the kind that controls the quantum theory. Among others it follows that the conformal invariance of the boundary theory (conformal field theory, CFT) is precisely encoded by a maximally symmetric {\em hyperbolic} ("anti-ball") geometry in the bulk, the Anti-de-Sitter space (AdS). One can view the correspondence  as a generalized symmetry processing machine. More than anything else, this leads to the striking feature that the extra "radial" dimension of the bulk is coincident with the "scaling direction" of the renormalization group of the boundary theory: the RG flow is "geometrized" in the bulk, the "general relativity (GR) = renormalization group (RG)" notion. 

\subsection{Finite density and the "covariant" RG flows of strange metals (Section  \ref{holoSM}).}

Up to this point in the text, I have just been setting up the stage and in Section  (\ref{holoSM}) the real work starts. What has AdS/CFT to say about finite density systems, "infested" by fermion signs? Different from the zero-density case, we have nothing to compare with. There is no doubt that the matter described by the boundary is densely many body entangled -- quantum supreme -- but the sign problem makes it impossible to address it in any other way theoretically. But the holographic outcomes are also at finite density controlled by the "GR = RG" principle that exerts its great power at zero density, now revealing a differently structured phenomenology. 

The common denominator with stoquastic quantum criticality is in the simplicity of observables governed by {\em scaling} behaviour, apparently as a consequence of the exponential complexity of the states. But the way that the scaling works is quite different.  

If successful, it would shed a penetrating light on the high Tc enigma and demonstrate that radically new forms of quantum supreme matter do exist in the physical universe. But it also works the other way around. It would be a most valuable piece of information alluding to the highly mysterious aspects of the correspondence, that may be of help for the string theorists in their quantum-gravity quest. When you are a condensed matter experimentalist, realize that there is a potential that with your humble equipment you may dig out facts alluding to the quantum gravity mystery that may turn out to be way more consequential than anything that will be delivered by the high energy physicists and astronomers! Yet again, although the jury is still out, and in a number of regards it is still quite confused, evidences have been accumulating in recent years that this is on the right track --  have a close look at  the final Section (\ref{highTcxep}). 

This finite density holography took off in 2008. The holographic "dictionary" is insisting that it is encoded in the bulk in the form of a charged black-hole like object. All along the progress was propelled by the richness of GR that is unleashed under these conditions, there is just much more possible than for the zero density case. The correspondence is a merciless computational device  and upon dualizing the precision bulk solutions a boundary world opened up that has striking similarities with what is observed experimentally in the high Tc style electron systems. But it is stronger than that -- there is an eerie similarity with the Fermi-liquid/BCS wisdoms as well considering the gross features of this physics. This was the main motive, to a degree subconscious, that fuelled the enthusiasm. I was myself not an exception. Although there is some mention of quantum information, the prevalent sentiment you will find in our book that we completed in late 2013 \cite{holodualbook} is of this "ain't it cool that fancy black holes encode for stuffs that are eerily similar to what electrons do?"

The present perspective that it reveals general phenomenological principle alluding to fermion-sign induced dense entanglement gradually cleared up in the intervening period \cite{JZsciPost19}.  This view may well be surprising (if not upsetting) for a string theorist that bailed out in the 2013 era. 

One way to phrase the outcomes is to call the holographic strange metals "generalized Fermi-liquids", actually in the specific sense that the strongly interacting stoquastic quantum critical states are generalizations of the free Landau critical state above the upper critical dimension. As for the latter, the Fermi-liquid is characterized by {\em power law} physical responses indicating some form of absence of scale which is yet quite different from the scale invariance controlling the quantum critical point affair. The origin is in the {\em degeneracy} scale: the Fermi- energy. The sign problem enforces nodes in the wave function that will have the universal consequence of inducing a huge zero-point motion energy. In a Fermi-gas this is easy: fill up the single particle states employing the Pauli principle finding a Fermi-energy that is in common metals easily of order $10^5$ Kelvins, while the ensuing Fermi-pressure keeps on the big stage of the universe neutron stars from collapsing into black holes. But this is universal: also the densely entangled incarnations have to deal with such a fermionic degeneracy scale.        

The Fermi-energy is just accommodated automatically in the Fermi-liquid power laws. For instance, its (Sommerfeld) entropy $S \simeq T/E_F$. Holography is in this regard an eye opener because it spells out the origin of this different scaling behaviour exploiting the geometrized renormalization group in the bulk. This very recent insight that I learned from Blaise Gouteraux \cite{Gouterauxcov} is spelled out in the most important passage of these lecture notes: Section (\ref{holoSM}), climaxing in Section (\ref{covariantRG}). Instead of the bulk isometry being {\em invariant} under scale transformations as for the CFT's, it is {\em covariant} instead. This  accommodates the degeneracy scale in a natural way and this principle enforces the properties of the Fermi-liquid in terms of a set of associated scaling dimensions of a kind that you don't find in a similar form in the statistical physics books.  

The Fermi-liquid takes now the role of the free critical fixed point and this is "deformed" in the densely entangled strange metal by turning the Fermi-gas "engineering" scaling dimensions  into {\em anomalous} scaling dimensions. The claim is that such scaling flows can be completely classified unleashing powerful gravitational means. This shows that physical theories exist where the anomalous dimensions can become very anomalous. The highlight is the dynamical critical exponent expressing the scaling relation between space and time becoming {\em infinite}. There is direct experimental evidence for such "local quantum critical" scaling in the cuprates  (Section \ref{EELSlocqucrit}) and together with the Planckian dissipation I perceive this as the leading evidence that we are on the right track. 

There is much more possible under the reign of scale-covariant scaling than under the highly constraining scale-invariant version of the stoquastic critical states. The bottom line is that one naturally reconstructs the portfolio of metal physics, that also includes the BCS-type instabilities of the Fermi-liquids. This turns out to be a greatly entertaining affair in the bulk. It includes spontaneous symmetry breaking in the boundary, requiring that the bulk black hole acquires "hair", a condition that is actually possible and in hindsight natural in a space-time which is asymptotically AdS. The ensuing superconductor is remarkably similar to the usual BCS variety, to the degree that the differences rooted in anomalous scaling dimensions are for practical reasons not measurable!

There are also warning signs. One has to be acutely aware that the IR "strong emergence" sector may still be constrained by the UV. At finite density one departs from the strongly interacting large $N$  CFT affair that one then pulls to a finite density. A basic form of such "UV sensitivity" one encounters in the context of transport (Section \ref{holotransport}) -- the UV point of departure is governed by ultra-relativistic, zero rest mass  matter while electrons have a rest mass that is a factor $10^8$ larger than the energy scales of interest. As I will explain, certain features in the finite density holographic transport are critically sensitive to this rest mass. 

But the real danger is in the large $N$ limit that we know too well is quite unphysical. Our Leiden group has a substantial IP in the discovery of the "Leiden-MIT" fermions \cite{Sciencefermions09}, an achievement that was a considerable stimulus early (2009) in the development. This implements the computation of  "holographic photoemission", revealing Fermi-surfaces and a lot more. Quite a bit later  this was however debunked as being an affair that is completely tied to the large $N$ limit (Section \ref{Holofermions}).

\subsection{The frontier: holographic transport and intertwined order (Sections \ref{holotransport},\ref{Intertwined}).}

All along transport has been on the foreground: I already alluded to the fluid-gravity duality. In Section (\ref{holotransport}) I will first remind the reader of the general principles underlying transport phenomena -- every physicist should know this but one may check it out since especially the solid state textbooks  tend to be infested by folklores (like "Drude conduction proves the Fermi-gas"). 

The take home message is that according to holography the finite temperature DC transport in the quantum supreme metals is invariably controlled by {\em hydrodynamical} flow behaviour. This is yet again rooted in the extremely fast thermalization in such systems -- local equilibrium is reached well before the translational symmetry breaking becomes noticeable, destroying the (conserved) total momentum. Given what we know experimentally, there is one ploy \cite{JZsciPost19} that explains the linear resisitivity of Fig. (\ref {fig:linresistivity}) in fact in terms of the minimal viscosity (Section \ref{sheardrag}). This is a quite predictive affair but yet again the type of experiments required to (dis)prove these assertions are for practical reasons quite hard to realize. This is a relatively mature affair and several experimental groups are trying to make this work. 

But we are now entering the {\em uncharted} part of the holographic portfolio. Electrons in solids are strongly influenced by the presence of an ionic lattice: the band structure affair. But this is also the case for the holographic strange metals and this translates to technical hardship in the gravitational bulk. This "Umklapp"  deteriorates the isometry having the ramification that the hardship of the Einstein equations as a system of highly non-linear partial equations has to be dealt with. This can be handled with state of the art numerical GR but only a couple of exploratory shots were fired, providing proof of principle that it can be done. This was actually an important motive for the string theorists to turn elsewhere -- the required computational effort is just not something that is in their genes.  Surely, the hardship of the computations goes hand in hand with the potential of finding yet other surprises. But we know very little, and this "computational holography" is still in its infancy. In Section (\ref{holoUmklapp}) I will present the little we know, referring to very recent work. 

The same theme is at work in the holographic symmetry breaking affair. I alluded to holographic superconductivity which is easy to compute. However, it was discovered that holographic strange metals are also subjected in a natural way to spontaneous breaking of space translations and rotations: crystallization(Section \ref{Intertwined}). But this kills Killing vectors and  the computation of these holographic crystals requires again numerical GR. The outcomes are actually the most complex stationary black holes that have been hitherto identified. This has everything  to do with the "hair" that becomes very structured. The ramification for the boundary is that this "Rasta black hole hair" dualizes in ordering phenomena that are eerily similar to the "intertwined order"  found below $T^*$ in the underdoped cuprates (Fig. \ref{fig:cupratephasedia}). 

\subsection{How about experiment?}

Of course, the 64K\$ question is: are the cuprate strange metals of the quantum supreme kind?  If so, do these give in to the general principles that are suggested by holography? As  I already stressed, this is as of yet undecided. It is actually right now in a rapid flux: the experimental community has taken up an intense research effort which is quite fertile. I will present what I perceive as clear "holographic signals" in the data, but there is also a lot that is not so easy to explain. This section is rather tentative, my guts feeling is that in a period of a year or so it will need already a thorough revision.     

\section{Landau's iron grip: when entanglement is short ranged.}
\label{SREproducts}

I assume that the reader will have succeeded in digesting the theoretical curriculum that is more or less standard in physics departments. Towards the end of it one learns the art of quantum field theory, the fundament of the standard model of high energy physics. It is presented as seemingly fundamental principle that one departs from a classical field theory, as formulated on basis of symmetry and locality. This is then re-quantized, in the canonical representation by lifting the classical field modes to quantum harmonic oscillators. These modes turn into particles identified by their quantum numbers. One then proceeds by incorporating the interactions using perturbation theory: the art of diagrammatics. Equivalently, in the path-integral representation one just inserts the Lagrangian of the classical theory in the action. The classical saddle point that minimizes the action coincides with the classical theory, and one requantizes the theory by expanding around the  classical saddle. 

In a parallel development that started in the 1930's,  it became clear that the same basic structure is also at the heart of the {\em collective} properties of mundane forms of matter under ambient conditions. This involves macroscopic numbers of particles that themselves belong to the realms of chemistry, as described by the Schr\'odinger equation: electrons, ions, spins. These were the subject of statistical- and later condensed matter physics. The  commonalities with the high energy realms were fully realized in the 1970's when visionaries like Ken Wilson, Sasha Polyakov and Gerard 't Hooft mobilized statistical physics wisdoms such as the renormalization group and weak-strong dualities to make progress with non-perturbative aspects as encountered especially in QCD.  This went back and forth, also fertilizing condensed matter physics. Profiting from the mathematical sophistication of the high energy community, books were written having in one or the other way "quantum field theory in condensed matter" in the title. 

All along the key principle in the condensed matter tradition has been what is nowadays called "strong emergence". The whole is so different from the parts that the parts can no longer be deduced from the properties of the whole. The fields of relevance to condensed matter physics only exist in a rigorous fashion in the thermodynamic limit where the number of parts goes to infinite. In the mean time there appears to be a community wide consensus that this strong emergence may also be governing principle in high energy physics. The standard model is now viewed as an effective, in essence phenomenological  description that does not reveal the nature of the underlying parts. Given the many free parameters of the model itself, but also the dark sector mystery and especially the need to unify with gravity: there {\em has} to be a new reality beyond the present high energy border but its signals are shrouded by strong emergence in the experiments that can be done. The present state of string theory may offer a vivid illustration of this way of thinking.

However, in this pursuit semiclassics appears as the first law. This also used to be the case in condensed matter physics. In the 1970's this discipline was presented to students like myself as a theatre forming a mirror image of the high energy realms, revolving around "elementary excitations" that are like the particles of particle physics: phonons, magnons, Fermi-liquid quasipartices and so forth. But under pressure of empirical developments that started with the discovery of high Tc superconductivity in the late 1980's, in a development that took 30 years or so the semiclassics paradigm  got increasingly into trouble. A myriad of mysteries were revealed by the experimentalists that appear to be completely detached from anything that can be explained by "particle physics". 

As announced, these lecture notes are  dedicated to yet a very different form of strong emergence giving rise to a collective quantum physics which is not governed by semiclassics. The recent progress is driven by the help arrived from the mathematical side -- this venture is propelled by the good work of string theorists culminating into equations relating to a quite different reality that may be realized in the electron systems of condensed matter -- I like the provocative nickname "unparticle physics" \cite{Phillipsunp}. 

For those who are intrigued by "fundamental physics", is there any reason to pay attention to this mundane stuff in the condensed matter laboratories? Precedent from the history physics suggests one better does so.  Once upon a time, van der Waals was studying the evaporation of liquids in gasses, thereby confirming that Boltzmann guessed it right. But the stochastic equations of Boltzmann turned in a much later era into the mathematical underpinnings of Yang-Mills gauge theory. Could it be that the dark sector or even the inflationary fields are unparticle physics? I would not dare to make any claim but it should be beneficial to realize that there is room for stuff that is radically different from a dilute gas formed from yet another kind of particle. 

\subsection{Short range entanglement.}

The other mathematical advance  has been the rise of  quantum information. In fact, this is still in its infancy dealing with thermodynamically large systems. Speaking for myself, I have experienced this mathematical thinking revolving around {\em information} as enlightening. All one has to realize is that {\em entanglement} is a unique quantum physical computational resource. The basic notion is that quantum computers can compute under certain conditions {\em exponentially faster} than any classical computer: the "quantum supremacy" that made headlines recently. 

Realizing that information matters one is urged to view matters from the standpoint of computational complexity theory. In first instance one just needs to appreciate the most basic notions and these are simple. All along we were somehow vaguely aware that we were cutting corners, but with the complexity theory at hand one can no longer stare away and one is forced  to face the challenge head on. As I will forcefully argue in the next section, the generic problem of {\em a thermodynamic system of strongly interacting fermions at a finite density} is presently {\em not computable} neither by any classical supercomputer that can be build, nor by any form of established mathematics. 

The first step is to appreciate the limitations of semiclassics in information language. The task is to address the question, what is the general nature of matter as formed from infinities of microscopic quantum degrees of freedom? This revolves around the nature of the zero temperature {\em ground state} ("vacuum" in the high energy language) -- the finite temperature properties ascend from the ground state as you will see. 

The information language reveals a single, completely general and very simple condition that a vacuum should satisfy in order for the physics to become semiclassical. When this is satisfied it prescribes the algorithm that tells us how to compute physical properties, a-priori ensuring success: the machinery of the  physics textbooks. This condition is: {\em the vacuum has the structure of a short-ranged entangled product state}, abbreviated as the "SRE product vacuum".  

For simplicity, let us depart from a macroscopically large system of $N$ qu-bits, where $N \rightarrow \infty$, in fact just two level systems that may be interpreted as microscopic $S =1/2$ spins. What follows is actually not depending in an essential way on the identity of these microscopic quantum degrees of freedom. Using this quantum computer language the Hilbert space is spanned by classical bit strings encoded quantum mechanically in a tensor product basis of the kind

\begin{equation}
| \mathrm{config.} \rangle_ k = \cdots \otimes | 0 \rangle_{i - 2}     \otimes | 1 \rangle_{i - 1}   \otimes | 1 \rangle_{i }   \otimes | 0 \rangle_{i +1}   \otimes | 1 \rangle_{i + 2}  \cdots 
\label{tensorprod}
\end{equation}

The dimension of this Hilbert space is set by all possible ways to distribute these two values over the $N$ parts: this amounts to $2^N$ dimensions and this is the origin of the a-priori exponential complexity. 

One expects that "typical" energy eigenstates are completely delocalized in this exponentially large Hilbert space, i.e. these are of the form

\begin{equation}
|\Psi \rangle_l = \sum_{k=1^{2^N}} a^l_k | \mathrm{config.} \rangle_ k
\label{qusupremestate}
\end{equation}  

where every amplitude $a^l_k$ is order $1 /\sqrt{2^N}$: these are exponentially small. 

This reveals the "quantum supremacy" troubles. As function of $N$ the Hilbert space grows exponentially. For example, when I entered the scene in the mid 1980's the  Cyber supercomputers of the day had a flop rate comparable to a cheap 2021 smartphone. Back then one could compute by brute force the exact ground state for a system (in fact $t-J$ model, see next section) of size $N \simeq 20$. Using the fastest 2021 supercomputer this has increased to $N \simeq 25$. The fun of the quantum computer is that the computational effort may scale in a polynomial fashion $\sim N^{\#}$ instead when sufficient computational qubits become available. 

Hence, for large $N$ such states are just not computable with classical means. Of course this also applies to analytical calculations that are particularly 'easy' viewed from the complexity angle. The teeny-weeny memory banks and millisecond flop rates of our human brains can even handle it. This intrinsic exponential complexity of the quantum many body problem is an insurmountable brick wall for our brains to comprehend how nature works in full generality. 

 Neverheless, the textbooks of physics are greatly successful in explaining large swathes of the matter as it occurs in the universe. How can this be? Apparently this stuff circumvents the exponential complexity brick wall.  We have arrived at the instance where I can reveal the hidden assumption on which semiclassics is resting, regardless the context.  Instead of the general wavefunction Eq. (\ref{qusupremestate}) matter that we understand is invariably characterized by the SRE product vacuum of the form,
  
\begin{equation}
|\Psi \rangle_0 =  A_0  | \mathrm{config.} \rangle_ {CL} +  \sum_{k} a^0_k | \mathrm{config.} \rangle_ k
\label{SREproductdef}
\end{equation}  

 There is a particular "classical state" (tensor product) $| \mathrm{config.} \rangle_ {CL}$ that "dominates" the vacuum: the amplitude $A_0$ is finite even in the thermodynamic limit.  If this is the case, the computation of the "dressing" $\sum_{k} a^0_k | \mathrm{config.} \rangle_ k$  will be of polynomial complexity. In fact, this is the familiar affair of computing the "fluctuations around the ground state" using the converging diagrammatic perturbation theory. 
 
It revolves around how the many body entanglement is organized and all what matters is that $| \mathrm{config.} \rangle_ {CL}$ is an unentangled tensor product. Pending the system this can be composed from any microscopic representation. To see what this means, let us consider a simple and overly familiar example: every day solids. Surely, solids are also at zero temperature on the macroscopic scale governed by classical field theory, in fact the one longest in existence: the theory of elasticity. As everything else, departing from the microscopic scale it should have a wavefunction and it is of course obvious how to stitch this together. We depart from real space wavepackets referring to the quantum mechanics  of single atoms. In second quantized representation, define  $Y^{\dagger} (\vec{R}, \sigma)$ creating an atom in such a wave packet where $R$ is the space coordinate and $\sigma$ the width. Since the particles are localized we can ignore the (anti)-symmetrization condition associated with indistinguishability (see next section). We now write   
 
 \begin{equation}
| \mathbf{R},\{ \sigma \} \rangle_ {CL}  = \Pi_{i=1}^N Y^{\dagger} (\vec{R}_{i}, \sigma_i ) | \mathrm{vac.} \rangle
\label{CrystalWF}
\end{equation} 

Where $\mathbf{R} = (\vec{R}_1, \vec{R}_2, \cdots  \vec{R}_N )$ is the configuration space specifying the positions of all atoms. When this forms a periodic lattice one recognizes immediately the crystal. But in order to keep the kinetic energy finite, these wave-packets require also a finite width. However, turning to a substance formed from large mass atoms and strong inter-atomic interaction potentials (like nearly all common solids) this may become so small that it is beyond observation.

We know of course that this is not yet the full story. We learn from the undergraduate text book that the lattice vibrations of the truly classical crystal have to be requantized by promoting them to harmonic oscillators, the general procedure defining semi-classics. This will give rise to zero point motions of the atoms (quantum Debye-Waller factor), and when the crystal becomes more quantal one has to address as well phonon-phonon interactions due to anharmonicities, to eventually consider virtual processes involving topological excitations: dislocation-antidislocation loops. One recognizes the perturbative gymnastics of the textbooks, and this just amounts in the wave function language of Eq. (\ref{SREproductdef}) to the computation of the amplitudes  $a^0_k$. 

The unique feature of the  SRE state is that the finiteness of $A_0$ acts like an "anchor in Hilbert space", preventing the vacuum to delocalize in the full, exponentially large Hilbert space thereby rendering the perturbation theory to be convergent. A complimentary perspective is to view it from a renormalization group perspective but now keeping an eye on the entanglement. The nature of physics is "running" with scale. The $| \mathrm{config.} \rangle_ k$ imply that the product $| \mathrm{config.} \rangle_ {CL}$ gets entangled but upon zooming out to larger scales this will disappear: invariably one can identify an "entanglement length", $L_{\mathrm{en}}$. This length signals that one has an effective "wave packet" which is dressed up by local entanglement,  but at length scales larger than $L_{\mathrm{en}}$ these form together a perfect product state. This is completely devoid of any form of quantum entanglement and can therefore be described by a theory that processes exclusively {\em classical} information: the "effective" classical field theory. 

This becomes later on a crucial insight. The "quantum supreme" vacuum states of form  Eq. (\ref{qusupremestate}) are the fundament for phenomenological theories explaining the observations that leave room for the exponential complexity -- we will see later how this works -- with however the ramification that these cannot possibly be captured by the un-entangled  semi-classical description. 

Dealing with typical atomic solids $L_{\mathrm{en}}$  is much less than a lattice constant. However, we know about quite a number of circumstances where $A_0 << 1$ such that $L_{\mathrm{en}}$ becomes quite large. But the miracle is that as long as $A_0$ is finite the macroscopic physics is described by classical fields, albeit typically characterized by strongly altered "renormalized classical" parameters. This principle is coincident with the text book  notion of adiabatic continuity. Departing from a limit where $A_0 \rightarrow 1$ where it is easy to derive the classical theory describing the collective state, one can deform the state (by truck loads of diagrams). The parameters of the classical theory describing the macroscopic system will be strongly affected but the structure of this theory will not change. This will end at a "thermodynamic singularity" -- the instance where $A_0$ becomes zero. 

\subsection{The classical states of matter: from crystals to Fermi-liquids.}  
\label{SREorder}

This is actually the title of lecture notes I wrote in the mid 1990's \cite{classicalcondensates} for a course where I had the marching order to explain the Bardeen-Cooper-Schrieffer (BCS) theory dealing with the superconducting state as found in normal metals. 

The discovery of this theory in 1957 was an epic event and in hindsight one may view it as the triumph of the "SRE product paradigm", pushed to its limits. The individual "qubits" correspond with single particle states living on the wave side of the particle-wave duality -- in this regard these reveal the workings of quantum physics.  The superconducting state is for this reason called a "quantum fluid" although it is "classical" in the sense of many body entanglement. However, the real leap forward embodied by the BCS revolution was in the demonstration of how to marry it with {\em fermion statistics}. The "Cooper logarithm" underlying the exponential gap function and the large coherence length is an exclusively fermionic affair that is way beyond the simple intuition that I alluded to in the above for the understanding of crystals. Eventually, the only way to comprehend it is through equations. I will review this portfolio in some detail in Section (\ref{Fermiliquid}).

Back then quantum information just started and I was myself not aware of even the simple notions that I just explained. But I had surely figured out the somewhat implicit recipe everybody was using to compute matters in main stream condensed matter physics. I reconstructed the SRE vacuum as the base line, with the ramification that next to crystals and magnets I was forced to also call the conventional quantum liquids "classical states of matter". Back then it was halfway understood that one can get away with this designation dealing with superconductors but to call a Fermi-liquid "classical" was quite eccentric.  Regarding the Fermi-liquid some physicists are still confused. But I got it right -- in Section (\ref{Fermiliquid})  I will outline the rock solid proof that the Fermi-liquid is devoid of any many-body entanglement while an order parameter is at work, albeit of an unusual kind: the Fermi-surface. 

This was inspired by pragmatism: departing from the SRE Ansatz it becomes easy to explain why we compute in condensed matter physics in the way it is done up to BCS theory. Step one: compute the energy expectation value of the product state  $| \mathrm{config.} \rangle_ {CL}$ as function of the classical configuration space ($( \mathbf{R},\{ \sigma \} )$ for the crystal) and minimize the energy. This coincides with the standard "Hartree-Fock" type mean field theory. Typically, but not always, one will find that quantities get expectation values corresponding with spontaneous symmetry breaking and one can identify the order parameter. In the crystal this describes the breaking of translations. It is then straightforward to develop the perturbation theory around this ordered state, and one continuous with the demonstration that this is converging. One then identifies  conditions for these fluctuations to be (nearly) ignored for disparate reasons (like heavy atom crystals, or weak coupling BCS superconductors). 
  
The issue is that pending the nature of the quantum problem different types of macroscopic "wave packets" are required to construct the "classical wavefunction". These choices are closely related to the notion of the "coherent state", quantum states that are as closely related as possible to their classical "descendents". The real space wave packets of the crystal are already case in point. Turning to spin systems, the point of departure is in the form of Heisenberg Hamiltonians $H = J \sum_{\langle i,j \rangle} \vec{S}_i \cdot \vec{S}_j$ where $\vec{S}$ are SU(2) operators with algebra $\left[ S^{\alpha}, S^{\beta} \right] = i \epsilon_{\alpha \beta \gamma} S^{\gamma}$ where $\alpha, \beta, \gamma = x, y, z$ dealing with microscopic spins characterized by a total spin quantum number $S$.  To construct the classical spin state, one maps the quantum spins onto classical spins (magnetic dipoles, being vectors) by using spin coherent states, 

\begin{equation}
|\hat{\Omega} \rangle = e^{- i \phi S^z} e^{i \theta S^y} | S, S \rangle_z
\label{spincohstates}
\end{equation}

where $| S, S \rangle_z$ is the max weight state in z-quantization, while $\theta, \phi$ are the Euler angles defining  points on the Bloch sphere. Such generalized coherent states are uniquely defined for microscopic constituents governed by any Lie group. The $\Omega$'s now take the role of $( \vec{R}_0, \sigma )$ in defining the classical configuration space and this maps the quantum Heisenberg models onto a classical spin system subjected to classical order. 

This is all still common intuition but it becomes perhaps less obvious dealing with the conventional "quantum liquids": superfluids, superconductors  and Fermi-liquids. Let us first zoom in on superfluids formed from bosons. The Bose-Einstein condensate formed by non-interacting bosons is obviously of the kind $\sim \Pi_1^N b^{\dagger}_{k=0} | \mathrm{vac} \rangle$ -- a product state "enriched" by Boson statistics. The big difference with the crystal is of course that now the product is formed from microscopic states that are on the wave side of the particle-wave duality and in this sense the state is "quantum". In the collective state the bosons are all completely delocalized. 

However, free bosons are completely pathological: physical bosons always interact. The next step is to switch on weak repulsive interactions as handled by the Bogoliubov theory, changing the Bose-Einstein condensate into a Bose condensate breaking the U(1) symmetry spontaneously such that the excitation spectrum is characterized by a genuine Goldstone boson (the phase mode or "second sound") with its linear dispersion relation. One encounters the Bogoliubov transformation for bosons where the phase mode is described by $b^{\dagger}_k = \cosh(u_k) a^{\dagger}_k - \sinh (u_k) a_{-k}$ where the $a^{\dagger}$ field operators creating the non-interacting bosons. The ground state is given by the condition $b_k | \Psi_{\mathrm{Bog} }\rangle = 0$ for all $k$: it follows that $| 0 \rangle$ is a Boson coherent state formed from the $ a^{\dagger}_k$ bosons of the form,

\begin{equation}
| \Psi_{\mathrm{Bog} } \rangle  \sim e^{\alpha a^{\dagger}_{k = 0}   - \sum_k \frac{1}{\tanh (u_k)}  a^{\dagger}_k a^{\dagger}_{-k}} | \mathrm{vac} \rangle
\label{Bogoliubovbosons}
\end{equation} 
  
in terms of the bare bosons.   

The bosons offers a vivid example of the adiabatic continuity. One can go all the way to the interaction dominated limit by considering hard core bosons defined on a tight binding lattice: per site $i$ one can either have no- or at most one boson.

\begin{equation}
| \Psi_{\mathrm{SF}}  \rangle= \Pi_i ( \cos (\theta_i)  + \sin (\theta_i) e^{i \phi_i} b_i^{\dagger} ) | \mathrm{vac} \rangle
\label{hardcorebosons}
\end{equation} 

The condensate is just formed from on-site Schr\"odinger cat states associated with zero and one boson.  In this strong coupling case it is not at all a good idea to depart from single particle momentum space as for the Bogoliubov problem. But on macroscopic scales both cases are described by the same "Mexican hat" Landau order parameter theory involving a complex scalar order parameter, although the numbers in the macroscopic theory are very different. 

In fact, it is much easier to derive this classical field theory describing the phenomenology of the macroscopic superfluid from this strong coupling limit as compared to the weakly interacting Bose-Einstein gas. Given Eq. (\ref{hardcorebosons}), it follows immediately that $\langle \Psi_{\mathrm{SF}} | b^{\dagger}_i | \Psi_{\mathrm{SF}} \rangle = \sin(2\theta_i) e^{-i \phi_i}/2$. Given that charge conjugation symmetry is broken the Euler angle $\theta_i$ will be fixed: this just represents an $XY$ spin with a length set by $\theta$ and a direction $\sim e^{i \phi_i}$. The microscopic Hamiltonian will include a hopping of the bosons and assuming this to be nearest-neighbour one finds immediately $ \langle \Psi_{\mathrm{SF}} | \mathrm{H}_{\mathrm{kin}}   | \Psi_{\mathrm{SF}} \rangle \sim \sum_{<ij>} \cos ( \phi_i - \phi_j)$: the Josephson Hamiltonian, the "Boson Schr\"odinger cats" are just described by a classical XY spin system!

The statistical physics textbook than insists that this coarse grains in a "$\phi^4$" thermal (Landau)  field theory for a complex scalar,

\begin{equation}
F =  | \nabla \Phi |^2 + m^2 |\Phi |^2 + w |\Phi|^4
\label{Landauphi4}
\end{equation}

where $m^2 \sim (T_c - T )/T_c$ and $\Phi (\vec{r})  = | \Phi (\vec{r}) | e^{i \phi (\vec{r})}$. This should be familiar "Mexican hat territory". The take home message is that the SRE-product maps the microscopic quantum system on a macroscopic {\em classical} effective field theory. 

These cases all involve ground states that break symmetry spontaneously -- a phenomenon in fact specific for {\em classical} field theories. But this need not to be the case,  there are plenty of "quantum disordered" ground state that do not break symmetry while these are still of the SRE product kind. 

The main story of these notes revolves around fermions at a finite density. Given the Pauli-principle this ESR-product affair acquires a "fermionic twist" that underlies profundity of both the Fermi-liquid and its BCS-type descendants. Given that these play a crucial role when we turn to holographic matter, I will take up this theme at length in Section (\ref{fermionsigns}) where I will explain why these states are yet again describing "classical matter" albeit endowed with properties that are quite different from this bosonic affair.   But let's first focus in on this bosonic matter that is lying in full view looking through the  "QFT 1.61 eyeglass".    

\section{Stoquastic matter and the quantum critical state.} 
\label{qucritical}

In this realm of physics, the stochastic mathematics underlying the success story of equilibrium statistical physics has been the main propulsion system. This started with Boltzmann formulating the partition sum, culminating in a rigorous understanding of the phases of classical matter at finite temperatures. Presently it is considered as a closed subject: the few loose ends are in the hands of mathematicians looking for mathematical proofs. In fact there is one millennium prize problem in this area:  to derive analytically the confinement mass gap in Yang-Mills theory. 

It cannot be stressed enough that the only truly general machinery available to deal with quantum many body physics is precisely this Boltzmannian affair. It is yet rather severely restricted: it can only be unleashed dealing with "sign-free" problems, and these are by themselves to a degree pathological. Special symmetries are required to "cancel the signs", and these symmetries are at best accidentally realized in nature. When the problem is "sign free" the equilibrium quantum problem can be mapped on an equivalent statistical physics problem that submits to the power of the stochastic methods. This subclass of problems was coined "stoquastic" in the quantum information community: a merger of "stochastic" and "quantum". 

In fact, the standard notion of the "quantum critical state" originating at the quantum critical point associated with a zero temperature quantum phase transition is entirely resting on this kind of mathematical technology. In condensed matter physics this was put on the map by Chakravarty and co-workers in the late 1980's \cite{CHN88} and subsequently charted with great precision by Sachdev.  This section is to be considered as an executive summary of his famous book \cite{QPTSachdev}. The take home message will be: be extremely aware of the "stoquastic tunnel vision" but at the same time you will meet a first example of "quantum supreme matter" in the form of the {\em strongly interacting quantum critical state}. In more than one regard this sets a template for matters to come.  

To understands how this works one has to familiarize one self with   "thermal field theory" as propelled by "Euclidean" path integrals. This is standard fare on the theory floors but it is taught very late in the course programs. In case you have not learned this, catch up first. To my strong opinion any physics graduate in the 21-th century should be aware of this box of tricks if not only to become acutely aware of the limitations of the "mathematical eye glass" in the hands of mankind. It is the mathematical pillar on which all of established field theory rests, metaphorically the "Quantum Field Theory 1.X".  This thermal field theory went through a long development. The "QFT 1.0" version  was formulated by Kubo in the 1950's: one of the unsung heroes of physics. Especially in the 1970's it went through a rapid progression of updates ("1.X") due to the groundbreaking insights of high energy theorists like Ken Wilson, Sasha Polyakov and Gerard 't Hooft that also when dealing with  QCD etcetera one should think like the statistical physicists. 

The mapping of the quantum problem to  an equivalent statistical physics exercise is actually a simple affair when you get used to the idea. It departs from the standard path-integral formulation leading to the expression for the zero temperature quantum partition sum

\begin{eqnarray}
{\cal Z}_{\hbar} & = & \sum_{\mathrm{histories}} e^{ \frac{i S_{\mathrm{history}}}{\hbar}} \nonumber \\
S_{\mathrm{history}} & = & \int dt \; {\cal L}_{\mathrm{history} }
\label{Pathint}
\end{eqnarray}

where $S$ is the action, and ${\cal L}$ the Lagrangian of the many body quantum system. One obtains the Euclidean incarnation by the seemingly simple "Wick rotation": continue real (Lorentzian time) analytically to "imaginary time": $t \rightarrow i \tau$. Kubo's discovery is that by assigning imaginary time to be defined on a circle with radius $R_{\hbar} = \beta \hbar = \hbar / k_B T$ the Euclidean path integral computes finite temperature equilibrium physics as well. This is remarkably powerful: as you will see, finite temperature {\em classical} statistical physics arises as a special case of the zero temperature quantum case. 

But the big deal is yet to come. Because of the $i$ in front of $S/\hbar$ in Eq. (\ref{Pathint}) one is dealing with the oscillating sum in Lorentzian time which is hard to address dealing with a complicated many body action.   But after the Wick rotation, the action picks up an $i$ coming from the time integral over the Lagrangian, $dt = i d\tau$. The effect is that $\exp{( i S/\hbar)} \rightarrow \exp{( - S_E/ \hbar)}$: at first sight the path integral turns into a Boltzmann thermal partition sum that is just living in a space with one extra dimension: the "Euclidean" time $\tau$. The Lagrangian of course also changes by the Wick rotation: time derivatives pick up the $i$, the standard relativistic kinetic term $ - (\partial_t \Phi) ^2 \rightarrow + (\partial_\tau \Phi) ^2$. The kinetic- and potential energy terms in the Euclidean Lagrangian add up. 

One now adds Kubo's time circle and the quantum partition sum becomes,

\begin{eqnarray}
{\cal Z}_{\hbar} & = & \sum_{\mathrm{histories}} e^{ - \frac{ S_{E, \mathrm{history}}}{\hbar}} \nonumber \\
S_{E, \mathrm{history}} & = & \oint_{R_{\hbar}} d\tau \; {\cal L}_{E, \mathrm{history} }
\label{Pathinteuc}
\end{eqnarray}
    
This has clearly the structure of a stochastic Boltzmann partition sum where $S_E$ and $\hbar$   take the role the potential energy and temperature, respectively, in setting the probability of a particular worldhistory. But this is deceptive: $S_E$ is typically {\em not} a real quantity and thereby the "Euclidean" Boltzmann weights may become negative, or even complex. Surely,  "negative probabilities" do not make sense and this is the sign problem. I will discuss this at length in the next section. 

There is a subclass of problems where it is possible to find representations where the Euclidean action is real. A first requirement is{ \em time reversal symmetry} -- a complex Euclidean action cannot be avoided when e.g. a magnetic field is switched on. Similarly, {\em charge conjugation invariance} is sufficient condition: zero density problems are stoquastic, the secret behind the stunning success of lattice QCD. Next, the signs in the path integral encode for sign changes in the ground state wavefunction. Dealing with bosons Feynman already proved that ground state wavefunctions of bosons are generically positive definite minimizing the kinetic energy. This is behind the success of the quantum Monte Carlo (the Metropolis algorithm unleashed in Euclidean space-time) in the description of $^4$He, see next section.  Other circumstances are typically associated with accidental, unphysical circumstances: unfrustrated Heisenberg spins on a bipartite lattice, electrons that only interact with phonons, and so forth.

But when it works it is astonishing powerful. Linear response is integral part of this equilibrium agenda and this is crucial to link it to experiment -- the "best" data is of the linear response kind because  per construction it is avoiding any complications associated with the experiment itself. One just computes the ($n$) point statistical physics correlation functions in Euclidean signature, to subsequently Wick rotate back to Lorentzian time and one obtains the ($n$ point, retarded) propagators that are central in the quantum theory. Let us put some beef on this affair by zooming in on some elementary examples. 

\subsection{The fruitfly: transversal field  Ising model.}

Ising spin systems have played a key role in the history of statistical physics, as the simplest models revealing the key principles. This is similar in the stoquastic quantum realms. The case in point is the Ising model in 2 space dimensions: this was exactly solved by Onsager in the 1940's demonstrating that phases of matter exist separated by a continuous phase transition. This integrability is actually a pathological condition but it does not matter for the big picture.  This model is playing a key role in Sachdev's book \cite{QPTSachdev}, and I just follow his exposition in this regard. 

This departs from the Ising problem of statistical physics,

\begin{eqnarray}
{\cal Z} & = & \sum_{\mathrm{config.}} e^{- H_{Ising}} \nonumber \\
H_{Ising} & = &  - \beta J \sum_{\langle i,j \rangle} \sigma^z_i \sigma^z_j
\label{IsingHam}
\end{eqnarray}

living, say, on a d+1-space dimensional hypercubic lattice with nearest neighbour couplings   while $\sigma_z$ can be taken to be the $z$ Pauli matrices. What matters is that the DOF's on the site can take two values like $\pm 1$. Every physicist has learned  what this is about. At high temperature ($\beta J < 1$) this describes a disordered state while at $\beta_c J \simeq 1$ a phase transition occurs to an ordered state which breaks symmetry spontaneously. For e.g. ferromagnetic couplings  ($J > 0$) one finds a twofold degenerate ground state, with the spins either all pointing up or down when $\beta J >> 1$.

By the path integral mapping in the reverse ("transfer matrix") this becomes in canonical formulation,

\begin{equation} 
H_{QuIsing}  =  -  J \sum_{\langle i,j \rangle} \sigma^z_i \sigma^z_j + B \sum_i \sigma^x_i
\label{transIsing}
\end{equation} 

The "transversal field" Ising (quantum) problem: $B$ is a magnetic field pointing in the $x$ direction and since $\sigma^x = ( \sigma^+ + \sigma^-)/2$ this is quantizing the classical Ising problem in d space dimensions.  

\subsubsection{The integrability in 1+1D.}

To set the stage further, let us zoom in on the special case in 1 space dimension. The key to the integrability is in a property of the Pauli matrices that is special to one dimensions: the operator identity called the Jordan-Wigner transformation,

\begin{eqnarray}
\sigma^z_j & = & 2 a^{\dagger}_j a_j -1 \nonumber \\
\sigma^+_j & = & e^{- i (\pi \sum_{k=1}^{j-1} a^{\dagger}_k a_k)} a^{\dagger}_j 
\label{JordanWigner}
\end{eqnarray}

where $a^{\dagger}_j$ creates a fermion at site $j$. Inserting this in Eq. (\ref{transIsing}) yields

\begin{equation} 
H_{QuIsing}  =  - J \sum_{j} \left(  a^{\dagger}_j a_{j+1} + a^{\dagger}_{j+1} a_j +   a^{\dagger}_j a^{\dagger}_{j+1} +a_{j+1} a_j - \frac{B}{J} ( 2 a^{\dagger}_j a_j -1 ) \right)
\label{transIsingferm}
\end{equation} 

This is a simple bilinear problem of non-interacting tight-binding fermions that is easy to diagonalize by a Bogoliubov transformation, thereby solving the problem exactly. But the relationship between the spins and the hidden free fermions is highly non-local: the Jordan-Wigner "strings" in Eq. (\ref{JordanWigner}) imply that the computation of a spin-spin correlator translates in an {\em infinite} point correlator in terms of the free fermions. Although equal time spin correlators are computable with quite some effort, it is extremely difficult to compute the unequal time (dynamical) spin susceptibilities for arbitrary couplings.

\subsubsection{The cohesive phases and quantum information.}
\label{Weakstrongduality}

Let us first address the "stable" or "cohesive" phases, away from the phase transition. It is very easy to convince oneself that the ground states are   of the SRE product kind. Let us consider first the limiting case $| J | >> B$. Depart from one of the two fully polarized states, product states that are eigenstates of the Ising Hamiltonian. Act once with the spin flip operators to find out that the flipped spin causes two wrong exchange bonds costing an energy $\sim J$. Hence, there is a large mass gap and $ | B/J |$ is a genuine small parameter regulating the perturbation expansion.  Hence, the short range entanglement "corrections"  in Eq. (\ref{SREproductdef}) $a^0_k \simeq B/J$,  falling off exponentially in higher order. In fact, we can increase $B$ further and further and as long as there is a mass gap, staying away from the critical point, this perturbation theory will converge. This implies that the $A$ will be finite any distance away from the "quantum critical point". The (short range) entanglement will fall off exponentially on a length scale that we will see can be identified with the correlation length/time associated with the phase transition happening in space-time. 

Obviously, the same logic will apply in the  the opposite limit $B >> | J |$, departing from the state where all spins are forced by the magnetic field to lie along the $x$ quantization axis. But departing from the classical interpretation something odd is going on: large $B$ corresponds with the high temperature limit and this is supposed to be a maximally random entropy dominated affair. How to understand this "orderly" nature revealed by the canonical formulation? The answer is in the notion of "weak-strong" duality that was discovered by Kramers and Wannier actually demonstrating that 2D Ising is self-dual: the high temperature phase is also described by an Ising model with an inverted coupling $\beta J \rightarrow 1 / \beta J$ and the dual Ising spins are therefore {\em ordered in the high temperature limit}. 

I am prejudiced that this duality notion is universal \cite{JZDuality}.  In full generality this works as follows. Departing from the ordered, symmetry broken low temperature state one can invariably identify the "operators" that are unique in destroying the order: the topological excitations, corresponding with domain walls (DW) for the $Z_2$ symmetry of the Ising model.  When one such excitation  "spans" the whole space time while it is delocalized it will destroy the infinite range correlations defining the long range order (LRO).  At low temperatures these DW will occur in the form of small closed loops (in $d=2$) but these grow in size when temperature is raised. Right at Tc these will proliferate and destroy the order. However, having the eyes fixed on the DW "disorder operators" one will find a strongly interacting system of DW's above $T_c$  that will in turn condense and break symmetry. In $d=2$ this is encoded by the dual Ising system of Kramers and Wannier. 

The gross principle appears to be that order cannot be avoided. What appears seemingly as a completely disordered entropy dominated affair is actually a prejudice based on just observing the original spins. When one would observe the system with machinery that  observes the disorder operators the conclusion would be the other way around. In turn it is a simplifying circumstance because it is rather easy to address the physics ruled by order. This reflects  the fact that in the quantum incarnation both the ordered and "quantum disordered" states are both governed by the tractable SRE products. 

The self-duality is special to Ising in $1+1$ D. In three (space-time) dimensions it is a rule that at least for simple symmetries ($Z_2$, $U(1)$, even Galilean symmetry) one runs into local-global dualities: the dual of Ising in $d=3$ turns out to be Ising Gauge theory. For XY (superfluids) one finds a neutral superfluid-charged superconductor duality structure (see next section). Even AdS-CFT appears to give to the rule, in a limited sense: global symmetry in the boundary turns into gauge symmetry in the bulk. 

This duality "confusion" has an image in the canonical language. The issue is that we learn from the two qubit "Bell pair" agenda that entanglement should be {\em independent} of the representation. For example, consider the two-qubit state: $| \mathrm{max} \rangle= ( | 0 \rangle_A \otimes | 0 \rangle_B + | 0 \rangle_A \otimes | 1 \rangle_B + | 1 \rangle_A \otimes | 0 \rangle_B + | 1 \rangle_A \otimes | 1 \rangle_B)/2$. One could jump to the conclusion that this is a maximally entangled state. However, it can also be written as $| \mathrm{max} = | + \rangle_A \otimes | + \rangle_B$ in terms of the single bit representation $ | + \rangle = ( | 0 \rangle + | 1 \rangle )/\sqrt{2}$ and this state is of no use for quantum information processing. For two bits there is a perfect measure for entanglement in the form of the bipartite von Neumann ("entanglement") entropy obtained by computing the entropy of the reduced density matrix of one of the bits. 

But there is no measure for infinite number of qubits -- I find the formulation of such a universal measure of many body entanglement the number one challenge in quantum information.  How can one be sure that a seemingly quantum supreme state turns into an ESR product upon identifying the appropriate microscopic representation? As an example, let us focus on the large B phase, a simple product in terms of $x$ quantized spins $\cdots | 1 \otimes | 1 \rangle^x_{i-1} \otimes | 1 \rangle^x_{i}\otimes | 1 \rangle^x_{i+1} \otimes \cdots$  (qubit notation). But let us now write this in terms of z-quantized spins $ | 1 \rangle^x_i = (  | 0 \rangle^z_i + | 1 \rangle^x_i)/\sqrt{2}$. Multiplying this out one finds $ 1/ \sqrt{2^N} \sum_{k=1}^{2^N} |\mathrm{config, k} \rangle$, a state that appears to be maximally entangled! In the next section  we will encounter a similar situation associated with superfluids. 
 
 \subsubsection{Particle physics needs the SRE vacuum. }
 \label{Isingparticles}
 
It is perhaps a provocative claim that the centre piece of conventional quantum physics, the quantum {\em particle} as organizing principle, is just a symptom of an "unentangled" SRE vacuum state. The logic is perfectly universal but the transversal Ising model is a particularly easy stage to see how this works.  

The way that linear response works is that in the measurement set up a local operator -- "flip a spin" (neutron scattering), "remove one electron" (photoemission) --  is infinitesimally sourced. This amounts to inserting a package of quantum numbers (energy, momentum, spin, $\cdots$) to measure the probability for this to succeed. This yields the spectral function which is then by causality (Kramers-Kronig) tied to the full response of the system. The specialty of the SRE product vacuum is that it keeps this package of quantum numbers "together" while this package delocalizes as a whole quantum-mechanically: we call this object a "particle". 
 
 But how to accomplish this in a quantum supreme state of matter where the ground state is already delocalized in the vast many body Hilbert space? By principle, when the ground state is quantum supreme so are all excited states: these have to be orthogonal to the ground state and this orthogonality pertains to the many body Hilbert space. Everything is entangled with everything and there is no room for the locality implicit to the notion of "particle". This is the no-brainer notion behind the statement that the presence of particles in the spectrum is a diagnostic for the SRE vacuum. Let us inspect  how this works in the stoquastic examples. 
 
\begin{figure}[t]
\includegraphics[width=0.9\columnwidth]{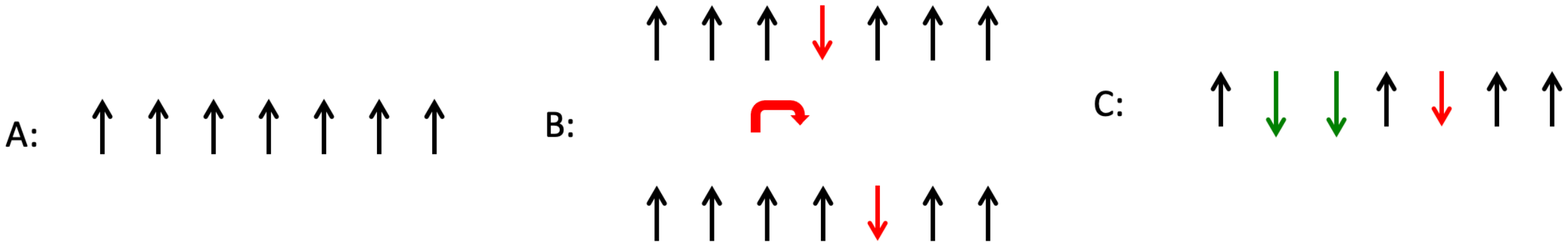}
\caption{ The transversal field Ising model in 1+1 dimensions is a simple context to understand the emergence of quantum particles in the spectrum of a short ranged entangled vacuum state. Consider the limit $B >> J$ and the ground state approaches (A) the "classical" tensor product state: all spin are aligned by the external field.  The excitation corresponds with a spin flip (B: red spin) that subsequently will hop ($\sim J$) resulting in a quantum mechanical particle characterized by a cosine dispersion relation. (C) Upon increasing $J$ the ground state is dressed by "vacuum polarization diagrams" (green spins) but eventually the spectrum will be characterized by an infinitely long lived "particle" at the bottom of the spectrum until the quantum critical point is reached where it will turn into a branchcut "unparticle" spectrum like in Fig. \ref{fig:fermionbranchcut}. }
\label{fig:ESRparticle}
\end{figure}

Let us focus in how this works in the transverse field Ising models, see Fig. (\ref{fig:ESRparticle}). Let us again depart from the large $B$ limit with its trivial product state vacuum. The experimentalist may source the $\sigma^z_i$ operator. Depart from the ground state where the spins are polarized in the $x$ direction, $\sigma^z = (\hat{\sigma}^+ +  \hat{\sigma}^-)/2$ where the "hat" refers to the x-quantization. The propagator $\langle 0| \sigma^z (i, t) \sigma^z (j, 0) | 0 \rangle$ turns into  $G_{zz} (q, \omega) =  \langle 0| \hat{\sigma}^- \hat{\sigma}^+ | 0 \rangle_{q, \omega}$ in the frequency-momentum domain. The spectral function telling where the excitations are is then $A_{zz} (q, \omega)= \frac{1}{\pi} \mathrm{Im} G_{zz} (q, \omega)$ according to the linear response lore. 

What happens when $J$ is finite, but small compared to $B$? At $t =0$ a spin in $x$ quantization is flipped (action of $\hat{\sigma}^+$) inserting a "triplet" $M_S =1$ quantum number. The Ising term becomes in $x$ quantization $\sigma^z_i \sigma^z_{i+1} \sim   \hat{\sigma}^+_{i} \hat{\sigma}^+ _{i+1}  + \hat{\sigma}^+_{i} \hat{\sigma}^- _{i+1}  + \hat{\sigma}^-_{i} \hat{\sigma}^+ _{i+1} + \hat{\sigma}^-_{i} \hat{\sigma}^- _{i+1}$. The two terms in the middle just correspond with simple tight binding hoppings of the flipped spin. The propagator becomes thereby, 

\begin{equation}
G_{zz} (q, \omega) = \frac{1}{\omega - \varepsilon_q}
\label{freespinflip}
\end{equation}

just describing a free quantum mechanical particle with dispersion $\varepsilon_q$, the usual "cosine" band with a minimum at the zone center  ($q=0$). We have identified the particle in this product state limit. 

But now we gradually increase $J$, what happens? From the form of the Hamiltonian it follows that the vacuum get dressed by increasing amounts of spin flips and the $A_0$ of Eq. (\ref{SREproductdef}) will decrease. We can keep track of it using diagrammatics and this becomes rapidly a tedious exercise. But we know the form of the outcome. As long as the perturbation theory is converging which is the case away from the critical point $G_{zz} (q, \omega) = 1 / (\omega - \varepsilon_q - \Sigma (q, \omega) )$ where $\Sigma$ is the self-energy. Invariably one will find that close to the zone center this turns into, 

\begin{equation}
G_{zz} (q, \omega) = \frac{A^2_0}{\omega - \hat{\varepsilon}_q} + G_{incoh} (q, \omega)
\label{freespinflip}
\end{equation}
 
I will specify $G_{incoh}$ underneath: what matters here is that this is characterized by a mass gap while close to the zone center a bound state will form below this cut-off and this disperses like the large $B$ particle albeit the dispersion is renormalized to  $\hat{\varepsilon}_q$ -- these renormalization ("mass enhancement") effects are in these kind of problems usually rather small. The big deal is that this quasiparticle emerges at the the very bottom of the spectrum. There are no states available for decay and this object is infinitely long lived -- a perfect particle. 

But the spectral weight of this quasiparticle in the spectrum is set by the product state amplitude $A_0$ in the SRE vacuum, becoming small upon approaching the phase transition. This is the "quasiparticle pole strength" or "wavefunction renormalization'" factor and this is ubiquitous. It works in the same way dealing with the quasiparticles of for instance the Fermi-liquid. This is a more complicated affair since there is no mass scale. But eventually the same infinitely long lived quasiparticle with its reduced pole strength and modified dispersion is realized in the zero energy limit, see Section (\ref{Fermiliquid}). 

In this particular example global symmetry is not playing a crucial role. This is different dealing with globally conserved currents. One meets here in addition principles associated with the "wholeness" per se. A first example are the hydrodynamically "soft" degrees of freedom rooted in {\em global} conservation laws, like the sound modes protected by {\em total} number- and  {\em total} momentum conservation in classical fluids. But these have their quantum counterparts, like the zero-sound of Fermi-liquids but also the protected zero temperature sound modes in the quantum supreme holographic liquids. Similarly, upon breaking symmetry spontaneously Goldstone bosons have to exist. Yet again, rooted in global symmetry principle, as far as we know these will all survive in the quantum supreme systems.    

Finally, let me allude to a phenomenon that is rather special for 1 space dimensions: the "fractionalization" phenomenon. On purpose I focussed in the above on the large $B$ phase. The same logic should apply to the large $J$ phase as well, and in fact it does. But there is a complication. The large $J$ phase breaks symmetry and this implies that besides the simple spin-flips there is yet another type of "particle" that can be identified departing from the classical limit: the "domain wall" corresponding with a point like kink. Under the influence of $B$ such kinks will also propagate quantum mechanically. It is easy to check that upon inserting a $z$ spin flip this will immediately fall apart in a kink-anti kink pair both propagating independently. The spin flip is actually in this vacuum a composite particle "fractionalizing" in two topological "particles" that both carry half of the spin triplet quantum number: the "spinons".  The spin-flip spectrum will be devoid of quasiparticle poles but this is just reflecting that the true particles cannot be sourced directly. 

This fractionalizing phenomenon is a manifest part of the semi-classical portfolio. This simple story is to quite a degree representative for physics in one space dimension in general. At first sight this may appear as confusing: lowering dimensionality is expected to have the effect that fluctuations increase such that quantum supremacy is becoming more natural. But this is not at all the case. Generic 1+1D physics was charted already in the 1970's in the form of the Luttinger- and the Luther-Emery liquids as well as a host of spin-like systems. At least for the former two, the key is that in 1+1D interactions are {\em always} relevant and everything turns into algebraic long range order (ALRO) involving spin-, charge density- and superconducting  order all at the same time. The bosons of bosonization are just the Goldstone bosons of this ALRO; a crucial aspect is surely that in 1+1D one can always transform away the sign problem, thereby avoiding quantum supremacy (next section).

The take home message of the above is that stable phases of stoquastic matter are invariably characterized by an energy scale that keeps the ground state to be a short ranged entangled product state being of polynomial complexity so that in principle it can be charted completely by classical computers. The discrete ($Z_2$) symmetry of the Ising systems is in this regard a simplifying circumstance. Systems characterized by a continuous symmetry may carry massless excitations but these are invariably Goldstone bosons, implied by the spontaneous symmetry breaking. But these are also a trait of classical field theory that can be safely re-quantized in a semiclassical guise. 

A final caveat is related to the analytic continuation of properties in Euclidean signature back to real time. This is not an issue dealing with the static (thermodynamical) properties but there is yet another difficulty dealing with {\em dynamical} linear responses. We learned that Euclidean signature is a greatly simplifying property, turning the stoquastic path integral in a stochastic affair. But we have to pay the prize when we want to deduce the real time dependences characterizing our quantum system. The essence is simple. Take a single, isolated excitation at energy $\epsilon$. The two point propagator revealing its existence will be an oscillating function in the (real) time domain $\sim e^{i \epsilon t}$. But upon Wick rotation this rapidly varying function turns into a smooth exponential function $\sim e^{- \epsilon \tau}$, the object that can be computed. 

This information can be retrieved dealing with a particle spectrum: this is the trick used by the lattice QCD community to find out about hadron masses, etc. However, dealing with any excitation spectrum that is more interesting than this one runs into the information loss problem. Euclidean correlators are smooth functions and tiny glitches may turn into sharp features  in Lorentzian time. There are pragmatic patches to deal with this, in the first place the "maximum entropy" algorithm. But it turns out that this is a lethal circumstance dealing e.g. with the finite temperature properties at long times. Any noise in the Euclidean computation which is impossible to avoid in the absence of closed analytical solutions  will amplify in an exponential fashion upon the continuation to real time.    

\subsubsection{Topological matter: signalling many-body entanglement. }
 \label{Topmatter}

SRE product vacua typically go hand in hand with spontaneous symmetry breaking, long range order captured by Landau order parameter theories. However, since the discovery of the fractional quantum Hall effects in the early 1980's a different kind of collective behaviour got into focus: {\em topological order}. I alluded to it in the first paragraphs of these notes: the triumph of Chern-Simons topological field theory  in 2+1 dimensions explaining the observations in very clean two dimensional electron systems in large magnetic fields. Since then this agenda expanded to a portfolio of  other systems and presently the research in "topology" is a dominating theme in physics. 

This is actually a quite diverse affair involving quite different kinds of phenomena. The common denominator is that it invariably rest on mathematical topology. This is the art of addressing global properties of things. The familiar story example is that for the eyes of a topologist a coffee cup and a doughnut are indistinguishable. Stick a finger through the handle of the cup and deform it without cutting and pasting the material and one finds a doughnut, while your finger is still sticking through the hole. Mathematical topology is about counting and classifying such "handles" for arbitrary manifolds. 

In the above I already implicitly alluded to the way it historically entered physics, in the form of what later became the "weak-strong" dualities that I just discussed (Section \ref{Weakstrongduality}). Long range order goes hand in hand with its "topological excitations" such as domain walls (Ising, $Z_2$), vortices (XY, $U(1)$) and so forth. These can be counted out using the means of homotopy groups. First identified by Burgers in the 1930's in the form of dislocations in solids, via the discovery of the fluxoid Abrikosov lattices formed in type II superconductors, it entered the main stream in the 1970's catalyzed by the Berezinskii-Kosterliz-Thouless theory of topological melting in 2D XY systems. But as I discussed in the above this is just part of the "classical matter" agenda -- it is just a matter of representation to recognize that "disordered" states are ordered states formed from the (topological) disorder fields. 

Adding to the confusion, a very different application of topology entered physics around 2006 by the discovery of {\em topological band insulators}. This rests on the  topological features occurring in wave mechanics. This started out in the context of single particle quantum mechanics defined in particular solids characterized by "band gap inversions" in particular band structures. This is all about non-interacting fermions  (next section), and this agenda expanded subsequently to include "Weyl fermions" and so forth highlighting the way that particular anomalies known from field theory can be realized in solids.  This diversified later to photonics and vibrational modes in meta-materials. But this has no relation to many body entanglement. 

There is however a third category of systems that is more closely related to the theme of these lectures: the "field theoretical" topological order. This includes the fractional quantum Hall states described by Chern-Simons, but also the deconfining states associated with discrete Gauge symmetry underlying e.g. Kitaev's "toric code". As early as 1992 Xiao-Gang Wen already pointed out how these should be understood in a quantum information language, see his book  \cite{XGWentop}. This topological order is encoded in their ground states in the form of a particular type of many body entanglement. Invariably these are {\em incompressible} states of matter where the ground state is separated from all excitations by an energy gap. We just learned that such gaps are supposed to be lethal for many body entanglement but the subtlety is that a particular form of very sparse entanglement can slip through. 

This translates  into the "quantum-weird" global properties of the system such as the ground state degeneracy pending the topology of the space, the edge modes and so forth. One can identify "topological quasiparticles" such as the (practical)  Majorana  zero modes that satisfy non-abelian braiding properties so that they can be used as topologically protected qubits. But the key is that an infinity of microscopic "qubits" is required to have a sharp definition of such topological qubits. It appears that this topological order can be captured quite generally  \cite{XGWentop} in terms of particular tensor network structures (see Section \ref{Compmethods}). 

Modulo the anomalies, topology is just not the kind of math that is required to shed light on {\em compressible} forms of densely entangled matter. I will therefore ignore it in the remainder and I refer the reader to the extensive literature of this fashionable field.

\subsection{The strongly interacting thermal critical state: universality and scaling.}  
\label{thermaluniversality}

We learned in the above that the polynomial complexity is protected by the stability (rigidity) characterizing a particular phase of stoquastic matter.  The only singular instance that this can fail is at the transition from one phase to the other. For this to happen the transition has to be {\em continuous}. In a first order transition one discontinuously jumps from one stable phase to the other. Let us first revisit the theory of the critical state at thermal transitions, an affair that flourished in the 1970's. This forms the fundament for the understanding of the quantum critical state.

The understanding of the "critical" states has been a triumph of the renormalization group (RG) theory in the 1970's. The RG language may well be familiar to the reader. In first instance it refers to the general conceptual notion that physics depends on the {\em scale} where the system is observed. Depart from the UV scale of the definition of  the standard model. Upon "renormalizing towards to the IR" (longer times, distances, etcetera) quarks confine in baryons, while the vector bosons and so forth acquire their Higgs mass. One then enters the realms where the protons, neutrons and electrons reign, forming nuclear matter. Descending further, atomic physics emerges, followed by chemistry, molecular biology and at some point we find ourselves, that then in turn form human societies, planets, galaxy's and eventually the whole cosmos. 

But this is not a particularly fertile context to find powerful equations.  RG as a mathematical theory triumphed particulary in dealing with the continuous phase transitions. Its powers rest eventually on a  principle that is not at all that obvious a-priori. Breaking symmetry by strong emergence is easy but making symmetry is a different affair. The secret of the critical state is that an immensely powerful symmetry {\em emerges} at the critical point: {\em scale invariance}, typically further enhanced  at stoquastic phase transitions to {\em conformal invariance}: "angles stay the same under scale transformation", implied by the effective Lorentz invariance realized at the quantum critical point.  

Although familiar from experimental observations involving systems as simple as water-steam mixtures, its profundity as emergence phenomenon cannot be stressed enough. Take the transversal Ising model: microscopically it is very scale full (the "$B$" and "$J$"). Onsager's exact solution of the 1+1D version played a crucial role in convincing man kind that it does  happen: at the critical point the Jordan-Wigner fermions turn into massless Majorana's, which is easy to see  by solving the simple problem Eq.(\ref{transIsingferm}). Absence of mass means scale invariance and this  is inherited  by the responses of the spins. 

This scale invariance  supplies the organizing principle that turns RG into a  tight framework to assess the critical state. Consider an infinitesimal change of scale. There are "operators" -- physical properties -- that stay precisely the same: these are the  {\em marginal} operators associated with the critical state per se. Upon deviating a bit from the critical temperature or critical coupling dealing with the quantum incarnation some operators turn  {\em relevant}. These grow upon moving to larger scales: in the context of the simple phase transitions these encapsulate the (dual) order parameters associated with the ordered state on both sides of the transition. Last but not least, one meets the {\em irrelevant} operators that fade out upon approaching the IR.  

The irrelevant operators may sound unimportant but the way these behave in the RG flow is actually crucial. This is highlighted in the statistical physics tradition, revolving around the notion of "free fixed point" versus "strongly interacting critical state" pending whether one is below of above the upper critical dimension. Let us consider again the coarse grained Landau-type action, like Eq.(\ref{Landauphi4}) -- in the case of Ising $\Psi$ is a real scalar. The simplest way to recognize the emergent scale invariance is in terms of "Landau mean field". Just ignore the self interaction term $\propto w$ in this action and what is left behind is just a massless, Lorentz invariant free field theory $\sim (\partial_\mu \Psi)^2 + m^2 \Psi^2$, $m^2 \rightarrow 0$. It is instructive to keep an eye on the canonical quantum interpretation: this is free field theory which is not entangled, the vacuum is a SRE product.  

But there is no a-priori reason that the self interaction can be ignored. Here the RG technology interferes. Especially in the statistical physics tradition, the role of {\em dimensionality} was highlighted. Given a symmetry one can identify an upper critical (space-time) dimension, $D_{uc}$. When  $D > D_{uc}$ it is easy to establish by power counting that $w$ is {\em irrelevant} towards the IR fixed point, meaning that it has completely vanished at infinite (Euclidean) times and distance. One can depart from the free fixed point and "climb up" the energy ladder using perturbation theory. But in doing so one directly starts to get information regarding the UV physics. Considering e.g. the critical point of the water-steam mixture or either the Ising spins that are both governed by a real scalar order parameter, at- and above its $D_{uc} = 4$ these perturbative corrections (wiring in short ranged entanglement) pick up immediately that H$_2$O molecules and simple Ising spins are very different objects. In the RG language one has to keep track of a myriad of irrelevant operators that switch on algebraically upon ascending from the IR fixed point. The outcome is a rather complicated affair.  

However, below  $D_{uc}$ the "self-interaction is finite at the IR fixed point" -- so far it is only a diagnosis of the trouble. Although I am not aware of a rigorous mathematical proof, such a "strongly interacting critical state" appears to be generically characterized by {\em exponential complexity}. It is impossible to enumerate it exactly: in the quantum incarnation it corresponds with a "quantum supreme" state of matter. 

The milestone of the 1970's is the discovery of {\em universality}: the properties of this strongly interacting critical state are completely determined merely by {\em symmetry and dimensionality}. Considering again the water-steam and Ising spin critical states, in both cases one is dealing with a real scalar: in 3D these fall in the same universality class. What is the meaning of this statement? The {\em irrelevant} operators haven been {\em completely diminished} in the IR, and only the order parameter field survives. This in turn implies a remarkable simplification: physical properties in the approach to- and at the critical point have to obey simple {\em scaling relations}. The only room that is left is in the form  of the various scaling dimensions that turn anomalous in the strongly interacting critical state, going hand in hand with various universal cross over functions. 

A phenomenological machinery is tied to universality, in the form of  {\em scaling relations}.  Their  power is in first instance tied to the scale invariance emerging at the critical point.  This imposes that all correlations have to be of the {\em powerlaw} kind. This is the familiar wisdom that only simple algebraic powerlaw  functions  are invariant under scale transformation. Change scale from $x$ to $y = \Lambda x$ and

\begin{equation}
f (x) = x^{\nu}, \; \; \; y = \Lambda x, \; \; f(y) = \Lambda^{-\nu} y^{\nu}
\label{powerlaw}
\end{equation}

But one can break the scale invariance weakly by e.g. tuning a bit away from the critical point, corresponding with switching on the relevant flow towards the stable fixed points. Widom introduced the notion that physical properties are {\em homogeneous} functions. In the one dimensional case, $f  (\Lambda x) = g(\Lambda) f (x)$, and e.g. in two dimensions

\begin{equation}
\Lambda^c f  (x, y) = f ( \Lambda^a x, \Lambda^b y)
\label{homofunctions}
\end{equation}

where $a,b,c$ are exponents, the (anomalous) scaling dimensions.  

Crucially, the free Landau critical theory is taken as a template to write down a homogeneous scaling form for the free energy density. The two control parameters are the reduced temperature $t = | T - T_c |/ T_c$ and the field $h$ coupling to the order parameter that both break the scale invariance ($d$ is space dimensionality),

\begin{equation}
f ( t, h ) = \Lambda^{-d} f (\Lambda^{y_t} t, \Lambda^{y_h} h )
\label{FREEENscaling}
\end{equation}

The free energy at the critical point is exhibiting the so-called {\em hyperscaling} -- it scales with the volume. 

From Eq. (\ref{FREEENscaling}) one can determine the behaviour of thermodynamics quantities in the approach to the critical point easily in terms of scaling expressions. Assume that $t$ sets the scale, such that  $\Lambda = t^{1/y_t}$. Resting on homogeneity, one can rewrite the scaling of the free energy as  $f (t, h) = t^{d/y_t} f ( \pm 1, t^{-y_h/y_t})$. By taking thermodynamic derivatives the scaling dimensions of the various thermodynamical quantities  upon approaching $T_c$ then follow as  $\alpha = d/y_t - 2$, $\beta = (d - y_h) / y_t$ and $\gamma = ( d - 2 y_h) /y_t$ for the specific heat ($C \sim 1/ t^{\alpha}$), magnetization ($M \sim t^{\beta}$) and susceptibility ($\chi \sim t^{\gamma}$), respectively. 

These scaling relations do apply also to the Landau free critical theory -- it forms a template for the construction of the scaling relations. It is easy to find that here the scaling exponents acquire simple "engineering" dimensions, such as  $\alpha=0, \beta = 1/2$ and $\gamma =1$ for a scalar order parameter. The revelation of the RG revolution was that scaling dimensions like $y_t$ and $y_h$ become {\em anomalous} in the strongly interacting critical state. These turn into a set of irrational numbers characteristic for the universality class. 

Given the strongly interacting critical state one can profit from the hyperscaling property. This implies that also the "geometric" scaling dimensions are set by the thermodynamic ones. This refers to the behaviour of correlation functions, as well as to the finite size scaling. Crucially, hyperscaling is violated at- and above $D_{uc}$. One just adds source terms to the free energy to compute two point correlations functions by taking functional derivatives. Exploiting homogeneity it follows,

\begin{equation} 
G ( r ) = \langle \phi ( r ) \phi ( 0) \rangle = \Lambda^{-2 (d -y_h)} G ( r/\Lambda, \Lambda^{y_t} t )
\label{Gscalingcrit}
\end{equation}

Near, but not at  the critical point,

\begin{equation}
G (r) \sim e^{- r/\xi}, \; \; \xi  \sim 1/ t^{\nu}
\label{corlength}
\end{equation}

$\xi$ is the correlation length, diverging at the critical point through $\nu = 1 /y_t$, the correlation length exponent. Right at the critical point, $G$ becomes a power law,

\begin{equation}
G (r) \sim 1 / r^{ d - 2 - \eta}
\label{corfunctionexp}
\end{equation}

characterized by an anomalous dimension, captured by the correlation function exponent $\eta = d+2 - 2y_h$.

Finally, we will see next that in the quantum interpretation the {\em finite size scaling} will have a crucial role in accounting for  temperature dependences. Given hyperscaling this will also become universal. One adds the finite size scale L to the free energy and assuming homogeneity, $f ( t, h, L ) = \Lambda^{-d} f (\Lambda^{y_t} t, \Lambda^{y_h} h, L /\Lambda)$.  As will become clear underneath, especially the finite scaling of the correlation function at the critical point will play a crucial role in the quantum-critical extension of the above. It follows immediately from the scaling Ansatz of the free energy, 

\begin{equation}
G (r) =   \frac {1} {r^{ d - 2 - \eta}} F ( \frac{r}{L} )
\label{finitesizeclas}
\end{equation}

with $F(x =0) =1$, i.e. when $L \rightarrow \infty$. The cross-over function $F(x)$ is also a universal quantity, determined solely by symmetry and dimensionality. The big deal is that this is a function exclusively of the ratio of $r$ and the system size $L$. 

The take home message is that because of the remarkable simplifications associated with the strongly interacting critical state, on this level one only has to compute the anomalous dimensions and the universal cross-over functions. In fact, this has evolved much further dealing with the theory right at the critical point in the form of {\em conformal field theory} that departs from the system being conformally invariant. This is constructed in a canonical language involving generalizations of harmonic oscillator bosons (Virasoro-, Kac-Moody algebra) becoming very powerful in $d=2$ due to integrability rooted in an infinite number of conservation laws. One learns that besides the scaling dimensions of the primary fields one also needs as extra data the 3-point operator product expansion coefficients required to construct arbitrary higher point correlation functions. 

I already emphasized that the full enumeration of the critical state appears to be  computationally  of exponential complexity.  In the classic literature universality appears as an empirical fact that is conceptualised in terms of the RG language with its scaling ramifications. The essence of the simplification embodied by the scaling relations is that only the marginal operator(s) and the relevant ones that switch on upon departing from the critical coupling describing the RG flow to the stable fixed points of both sides of the phase transition take part. Dealing with the free unstable fixed point above the upper critical dimension one also has to deal with the myriad of {\em irrelevant} operators that kick in directly upon ascending from the IR fixed point. These have to be accounted for as well in a scaling language, having the effect that scaling is no longer simple: the "difference between water molecules and Ising spins" becomes noticeable immediately. Last but not least, all that is remembered in the strongly interacting quantum critical state is symmetry and dimensionality, entering through a unique set of  anomalous scaling dimensions. 

In the classic literature this universality is observed as "emperical fact" but it is not quite  explained {\em why} this simple structure emerges. The following claim is conjectural. It seems however to be so obvious that I find it hard to see how it can fail: {\em the simplicity of the observable physics  is a consequence of the exponential complexity of the state.} 

This may appear at first sight as paradoxical. Let us focus on the quantum version of this affair. The strongly interacting quantum critical state is apparently quantum supreme, a coherent superposition delocalized in an extensive part of the many body Hilbert space. The observables are however VEV's associated with at best a few point correlation function. Extracting this minute amount of information relative to the exponential complexity of the state itself amounts to the {\em best possible averaging} of the former.  It is then not surprising that all the specifics of the UV physics are completely averaged away, wiping out the irrelevant operators.  Only the bare essentials of symmetry and dimensionality are then surviving in this "optimal average". This "scaling simplicity born from quantum supremacy" may be an important key to appreciate the outcomes of finite density holography which appears to be very similar in this regard. 

Finally, to what extend is this critical state computable using classical means? This boils down in first instance to the determination of the anomalous scaling dimensions. This has been more successful than one could have anticipated before hand. In the first place, various perturbative methods departing from the free critical state were devised organized to capture the scaling dimensions. The $\epsilon$ expansion is case in point; one departs from the upper critical dimension lowering it by an amount $\epsilon$ computing the corrections in a power series in the small number $\epsilon$. This is showing the right trends but it is at the same time the proverbial asymptotic expansion. For any finite $\epsilon$ it will eventually diverge, signalling the impossibility to compute the critical state in full. Similarly, dealing with states of polynomial complexity the Metropolis (Monte-Carlo) algorithm is very powerful. It was early on established that it is failing upon approaching the critical point and special "loop" and "cluster" algorithms were devised to improve these matters. The outcome is that the critical exponents can be computed with a respectable accuracy.  My understanding is that despite the exponential complexity it is possible with polynomial means to capture the scaling dimensions with an accuracy of a few digits while every extra digit takes an exponentially larger computational effort.  

\subsection{The quantum critical state: criticality with a Lorentzian twist.}
\label{QuCritWick}
  
Let us return to the main theme of the quantum systems. By definition, a stoquastic system is in one-to-one correspondence with a thermal statistical physics living in the Euclidean path integral. I just explained how to capture the strongly interacting critical state of the latter in terms of the scaling expressions. All we need to do this is to re-interpret these in Lorentzian signature. Temperature of the thermal incarnation turns into the coupling constant of the quantum system that we tune through the critical coupling where the phase transition takes place in space time. Physical temperature on the other hand relates to a finite size: the imaginary time circle with radius $\hbar/k_B T$.

Departing from the thermal classical state, there is one more scaling dimension that is really special for the quantum critical state: the {\em dynamical critical exponent} $z$ expressing how space and time are scaling relative to each other. A similar quantity can be identified at thermal transitions, but the meaning in the quantum case is completely different. 

By asserting that the quantum critical state is obtained by inserting the thermal state in Euclidean space time implicitly assumes that the Eucidean space-time is isotropic: the imaginary time direction is just like a space dimension. This coincides with Lorentz invariance: the Euclidean rotations turn into Lorentz boosts in real time. However, this does not have to be the case. In a stoquastic setting one can encounter for instance the situation that the order parameter fluctuations can decay in a heat bath of other excitations. Integrating these out yields an Euclidean action where the space- and time gradient terms do no longer appear on the same footing, see for instance Section (\ref{hertzmillis}). This will have the ramification that the scaling of time relative to spatial scale transformations acquires the form $t = l^z$.  For instance, diffusion means $z=2$. This is equivalent to asserting that there is not one but $z$ time directions: the total dimensionality of space-time is $d + z$.  We will see that $z$ is in the lime light dealing with the non-stoquastic critical matter, where $z$ can take seemingly absurd (from a stoquastic viewpoint) values,  including infinity.

\subsubsection{Anomalous scaling dimensions and the branch cuts.}
\label{branchcuts}

Let us first focus on what happens at zero physical temperature in the quantum system.  The first question is, what happens with the two point propagators right at the quantum critical point? We learned in the above that the Euclidean correlators have to be power laws in a system that submits to scale invariance. To keep track of time let us focus in on the Lorentz invariant (conformal) $z=1$ case. Define a distance $r_E = \sqrt{\omega_n^2 + c^2 q^2}$ in the Euclidean (Matsubara "$\omega_n$") frequency-momentum space.  A 2-point correlation function has to be of form,

\begin{equation}
\langle \Psi (r)  \Psi (0) \rangle \sim \frac{1}{r^{2 \Delta_{\Psi}}_E} 
\label{euclconfromal}
\end{equation}

using the "string theory" (conformal field theory, CFT) convention, involving the "anomalous" scaling dimension $\Delta_{\Psi}$ of field $\Psi$ (the $\eta$ of the stat. phys. literature). Fourier transform first to momentum-energy ($q$, $\omega$); the system is assumed to be Lorentz invariant characterized by a "velocity of light" $c$.  
 Upon Wick rotation to Lorentzian time $\omega_n^2 \rightarrow - \omega^2$ and it follows, 

\begin{equation}
\langle \Psi \Psi \rangle_{\vec{q}, \omega} \sim \frac{1}{(c^2 q^2 - \omega^2)^{d/2 + 1/2 - \Delta_{\Psi} }} 
\label{Lorentzconforomal}
\end{equation}

This is referred to as "kinematics alone fixes the form of the two-point functions" in the CFT literature. The outcome is very simple but iconic: response functions of this form  we refer to as "branch-cuts".    

Considering the free critical theory which is characterized by $ 2\Delta_{\Psi} = d -1$ and the propagator takes the form

\begin{equation} 
\langle \Psi \Psi \rangle_{\vec{q}, \omega} \sim \frac{1}{c^2 q^2 - \omega^2}
\label{freecrtitfield}
\end{equation}

One discerns immediately that this corresponds with sharp poles on the light cone $\omega = c q$: this describes a system of free massless particles.

\begin{figure}[t]
\includegraphics[width=0.9\columnwidth]{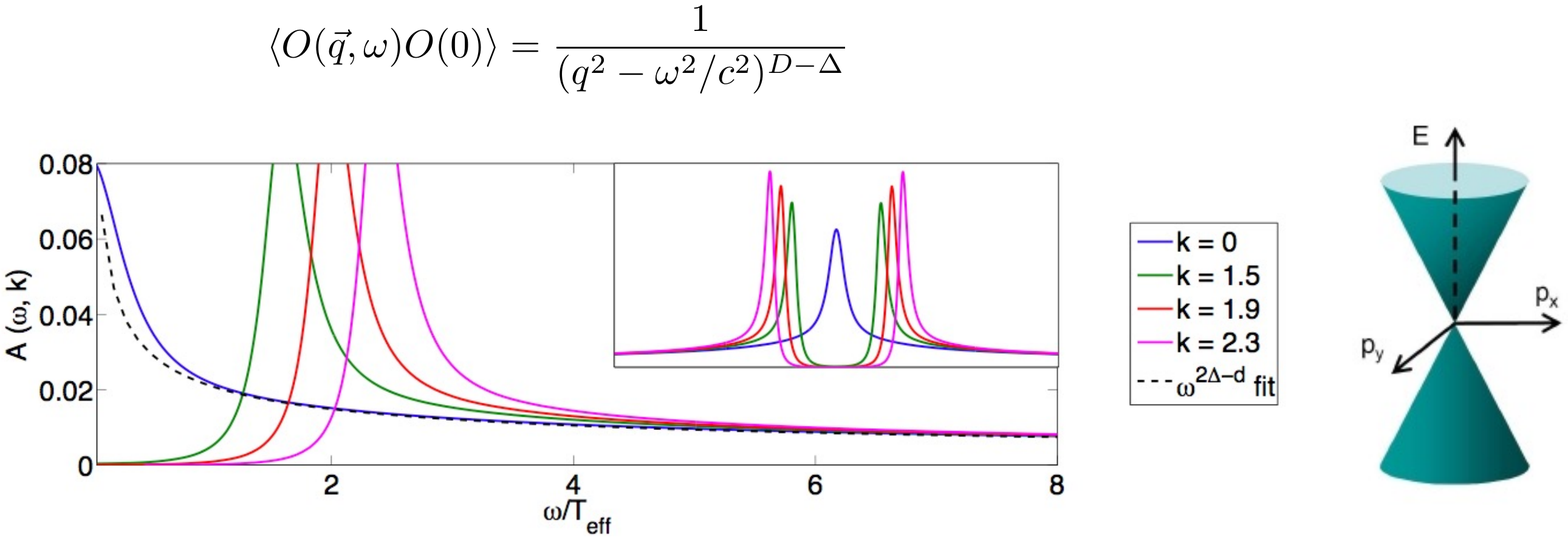}
\caption{A typical example of the spectral function of a strongly interacting stoquastic, conformally invariant critical state. This is the one associated with the two point fermion propagator actually computed using AdS/CFT at a small but finite temperature \cite{Sciencefermions09} , as would be measured by photoemission dealing with electrons. At zero momentum ($k=0$) this is just a powerlaw  as function of frequency. At finite momentum one first to enter the light-cone (right) before the spectral weight becomes finite.}
\label{fig:fermionbranchcut}
\end{figure}

But we learn that in the strongly interacting critical state the exponent is anomalous, it is an irrational number that is bounded by the free case (the "unitarity limit").   How does the spectrum of such a system looks like? This is illustrated In Fig.  (\ref{fig:fermionbranchcut}) dealing with Dirac--like fermions. This is actually computed using the "plain vanilla" Maldacena AdS/CFT associated with zero density. As we will see later this is just tailored to describe the {\em structure} of the strongly interacting quantum critical state, where one can in fact continuously tune anomalous dimensions by the free parameters of the "bottom up" phenomenological incarnation of holography. 

How do these branch-cut propagators work? Let us first consider zero momentum and abbreviate $\alpha_{\Psi} = d + 1 - 2 \Delta_{\Psi}$. The propagator becomes,

\begin{equation}
\langle \Psi \Psi \rangle_{\vec{q}=0, \omega} \sim \frac{1}{(  i \omega)^{\alpha_{\Psi}}} = \frac{1}{ | \omega |^{\alpha_{\Psi}} }\left( \cos \phi_{\Psi} + i \sin \phi_{\Psi} \right)
\label{Phaseangle}
\end{equation}

with the phase angle $\phi = \pi \alpha_{\Psi} /2$. The imaginary part is the spectral function corresponding with a simple power law $1/  | \omega |^{\alpha_{\Psi}}$. But the analytical structure is very tight: it follows that also the real part of the response shows this powerlaw while the relative weights of the real- and imaginary parts are fixed via the anomalous dimension as well. When both can be measured the phase angle should be independent of frequency.  In fact, the most convincing experimental evidence for such "strongly interacting emerging scale invariance" (there is a lot more going on than this stoquastic affair) in the cuprate strange metal is in the form of such a branch cut behaviour in their optical conductivities as I will discuss in Section (\ref{optbranchcutcupr}).

One can now boost this to finite momentum. One observes that the propagator for frequencies smaller than $cq$ has no longer an imaginary part. The spectral weight is vanishing outside the lightcone in the frequency-momentum domain obeying the causality requirements (see Fig. \ref{fig:fermionbranchcut}). At $\omega = cq$ the imaginary part jumps up to then again decrease in the power law fashion. 

A crucial observation is that from the analytical structure of the propagator one can directly discern whether the problem is perturbative in a free limit  (the "particle physics") or whether one is dealing with a quantum supreme state: the branchcut is a first example of a signature of "unparticle physics". The belief is widespread that the self-energy form of a propagator $G (\omega, k) \sim (\omega - \varepsilon_k - \Sigma (\omega, k))^{-1}$ is universal but this is actually particle-physics folklore.   Surely, the self-energy $\Sigma$ may be very fanciful and difficult to compute when one has to keep track of high-order "diagrams". This is however not what matters. The issue is that the analytical structure of the self energy expression implies that it should be possible to reconstruct it  departing from a free Hamiltonian ( the $\varepsilon_k$ factor) that is remembered in the full response given the convergent perturbation theory/SRE ground state. It is obvious by comparing the branchcut expression with the self energy form that the information regarding a free limit has completely disappeared  from the former.

This amounts to a warning for experimental spectroscopists. With the eyes half closed, one sees in Fig. (\ref{fig:fermionbranchcut}) peaks dispersing in a way that looks like free Dirac fermions. It is a community habit to immediately jump to the conclusion that this reveals band-structure electrons in the case of the most abundant form of spectroscopy: photoemission. But the information telling us whether it is particle- or unparticle physics is actually encoded in the analytical form of the propagators, i.e. the lineshapes. In fact, when one has access to both the real and imaginary part of the propagators it is quite easy to construct a "quasiparticle detector" in the form of a rigorous algorithm that will tell whether any information regarding a free limit is hidden in the experimental outcome. Although the experimental information falls short for a definitive proof, line shapes extracted from very recent high resolution ARPES data  indicate that the most "particle physics" like features seen in cuprate strange metals (the "nodal fermions") are actually not passing the quasiparticle detection machinery, see Section (\ref{Cupratefermions}). 

As a final remark, it is a habit to call responses characterized by quasiparticles "coherent" while the "unparticle" spectra are called {\em "incoherent"}. This is actually a rather awkward semantics. Coherence refers to wave phenomena, like a laser beam is a form of coherent light. This is rooted in the quantum mechanical coherence of single (non interacting) particles. But Branchcuts etcetera are also rooted in quantum physical coherence but now present in the many body Hilbert space. This terminology gets really painful in the context of transport. I already alluded to hydrodynamically modes protected by global symmetry  like sound waves. Such "particles" have nothing to do with quantum-mechanical coherence, they are just rooted in global conservation laws. As I will explain later, these are behind the ubiquitous Drude transport. There is a habit to call these also "coherent" which is just irritating, reflecting the effect of 100 years living in the particle physics tunnel. For this reason, I will be obnoxious: I will call "coherent" and "incoherent" instead "particle" and "unparticle" when this is not governed by global conservation law. When global symmetry is at work I will name the modes referring specifically to the conservation laws, like "zero sound" giving rise to Drude phenomenology.      

\subsubsection{The quantum critical wedge.} 
\label{QPTgen}

What  else can an experimentalist expect to observe dealing with stoquastic quantum criticality?  

This affair revolves around the special, singular "quantum critical point" (QCP). The experimentalists should be in the position to tune the zero temperature coupling constant $g$ controlling the quantum fluctuations. This may be a magnetic field, pressure or a chemical potential. Away from the critical coupling $g_c$ he/she will be in the position to identify a (dual) order parameter. Upon approaching the QPT the order parameter rigidity (elastic moduli, superfluid stiffness etc) will decline but at sufficiently long times, low temperatures and so forth it will give in to the thermal physics described by classical theory: the SRE product moral, called the "renormalized classical regime". 

Dealing with quantum physics, space and time are "intertwined" and to get a clear view one needs {\em dynamical} response functions: the ideal observables are the dynamical linear response functions that at zero temperature and precisely at the critical coupling will reveal the branch cuts. But what to expect when $\delta g$ is small but finite? The crucial quantity is the correlation length/time: at distances larger than $\xi_{\mathrm{cor}}$ the system is in the renormalized classical regime but the critical state is re-entered on scales smaller than the correlation length. 

In fact, time matters most and it is just convention to define $\nu$ to be associated with  length, such that the correlation time $\tau_{\mathrm{cor}} = 1/ (\delta g)^{\nu z}$. Hence, at an energy $\hbar/ \tau_{\mathrm{cor}}$ the response crosses over from the dynamics of the classical theory (e.g., exhibiting Goldstone bosons) back to the critical branch cut affair at higher energy. In fact, the crossover functions are also supposed to be governed by universality but they are in practice hard to compute. The take home message is that in the idealized situation of zero temperature, full control over $\delta g$ and spectroscopic machinery giving full access to the dynamical susceptibilities, the experimentalist could measure the {\em quantum critical wedge}, see Fig.'s (\ref{fig:PlanckiandissCFT},\ref{fig:QCPwedge}).  Upon moving away from the QCP he/she will see that the crossover line governed by $(\delta g)^{\nu z}$ is gradually moving to higher energy. 

\subsection{The finite temperature physics near the quantum phase transition.}
\label{finiteTQCP} 

Up to this point I addressed the physics at strictly zero temperature. Using the Euclidean path integrals it is actually quite easy to find out what happens at {\em finite} temperature: we just have to roll up the imaginary time direction in a circle with radius $R_{\tau} = \hbar / (k_B T)$ where $T$ is the physical temperature. This in fact a geometrical operation and it is  well understood what has to be done in the statistical physics incarnation: the art of finite size scaling ruled by universality, being integral part of the RG portfolio! All one has to do is to compute the influence of the finite time circle on the statistical physics correlators, to then analytically continue to real time to establish the physical linear response functions. 

This is easier said than done because this is a first instance where the analytic continuation information loss problem surfaces. Let us depart again from the statistical physics of the strongly interacting critical state  in Euclidean space time. Right at the critical point the system acquires the conformal invariance and one can now ask the question how correlation functions are influenced when one of the dimensions (i.e., Eucidean time) becomes finite, thereby lifting the scale invariance.  Correlations at a distance $r$ should become scaling functions of $r/L$ where $L$ is the finite size. Associating $r$ with the Euclidean time direction, and realizing that the "duration" of imaginary becomes $L_{\tau} = \hbar / k_B T$, the issue is that  the Wick rotation maintains locality in the sense that a certain duration of imaginary time is in one-to-one correspondence to the same duration in Lorentzian time. Dealing with the strongly interacting critical state, the Euclidean propagators should give in to to the simple finite size scaling Eq. (\ref{finitesizeclas})
This implies that  real frequency linear response functions should submit to the scaling behaviour,

\begin{equation}
\chi_{\Psi, \Psi} (\omega, T) = \frac{1}{T^{\Delta_{\Psi}}} F_{\Psi} ( \frac{\hbar \omega}{k_B T} )
\label{ETscaling}
\end{equation}

dealing with non-conserved quantities -- 'hydrodynamically protected" quantities are more tricky and we will return to this at length in Section (\ref{holotransport})

This is called "energy- temperature scaling", exquisitely associated with the way that the quantum-physical scale invariance manifests itself in experiment.  We learned in the above that according to the statistical physics the cross-over function  $F_{\Psi} ( x )$ is universal as well, set by symmetry and dimensionality. But these are {\em not} available in closed analytical form dealing with non-integrable critical states -- AdS/CFT is an exception (see next section). 

Why is there a problem of principle? The Wick rotation back to Lorentzian signature is the culprit.  Tiny  "bumps" in the smooth Euclidean correlators may turn into rapid variations in the Lorentzian responses. This information-loss problem becomes particularly severe at frequencies small compared to temperature; small noise in the numerics of e.g. the QMC explode upon analytical continuation. 

Resting on elementary finite size size reasoning it is however still possible to arrive at a variety of generic behaviours associated with the finite temperature physics -- this was first realized by Chakravarty et al. \cite{CHN88}  in the specific context of magnetic systems close to a quantum phase transition. This was further elaborated by Subir Sachdev -- it is a central theme in his  book \cite{QPTSachdev}. 

\subsubsection{The $E/T$ scaling and the quantum critical wedge.}
\label{qucritwedge}

\begin{figure}[t]
\includegraphics[width=0.9\columnwidth]{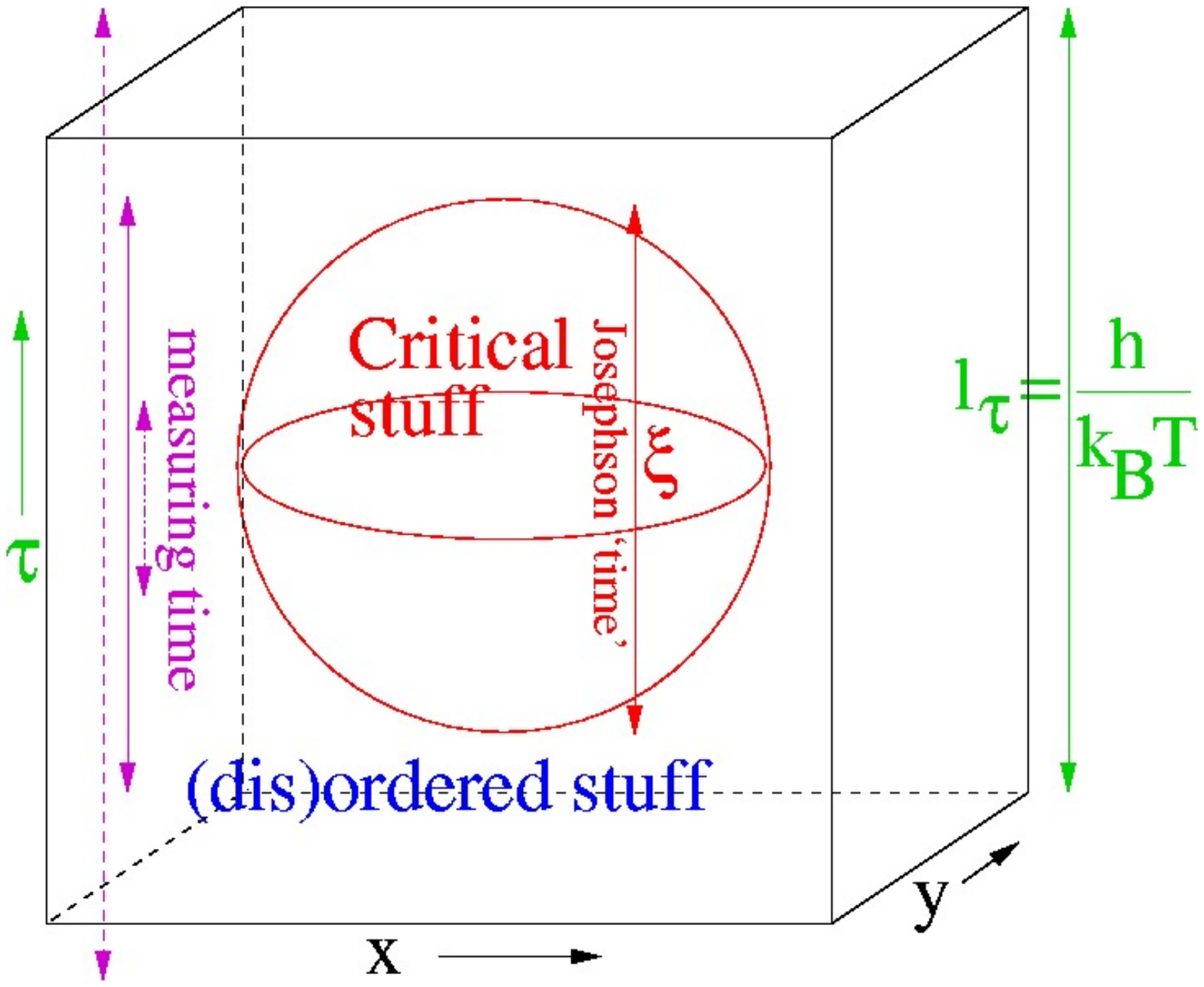}
\caption{ Euclidean space time "road map" dealing with a stoquastic critical state. Upon approaching the quantum phase transition the "Josephson correlation time" $\xi_{\tau}$ increases algebraically in terms of the reduced coupling $\delta g = | g - g_c | /g_c$ to diverge at the QCP. The meaning of this is that  at times $l_{\tau}$ shorter than $\xi_{\tau}$ the system behaves as if it is right at the critical point (red "ball") characterized by the Planckian dissipation, branch cuts and so forth. When the time associated with the measurement is short compared to  $l_{\tau}$ the response will be as if temperature is zero.  However, when $\delta g  \neq 0$ upon raising temperature the system will be first governed by the stable (ESR product) state and instead one will find the "renormalized classical response" until $l_{\tau} \simeq \xi_{\tau}$ where it will crossover to the quantum critical "Planckian" regime at macroscopic (DC) measurement times. This is the explanation for the "quantum critical wedge",  Fig. (\ref{fig:QCPwedge}).}
\label{fig:PlanckiandissCFT}
\end{figure}

Let us depart from the energy-temperature scaling form Eq. (\ref{ETscaling}). Experimentalists can measure the response function over a large frequency range at many different temperatures. By plotting $T^{\Delta_{\Psi}} \chi_{\Psi, \Psi} (\omega, T)$ as function of $\hbar \omega / k_B T$ all the data will collapse on a single curve (representing $F$) for the correct value of $\Delta_{\Psi}$. This $\Delta_{\Psi}$ should then in turn be consistent with the branch-cut setting in at $\hbar \omega > k_B T$. This is very powerful: scaling collapses are the primary weaponry dealing also with the thermal critical state but in the quantum context the "temperature" finite size scaling becomes very convenient. Just dial up the heating! When one finds this $\omega /T$ scaling one can  be sure that one is dealing with a quantum critical system. Yet again, in practice this is less glorious because of the difficulties to measure dynamical response functions over the required dynamical range. 

Let us now turn to the regime where the measurement time is {\em long} as compared to $R_{\tau}$ (Fig. \ref{fig:PlanckiandissCFT}): in other words $\hbar \omega << k_B T$, which includes DC measurements ($\omega = 0$). We imagine that the experimentalist can vary the coupling constant $\delta g$ through the QCP while he/she has access to a large temperature range. He/she will encounter yet another fingerprint signalling that a quantum phase transition is at work. This "quantum critical wedge" is in everyday laboratory practice actually the most important indication for the claim that such physics is at work. 

I already alluded to the role of the correlation "time" that is decreasing algebraically with $\delta g$ upon moving away from the QCP. I explained that this will be visible in the zero temperature dynamical response functions. But the principles of finite size scaling imply that it should be visible as well at zero frequency when temperature is varied. This is again very simple. When one is close but not at the critical point, at Euclidean times smaller than the correlation time $\tau_{\mathrm{cor}}$ the system behaves as if it is precisely at the critical point. Upon increasing temperature the time circle is shrinking and when its radius becomes of order of $\tau_{\mathrm{cor}}$ the finite temperature system re-enters the critical state! Hence, one expects a cross over from the long time finite temperature renormalized classical physics to the finite temperature physics one finds right at the QCP.  A cross over will become detectable in DC physics at a temperature,

\begin{equation}
k_B T_c \sim (\delta g)^{z\nu}
\label{qucritwedge}
\end{equation}

\begin{figure}[t]
\includegraphics[width=0.9\columnwidth]{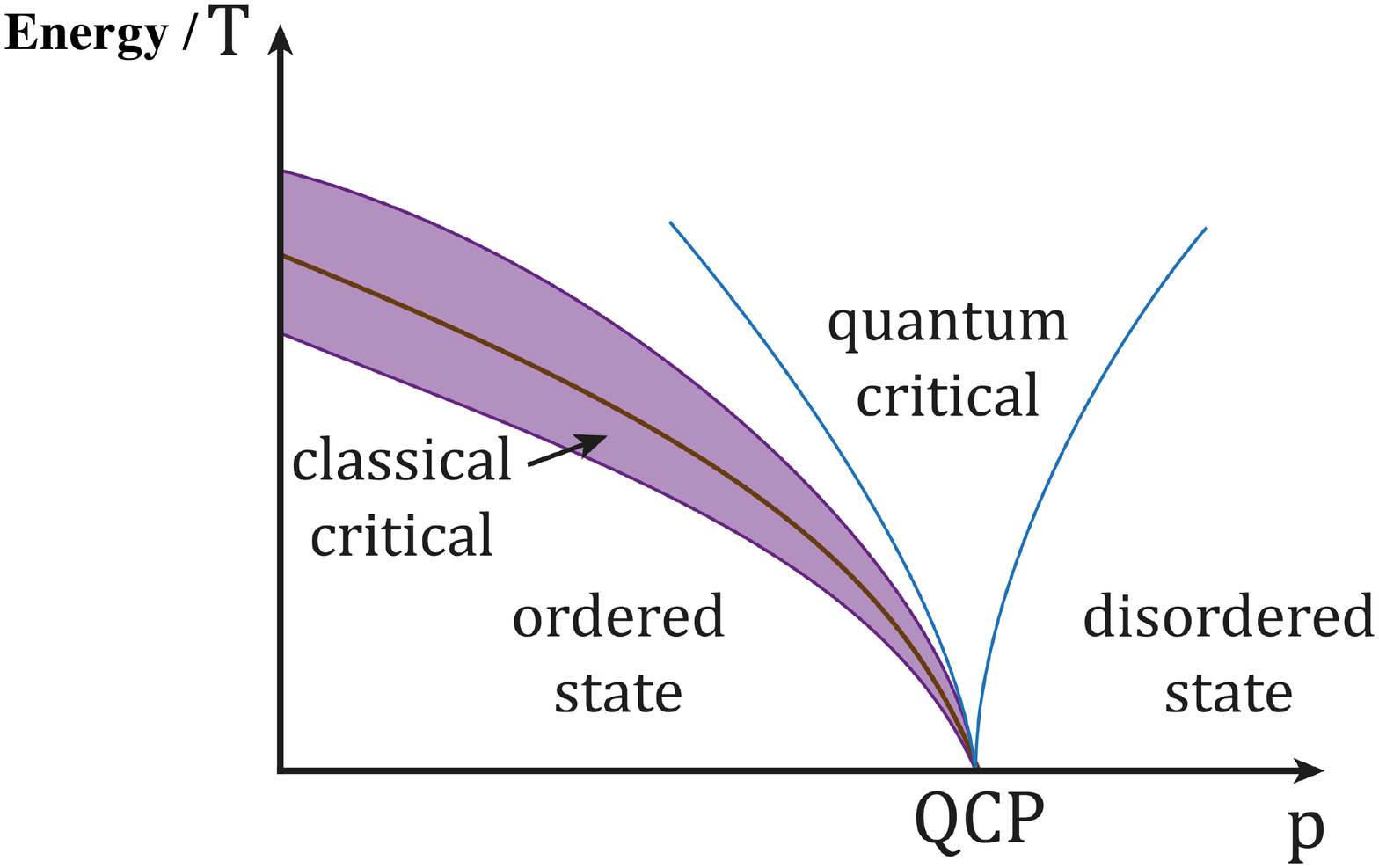}
\caption{ The "quantum critical wedge": upon tuning the system with a zero temperature control parameter ($p$, like a magnetic field, pressure, thermodynamical potential, etc) one encounters a quantum phase transition at a critical coupling $g_c$. On both sides the physics of the stable (semi) classical states will be in charge at sufficiently low energy and/or temperature. This includes the possibility of a {\em thermal} phase transition (purple). However, as explained by Fig. (\ref{fig:PlanckiandissCFT}) upon raising temperature the cross over to the quantum critical state will take place that is in turn characterized by the Planckian dissipation. The crossover lines in the ($p, T$) plane are determined by the correlation length exponent.}
\label{fig:QCPwedge}
\end{figure}

This defines a "wedge" in the coupling constant - temperature plane (Fig. \ref{fig:QCPwedge}), and one can  read off the combination of the anomalous dimensions $z \nu $.  

Although quite simple, this is empirically very powerful. In the mean-time there is a plethora of such quantum critical wedges identified in a variety of experimental incarnations (heavy fermion systems, pnictides, structural transitions) while for a long time it was taken for granted that such a quantum phase transition occurring at optimal doping would also be responsible for the "strangeness" of the cuprates. In a very recent development hard evidence appeared that this is not quite the case as I will discuss in Section (\ref{qucritphase}).

When I discussed the Kadanoff scaling thermodynamical quantities were on the foreground. The above discussion concentrates however on the geometrical exponents ($\eta/\Delta_{\Psi}, \nu, z$). As it turns out the workings of the thermodynamics at the quantum critical point did not attract much attention, largely because it is not easy to get an experimental handle. However, this is an entertaining affair by itself, and it has been completely charted  for  a density driven QCP in Ref. \cite{JZHossein04}.   

But now the question arises, what is special about the finite temperature -- low frequency physics of the quantum critical state itself? The revelation is in the realization that the classical dissipative state that ascends from the zero temperature strongly interacting critical state appears to be the {\em best heater} that can be realized in nature. This is the notion of "Planckian dissipation". 

\subsubsection{Eigenstate thermalization: the origin of heat and the collapse of the wavefunction.}
\label{ETHsummary}

The word "dissipation" refers to the origin of heat -- eventually, where does the second law of thermodynamics come form, insisting that entropy is always increasing towards the future? Conventionally, this was discussed in terms of the {\em stochastic} equations of motions underlying irreversible thermodynamics. Rooted in Brownian motion, these involve random forces and damping terms that adequately describe  steam engines and so forth. These are the fundaments behind the usual second law arguments, e.g. the probability that eggs that shattered falling of the table will have a vanishing probability to re-assemble. This is then underpinned by reference to chaos, the Lyaponov exponent affair that insists that uncertainties in initial conditions will amplify exponentially. 

However, since the early 1990's the view on this has been on the move again through developments in the quantum information community. This eventually amounts to the insight that this stochastic nature of reality can be traced back to a more fundamental origin: {\em probability appears when the wave function collapses and dealing with thermodynamically large, closed systems irreversibe thermodynamics is the consequence.} 

This insight is labelled by "Eigenstate thermalization hypothesis" (ETH) referring to seminal contributions by Deutsch and Srednicki  in the early 1990's. This refers to the following hypothesis: prepare a pure state that is a coherent superposition of highly excited energy eigenstates (that are by default "quantum supreme") narrowly distributed around the mean energy. The outcome for the VEV of any operator at a sufficiently late time is then {\em identical} to what one would find in a thermal, mixed state at a temperature corresponding with the expectation for the fully thermalized state. 

In a way, it just overcomes a psychological bias. The "collapse" was historically discussed in the context of highly contrived baby models describing tiny quantum systems, typically two level problems. But the rules change dealing with generic systems that are intrinsically many body: the cocktail of exponentially large Hilbert spaces together with the fact that for whatever reason mankind can only measure VEV's is the natural theatre for the emergence of the heat as understood by the statistical physicists. In this regard, I recommend the reader to have a close look at Ref. \cite{dAlesio}, where the Master equation governing  stochastic dynamics is {\em derived} from the quantum system undergoing repeated spontaneous collapses. 

Another development is next door to the main theme of these lecture notes: mobilizing the AdS/CFT correspondence to study the {\em non-equilibrium} physics associated with CFT's. A whole slur of novel quantum non-equilibrium phenomena were identified, referred to as "rapid scrambling", "quantum Lyaponov time", "Pole skipping" to name a  view. This is beyond the scope of these lectures and I refer the reader to a recent review, ref. \cite{SonnerLiuRev}. 

I will stay focussed  here on linear response, which is a simplifying circumstance. One just injects an infinitesimal amount of energy that converts into an infinitesimal amount of heat. Although technically much easier to handle -- one can just rely on the simple equilibrium path integrals -- eventually the reason that it reveals dissipation is the same. It is what remains of the unitary time evolution after collapsing the wave function.  Among others, ETH "logic" shows that the rate of entropy production is tied to the many body entanglement -- the more strongly the states in the game are entangled, the faster it goes. In linear response this affair takes the shape of the Planckian dissipation. 

\subsubsection{The finite temperature quantum critical fluid: the Planckian dissipation.} 
\label{Planckiandiss}

 Chakraverty {\em et al.} \cite{CHN88} were the first to realize that this linear response heat production in the quantum critical fluids submits to a principle of extraordinary simplicity which is at the same of an extraordinary weirdness within the reference frame of a "particle physicist". It is an instance where wording matters to help the human mind to get acquainted to the mathematical concept: I did something good for physics when I invented  myself \cite{Planckiandiss} the quote "Planckian dissipation."

I already discussed the way that this can be computed in principle using the thermal field theory machinery. Compute the statistical physics correlation functions in Euclidean space time as these get modified by the time-circle finite size scaling. Subsequently continue to Lorentzian time. But I already stressed the trouble that here the Euclidean "information loss" problem becomes insurmountable when analytical formulae are not available upon entering the regime where $\hbar \omega << k_B T$. 

Given the Eigenstate thermalization principle the system has to behave as a classical, dissipative system at finite temperatures and macroscopic times.   But such systems also give in to universal principle. We are typically interested in a {\em non-conserved} order parameter, associated with the spontaneous breaking of symmetry on the "ordered" side of the transition. Since such a field is not protected by a global conservation law it should be subjected to the phenomenon of {\em relaxation}. Having it switched on by applying an external order parameter stabilizing field (in infinitisemal form in linear response), by switching off this field the order parameter will {\em disappear}, relax on a characteristic time scale called the relaxation time. The work exerted by the external source will thereby turn into heat: this relaxation time is therefore linked generically to dissipation, the time it takes to generate entropy. 

Such relaxational processes may be quite complicated, involving a hierarchy of relaxation times -- see the discussion on the memory matrix in  Section (\ref{holotransport}). However, when it is governed by a single relaxation time it gets very easy. The EOM of the order parameter field will then be of the simple form,

\begin{equation}
\frac{ d \Psi}{d t} + \frac{1}{\tau_{\Psi}} \Psi  =  M_{\Psi}
\label{relEOM}
\end{equation}

$ M_{\Psi} $ is the force exerted by the external  source term, while $\tau_{\Psi}$ is the relaxation time. 
Inserting an oscillating source $M_{\Psi} (t) = M^0 e^{-i \omega t}$ it is an elementary exercise to derive the form of the dynamical susceptibility,

\begin{equation}
\chi_{\Psi} (\omega)  =  \chi^0_{\Psi} \; \frac{ i}{ \frac{1}{ \tau_{\Psi}} - i \omega}
\label{relpropagator}
\end{equation}

the absorptive part describes a Lorentzian peak with width $1/\tau_{\Psi}$ centred at $\omega = 0$, This may be familiar from the textbook Drude theory, a typical example of a simple relaxation response (Section \ref{holotransport}). 

Back to the Euclidean field theory. We are interested in the behaviour of the Euclidean correlators at imaginary times long compared to the radius of the time circle, representing the $\hbar \omega << k_B T$ regime. As I already stressed, it is practically impossible to compute them. But now we can rely on simple reasoning: (a) assert that in this regime one is encountering classical, dissipative physics, (b) observe that on the quantum side there is only {\em one} characteristic time that enters the $\omega/T$ cross over function: the radius of the time circle itself! Hence, the outcome should be that the classical relaxational dynamics can only know about a single time, and the high school theory Eq. (\ref{relEOM}) should therefore apply. 

This has the unavoidable consequence that,  

\begin{equation}
\tau_{\Psi}= A_{\Psi} R_{\tau} = A_{\Psi} \frac{\hbar}{k_B T}
\label{planckiandis}
\end{equation}

where $A_{\Psi}$ is a universal amplitude associated with the universality class and this a parametric factor that has to be of order unity. This incredible simple result is the Planckian dissipation. 

The issue is that by importing Planck constant scales appear in the dimensional analysis that otherwise do not exist. A case in point are the UV Planck scales of quantum gravity, obtained by combining $\hbar$ with Newton's constant. The Rydberg scales of atomic and molecular physics are yet another example. But dissipation is associated with temperature, $k_B T$ having the dimension of energy. Planck's constant has the dimension of action which is energy times time. Henceforth, $\tau_{\hbar} = \hbar / (k_B T)$ is a  characteristic time associated with temperature $T$. This time now becomes the time ruling the dissipative dynamics but only so dealing with a quantum system that is scale invariant (modulo the time circle) submitting to universal finite size scaling (meaning it is quantum supreme) in Euclidean space time. Anybody can imagine something simpler than this wisdom? 

On a side, a strange folklore appeared trying to argue this in another way, even blaming at least implicitly my person for the copyright. This departs from the {\em quantum mechanics} wisdom called the energy time uncertainty relation: $\Delta E \Delta t \ge \hbar$. Assert that $\Delta E = k_B T$ suggesting a characteristic time $t = \tau_{\hbar}$. But this is ludicrous, quantum mechanical systems of a few degrees of freedom to which this applies are of course not subjected to the genuine quantum field theory affair in the above! To simplify is good but do not take it too far: after all we have learned something since 1927.

I am also responsible for the conjecture that $\tau_{\hbar} $ is the shortest possible relaxation time in a linear response context, permissible by the laws of physics. This is entirely motivated by the Euclidean field theory reasoning in the above. To which degree it is absolutely certain that at times long compared to the Euclidean time circle one meets invariably classical dissipative physics, assumption (a) in the above? I am not aware of a mathematical proof but it is invariably the case in 1+1D integrable systems including the transversal field Ising model playing a key role in Sachdev's book. Better testimony follows from AdS/CFT dealing with field theories in the boundary that are a-priori not integrable where it is hardwired in the bulk geometry.  

The other crucial condition is the scale invariance in Euclidean space time, assumption (b) in the above. This relies on generic behaviour associated with the finite size scaling in a scale invariant system -- this may even not rely on the stoquastic nature of the problem where the rule is well established. When the system is characterized by internal scales as for the stable SRE phases away from the critical point, these scales will interfere in the finite size scaling. Euclidean correlators are no longer only dependent on the radius of the time circle but also on the rigidity scale having the effect that relaxation is delayed: this is the reason that the Planckian dissipitation limit is {\em never}  reached in particle physics.  

A familiar example is the Fermi-liquid. In a way it is a most extreme example of a "classical" (SRE) system that is as quantum-mechanical as can be with the Fermi-energy representing a gigantic quantum zero point motion energy, see next section. The relevant relaxation time is the time it takes for the finite temperature quasiparticles to collide: $\tau_{\mathrm{col.}} \simeq  E_F \hbar /(k_B T)^2$. This relaxation time is stretched by the factor $E_F /(k_B T)$ where $E_F$ is the scale associated with the "rigidity" of the Fermi liquid. 

Finally, scale invariance by itself does not suffice. {\em Universality} is required; departing from the free critical theory above the upper critical dimension the finite size scaling is no longer {\em simple} given the myriad of irrelevant operators that rear their head. Perturbative corrections become important that will wreck the simple energy-temperature scaling as well as the ruling of  $\tau_{\hbar} $. I will illustrate this later in the context of holographic superconductivity with a colorful example (Section \ref{HoloSC}). 

The interest in Planckian dissipation is presently flourishing especially in the experimental condensed matter community. It appears to be quite ubiquitous in governing {\em transport} properties, with as icon the famous linear resistivity in cuprate strange metals that is experimentally proven to be governed by $\tau_{\hbar}$ (see Section \ref{linresemp}). But this is actually a context where the Planckian dissipation conjecture does not apply directly. The difficulty is that  this transport is controlled by a nearly conserved hydrodynamical soft mode, rooted in the {\em total momentum} being conserved in a translationally invariant system. The current relaxation can therefore not be universal in the Planckian sense: it eventually will reflect the way that the translational symmetry breaking survives in the deep IR. I will come back to this subject at length in Section (\ref{holotransport}).  

\subsection{Improvising quantum-critical metals: the Hertz-Millis model.}
\label{hertzmillis}
 
 The take home message is that we understand so much of the stoquastic quantum critical state that a tight phenomenological framework is available to test it in experiment. This is all rooted in the powers of scaling; this is in close parallel to the history of the thermal critical state where Kadanoff developed the phenomenological scaling theory well before the more "microscopic"  Wilson RG became available.  The charm is just in the Wick rotation ploy where the familiar properties of the thermal critical state are twisted into the branch cuts, the quantum critical wedge and the Planckian dissipation. 
 
 The question that immediately arises, where is the experimental realization of such a strongly interacting stoquastic quantum critical state?   One would like to find a no-frills version, in particular a sign-free spin system with an Euclidean version that is well charted in its thermal incarnation. Although experimentalists tried hard, no such case was until now identified. 
 
One may have a second thought. Presently a grand pursuit is unfolding in the form of the creation of the quantum computer. The difficulty is in keeping this contraption isolated from the outside world to a degree that the coherence of states delocalized in the vast many body Hilbert space can be maintained. Right at the quantum critical "singularity" one is dealing with such states and it is as if it is impossible for nature itself to avoid circumstances that spoil this quantum supremacy. In fact, the same day that I was writing this passage the news broke that two groups managed to "program" a cold atom quantum simulator to deal with the transversal field Ising model -- it is the closest approach I have seen coming out of experimental laboratories \cite{QuSimIsing}. 
 
 A case in point is the experimental realization of a literal transversal field Ising model. Ising spin systems are quite abundant, typically insulating salts containing rare earth elements. There is nothing easier than applying an external transversal magnetic field. Such a system was explored by Aeppli and Rosenbaum and coworkers in the late 1990's \cite{AeppliQuAn}. A complicating circumstance was encountered. Because of the dipolar interactions the classical (Ising) spin system is frustrated, exhibiting spin-glass properties.  Glasses are generically characterized by extremely slow relaxation. The authors demonstrated that this speeds up significantly by the transversal field induced tunnelling in the complex energy landscape characteristic for glasses. This was actually the birth of quantum annealing, exploited among others by the d-wave quantum computer.  Similarly, upon going to even lower temperatures one encounters the interactions with nuclear spins that will act as a heat bath decohering the electron spin system. 
 
 Nevertheless, quite a large number of systems were identified exhibiting quantum phase transitions but invariably these cannot be related straightforwardly to well understood thermal analogues. Invariably, these occur in metallic-like systems and the sign problem discussed in the next section is in the way of a rigorous understanding. Perhaps the best examples are the two dimensional  systems exhibiting  superconducting-insulator transitions that were conjectured to reflect the $U(1)$ (XY) universality class associated with the fluctuations of the superconducting phase as early as the late 1980's \cite{MatFisch89}. But these examples are further complicated by disorder, the role of fermion excitations as well as the fact that the experiments in first instance rely on transport measurements where conservation laws may obscure the quantum critical features. Another example are the "plateau transitions" observed in the quantum Hall effects \cite{SondhiRMP}. These are obviously of the non-stoquastic kind (fermions and time reversal symmetry breaking) and are generally perceived as not understood. 
 
 Starting in the early 1990's a host of metallic quantum critical systems were discovered, the great majority of them in the context of the "heavy-fermion" lanthanide- and actinide intermetallic compounds \cite{LohneysenRMP}. These typically show magnetic order and applying influences like pressure or magnetic fields these can be tuned through a zero temperature transition  where the magnetic order disappears at zero temperature. These exhibit the characteristic quantum critical wedges in the temperature-coupling constant plane, bordering a region characterized by anomalous transport properties including linear resistivities that appear to reflect the Planckian dissipation. In addition, at low temperatures one finds often a superconducting "dome" centred at the quantum critical point. 
 
The community standard is to view  such transitions departing from the very early (1975) work by John Hertz,  while Andy Millis ironed out some minor flaws more recently:  the "Hertz-Millis" theory \cite{HertzMillis}. Hertz departed from a Fermi-gas subjected to weak interactions (Section \ref{Fermiliquid}). These may then exhibit thermal transitions where either weak ferro- (Stoner-like) or antiferromagnetic order of the "nesting" spin density wave kind may appear. Asserting that the transition is mean field which is a good idea given the large BCS-style coherence lengths suppressing the fluctuations one can employ time dependent mean field ("RPA", "bubble sum") to obtain the relevant order parameter susceptibility describing the approach to the transition (see  Section \ref{Fermiliquid}),

\begin{equation}
\chi_{\Psi, \Psi}  (\omega, q, T) = \frac {\chi^0 ( \omega, q, T )} { 1 - J_q \chi^0 ( \omega, q, T )}
\label{RPAsusc}
\end{equation}

where $\chi^0 ( \omega, q, T )$ is the Lindhard function of the Fermi gas (next section) and $J_q$ the momentum dependent interaction strength.

The vanishing of the denominator than signals the onset of the order: $\mathrm{Re} \chi^0 (\omega=0, q, T ) = 1/ J_q$. When this condition is reached as function of decreasing temperature at $q=0$ ferromagnetism will set in, or either  an antiferromagnetic  spin density wave when it happens at a finite momentum $q$.  I will show how this works in more detail in the discussion of holographic superconductivity (Section \ref{HoloSC}): a peak start to develop in $\chi"$ in the normal state centred at an elevated energy having a comparable width. Upon lowering temperature this peak moves down and narrows to turn into a delta function at zero energy right at the quantum critical point. 

Hertz addressed this as a zero temperature transition using the thermal field theory language that was in the 1970's not yet widely disseminated in the condensed matter community. One can recast it in the form of a $\Psi^4$ type effective action, to add a Yukawa coupling to the Fermi gas of the form $\vec{\Psi} \cdot \psi^{\dagger} \vec {\sigma} \psi$ where $\vec{\sigma}$ are Pauli matrices and $\psi$ the fermion field operators. For a non-conserved order parameter the Landau damping of the order parameter by the electron-hole excitations yields after integrating out the fermions using second order perturbation theory a modification of the kinetic term ${\cal L} \sim (| \omega| + q^2 + m^2 ) | \Psi_(\omega, q) |^2$.  This implies a "diffusion" dynamical critical exponent $z=2$. Given that the effective dimensionality of Euclidean space-time is $d+z$ for $d > 2$ one is above the upper critical dimension (typically $d_{u.c.} = 4$ for simple magnetic transitions) and the self interaction $w$ can be ignored at the IR fixed point: these transitions are supposed to be of the non-interacting kind.

The difficulty is that "integrating out" requires that the fermions are fast as compared to the order parameter fluctuations. For any finite correlation length this is not quite true since particle-hole excitations occur at lower energies than the mass of the order parameter field. One has to address the "backreaction" of the critical order parameter fluctuations on the fermions. In higher dimensions it is argued that these are strongest in the pairing channel: the critical fluctuations act as an efficient pairing glue, driving a superconducting state \cite{LohneysenRMP}. The superconducting gap then protects the order parameter from potential danger coming from massless fermions. Specifically in two space dimensions it was found that there are yet other, more subtle IR divergences associated with the fermions and an effort developed addressing this affair with quite sophisticated means \cite{SungSik2D}. To the best of my understanding this is still not quite settled. 

A vast literature, both experimental- and theoretical, evolved \cite{LohneysenRMP}. Resting on the mean-field nature of Hertz-Millis many properties can be computed and it was argued that a subset of the Heavy-Fermion style quantum phase transitions appear to be consistent. However, also a group of "bad actors' were identified where this is not the case. This trouble is rooted in the fact that the physics of the metal is at the scales of interest not quite like a Fermi-gas. Surely in the case of the heavy fermion "QCP's" the interactions in the UV are very strong, encapsulated by Anderson lattice models \cite{LohneysenRMP}: "Hubbard electrons" hybridizing with weakly interacting band structure state. Because of the fermion signs (next section) these escape a controlled mathematical description. Likely the metallic states are of the densely entangled "quantum supreme" kind but upon cooling down at low temperatures the heavy Fermi-liquids {\em emerge} as "instabilities" of this uncharted metallic stuff, as well as the magnetic order.  

Recently a step in this direction was taken in the form of artificial fermion models characterized by a fine tuning making it possible to cancel the fermion signs (see e.g. \cite{Signfreenematic}). The effective actions descending from such strongly interacting fermions are yet very different from the Fermi-gas kind and these can only be addressed by quantum Monte Carlo. Far from settled, this work does indicate that even in sign free models such fermionic QPT's behave differently. But nature does not give in naturally to sign cancellations and the general nature of such metallic quantum phase transition is shrouded behind the sign problem brick wall. 

There is much more to be said regarding these metallic quantum critical points but perhaps surprisingly it is in the present context a side-line. We are heading towards what holography has to tell about quantum supreme matter. Contrary to some folklores, it has little to say about such quantum phase transitions. Within the established condensed matter paradigm such a transition is a necessary condition to avoid the SRE product states. But holography is insisting on the existence of "quantum critical phases", in essence densely entangled generalizations of the Fermi-liquid as I will highlight in Section (\ref{holoSM}). 

It became a community consensus in the cuprates that the strange metal physics around optimal doping also should be rooted in a QPT taking place right at optimal doping.  At the time that we wrote the Nature "consensus document" \cite{Naturecons15} we also took this for granted. Accordingly, in the phase diagram Fig. (\ref{fig:cupratephasedia})  we indicated a quantum critical wedge -- its fingerprint par excellence  in experimental data, obviously with doping level as zero temperature tuning parameter. But as in the other cases one should be able to identify the order parameter that is disappearing. The experimental machinery of condensed matter physics is just tailored to detect order but despite an intense effort it could not be detected. The "intertwined" stripe etcetera orders appear to have vanished already at doping levels that are significantly lower than where the QPT should be located at a critical doping $p_c \simeq 0.19$. The idea was born that some form of "hidden order" was in charge, a symmetry breaking that is difficult to detect with the experimental machinery. 

One such form of hidden order was proposed  that although still controversial deserves to be taken quite seriously: the spontaneous diamagnetic "loop currents" proposed by Varma (\cite{Varmaloops},  Section \ref{Intertwined}). This breaks time reversal invariance  but not translations and is hard to detect -- evidence was claimed for it in neutron scattering measurements but is not seen unambiguously in muon- and NMR measurements that should be exquisitely sensitive for such order.  But in addition it has attractive ramifications such as the absence of a specific heat anomaly at the thermal transition, an induced pairing interaction favouring d-wave superconductivity and even a controversial claim that it could drive $z \rightarrow \infty$ local quantum criticality.   

But this view changed drastically by the very recent experimental evidences for a highly anomalous {\em first order} like zero temperature phase transition residing at $p_c$. Similarly, in the last few years evidences have been accumulating that the {\em overdoped} metal realized at $p > p_c$ is a quite strange metallic phase as well. As I will discuss in Section (\ref{qucritphase})  these are all evidences supporting the notion that both the under- and overdoped metals are in fact "quantum critical phases" that may well be governed on by general physics principle as suggested by holography.

\section{ Quantum supremacy and the (fermion) sign problem.}
\label{fermionsigns}

At this instance we have arrived at a frontier of human knowledge. Our understanding of matter eventually hinges on the availability of mathematical machinery that is a prerequisite for this understanding. We learned that the material in the textbooks of physics hinges on the existence of the SRE product vacuum states. In the previous section we learned that by resting on the machinery of statistical physics we can extend this to the strongly interacting quantum critical point, but this requires that the problem is stoquastic. 

But stoquastic systems are in fact very rare in nature -- only real boson systems like $^4$He satisfy the conditions without fine tunings of various kinds.  {\em Generic finite density quantum systems are characterized by Euclidean path integrals that are not of the statistical physics kind:} nature is intrinsically non-stoquastic. 

The bottom line is that no mathematical machinery is available to tackle such problems. It is of course not pleasant for theoretical physicists to admit that their hands are empty, and accordingly in the mathematically inclined part of the community this was just brutally worked under the rug. In such a sociological circumstance it is typically beneficial  to refer to quotes in a remote past by the prophet  Feynman. In fact, in the illustrious Feynman-Hibbs path integral book \cite{FeynmanHibbs}  you will find reference to the failure as perceived by Feynman of the path integral to shed light on fermion problems. 

Until rather recently, the (fermion) sign problem was often put away as a software engineering problem faced by computational people, not of interest to serious theoretical physicists. It is still a prevalent attitude in the string theory community, annoyingly so because it is quite obvious that the charm of AdS/CFT is in that it makes possible to have a look on what is going on behind the "fermion sign brick wall."  The best therapy is to interact with the brave part of the computational community that has been resiliently chasing the sign problem  for the last thirty years or so. Computers keep us honest for the simple reason that these do not offer the opportunity to look away from the trouble. 

This first half of this section is intended to offer a pedestrian introduction in the sign problem, to get some intuition of where the trouble is. All we can do is to diagnose the problem, as an introduction to the potential cure offered by holography.  I will first review a story that is for no good reason not found in the  quantum field theory textbooks: the way that fermion- and boson statistics is encoded in the first quantized (worldline) path integral (section \ref{worldlinePI}). This yields a basic insight in why fermion problems are so much harder than their bosonic, stoquastic counterparts.  

Section  \ref{TroyerWiese} is the crucial passage:  I will shortly review a  mathematical theorem stating that non-stoquastic problems are in general of exponential complexity: "NP-hard".  I will then focus on the typical circumstances that are at work in the strongly correlated electron systems:  the "Mottness" condition, associated with the Mott insulating state that is turned into a metal by doping and/or the coexistence with weakly correlated electrons states  (Section \ref{Mottness}). I will highlight here yet another unfamiliar story that allows us  to  diagnose the nature of the horrid sign problem as encountered in this specific context that is invariably at work in the condensed matter mystery systems. For completeness I will finish this discussion with a very concise summary of the recent progress in the computational community.   

The second half of this Section is devoted to a discussion of the only state of fermionic matter that is thoroughly understood: the Fermi-liquid and its BCS style descendants (Section  \ref{Fermiliquid}). A thorough understanding of this physics will be a prerequisite to appreciate the nature of the holographic strange metals that can be viewed as a generalizations of the Fermi-liquid portfolio. I will spent quite some time on this subject, suspecting this o be material that may be less familiar to part of the readership. Although the expert reader will not find any news in these passages, the conceptual framework I have in the offering is quite different from the traditional one.  

 It is yet again "quantum information modernity". I will first explain in which sense the Fermi-liquid is a "classical state of matter". Its vacuum is like an SRE product state, but there is a twist compared to the stoquastic agenda: the classical "anchor" is no longer a simple "bit string" but instead it is "infused" with fermion statistics. For the (interacting)  Fermi-liquid it is just the Fermi gas, but what really matters is that it is "classical" in the sense of being devoid of many body entangled, the rigorous proof will be outlined in Section (\ref{coldatomFL}). As stoquastic SRE product vacua go hand in hand with spontaneous symmetry breaking and order parameters, so does the Fermi-liquid. But the associated "order parameter physics" is of a quite different kind --  it is very "quantum-like" in the classic sense of "quantum fluids": the Fermi-surface takes this role. Instead of the Goldstone bosons, one finds besides modes associated with the fluctuations of the Fermi surface as a whole (the "zero sounds") also "fermionic Goldstone modes": the quasiparticle-anti quasiparticle excitations spanning up the Lindhard continuum (Section  \ref{RGFermiliquidzeroT}). The thermodynamic rules are also quite different. This Fermi liquid "order" only exists at strictly temperature regardless the dimensionality of space. At any finite temperature the Fermi-liquid is behaving as a classical dissipative kinetic gas (Section \ref{FermiliquidfinT}).
 
 The take home message is that the {\em weakly} interacting Fermi liquid obeys {\em scaling laws} that are in some regards  remarkably similar to those governing the strongly interacting quantum critical state, as for instance the energy temperature scaling. But these are yet quite different -- in a way the scaling structure behind the Bardeen-Cooper-Schrieffer superconducting instability is a vivid example showing the similarities and the differences (Section \ref{BCSbasics}). But these simple scaling laws are quite fragile -- when the interactions get stronger these are compromised by perturbative corrections (Section\ref{FLdiagrams}), as the effect of "irrelevant operators" comparable to the non-universal "free" quantum critical state above the upper critical dimension. 
 
 When we arrive at the holographic strange metals we will find that the holographic duality reveals a particular form of "covariant" renormalization group flow that is different from the way that the RG works at stoquastic quantum critical points. As it turns out, the Fermi-liquid is governed by this kind of RG while it takes the role of template for the quantum supreme generalization in a similar way as the Landau mean field critical theory "organizes" the thermal critical state.

\subsection{Exercising the fermion signs: the path integral of the Bose and Fermi gas.}
\label{worldlinePI}

In the canonical formalism fermion statistics enters through the Pauli principle.  To acquire some intuition regarding the way that it translates into the sign problem, it is entertaining to consider the non-interacting Fermi- and Bose gas  in the first quantized (worldline) path integral formalism. This should be part of the basic training of any physicist but it seems that for accidental historical reasons it did not enter the textbooks. An exception is found in the book "Path Integrals" by Kleinert \cite{KleinertPI} and let me outline the way it works.  

The worldline representation is a pleasant arena to familiarize oneself  further with the Euclidean path integral that we already encountered in the previous section.  Let us first exercise this with the elementary problem of a single free quantum mechanical particle in first quantized representation. Depart from the "transition amplitude", the probability that a particle with mass $M$ in $D$ space dimensions  that is created at $x_a, t_a$ will arrive at the space-time coordinate $x_b, t_b$. As explained in elementary texts,

\begin{equation}
\langle x_b t_b | x_a t_a \rangle = \frac{1}{\sqrt{ 2 \pi i \hbar (t_b - t_a) / M}^{D} }e^{ \frac{i}{\hbar} \frac{M}{2} \frac{ (x^a - x_b)^2} {t_a - t_b}}    
\label{Lorentzanpart}
\end{equation}

Upon Wick rotation to imaginary time $\tau$,

\begin{equation}
\langle x_b \tau_b | x_a \tau_a \rangle = \frac{1}{\sqrt{ 2 \pi \hbar (\tau_b - \tau_a) / M}^{D}} e^{ - \frac{M}{2 \hbar} \frac{ (x^a - x_b)^2} {\tau_a - \tau_b}}   
\label{Euclideanpart}
\end{equation}

Here one already infers the magic: this is identical to the two point function of a "polymer", an elastic line, subjected to Gaussian thermal fluctuations after associating the line tension with $M$ and temperature with $\hbar$. At finite physical temperature the imaginary time direction turns into the circle and all that can happen in the vacuum state is that the "polymer" forms a closed loop "lassoing" the time circle. The quantum partition sum is then determined by  the effective classical partition sum of this  "ring polymer". An elementary computation yields, 

\begin{equation}    
Z_{\mathrm{part}} = \frac {V_D}{\sqrt{ (2 \pi \hbar \tau_{\hbar})/M}^{D}}
\label{partsumpart}
\end{equation}

which you may recognize as the partition sum of a single quantum mechanical particle in the spatial continuum.  

Let us now consider an ensemble of such particles. We have  to add the postulate that particles are either fermions of bosons characterized by (anti-)symmetrized Fock space. Depart from the first quantized configuration space where $\bR (\tau)$ specifies the coordinates of all particles at (imaginary) time $\tau$, $\bR = (\br_1,\ldots, \br_N) \in \mathbb{R}^{Nd} $. The partition sum can be expressed as the integral over this configuration of the diagonal density matrix evaluated at $\tau_{\hbar}$, the wisdom we saw already at work in the previous paragraph, 

\begin{equation}
\label{partition-function-1}
\mathcal{Z} =  \textrm{Tr} e^{- \beta H} = \int \ud\bR \rho (\bR,\bR;\beta).  
\end{equation}

But dealing with indistinguishable fermions or bosons one has to (anti) symmetrize the wave function and in this worldline path integral representation this translates into summing over all $N!$ {\em permutations} ${\cal P}$ of the particle coordinates,
\begin{equation}
\rho_{B/F} (\bR,\bR;\beta ) = \frac{1}{N!} \sum_{\cal P} (\pm 1)^p \rho_D (\bR, {\cal P}\bR; \beta ), 
\label{pathstat} 
\end{equation}

where

\begin{subequations}
\begin{eqnarray}
\rho_D (\bR,\bR';\beta) & = &  \int_{\bR\to\bR'} \mathcal{D}\bR  \exp(-\mathcal{S}[\bR]/\hbar),\\
\mathcal{S}[\bR] & = & \int_0^{\hbar \beta} d \tau \left( \frac{M}{2} \dot{\bR}^2 (\tau) + V (\bR(\tau) ) \right),\quad  
\end{eqnarray}
\label{pathdist} 
\end{subequations}

$ V (\bR(\tau) )$ is the potential energy which can be due to either interactions of external potentials that we will ignore for the time being. 

\begin{figure}[t]
\includegraphics[width=0.7\columnwidth]{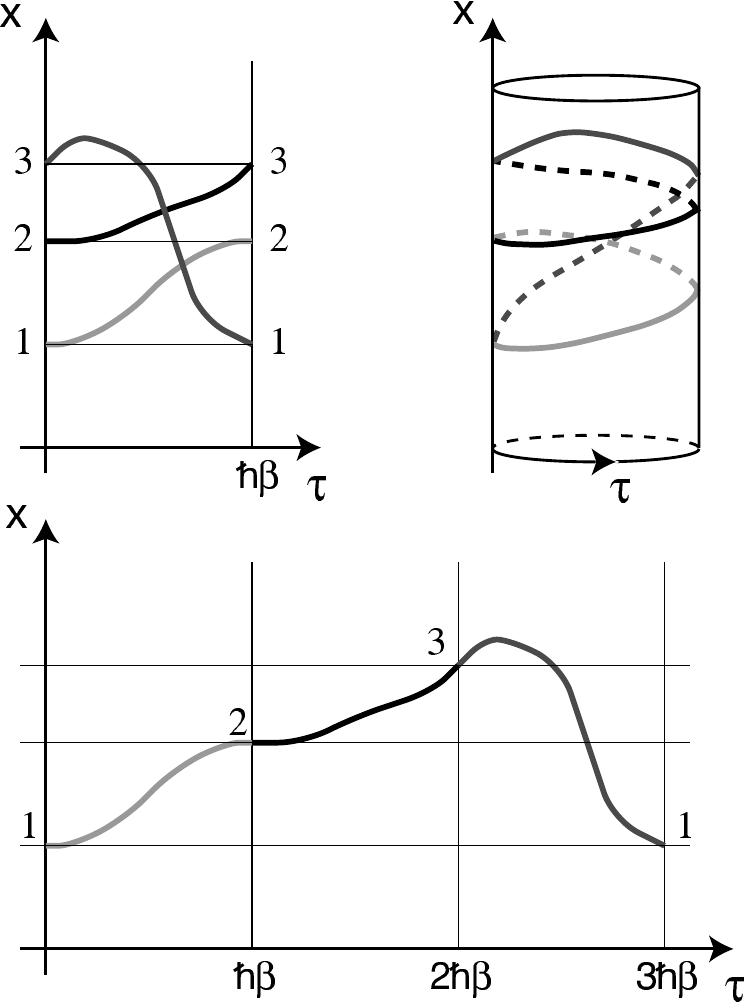}
\caption{In Euclidean signature worldlines are wrapping around the imaginary time circle \cite{devioussigns} in this example, a single worldline wraps three times along the Euclidean time circle ($w = 3$) representing a permutation of three particles like $123 \rightarrow 312$ in the (anti)symmetrized wave function. In the full partiton sum one has to account for all possible winding configurations, for three particles this includes $123  \rightarrow 123 \; (w=1)$, $123 \rightarrow 213 \; (w=2)$, etcetera.}
\label{fig:winding}
\end{figure}

Remarkably, the quantum statistical requirement turns into a topological affair. The way this works is as follows. Depart from a bunch of worldlines at imaginary time $\tau_0$. One follows the "word history" by evolving along the imaginary time direction until one has traversed the time circle arriving again at $\tau_0$: at this instance one has to connect the worldlines again but the (anti) symmetrization requirement insists that this can be accomplished a-priori in any way.  Dealing with $N$ particles this can be accomplished by having $N$ single particle "loops". But one can also wrap one worldline $N$ times around the time circle, see Fig. (\ref{fig:winding}). For the latter case, on a single time slice one discerns $N$ particles, worldlines piercing through through, but these share all {\em one worldline}. 

Let's inspect how this works as function of temperature. The worldlines will meander along the imaginary time direction like polymers with fixed string tension $\sim M/\hbar$.  When temperature is high the time circle is short and upon traversing the time circle the net average distance that is traversed will be small with the effect that every particle line connects with itself. This is the path integral way to understand that above the degeneracy  temperature the system becomes a classical gas. However, upon increasing the time circle upon lowering temperature this meandering distance becomes of order of the interparticle distance: the de Broglie thermal wavelength becomes of order of the interparticle distance and degeneracy sets in. 

Consider the probability in the case of two particles that a single worldline wraps twice around the time circle: this may become of the same order as two individual "lasso's". But now the difference between fermions and bosons becomes also consequential through the $(\pm 1)^p$ factor in Eq. (\ref{pathstat}). For bosons both the single- and two wordlines contribute with a probability factor $> 0$ to the quantum partition sum. However, for the fermions the two single loops have a positive weight but the one wordline wrapping twice around the time circle carries a "negative probability": you are facing the sign problem!

The next step is a bit of an intricate combinatorics exercise \cite{KleinertPI,devioussigns} that only works for the free problem. The outcome is that one can write the partition sum entirely in terms of  a sum over the {\em winding numbers} $w$, see Fig. (\ref{fig:winding}).The grand canonical partition sum becomes, 

\begin{eqnarray}
Z_{G}(\beta,\mu) &  = &  \sum_{N=0}^{\infty}Z_{B/F}^{(N)}(\beta)e^{\beta\mu N}\nonumber\\
& = & \exp\left( \sum_{w=1}^\infty (\pm 1)^{w-1}\frac{Z_0(w\beta)}{w}e^{w\beta\mu}\right),                   
\label{ZG}
\end{eqnarray}

corresponding to a grand-canonical free energy
 
\begin{eqnarray}
F_G (\beta) & = & - \frac{1}{\beta}\ln Z_{G}(\beta,\mu)\nonumber\\
 & & = - \frac{1}{\beta} \sum_{w=1}^{\infty} (\pm 1)^{w-1} \frac{ Z_0 (w \beta)} {w} e^{\beta w\mu},
\label{signfreeen}
\end{eqnarray}
with the $\pm$ inside the sum referring to bosons ($+$) and fermions ($-$), respectively. This is quite elegant: 
one just sums over worldlines that wind $w$ times around the time axis; the cycle combinatorics
just adds a factor $1/w$ while $Z_0 (w \beta) \exp{ ( \beta w \mu ) }$ refers to the return probability of a single worldline of overall length $w \beta$. This  further simplifies in the spatial continuum to,  

\begin{eqnarray}
Z_0 (w \beta) & = &  \frac{V^d}{\sqrt{ 2\pi \hbar^2 w \beta / M}^d} \nonumber \\
                         & = & Z_0 (\beta) \frac{1}{w^{d/2}},
\label{zodef}
\end{eqnarray}

Yielding the free energy and average particle number $N_G$, respectively,

\begin{subequations}
\begin{eqnarray}
F_G & = & - \frac{Z_0(\beta)}{\beta} \sum_{w=1}^{\infty} (\pm 1)^{w-1} \frac{e^{\beta w\mu}}{w^{d/2+1}}, \\
N_G & = & -\frac{\partial F_G}{\partial \mu}=Z_0 (\beta) \sum_{w=1}^{\infty} (\pm 1)^{w-1} \frac{e^{\beta w\mu}}{w^{d/2}}.
\end{eqnarray}
\label{freefreeen}
\end{subequations}

Finally, the sums over windings can be written as, 

\begin{equation}
\sum_{w=1}^{\infty} (\pm 1)^{w-1} \frac{e^{\beta w\mu}}{w^{\nu}} = \frac{1}{\Gamma(\nu)} \int_0^{\infty} d \varepsilon 
\frac{\varepsilon^{\nu-1}} {e^{\beta (\varepsilon - \mu)} \mp 1},
\label{fdwinding}
\end{equation}

and we recognize the textbook expression   involving an integral of the density of states ($N(\varepsilon) \sim  \varepsilon^{d/2}$ in $d$ space dimensions)  weighted by Bose-Einstein or Fermi-Dirac factors!

I presume the reader has captured the physics entertainment that is going on in the above. The indistinguishability of the quantum particles translates into the "wind around the time axis as often as is allowed" -- this is a topological affair, associated with "lassoing the cylinder" (homotopy group $\Pi_1 (S_1) = Z$). By subtle combinatorics this then eventually translates in the texbook Bose-Einstein and Fermi-Dirac distributions familiar from the canonical formalism. The big deal is that one immediately infers that for the bosons  the sum over the winding numbers is also a stochastic affair -- every winding configuration is characterized by a Boltzmann weight probability, and this sums up to the Bose-Einstein distribution. The problem can actually as wel be interpreted in terms of the classical problem of "ring polymers" winding around a cylinder in space. 

But how  about the fermions? In the "winding sum"  Eq. (\ref{fdwinding}) one observes the $(-1)^{w-1}$ factor: uneven windings contribute with a positive "Boltzmann weight", but even windings are like "negative probabilities". The overall contributions of adjacent $w$'s are nearly the same when $w$ gets large and these nearly cancel each other. The bottomline is that the free energy is pushed upward, and these "destructive interferences" are the origin of the Fermi-energy -- the fermion system at finite density is characterized by an enormous zero point motion energy.  

Dealing with free fermions this alternating sum just turns into the Fermi-Dirac distribution but when interactions get important this sum becomes ill defined. This is the first quantized way to appreciate the origin of the sign problem. 

To appreciate this a bit more, let's focus in on a context where the first quantized representation is the natural one:  the Helium quantum liquids.  At $\simeq 3$ K both the fermionic $^3$He and the bosonic $^4He$ form a dense {\em classical} van der Waals liquid. The description of such fluids has been a central challenge in the classical fluid community. One can picture it as a form of dense traffic. The He  atoms are in first instance like hard, impenetrable balls that occur at a high density so that they all touch each other, and this cohesive fluid is kept together by the weak, long range van der Waals attractions. The motions are like stop and go traffic, it is extremely collective. For your car to move, a mile away or so a car moved and a string of cars followed to eventually make it possible for you to get your foot from the brake pedal. It is the same affair for the Helium balls and the classical van der Waals fluid was tackled by brute force numerics, based on solving Newton's equations of motion with added noise terms -- "molecular dynamics". 

But one now lowers temperature further so that quantum degeneracy becomes on the foreground. Already in the 1950's or so the liquid form factor was measured by neutron scattering showing that down to the lowest temperatures the local physics does not change at all: microscopically it is for both isotopes the same van der Waals traffic jam, eventually kept fluid by quantum zero point motion. 

This sounds like an insurmountable problem: the UV "coupling constant" is in a way close to infinite given the hard steric repulsions. But nevertheless the bosonic $^4$He problem can be regarded as completely charted theoretically at least for equilibrium properties. The reason is that the worldline quantum Monte-Carlo just gets it done as demonstrated by David Ceperley in the 1990's as one of the great triumphs of QMC \cite{Ceperley4He}. 

How does this work? Let  us first revisit the free bosons: we know from the canonical formalism that at some temperature Bose-Einstein condensation will take over, where a finite density of Bosons will occupy the $k=0$ single particle momentum states. The way this works in the worldline representation is that by lowering temperature at average the number of windings will increase. At some temperature worldlines with $w \rightarrow \infty$ will acquire a finite probability and this corresponds with the Bose-Einstein condensation temperature. In fact, at $T=0$ the probability to find a particular winding $P(w)$ becomes completely $w$ independent. The value of $P(w)$ at $w \rightarrow \infty$ is associated with the condensate density $\rho_{SF} (T=0)$ and the flat distribution means that all particles are in the condensate, $\rho_{SF} = 1$. This is in fact an example of weak-strong duality  that I touched upon in Section (\ref{SREproducts}). The winding of the worldlines in the path integral is actually encoding for the many-body entanglement and a flat winding distribution implies a form of maximal entanglement in position (number) representation. But in momentum (phase) representation it is just a simple tensor product. 

Let us now consider the van der Waals fluid version. Because of the hindrance coming from the steric repulsions, the contribution from long windings are suppressed relative to the short windings. But given that eventually an infinitely long worldline can get through there will be a temperature where as for the free system $P$ becomes finite at infinite $w$: this is the transition to "Helium II", the superfluid! The bottom line is that at zero temperature $P(w)$ is no longer $w$ independent as in the free Bose gas: it is skewed towards small $w$ but above a critical $w_{crit}$ it will again become $w$ independent although at a lower value than in the free system. This expresses that the zero temperature condensate fraction is reduced significantly compared to the Bose-Einstein case, one of the well known properties of Helium. Surely, every equilibrium property can be computed, being right on top of experiment as Ceperly showed \cite{Ceperley4He}.

But let us now turn to $^3$He, the fermionic version -- a classic example of the difficulties associated with the "non-stoquastic" quantum physics. Departing from the van der Waals liquid conditions in the UV, it is obvious that the winding summation with the alternating signs turning into the Fermi-Dirac distribution is no longer working. What happens instead? The answer is an empiricism that added to the fame of Landau. He realized that at temperatures well below 1K all experimental properties are consistent with the emergence of a Fermi-gas formed from renormalized quasiparticles. In turn he formulated the Fermi-liquid theory to describe these renormalizations, see Section (\ref{Fermiliquid}). 

One has to realize what is going on here. One departs also in the case of $^3$He from the UV "quantized traffic jam physics" of the dense van der Waals liquid  as proven by the liquid form factor measurements. Upon renormalization towards the IR a miracle happens: the impenetrable hard balls of the UV turn into perfectly non-interacting quasi-$^3$He particles that have just increased their mass by a factor of 10 or so (Fermi-liquid phenomenology) communicating only by the Pauli-principle! It is one of the dirty secrets of physics that this painful problem got worked under the rug completely: nobody has even the faintest clue how this can happen! 

To further demotivate wishful theoretical dreaming, as for $^4$He one can unleash all the supercomputing power available to attempt  to compute this brute force. But one then finds that when temperature gets well below the bare Fermi-energy the required computational resources are growing exponentially -- this exponential complexity rooted in quantum supremacy appears to be fundamental: a no-go theorem has been claimed, see next section. 

The take home message is that experiments show that the Fermi-liquid fixed point is amazingly stable: the $^3$He  example was later followed by a zoo of "heavy Fermi liquids" in electron systems characterized by up to a 1000 fold quasiparticle mass enhancements \cite{LohneysenRMP}. But the explanation of this excessive stability is completely in the dark: it is shrouded by the sign problem! This is the problem I alluded to discussing the metallic quantum phase transitions in Section (\ref{hertzmillis}). Eventually very heavy Fermi-liquids are realized at very low temperatures surrounding the quantum critical point in the heavy fermion systems. But the UV is similar as in Helium, with interactions that are much larger than the bare Fermi-energy. By miracle somehow this manages to turn at very low temperatures in a Fermi-liquid but actually nobody can tell why this seemingly unreasonable feat is accomplished. 
  
\subsection{The Troyer-Wiese no-go theorem.}
\label{TroyerWiese}

Computational complexity theory is a corner stone of mathematical computer science. It departs from the universal classical computer -- the Turing machine -- and asks the question how the amount of computer time scales with the size of the problem. Remarkably, it is then possible to identify various complexity classes and decide at least in principle in which class a particular algorithmic problem belongs. Consider a problem characterized by $N$ bits of information; this is considered as benign when it can be solved in a time that scales polynomially with $N$ (computer time  $t \sim N^{\alpha}$) defining the class "{\bf P}". However, there is also the class of "non-polynomial" ("{\bf NP}") problems that are typically associated with an exponential growth of the computational effort, the "exponential complexity" $t \sim e^N$. Obviously, for large $N$ this amounts to a no-go theorem.  As an added subtlety, one can identify a subclass of {\bf NP} problems characterized by the property that when an algorithm would be found that can solve it in polynomial time this can be used to solve {\em all} {\bf NP} problems: the {\bf NP}-hard class. This is obviously the holy grail  of computation:  when such an algorithm would be found one could dream that e.g. chaotic problems like weather forecasting could be extended over much larger periods. 

Although often not emphasized, this also encapsulates the glass ceiling of the "unreasonable effectiveness of mathematics in natural science". Scribbling mathematical equations and solutions on a piece of paper is also a form of computation. When this works one is obviously dealing with problems that viewed from the computational complexity angle represents the {\em simplest} problems altogether. In the daily practice, the effective/phenomenological  theories of physics are typically of the kind but in order to get the numbers right one has to mobilize a "polynomial effort" of a computer. To say it in a more provocative way, physics has been a cherry picking affair, filtering out those problems that are of polynomial complexity  that can be cracked by mathematics. The vast stretches of reality that are of exponential complexity are left to the other sciences which are inherently entirely empirical with theories that are to physicists standards no more than heuristic hand waives.  This is eloquently captured by the classic wisdom among young male physicists that "girls are way more complicated than quarks".      

Troyer and Wiese presented in 2004 the remarkable claim \cite{Troyerwiese} that the {\em generic equilibrium problem of strongly interacting fermions at finite density is {\bf NP}-hard.} This departs from the now familiar territory that a-priori the Hilbert space of the quantum many body system is exponentially large, while you already learned that the alternating signs that cannot be avoided dealing with a path integral of a strongly interacting non-stoquastic problem are causing headaches. But are there loopholes? The take home message of Troyer and Wiese amounts to a no-go theorem.  

Let me present here a very short sketch of the strategy behind the derivation of this theorem -- see the paper to get the fine touches \cite{Troyerwiese}. It departs from a {\em classical} problem for which it is well established to be  {\bf NP}-hard: the Ising spin glass problem.  

\begin{equation} 
H = \sum_{\langle j, k \rangle} J_{jk} \sigma^z_j \sigma^z_k
\label{TWIsingglass}
\end{equation}

where $\sigma^z_j$ are the usual Pauli matrices, while  the nearest-neighbour couplings $J_{jk}$ take values randomly from the interval bounded by  $+J$ and $-J$, crucially the couplings have to take both positive and negative values for it to be  {\bf NP}-hard. This is the well known affair with its exponentially large number of local minima that can be related to the travelling sales man problem and so forth. 

By a very simple, familiar operation this can be shown to be equivalent to a generic quantum physical problem that is claimed to be representative for a typical non-stoquastic problem.  Just re-identify $\sigma^z \rightarrow \sigma^x$ and in $x$ quantization one finds that the Hamiltonian in this representation becomes an "all to all" random matrix like affair where the off diagonal entries are characterized by random signs -- a generic sign-full problem.  

\subsection{Mottness as the sign problem amplifier.}
\label{Mottness}

In the spatial continuum electrons exclusively interact via the Coulomb interactions. Its long range nature is yet another, independent factor that adds to the stability of the Fermi-liquid. This is in essence not different from the wisdom in stoquastic systems that long range interactions are less "dangerous" since these are quite well captured on the mean field level. Although there is theoretical bluff involved, there are good indications that this "jellium" Fermi-liquid has quite some extra stability given the long range  Coulomb interactions in the UV.

But it was realized perhaps as early as in the 1930's that this logic is severely impeded by the presence of a strong external potential coming from the ionic lattice. In chemistry this is known as the "Heitler-London" description of e.g. the $H_2$ molecule. One departs from two neutral hydrogen atoms a relatively large distance apart. The $1s$ atomic states do however overlap and these tunnel with a rate "$t$". When an electron tunnels one obtains a $H^-$ atom and a proton. But one can now just measure how much energy this takes and one discovers that in $H^-$ the two electrons are close together paying a Coulomb energy $U$ which is actually very large, typically of order $10$ eV.   

Different from the single particle  description, the fluctuations that  exchange the electrons are now hampered by the large interaction energy $\sim U$.  The ground state becomes that of two protons both with an electron tied to it: all what remains are localized spins on each neutral hydrogen atom. But virtual exchanges are still possible but only when these spins are antiparallel because of the Pauli blocking. This turns into an  antiferromagnet spin-spin "superexchange".  The ground state of the $H_2$ molecule  is a two spin spin-singlet stabilized by an exchange interaction $J = - t^2/U$ as follows from second order perturbation theory controlled by the small "strong coupling" parameter $t/U$.
 
Consider now a lattice of such atoms -- this is the Hubbard model;

\begin{equation}
H = \sum_{\langle i j \rangle \sigma } t c^{\dagger}_{i \sigma} c_{j \sigma} + U \sum_i n_{i \uparrow} n_{i\downarrow}
\label{Hubbardmod}
\end{equation}

This describes a lattice of tight binding electrons with a hopping $t$ with a "H$_2$" repulsive interaction: when two electrons are on the same site they have to pay a Coulomb penalty $U$. Despite its simple appearance more papers appear to have been published dealing with it than even for the standard model of high energy physics!

There is now a general consensus that this is the guinea pig of the sign problem. This model became the focus of attention by the discovery of superconductivity in copper oxides in the late 1980's. It became clear soon after the discovery that these "cuprates"  are {\em doped Mott insulators}, see Fig. (\ref{fig:cupratephasedia}). The insulating parent materials are well established to be Mott insulators, that  are subsequently doped in the same guise as doping semiconductors. 

 Back to the Hubbard model. At a density of one electron per site ("half-filling") and large $U$ it describes an insulator.  Notice that this "Mott-insulator" is nothing else than a traffic jam occurring in the electron traffic. It is potential energy dominated and no funky quantum mechanical wave function effects are required as in band insulators which are rooted in quantum mechanical interference.  
 
 Although worked under the rug for a long time by the (single particle) band structure community, this "large U" physics is actually quite ubiquitous in the zoo of solids. In essence, anything that is insulating having a nice colour is of this kind: the Mott insulators formed in transition metal and rare earth 'salts' (like FeCl$_2$, Mn$S$, $NiO$, La$_2$CuO$_4$, Ce$_2$O$_3$ ...). It turns out that for 3d- or 4f electrons the "effective U's" after incorporating the additional screening processes in solids are typically still larger than the bandwidth. The Hubbard model is just the  minimal model capturing the essence of this "big number" physics. 

In the Mott insulator the spins of the electrons are "left behind" at low energy and these are subjected to the antiferromagnetic superexchange interactions. Accordingly, an antiferromagnetic insulator is expected to form and this was indeed confirmed  in no time in the parent cuprates.   These Mott-insulators are in fact the only aspect of the cuprates that is thoroughly understood. Hell breaks loose for the theorists when these are doped.  After countless attempts to get a mathematical  handle on the doped systems, it gradually turned into the exercise ground for the "big machine" computational efforts aimed at hacking the sign problem. 

Before turning to the specifics of these efforts, so much is clear that the computers tell us that the sign problem in doped Mott insulators is particularly severe. It is not widely realized that the origin of this trouble can be {\em diagnosed} although this does not help at all to find a cure. This diagnosis is an achievement of the Tsinghua theorist Zheng-Yu Weng who already started to realize the quantum statistical troubles in the mid 1990's with his "phase strings" -- I found it myself  remarkable to the degree that I gave it a name: "Weng statistics"\cite{Motnesscoll} .

This becomes transparent using once again the first quantized path integral language with its winding number encoding of the quantum statistics. The root of the trouble is that in the Mott-insulator the low energy physics involves exclusively  (Heisenberg) spins. Spins are distinguishable objects, having  a Fock space spanned by tensor product states that are {\em not} anti-symmetrized. The reason is that spins are localized fermions which do not exchange with each other -- the virtual exchanges turn into the effective spin-spin interactions. However, when an electron is removed by doping the hole can move around and for this reason the electrons in the immediate vicinity start to remember that they are indistinguishable fermions that need anti-symmetrized wavefunctions. One is dealing with "part time" fermions that surely do not submit to the statistical winding rules of non-interacting fermions!

One can actually find out what takes the place of the free windings. This is quite straightforward employing the high (physical) temperature expansion in the worldline language \cite{KaiwuhighT}. One departs from the $t-J$ model, the effective theory at energies $<< U$, describing holes hopping (with $t$) through the Heisenberg spins system (with super-exchange J) avoiding double occupancy in the process.  For counting purposes one appoints the up spins as reference vacuum state, while the world histories of the down spins ("spinons") and holes ("holons") are  tracked keeping the sign associated with a particular overall worldhistory as a separate entry in the accounting. One then derives that the quantum partition sum  has the structure,      

\begin{equation}
Z_{t-J}=\sum_{c}{\tau }_{c}\mathcal{Z}[c]  
\label{PhasestringZtj}
\end{equation}
  
Here $\mathcal{Z}[c] $ is the absolute value ("positive probability) of the contribution of a particular world history $c$ formed from spinon- and holon worldlines.  But the {\em sign} of this contribution is given by 

\begin{equation}
{\tau }_{c}\equiv (-1)^{N_{h}^{\downarrow }[c]+N_{h}^{h}[c]} 
\label{Phasestringsign}
\end{equation}

$N_{h}^{h}[c]$ is  business as usual: this counts the number of mutual windings (exchanges) of the holon wordlines, these behave as normal fermions relative to each other. But the surprise is in the quantity $N_{h}^{\downarrow }[c]$. When a hole hops to a site where a down spin ("spinon") is present the corresponding electron hops in the opposite direction, a "holon-spinon collision". For a given world-history  $N_{h}^{\downarrow }[c]$ counts the total number of such collisions.  One adds them up, and the parity of this overall integer than determines whether the worldhistory contributes with a "positive" or "negative" probability to the partition sum. 

One discern immediately the difference with the sign structure of the free fermion problem. There we found that the winding distribution is statistically hardwired, making it possible to sum the alternating series into a Fermi-Dirac distribution. But in the doped Mott-insulator it becomes a {\em dynamical} affair. The negative probabilities are like destructive interferences pushing up the ground state energy -- the origin of $E_F$ as the zero point motion energy of the Fermi-gas. But in the doped  Mott-insulator one can organize worldhistories in such a way that the negative signs are avoided as much as possible. This  may however in turn have the effect that the "sign-free"  $\mathcal{Z}[c]$ is reduced, having also the effect of pushing up the energy. There is just no way that one can keep track of this balance having only polynomial complexity (Metropolis) means.

This "minimization of negative signs" does to a degree depend on the way that the spin system is organized. It is easy to check that in an ordered, conventional antiferromagnet that can be stabilized by adding an external staggered magnetic field departing from a bipartite lattice {\em all}  $N_{h}^{\downarrow }[c]$'s are even. The sign is then determined by  $N_{h}^{h}[c]$. The outcome is that small "Fermi-pockets" are formed  \cite{Motnesscoll} with a Fermi surface spanning up a Luttinger volume proportional to $p$, the hole density. Departing from $U =0$ one would find instead a volume $\sim 1 + p$ since one is dealing with a nearly half filled band. This is precisely coincident with what is found using straightforward Hartree-Fock mean field theory in the presence of antiferromagnetic order also for large $U$. This sign counting is completely general and therefore the hole pockets of conventional mean-field theory are no more than a special case. 

But there are different games one can play. For instance, one can contemplate that the spin system is in a resonating valence bond state ("RVB"), introduced early on by Phil Anderson merely on basis of his intuition. Thirty years down the line I am still waiting for either solid mathematical- or experimental evidences that it makes any sense. Whatsoever, the idea is to form singlet pairs of neighboring spins ("valence bonds") to then form a "maximal" coherent superposition of all possible tilings of the lattice by such valence bonds. The trouble is in the assertion that everything quantal in magnetism should be stitched together from the two spin "Bell pairs", the same kind of "two-ness" intuition that has been clouding the general understanding of many-body entanglement more recently in the quantum information community. 

The original RVB  idea was that these valence bonds stay intact when the system is doped. In the presence of holes these turn into charged electron pairs that can delocalize and Bose condense, explaining superconductivity-at-a-high temperature. Again, no shred of evidence ever appeared substantiating this claim. But Zheng identified an interesting loop hole: it turns out that departing from such an RVB in the form of an Ansatz the number of negative signs is strongly reduced lowering the ground state energy.  This suggests that RVB is perhaps not such a bad idea, although one better be aware that the sign problem may be the culprit. 

\subsection{ The big computational guns.}
\label{Compmethods}

Computers keep people honest -- for quite a long period in history the sign problem was worked under the rug by a large part of the physics community. But in the computational community there has been a consensus all along that the problem is foundational. 

In the stoquastic context quantum Monte Carlo has been victorious -- I alluded to $^4$He and the role it played in stoquastic quantum criticality. The ultimate benchmark is perhaps the staggering success, computing the difference of the proton and neutron mass by the "lattice QCD" community. However, in a concerted effort of admirable perseverance that has been going for some 50 years computational physicists have been chasing the sign problem with only very limited success.

In recent years there has been noticeable progress. A first venue is meat-and-potatoes QMC. The representation of choice is not the first quantized affair that I highlighted in the above. Instead, it turns out that the so called "determinant QMC" is much more efficient.  This departs from the second quantized path integral -- standard material in the advanced text books. One spans up the Hilbert space with generalized coherent states, mapping the second quantized canonical theory to the field-theory style path integral where the fermions are encoded by Grassmann (anticommuting) numbers. The fermion interactions are taken care of by the Hubbard-Stratonowich auxiliary fields such that the fermions can be integrated out. At the classical saddle point this turns into the standard Hartree-Fock (BCS etcetera) mean field theory, where the auxiliary fields are recognized as the order parameter. The effective action for these fields are however sign-ful through the fermion determinants. As the first quantized PI's, these are finite temperature methods defined on the Euclidean space-time cylinder. 

The algorithms were developed already in the mid 1980's but with the computational resources available back then one could not get at temperatures well beyond the (bare) Fermi temperature before the exponential complexity brought it to a hold. But this has been partially off-set by the exponential growth of these resources -- Moore's law. In the present era it is possible to get to much lower temperatures, that have started to overlap with the temperatures relevant for experiment, like 600 K or so dealing with cuprates \cite{DeverauxQMC}. Interestingly, the outcomes for the Hubbard model do not seem to connect that well to experiment: it is hanging in the air that plain-vanilla Hubbard models may fall short to explain the physics of the strange metals. 

The other progress has been in the arrival of algorithms that are inspired by quantum information, revolving around the many-body entanglement. These are the "tensor network" methods. This departs from the canonical formalism. For simplicity imagine a problem that is casted in terms of qubits, or equivalently $S=1/2$ spins. The Hilbert space is spanned by "bit strings", tensor products of these local DOF's and one can write any energy eigenstate state as
 
\begin{equation}
| \Psi \rangle_k = \sum_{i_1, i_2, \cdots i_N} C_ {i_1, i_2, \cdots i_N} | i_1 \rangle \otimes  | i_2 \rangle \otimes \cdots \otimes  | i_n \rangle
\label{wavefietensor}
\end{equation}

where every local bit $| i_k \rangle$ can take two values. The amplitudes $C_ {i_1, i_2, \cdots i_N}$ form a set of  $2^N$ numbers -- this is of course the quantum supremacy trouble.  But these are evidently also the coefficients of a tensor $C$ with $N$ indices $i_1, i_2, \cdots i_N$ where each of them can take 2 values. It is a rank $N$  index with $2^N$ coefficients.  

The idea of tensor networks departs from the mathematical fact that the rank $N$ tensor can be replaced by a "network" of smaller rank tensors. For instance, consider a chain of sites $i$. We can write in full generality $C$ in terms of local rank 2 tensors (matrices) $A$ as  $ C_ { \cdots i-1, i, i+1  \cdots }   =  \sum_{k = 1}^{2^N} \cdots  A^{k_{i-2} k_{i-1}}_{i-1}  A^{k_{i-1} k_{i}}_{i}  A^{k_{i} k_{i+1}}_{i+1} \cdots$. In a spatially homogeneous system the $A^k_i$ are the same, and all the data can be stored in this local tensor. However,  the extra $k$ label is referred to as the bond dimension: when one contracts the $k$'s over the full $2^N$ Hilbert space the local tensors $A$ contain all the information required to reconstruct the full tensor $C$. The rank of $A$ is determined by the connectivity of the Hamiltonian -- dealing with a e.g. a square lattice with nearest neighbour interactions a rank three tensor is required, etcetera.

The crucial ingredient is that the {\em bond dimension encodes the "range" of the entanglement}. Take only $k=1$ and you get a product state. When $k$ takes two values you wire in nearest pairwise entanglement and so forth. Hence, these can be viewed as a particular variational Ansatz that is constructed to regulate the entanglement. For small bond dimensions it describes SRE product states with only very local entanglement. By gradually increasing the bond dimensions an increasing degree of many body entanglement  is wired in and one tracks how the outcomes are evolving. Of course, there is just "conservation of misery" dealing with quantum supreme states -- to capture it in full the bond dimension has to be exponentially large. 

Tensor networks were born as early as 1993 when Steve White designed the matrix version as of relevance to one space dimension in combination with an efficient algorithm to optimize the Ansatz in the form of the  "Density Matrix Renormalization Group" (DMRG) \cite{WhiteDMRG}. The name is not quite appropriate -- it has nothing to with the renormalization group in the usual sense, and only later it became clear that it revolves around the degree of many body entanglement. Although limited to systems having eventually a 1D connectivity this turned into the intervening period in a success story (e.g., \cite{tJladders}).  In addition, a substantial portfolio of other tensor networks was constructed, designed to deal with specific circumstances that are harder to handle for all kind of computational reasons (e.g., "IPEPS" for two dimensions, "MERA" for scale invariant problems). 

QMC and the tensor networks do have the benefit that they reveal their range of applicability.  In addition, there are more ad-hoc methods that do invoke uncontrollable assumptions. A first category are implementations of Ceperley "fixed nodal surface", see Section (\ref{coldatomFL}). Another important category is resting on the  "dynamical mean field theory"  (DMFT) idea. Dealing with short ranged models (like Hubbard) one can show that in {\em infinite} dimensions the problem reduces to a single strongly interacting ("Kondo") impurity in an effective medium -- although quite un-trivial such impurity problems can be solved using e.g. Quantum Monte Carlo. There is literally nothing lying on the shelf telling that these solutions have anything to do with the physics in 2 or 3 dimensions. However, one can just implement a cluster of increasing size feeling the effective medium boundary conditions and see how it evolves: the "cluster" DMRG. 

What is the state of the art? This is encapsulated by a recent large scale effort (the "Simons collaboration") where the whole repertoire of different methods as discussed in the above was unleashed on the Hubbard model for large $U$ and a doping around $1/8$ \cite{Simonscol}. Although quite different systematic errors are involved, eventually all these models arrived at the same answer: the "spin stripes" of the kind that I discovered in 1987 using plain-vanilla mean field \cite{ZaGun}. It is a sideline in the big picture story of these lectures, and I  will discuss them in a bit more detail in Section (\ref{Intertwined}). 

Such "stripes" are insulating and SRE products -- this just reveals that there are strong ordering tendencies in the Hubbard model. But such stripy states are just a small part of the whole story. Presently it appears that the most controlled results are produced by DMRG for Hubbard type Hamiltonians defined on a ladder geometry: the legs are associated with one dimensions where DMRG works very well. One can then gradually increase the width (putting "rungs" between the legs) paying the prize that the computational complexity grows exponentially in the width. Even for a width $4$ it turns out that one has to invoke a very large bond dimension (like $\sim 10^4$) to get convergence. But yet again, it does not shed light on the strange metals \cite{tJladders}. One finds at long distances that Luttinger liquid universality takes over, an affair which is completely tied to one space dimension. Quantum statistics is meaningless in 1+1D and the ramification is that one can always identify a "sign free" representation. In essence, in one dimension everything scales to strong coupling and conventional algebraic long range order takes over: the secret of one dimensional physics as revealed in the 1970's.    

\subsection{The remarkable Fermi liquid.}
\label{Fermiliquid}

As I emphasized in the beginning of this section, the only fermionic state of matter that is thoroughly understood is the Fermi-liquid.  Those readers who are at home with this substance will not find anything new in this exposition. However, I will frame it in a different way compared the conventional canon. It is yet again exploiting the modern quantum information view, that is just making explicit the underlying principles that are kept implicit in the classic literature. Once again, the notion of "classicality" resting on the ESR product vacuum is generalized seamlessly to fermionic matter: the Fermi gas as "classical anchor" devoid of many body entanglement (Section \ref{coldatomFL}), that actually implies the fermionic version of classical {\em order} governed by the Fermi surface that only exists at zero temperature (Section \ref{RGFermiliquidzeroT}). At finite temperature this turns on the macroscopic scale into a classical fluid endowed with special properties: at low temperatures it turns into the closest realization of an ideal gas realized
in nature (Section \ref{FermiliquidfinT}).

However, the Fermi-liquid is characterized by the Lindhard continuum of particle-hole excitations. In the weakly interacting Fermi-liquid this gives rise to physics that is to a degree a mirror image of the quantum critical physics that you learned to appreciate in the previous section. Dynamical susceptibilities acquire simple scaling behaviour, including the energy-temperature scaling. This now happens in a stable phase matter, while the organization of the underlying RG flow is different from the quantum phase transition affair. The principle will be revealed by strange metal holography in the form of the "covariant RG". 

The take home message will be that the weakly interacting Fermi-liquid takes the role of the free Landau critical point, hard wiring the "structure" of the RG, which then "morphs" into the quantum supreme generalization by invoking anomalous scaling dimensions. At the same time, the onset of "quantum supremacy wipes out the irrelevant operators" according to holography -- as  for the free critical state, scaling is  destroyed in the Fermi liquid itself by perturbative corrections (Section \ref{FLdiagrams}). The idea of strange metals as "generalized Fermi-liquids" is perhaps best examplified by the icon of Fermi-liquid physics: the BCS mechanism of superconductivity. As a preliminary for a spectacular holographic story in this regard (Section \ref{HoloSC}), I will present BCS in its scaling incarnation (Section \ref{BCSbasics}).  

\subsubsection{The Fermi-gas is not entangled: the nodal surface representation.} 
\label{coldatomFL}

The precision argument identifying the Fermi gas as being strictly unentangled involves a less well known representation of the fermionic path integral. Ceperly discovered that one can re-shuffle the sign problem in the first quantized fermion path-integral into an object called the "nodal surface", see ref. \cite{devioussigns}) for a concise review. This is defined as the hypersurface where the full dynamical density matrix is vanishing, closely related to the nodes of the Slater determinant in configuration space. It is a theorem that given a full knowledge of this nodal surface one can reformulate the path integral in a stoquastic form: the nodes turn into reflective boundary conditions "trapping" the bosonic world histories. 

\begin{figure}[t]
\includegraphics[width=0.7\columnwidth]{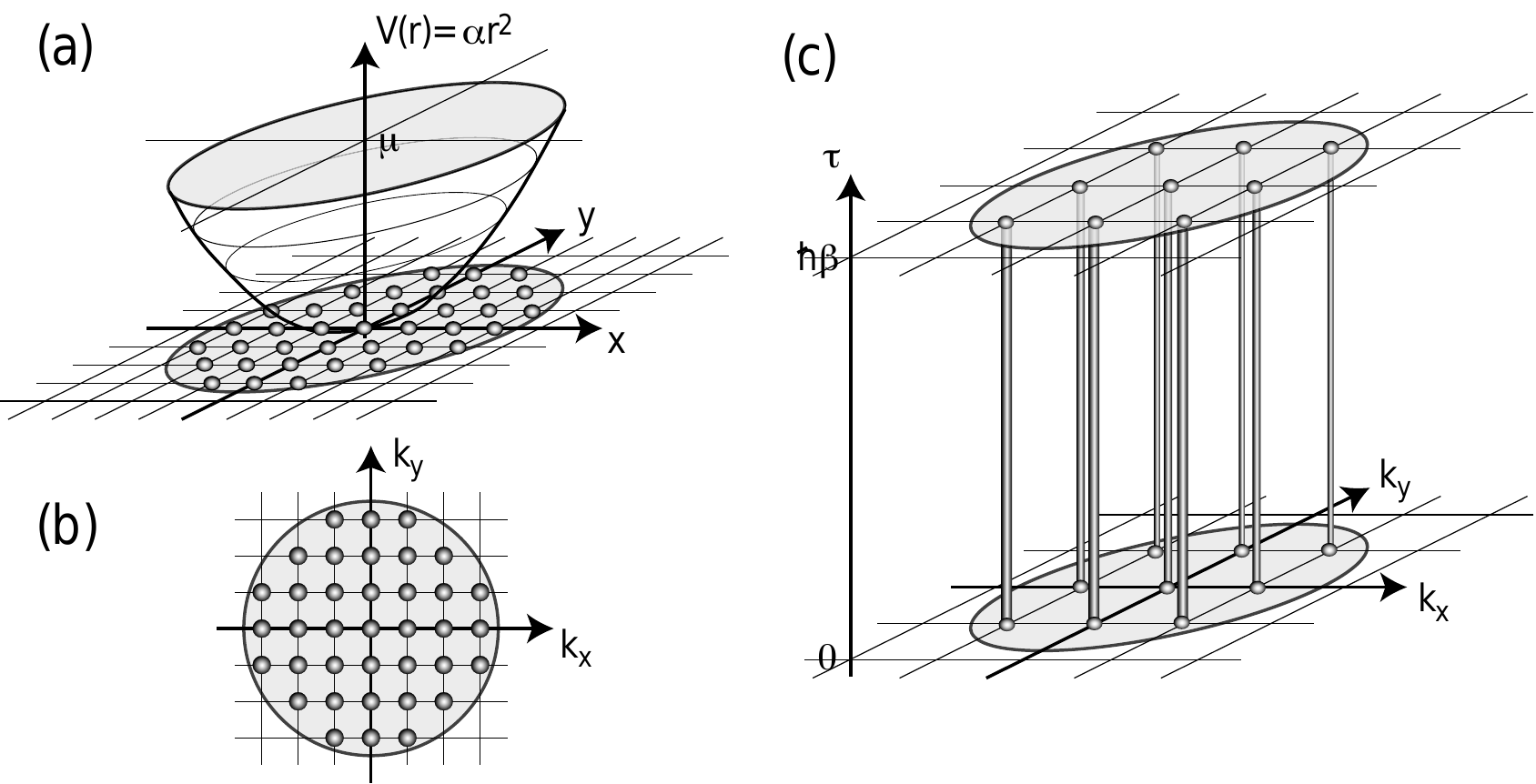}
\caption{Using Ceperley's  nodal surface representation, the Fermi-gas maps on the problem of classical hard-core particles residing in a momentum-space "optical lattice" being trapped in a "harmonic crucible".}
\label{fig:fermi}
\end{figure}

Sergei Mukhin \cite{devioussigns} realized that this becomes a surprisingly simple affair dealing with the Fermi gas in momentum representation where the nodal surface turns out to become equivalent to a {\em classical} "Mott insulator" in momentum space. The partition sum turns out to be identical to a system of classical particles living on the lattice formed by the allowed momenta.  The nodal surface acts as a hard core condition such that only one particle can reside at any momentum space "site". In turn these "classical particles in the momentum space optical lattice" live in a harmonic potential corresponding with the free particle dispersion relation $E(k) = (\hbar k)^2/(2m)$. The vacuum state is obtained by "filling the cup" with the Fermi surface corresponding to the rim, see Fig. (\ref{fig:fermi}). 

This mapping shows unambiguously that the Fermi-gas is one-to-one realization to a classical problem devoid of entanglement-- the hard core classical  particles in the momentum space crucible. It is convenient to rely on this precise analogy in addressing the properties of the Fermi-gas "anchor" as an entirely classical reference state. One immediately infers that the Fermi-surface (the "rim") is only sharply defined precisely at zero temperature. The  issue is that regardless the dimensionality of the Fermi gas the rim looses its sharp definition in the form of a step function in the density distribution in momentum space $n_k$ at any finite temperature. The co-dimension of the thermal excitations is always zero regardless dimensionality. Excite one particle from below to above the rim; this takes a finite energy $\Delta E$ and the Boltzmann weigth $\exp{( - \beta \Delta E)}$ is always finite, while $\Delta E$ can become arbitrarily small. The system is in this regard "critical", these excitations are massless. 

This is of course nothing else than the simple wisdom taught in undergraduate courses that one can count out the Fermi gas properties by just shuffling the particles between their quantum mechanical eigenstates relying on the Pauli principle. But by using the nodal surface representation one gets entertained by the precision classical analogue.  

\subsubsection{The zero temperature Fermi-liquid as an ordered state of fermions.}
\label{RGFermiliquidzeroT}

The theory of the normal metallic state including the neutral version realized by $^3$He is a well charted subject; it should be appreciated as a monumental achievement by Landau and his Soviet school of physics. It has been cooked to perfection, and in this regard there is nothing new to be learned.  But in hindsight the underlying physical principles that rendered the mathematical control were to a degree implicit. In the present context, aiming at its generalization to the holographic strange metals, it is clarifying to spell out what the principles are. 

Lacking insights in the nature of states that are {\em not} Fermi-liquids a tunnel vision developed that is widespread, misidentifying traits that are singularly special to the Fermi liquids to be governing principle instead. In particular, the low temperature Fermi-liquid is the closest approach to a non-interacting low density gaseous physics realized by nature. Among others, in such a system the vacuum structure as a whole is encoded in the information that can be retrieved from the single fermion propagators -- e.g., by analyzing the self-energy of electrons measured by photoemission the properties of the collective state can be deduced in full. But this is not at all the case dealing with the "unparticle" quantum supreme metals. For instance, we will find out much later that albeit for rather special reasons the fermion propagators  contain literally no information {\em at all} regarding the vacuum structure of the holographic systems. Even when knowing the fermion propagators with infinite precision nothing is then learned regarding  collective properties like  macroscopic transport.   

The central principle governing the Fermi-liquid as a special state formed from {\em interacting} fermions will ring a bell. The ground state wavefunction controlling  the {\em zero temperature} state is,

\begin{equation}
| \Psi \rangle_{\mathrm{FL}} = A_{\mathrm{FL}} \Pi_{k=0}^{k_F} c^{\dagger}_k | \mathrm{vac} \rangle + \sum_{i} a^0_i | \mathrm{config}, i \rangle
\label{FLgroundstate}
\end{equation} 

ignoring spins and other internal quantum numbers to keep it concise.

At first sight  this is the familiar short ranged entangled product vacuum. However, the classical "anchor" is no longer a simple bit string but instead the antisymmetrized Fermi  gas state. Having the mind focussed on the quantum information, this does not matter since we just learned that the Fermi gas is devoid of many body entanglement. But the fact that the anchor state is formed from fermions does change the rules dramatically as compared to the stoquastic "bosons". There are now large amounts of massless excitations (the electron hole pairs) and the perturbation theory becomes a much more laborious affair. But the IR fixed point stability theorems rely entirely on the fact that the perturbation theory can be mathematically proven to be convergent when the proper conditions are met.  This implies that  $A_{\mathrm{FL}}$ is {\em finite}.

We learned that it is an SRE product goes usually hand in hand with a (dual) order parameter breaking symmetry spontaneously dealing with stoquastic matter. This is also the case for the Fermi-liquid although the sense of "order" is very different. The {\em Fermi-surface} now takes the role of order parameter. The easiest way to appreciate this is by realizing that the Fermi-surface is a geometrical object ("surface") with a precise locus in the space formed by {\em single particle momentum} quantum numbers. But according to the {\em exact} quantum rules this space does not exists when the fermions submit to any form of physical interactions! This is simple: the single particle momentum operator $n_k = c^{\dagger}_k c_k$ does not commute with the interactions generically of form $\sim \sum_{k, k', q} V( \cdots )  c^{\dagger}_k c_{k+q}  c^{\dagger}_{k'} c_{k'-q}$. Total momentum is conserved in a homogeneous background but single particle momentum is not and therefore it is not quantized in an exact fashion. 
 
But this is the same phenomenon that is underpinning the most familiar of all forms of conventional spontaneous symmetry breaking: the single particle position operator is not conserved either but in a crystal it {\em emerges} as a constant of motion. We know where the atoms are in space. This of course ascends from the fact that the crystal vacuum is an SRE product controlled by stitching together atoms in real space wave packets. This role is now taken by the unentangled Fermi gas.

At first sight this may be confusing. What has order to do with the Fermi-gas? There is actually a direct analogy to be found in stoquastic physics. As for the Fermi-gas, there is no such thing as a strictly non-interacting system of bosons to be found in nature. But for reasons of pedagogical convenience we learn as undergraduate students much about the physics of the free systems that are completely governed by quantum statistics. Accordingly, free bosons should be subjected to Bose-Einstein condensation. However, in the presence of any finite repulsive interaction the rules change drastically. As the Bogoliubov theory of weakly interacting bosons demonstrates directly, the system turns into a 
{\em superfluid} (or superconductor) in the scaling limit. This breaks $U(1)$ spontaneously (in contrast to the BEC ground state which is an energy {\em eigenstate}) having among others as consequence that rigidity emerges expressing itself in the form of a Goldstone boson (phase mode) with linear dispersion. What I just argued is the fermionic incarnation of the same affair, the non-interacting limit is just deceptive.     

Goldstone bosons of conventional order are expressing  that the ordered system carries excitations that act in the "direction of symmetry restoration". These behave in a way as a critical system. For instance, the entropy will be algebraic exhibiting a Debye law $S \sim T^d$ as in a CFT.  However, given the Goldstone protection the system is as exquisitely non-interacting as possible. But in Fermi liquids one finds in addition the continua of electron-hole excitations. These also fluctuate the Fermi-surface order -- one could call them "Goldstone fermions" -- and as a consequence  the ensuing phenomenology is much richer. This is captured by Landau's phenomenological (IR fixed point) Fermi-liquid theory. 

To find out how this works let use the same strategy as for the stoquastic SRE product states. Let us first focus on the "classical limit"  $A_{\mathrm{FL}} \rightarrow 1$ to find out how the fixed point physics is organized. Subsequently we can load this up with any amount of required perturbative corrections, realizing that on scales larger than the entanglement length this system will be governed by the "classical" theory (which is Landau's phenomenology). Let us specialize to the Fermi-liquid formed in homogeneous space, realized in nature in the form of the $^3$He incarnation. Electrons in solids "suffer" from complications associated with their single particle band structure, but these give rise to no more than a number of special effects that are not essential for the big picture.

\begin{figure}[t]
\includegraphics[width=0.7\columnwidth]{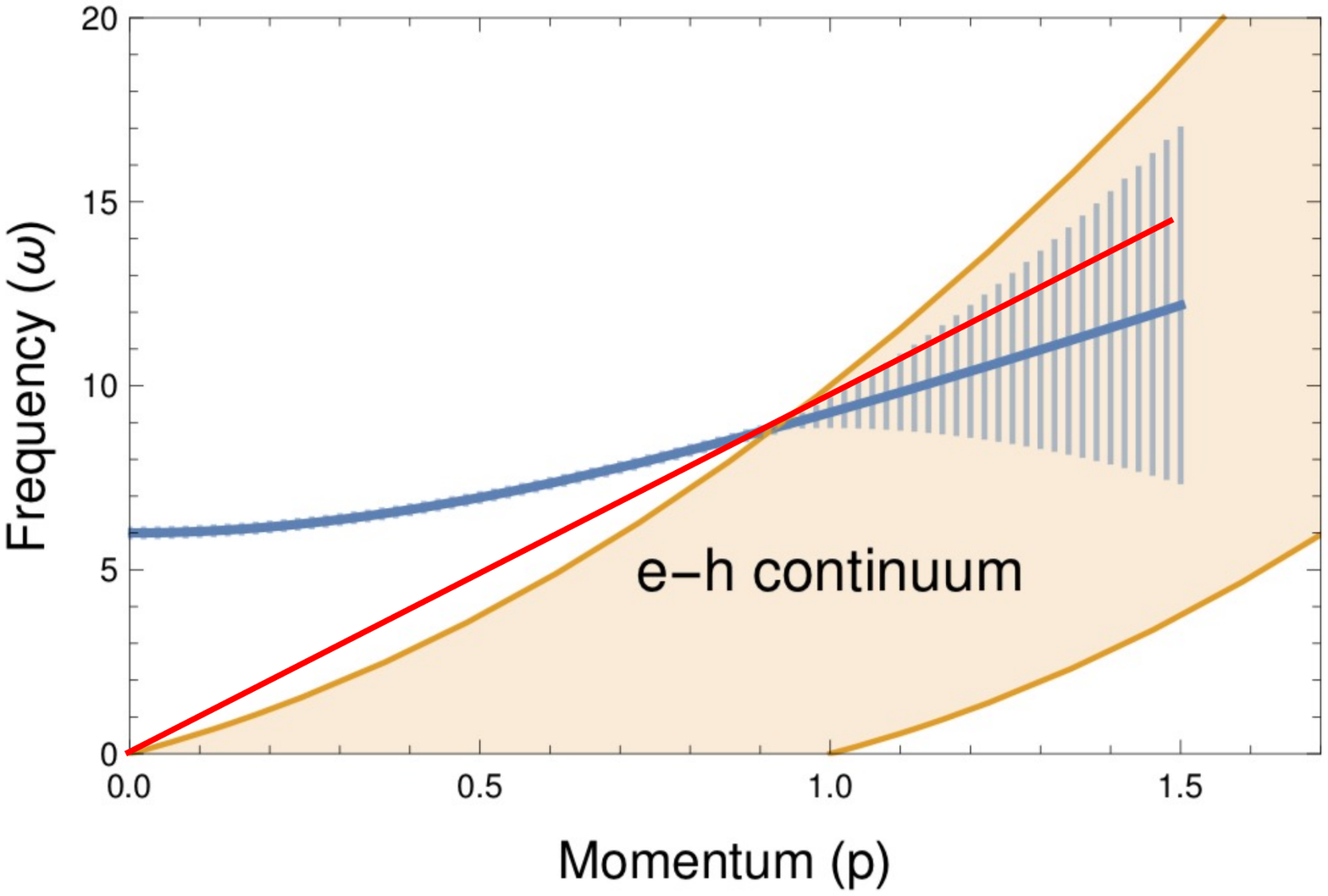}
\caption{The fermion-kinematics of the free Fermi gas in the momentum-energy plane encapsulated by the Lindhard function associated with the density response function of the Fermi gas. The red line indicates the zero sound mode of the neutral Fermi-liquid, decaying at large momenta in the Lindhard continuum (Landau damping). In the charged Fermi-liquid this zero sound mode gets "promoted" to a plasmon in the density response (blue line). }
\label{fig:LindhardFG}
\end{figure}

The excitation spectrum of the Fermi gas itself  is determined by just counting the number of particle-hole excitations that can be created for a given centre of mass momentum and energy.  The first quantity of interest is the spectrum of density excitations. The density operator in momentum space $\rho_q = \sum_k c^{\dagger}_{k + q} c_k$ and a simple but tedious  counting exercise shows that the dynamical density susceptibility is governed by the {\em Lindhard} functions. 

Their precise form  in various dimensions can be looked up. The take home message is that these span up a bounded region in the momentum-energy plane, see Fig. (\ref{fig:LindhardFG}).  What matters most in the present context is that at low energy the imaginary part (spectral function) of the dynamical density susceptibility $\chi_{\rho}^0 (q, \omega ) = \langle \rho \rho \rangle_{q, \omega}$   is of the form, 

\begin{equation}
\mathrm{lim}_{\omega \rightarrow 0} \mathrm{Im} \chi_{\rho}^0 (q, \omega) \sim \frac {1}{q} \; \frac{\omega}{E_F} 
\label{Lindhardasymptotics}
\end{equation}

This is a {\em powerlaw}, reminiscent of the scaling functions we encountered at quantum critical points. But the Fermi gas is a stable phase of matter. As we will learn to appreciate when we turn to the holographic strange metals, a difference of principle with the "critical point" scaling forms is however that these are "UV sensitive" in the sense that the Fermi energy is remembered in the deep IR. By measuring $\chi_{\rho}$ at low frequency one can determine the Fermi energy in principle.
 
 In fact, this is just the tip of the iceberg. Regardless the property one measures the response of the Fermi-liquid is a {\em powerlaw}. In addition, at least sufficiently close to the fixed point it exhibits energy-temperature scaling. Also the finite temperature properties are power laws, as for instance the Sommerfeld entropy of the Fermi-gas $S \simeq T / E_F$ regardless the number of dimensions. In a way it behaves like a critical state, but not quite: the structure of the theory is entirely different from the genuine quantum critical states. The scaling dimensions are entirely detached while crucially one always finds the Fermi-energy to be present. Obviously, the greatest difference is that the Fermi-liquid is a stable phase of matter that does not need fine tuning to a critical point. Keep this in mind, this theme will re-enter in the strange metal section with a  loud bang!
 
The next step is to find out what happens with this excitation spectrum when the interactions are switched on. One would first like to infer what happens in the "classical limit", the analogue of the classical equations motion describing the lattice vibrations dealing with a crystal. In case of the Fermi-liquid such a "normal" classical limit is a bit obscure but there is a universal procedure to derive it directly from the quantum physical time evolution, resting on the Ehrenfest theorem (e.g., \cite{classicalcondensates}). In Heisenberg representation the time evolution of the operator $O$ is governed by $ i \hbar ( d O /dt ) = \left[ H, O \right] $. Bt taking VEV's on both sides one will recover the classical EOM's when the ground state is a classical product vacuum. The r.h.s. will then factorize in a vacuum expectation value associated with the order parameter and the VEV of an operator. Together with the VEV's of the l.h.s. this coincides with the  EOM's describing the classical modes. This corresponds in diagrammatics with the "Wick theorem" and departing from $A_{\mathrm{FL}}  \rightarrow 1$ this is just "time dependent mean field". This is called for historical reasons the "random phase approximation" (RPA) in the Fermi-liquid literature, or either the "bubble resummation" in diagrammatics.   

In this minimal setting, the result  for the "full" density susceptibility associated with a Fermion system subjected to a repulsive  density-density interaction $H_{\mathrm{int}} = \sum_q V_q \rho_q \rho_{-q}$ becomes,  

\begin{equation}
\chi_{\rho} (q, \omega) = \frac{\chi^0_{\rho} (q, \omega)} { 1 - V_q \chi^0_{\rho} (q, \omega)}
\label{RPAdens}
\end{equation}

where $\chi^0_{\rho} (q, \omega)$ is the Lindhard susceptibility of the free Fermi gas. This yields a first Fermi-liquid novelty. When the denominator is vanishing because $\mathrm{Re}  \chi^0_{\rho} (q, \omega) = 1/ V_q$ while $\mathrm{Im}  \chi^0_{\rho} =0$ it signals that the interactions cause a new excitation. Invariably, one finds in this way an "antibound state" lying above upper bound of the Lindhard continuum at small momenta. Assuming a short range interaction as of relevance to a neutral system (like $^3$He), this "interaction pole"  exhibits a {\em linear} dispersion: the red line in Fig. (\ref{fig:LindhardFG}). This describes propagating density oscillations: the {\em zero} (temperature) {\em sound mode}.  The physical interpretation is that the interactions "harden" the Fermi surface which starts to behave as a soccer ball, submitting to coherent breathing (s-wave) vibrations. Given the Luttinger Volume theorem insisting that the volume enclosed in momentum space by the Fermi-surface is in one-to-one correspondence to the microscopic number density this Fermi-surface fluctuation is in one-to -one correspondence with a density fluctuation: it is a sound wave.   

Neatly formulated as a phenomenological theory in terms of a free energy relating to momentum-space distribution functions, Landau's formulation of the Fermi liquid revolves precisely around these "Hartree" mean field terms (see e.g. \cite{baymFLbook}) . The RPA in the above is subjected to an angular momentum ($L=0,1,2, \cdots$) and spin ($s$, $a$, singlet and triplet) decomposition. Zero sound is associated with the scalar ($F^s_0$ Landau parameter). Perhaps the most spectacular prediction (confirmed in $^3$He) is that a propagating mode develops as well in the $L=1$ channel when the interaction parameter $F^1_s$ exceeds a critical value. This parameter also governs the mass enhancement of the quasiparticles (see underneath). This mode corresponds with a {\em shear} wave -- it is indistinguishable from the transversal acoustic phonon of a solid. At least in the homogeneous case it will appear when the quasiparticle mass enhancement $m^*/m \ge 7$, 

How can this be? A reactive (elastic) response to a shear stress should be universally association with the formation of a crystal -- in classical liquids this is exclusively dissipative encapsulated by the shear viscosity. But in the zero temperature "heavy" Fermi liquid the Fermi-sphere "hardens" also with regard to dynamical  shear deformations, although it continues to behave like a liquid with regard to static shear stress.  This static response is associated with the $L=2$ channel. This  highlights the unique nature of the physics controlled by the Fermi-surface "order parameter".  In a very recent paper these matters are discussed in the context of the metallic  heavy Fermi-liquids formed from electrons \cite{valentinisshear}.    

I already stressed the singularly special role played by the single fermion response in a Fermi-liquid. This is easy: there is no better way to directly observe the Fermi-surface "order" than by photoemission in electrons system.  As for the transversal Ising system (Section \ref{Isingparticles}) the SRE product "keeps the quantum numbers together" as required for the particles in the spectrum, but these are now associated with the Fermi-gas. Right on the Fermi-surface particles carrying the quantum numbers of the non-interacting Fermi-gas fermions will occur, having a pole strength equal to $A^2_{\mathrm{FL}}$: these are the quasiparticles, and I have just rephrased Landau's principle of adiabatic continuity in the modern quantum information language. Adiabatic continuity just means the the Fermi-gas "anchor" has a finite amplitude $A_{\mathrm{FL}}$ in the SRE product.

The complicating circumstance compared to the stoquastic case is however that a gap protecting the particles is now absent. Yet again the massless Lindhard continuum is in the drivers seat, but its scaling dimensions conspire to offer sufficient protection of the "fermionic Goldstones"  (quasiparticles) to live infinitely long right at the Fermi-surface. At  finite energy a quasiparticle decays by exciting quasiparticle-antiquasiparticle pairs through the residual ("$F_1^s$") interactions but this gives rise to a decay ("collision") time 

\begin{equation}
\tau_{QP} \sim E_F / (\hbar \omega^2) 
\label{FLcolltime}
\end{equation}

becoming infinite for $\omega \rightarrow 0$. Technically this follows from the usual "loop" self-energy diagram, that is expressing how the excited electron decays in the Lindhard continuum.  This sets the imaginary part of the self energy, and it is easy to show that the real part of this self-energy translates into the mass enhancement of the quasiparticles alluded to in the above.     

\subsubsection{The Fermi-liquid at finite temperature.}
\label{FermiliquidfinT}

Appreciating the Fermi-surface as the order parameter of the {\em zero temperature} Fermi-liquid is clarifying with regard to the finite temperature physics. As  is obvious  from the "rim of the cold atoms" (Section \ref{coldatomFL}), this order parameter is destroyed at any finite temperature since the "particle-hole" fluctuations are always zero dimensional having a finite Boltzmann when temperature is finite. This has the ramification that on macroscopic scales the state is indistinguishable from a classical fluid: one can adiabatically continue to the high temperature limit without encountering any thermodynamic singularity (phase transition). Yet again, $^3$He is the best example: its flow properties in the very low temperature Fermi-liquid regime is governed by Navier-Stokes hydrodynamics albeit its hydrodynamical coefficients are quite anomalous. 

In first instance one can depart from the Fermi-gas wiring in finite temperature through the Fermi-Dirac distribution functions: an undergraduate textbook affair. This is so straightforward that a rather profound outcome is easily overlooked: {\em the finite temperature weakly interacting Fermi-liquid exhibits impeccable energy-temperature scaling}. We learned that this is in the stoquastic arena exquisitely special to the strongly interacting  quantum critical state, and in the Fermi gas we get it for free!  This will be a crucial motive in the context of the holographic strange metals. 

I already alluded to the thermodynamics exhibiting power law behaviours suggesting some form of scale "invariance", the Sommerfeld law for the entropy $S \sim T$ regardless dimensionality. In Section (\ref{holoSM}) we will learn how this emerges from a particular form of $E/T$ scaling. But it pertains to all responses, including the Lindhard excitations -- see e.g. the BCS "particle-particle" channel that I will discuss in the next subsection. 

The best known example is associated with the quasiparticle life time $\tau_{QP}$ (Eq. \ref{FLcolltime})  in the finite temperature Fermi-liquid. This exhibits the simple energy-temperature scaling behaviour, 

\begin{equation}
\frac{\hbar}{\tau_{QP}} \sim \frac{( \hbar \omega)^2}{E_F} \left (1 +  ( \frac{ 2\pi k_B T} {\hbar \omega} )^2 \right)
\label{fLfiniteTQP} 
\end{equation}

At macroscopic times the Fermi-liquid at low but finite temperatures should be viewed as the closest approach to a classical kinetic gas that can be realized in nature. The central quantity in Boltzmann kinetic gas theory is the collision time $\tau_c$. In this regime this is of course coincident with the quasiparticle life time  $\tau_{QP}$. From Eq. (\ref{fLfiniteTQP}) it follows that at macroscopic times $\tau_c \simeq  E_F /(k_B T) \times \tau_{\hbar}$ where $\tau_{\hbar} = \hbar / (k_B T)$, the familiar Planckian time. In a good metal $E_F \simeq 10^5$ K and one infers that $\tau_c$ is many orders of magnitude larger than the Planckian value in the Fermi liquid regime where the (renormalized) $E_F >> k_B T$.  

As I will discuss in more detail in the transport Section (\ref{holotransport}), in nearly all metals the collision time is typically much larger than the single particle momentum relaxation time associated with the breaking of translations and this momentum has decayed well before the local equilibrium can be established at times longer than $\tau_c$ as required for the  hyrodynamical collective flow behaviour.  To see the hydro at work one can turn again to $^3$He. The quantity of interest in Navier-Stokes is the {\em viscosity} responsible for the dissipation. By dimensional analysis the (shear) viscosity $\eta$ is set by the free energy density $f$ and the momentum relaxation time $\tau_c$,

\begin{equation}
\eta_{FL} \simeq f \times \tau_c = n E_F \times  \tau_c = n \hbar \frac{E^2_F}{ (k_B T)^2}
\label{FLeta}
\end{equation}

At low temperatures $^3$He becomes extremely viscous, a challenging complication for the design of dilution fridges. We will see much later that this actually offers an extreme contrast with the quantum supreme finite temperature fluids, offering an unique opportunity for the experimentalists to find out whether one is dealing with a quasiparticle or "unparticle" system  

What happens to the dynamical responses at long wavelength when temperature becomes finite? As a ramification of the energy-temperature scaling the dividing line is set by $\hbar \omega \simeq k_B T$. As in the quantum critical systems, when $\hbar \omega > k_B T $ the system behaves as if it is at zero temperature and the response is dominated by the zero sound, and one recovers when the conditions are right even the propagating shear mode. This is called, somewhat mystifying, the "collisionless regime" in the classic literature.  However, when $ \hbar \omega < k_B T$ the hydrodynamics of the thermal fluid takes over (the "collision-full regime"). Instead of the zero sound one now encounters the hydrodynamical "first sound"; a famous result is the attenuation  maximum at $\omega \simeq k_B T$ measured by ultrasound signalling that the protection mechanism of sound is qualitatively different in both regimes. Similarly, shear submits to the viscous rules in the hydro regime, and the propagating shear recovers above $\hbar \omega \simeq k_B T$ as spectacularly confirmed in $^3$He in the 1970's (see ref. \cite{valentinisshear}).  

\subsubsection{The stability of the Fermi-liquid: the BCS theory.}
\label{BCSbasics}

Despite its pseudo-critical attitudes the Fermi-liquid has an existence as a {\em stable} state of matter. By exposing it to particular microscopic conditions it can get destroyed but it is characterized by a basin of stability. For instance, the Fermi-liquid in copper metal continuous to persist down to the lowest temperatures that can be realized in the laboratory.

However, when it becomes unstable the state that takes over is characterized by unique "fermionic" traits. The case in point is the conventional superconducting state that was explained in 1957 by the Bardeen-Cooper-Schrieffer (BCS)  theory, setting the paradigm. I assume the reader has familiarity with this famous story. For the canonical textbook explanation, see for instance Ref. \cite{classicalcondensates}. Perhaps more than anything else, the BCS mechanism illustrates the underlying commonality between the Fermi-liquid and the holographic strange metal, as I will highlight in Section (\ref{holoSM}).

To appreciate this similarity, it is informative to discuss BCS in a scaling language which is well known but perhaps not for all readers. This departs yet again from the time dependent mean field (RPA) formula Eq.  (\ref{RPAdens}). Dealing with repulsive interactions in the density channel, I pointed out the occurrence of anti-bound states corresponding with the zero sound excited states. But when bound states would like to form at {\em negative} energies these would signal the instability of the Fermi liquid -- this is just about linear instability, referred to as "tachyons" by the string theorists. 

 As a first try one could reverse the sign of the interaction $V_q$ in the density channel to find out whether such negative energy states would form.  However, for this to happen $\mathrm{Re}  \chi^0_{\rho} (q, \omega =0) = 1/ V_q$ has to be satisfied:  the left hand side has to become large enough at some $q$. But we learn that the spectral density is {\em decreasing} with energy according to Eq. \ref{Lindhardasymptotics}: it is "irrelevant" in the (covariant) RG language explained in Section  (\ref{holoSM}). This has the consequence that  $\mathrm{Re}  \chi^0_{\rho}$ is small at zero frequency and large interactions are required. The Fermi liquid is accordingly stable to such density instabilities when the effective attractive interactions are weak. In fact, dealing with special fine tuned band structures one can encounter "nesting conditions" in this "particle-hole channel" that may change this into a BCS like affair, see Ref. \cite{classicalcondensates}.

However, let us consider instead  the "particle-particle" channel. Define a zero center of mass pair operator, 

\begin{equation} 
b^{\dagger} (q=0) = \sum_k c^{\dagger}_{k \uparrow} c^{\dagger}_{-k \downarrow}
\label{pairopBCS}
\end{equation}

to subsequently "count out" the dynamical pair propagator of the Fermi-gas which is again originating in the continuum of Fermi-surface excitations, 

\begin{eqnarray} 
\chi^0_{\mathrm{pair}} (\omega, T) & = &  \langle  b^{\dagger} b \rangle_{q=0, \omega} \nonumber \\
\mathrm{Im}  \chi^0_{\mathrm{pair}} (\omega, T =0 ) & = & \frac{ \omega^{\Delta_{\mathrm{pair}} = 0}}{E_F} \nonumber \\
\mathrm{Re} \chi^0_{\mathrm{pair}} (\omega, T =0 ) & = & - \frac{1}{E_F} \ln \left( \frac{\Lambda_{\mathrm{UV}}}{\omega} \right)
\label{pairsuscFG}
\end{eqnarray}

The spectral density is frequency independent -- the pair operator is "marginal" with scaling dimension $\Delta_{\mathrm{pair}} = 0$ in the RG language. Kramers-Kronig consistency then implies that the real part is characterized by an IR {\em logarithmic divergence}. In principle the UV cut-off $\Lambda_{\mathrm{UV}} \simeq E_F$ but for future purposes we keep it as a parameter. 

Departing from a system characterized by generic {\em attractive} interactions, the interaction Hamiltonian will contain terms $H_1 = V_{\mathrm{BCS}} b^{\dagger} (q=0) b (q =0)$ originating in attractive interactions "filtered" in the pair channel, the "BCS Hamiltonian".  Applying the RPA, it follows that the condition for the formation of new poles becomes $1/E_F = 1 / V_{\mathrm{BCS}} \ln \left( \frac{\Lambda_{\mathrm{UV}}}{\omega} \right)$. Because of the $\omega \rightarrow 0$ IR logarithmic divergence this cannot be satisfied for any finite $V_{\mathrm{BCS}}$ but one can deduce immediately the energy scale (gap) associated with the (superconducting) state that takes over by cutting off the lower bound of the "Cooper logarithm" by the scale $\omega = \Delta$.  Accordingly, 
$1/E_F = 1 / V_{\mathrm{BCS}} \ln \left( \frac{\Lambda_{\mathrm{UV}}}{\Delta} \right)$, and the famous gap equation follows,

\begin{equation}
\Delta \simeq \Lambda_{\mathrm{UV}} e^{- E_F / V_{\mathrm{BCS}} }
\label{BCSgap}
\end{equation}

By employing the energy-temperature scaling it follows immediately that the transition temperature $T_c \simeq \Delta$ -- computing this explicitly yields the parametric factors in the form of the zero temperature  gap ($\Delta$) to $T_c$ ratio as $ 2 \Delta /(k_B T_c) = 3.51$. 

Below $T_c$ one of course encounters the "new vacuum" associated with the "off diagonal long range order" characterized by the VEV $\langle   b^{\dagger} (q=0) \rangle \neq 0$ and the $T=0$ BCS "product" wave function $| \Psi \rangle_{\mathrm{BCS}} = \Pi_k (u_k + v_k   c^{\dagger}_{k \uparrow} c^{\dagger}_{-k \downarrow} | \mathrm{vac.} \rangle$. 

The magic revealed by this simple consideration is responsible for the iconic status of BCS. The attractive interactions are in conventional superconductors due to the virtual exchange of phonons. These are only present below the characteristic phonon frequency, $\Lambda_{\mathrm{UV}} = \omega_{\mathrm{ph}}$ while in metals like aluminium these are quite weak: $\lambda =  V_{\mathrm{BCS}}/E_F << 1$. But given the Cooper logarithm the Fermi-liquid {\em has} to become unstable be it at a temperature that is extremely (exponentially)  small compared to the UV scales. The Fermi energy of aluminum is $\sim 10^5$K and $T_c \sim 1$K. In addition, $\Delta$ implies a length scale via $\xi \simeq \hbar v_F/\Delta$: the coherence length ("pair size") that can easily be of order of microns.

\subsubsection{Perturbation theory: the AGD diagrams and functional RG.} 
\label{FLdiagrams}

The take home message of the above is that the Fermi gas subjected to {\em weak} interactions is subjected to a scaling logic which is reminiscent of, but yet different from what we encountered in quantum criticality. By "listening to the black holes" of holographic duality in Section  (\ref{holoSM}) we will learn where this difference resides, finding out how this scaling structure generalizes to the quantum supreme realms. In this context the Fermi-gas has a similar status as the free critical theory above the upper critical dimension in the stoquastic realms. It defines a template of how the theory is organized, but it gets endowed with anomalous scaling dimensions when the quantum supremacy is in effect. 

In the stoquastic realms we learned that simple scaling behaviour is destroyed above the critical dimension because of the manifold of irrelevant operators that give rise to perturbative corrections. This is a disease of SRE "products" and the Fermi-liquid suffers from the same pathology. When the interactions become stronger the elegant scaling properties of the Fermi gas are compromised. The expression of this messiness is in the intricate nature of the diagram "gymnastics" embodied by "AGD" (the early book by Abrikosov, Gorkov and Dzyaloshinkii \cite{AGD}). 

The best formulation is in terms of the perturbative functional renormalization group pioneered by Shankar and Polchinski (see \cite{metznerfrg}).  At every point on the Fermi-surface one can define a bunch of coupling constants and these all run  a-priori in their own ways under renormalization, one needs {\em functions} of coupling constants. Modulo band structure effects (rooted in nesting) one can then show that {\em all} operators associated with these couplings are IR irrelevant, except for the pair channel. This yields a precision story for the stability of the Fermi-liquid singling out the superconductor as the only dangerous competitor. As not always realized, it is however severely limited. It departs from a UV Fermi gas basis, obviously required to organize the RG resting on the Fermi surface. One than switches on {\em weak} interactions in the UV, to  find out how these extinguish in the flow to the IR. It is a superior formalism to assess the stability of the IR fixed point but it has {\em nothing} to tell about the situation encountered in $^3$He or the strongly correlated electron systems (Section  \ref{Mottness}) where the interactions overwhelm the bare Fermi energy in the UV. 

Although less flexible the conventional "AGD" diagrams yield similar outcomes. Departing from the Fermi gas, interactions are switched on and when these get stronger one is forced to compute higher and higher order perturbative corrections. This hits a glass ceiling in the form of the vertex corrections, boiling down to hard to compute integrals. To maintain control "retardation" may come to help. Consider the electron-phonon coupling problem. I already alluded to the fact that the characteristic phonon energy is small compared to $E_F$, implying a different small "Migdal Parameter" $ \omega_{\mathrm{ph}}/E_F << 1$. One can show that vertex corrections are suppressed by this quantity and the ensuing "self-consistent" theory can in principle be solved, at least when the coupling $\lambda$ is not becoming too large \cite{Eliasbergkiv}. Applied to the electron-phonon problem, the ensuing "Migdal-Eliasberg theory of strong coupling superconductivity" has been quite successful in describing for instance the superconductivity of $Pb$. 

If not only for its iconic status, the superconducting instability forms an interesting stage to confront this Fermi-liquid controlled affair with its quantum supreme holographic generalizations. In Section (\ref{HoloSC}) I will present a comparison between the way that superconducting correlations develop in holographic strange metals versus both the weak- and strong coupling BCS varieties, showing how the impeccable simple scaling behaviour in the former is completely wrecked by the perturbative corrections according to  Migdal-Eliasberg.

\section{What you absolutely need to know about the AdS/CFT correspondence.}
\label{AdSCFTgen}

We have arrived at the last collection of preliminaries as required to appreciate the substance of these lecture notes. This is not intended to be a primer on the AdS/CFT correspondence. This mathematical contraption rests on roughly 40 years of hard work by string theorists, being so remarkable that a large part of the contemporary activity in this community revolves around the "correspondence". In the 20+ years after the discovery by Maldacena in 1997  thousands of papers have been published turning it into a vast subject. In the particular corner of interest here -- the Anti-de-Sitter to Condensed Matter Theory, "AdS/CMT" -- three bulky textbooks are available which are complementary in their focus \cite{Erdmengerbook,holodualbook,lucasbook}.  

To learn how this works may take a year of your life. Stronger, to operate at the frontier of the development you better engage in it for your PhD thesis in order to build up a big repertoire of all the technical tricks of the trade. I learned it an elevated age: I do believe that I have a fair understanding of how it all works but I am heavily dependent on skilled  coworkers to get the computations done. 

I will explain in this concise section what you really need to know in order to appreciate the power it exerts on thinking out of the established condensed matter box that I exhibited in the above. Handle it in the same way as a theorist deals with experiment. You don't want to know how the op-amps and vacuum pumps of the laboratory rigs precisely work, but you do need to know what kind of specific information the experimental colleagues can deliver, and especially the caveats and  restrictions that are invariably present in the real world. 

\subsection{The allure: fancy black holes as quantum computers.}

AdS/CFT was not meant to shed light on electron systems. All along the development of string theory was propelled by the promise that it could shed light on quantum gravity. It is really an outgrowth of quantum field theory, dealing with extended objects (strings, branes) that are infused with great amounts of symmetry (supersymmetry, Weyl-invariance, $\cdots$).  In a  mathematical miracle, results dropped out hinting at surprising relations with "gravity", Einstein's theory of general relativity (GR). This really landed on its feet with Maldacena's discovery. 

We are witnessing  presently the second youth of GR. It started with the cosmology revolution in the 1990's, where this somewhat flaky affair turned into a quantitative science due to the high resolution mapping of the CMB and so forth. More recently the gravitational wave detectors got on line, while the astronomers pulled off the observation of the supermassive black hole shadow with the event horizon telescope. This is testimony of the mathematical quality of GR: the math forced upon us the wisdom that black holes are like the "atoms" of space time and half a century later the hardware was developed to a degree that it could be confirmed by observations. Black holes are presently at the centre of attention. 

You have to be quite literate both in GR and quantum theory to recognize the allure that is eventually responsible for the prominence of  the correspondence in the string theory community. The equations that tell you that black holes exist are completely different from anything in the quantum theory, not to speak about the phenomena they describe. It is then a revelation to find out that  the black hole equations can be used to reconstruct anything that really matters in the established condensed matter  agenda. But it reaches much farther. 

My claim is that {\em state of the art black hole mathematics acts as  a quantum computer revealing  general phenomenological principles governing  observable properties of quantum supreme matter.} In fact, the ultimate promise is that by establishing this connection more thoroughly -- it is still under construction -- there is a potential that experimental work will reveal surprises that may impact in the "reverse gear" of holography: shed light on the deep mystery of quantum gravity. Frankly,  explaining why the superconducting Tc is high (the holy grail in traditional condensed matter) is worthy no more than a footnote as compared to this challenge.  

A large part of the technical hardship I just referred to is due to the fact that the gravitational physics that is "dual" to the quantum matter is state of the art. The black holes at the centre of public attention originate in the 1960's -- the Kerr solution. These are quite simple for the reason that it was well into the 21-th century taken for granted that "black holes cannot have hair". In an asymptotically flat space time stationary black holes have to be featureless according to the no-hair theorem: a black hole is like an elementary particle that can only be characterized by its overall energy, (angular) momentum and electromagnetic charge. Admittedly, dealing with dynamical circumstances such as black hole mergers the Einstein equations representing a system of non-linear partial differential equations (PDE's) come to life. It took actually half a century or so to tame these horrible equations to a degree that supercomputers can handle them -- the numerical GR that plays a crucial role in interpreting the GW detector signals. 

Spurred actually by the AdS/CMT agenda, it was a decade ago realized that under a general condition tied to the correspondence -- the gravitational space time has to be asymptotically Anti-de-Sitter -- there is no such thing as a no-hair theorem. In a flurry that followed the string theorist pulled off a hitherto unexplored new area in GR. Equilibrium in the quantum theory corresponds with stationary gravitational solutions, and the holographists discovered  a zoo of fanciful black hole hairs. The state of the art is that one needs the numerical  GR technology running on supercomputers to find out how such "rasta hair" black holes look like. This present frontier will be highlighted in particular in  Section (\ref{Intertwined}).

\subsection{The plain vanilla AdS/CFT correspondence.}
\label{Maldacenaholo}

To get an idea of how the correspondence works, let us start focussing in on the bare bones version: the "set up" discovered by Maldacena. This was in fact a spin-off of the so-called "second string revolution".  This is a remarkable story involving a series of profound mathematical discoveries  that is beyond the scope of these notes. What you need to know is that this correspondence is regarded as a mathematical fact, proven at least to physicists standards.  This theorem is as follows: "Maximally supersymmetric ${\cal N} = 4$ Yang-Mills in $\mathrm{D} =4$ space-time dimensions in the large $N$ limit for infinite 't Hooft coupling is {\em dual} to classical supergravity on AdS$_5 \times$S$^5$."  What has this string-speak to do with electrons in solids? 

Let's start out with the  ${\cal N} = 4$ Yang-Mills part. Quantum chromo-dynamics with its quarks and gluons is an example of a Yang-Mills theory. The quarks carry three colour charges ($N=3$) and there are a total of $3^2-1 = 8$ gauge bosons (gluons) exchanging the colours between the quarks. The mapping to {\em classical} (computable) gravity claimed by the correspondence requires  that this number of colors is sent to infinity. The crucial part is that one is dealing with a {\em matrix} field theory (the $N^2$ gluons) in the  large $N$ limit. 

Supersymmetry on the other hand is not crucial: a zoo of non-supersymmetric correspondences were identified later. But there is one particular feature that is important. Maximal supersymmetry means that the fermionic- and bosonic fields are in perfect balance at zero (quark) density  and this leads to the "non-renormalization theorems": departing from a scale invariant free theory the fermion- and boson contributions in the RG equations cancel each other exactly and this has the implication that one no longer has to fine tune to a phase transition to obtain a perfect conformal invariance. This just describes a quantum critical state as discussed in the previous section, where "large 't Hooft coupling" has now the meaning of "strongly interacting critical". The take home message is that one does not have to fine tune to a phase transition: the state in the boundary is automatically of the  (extremely) strongly interacting quantum critical kind. 

Such theories are giving in to the rules explained in Section (\ref{qucritical}): two-point propagators are branch cuts characterized by anomalous dimensions and these may become  {\em very} anomalous. But the large $N$ CFT's  are actually not at all understood in the canonical language used in quantum information. There are however good reasons to believe that the combination of large $N$ and infinite 't Hooft coupling implies that the many-body entanglement is pushed to its very limit. This is the important reason to pay attention: the string theory speak is just coding for a maximally "quantum supreme" stoquastic state of matter -- as always in physics one wants to know first the limiting cases and this Yang-Mills affair is about the dense entanglement limit.

The crucial word in the definition of the correspondence is "dual". Metaphorically this is similar to the quantum-mechanical particle-wave duality. "Particle" is in a way opposite to "wave" but we understand too well that these are two sides of the same coin. There is the quantum-mechanical "wholeness" and pending the way one observes the system one percieves it as a particle- or wave physics. In fact, AdS-CFT is more closely related to the field-theoretical weak-strong ("S") duality: the Kramers-Wannier type of affair where the "strongly coupled" (disordered) state corresponds with an ordered state of the disorder operators (topological excitations) of the weakly coupled (ordered) state. The Yang-Mills theory is like the strongly coupled affair, and the gravity side is like the "order by disorder operators" that we encountered in the previous sections. But there is one extra feature that is beyond anything that is found in the field-theoretical dualities. It is  a {\em holographic} duality. 

The word refers to the familiar "holograms": one has a two dimensional photographic plate with complicated interference patterns, and upon shining through a laser beam it reconstructs into a recognizable three dimensional image. The big deal is that all the information for the 3D image is encoded in one lower dimension, be it in the form of completely incomprehensible interference fringes. In this comparison, the Yang-Mills theory is like the photographic plate and gravity is the three dimensional figure. It is in more than one regard an effective metaphor. Somehow, the gravitational side is after some training becoming intuitively comprehensible  for our ape-brain: black holes  are employed with great effect in Hollywood movies. But the quantum world is inherently abstract, like the interference fringes. 

The real depth is however in the "holographic principle" formulated in the early 1990's by 't Hooft and Susskind stating that the number of degrees of freedom in a gravitational theory relates to that in a quantum field theory as if the former lives in a space-time with one less dimension compared to the latter. This is rooted in semi-classical black-hole wisdoms: the famous black hole radiation by Hawking. Depart from a classical (Schwarzschild) black-hole space-time and just insert quantum fields, to discover that the black hole turns into a black body radiator according to external observers. Similarly one can associate a Bekenstein-Hawking entropy to the black hole and this scales with the {\em area} of the black hole horizon. Field theoretical objects have an entropy scaling with the {\em volume}: here is the 'missing dimension'. 

Part of the 1997 excitement was due to the fact that AdS/CFT was the first mathematically controlled ploy that confirmed the holographic principle. The field  theory lives on the 3+1 dimensional "boundary" while the gravitational dual lives in the "bulk" 4+1 dimensions: the "$5$" in AdS$_5$ while S$^5$ refers to 5 extra dimensions that are rolled up (compactified) in tiny little circles forming a 5 dimensional "ball".

\begin{figure}[t]
\includegraphics[width=0.7\columnwidth]{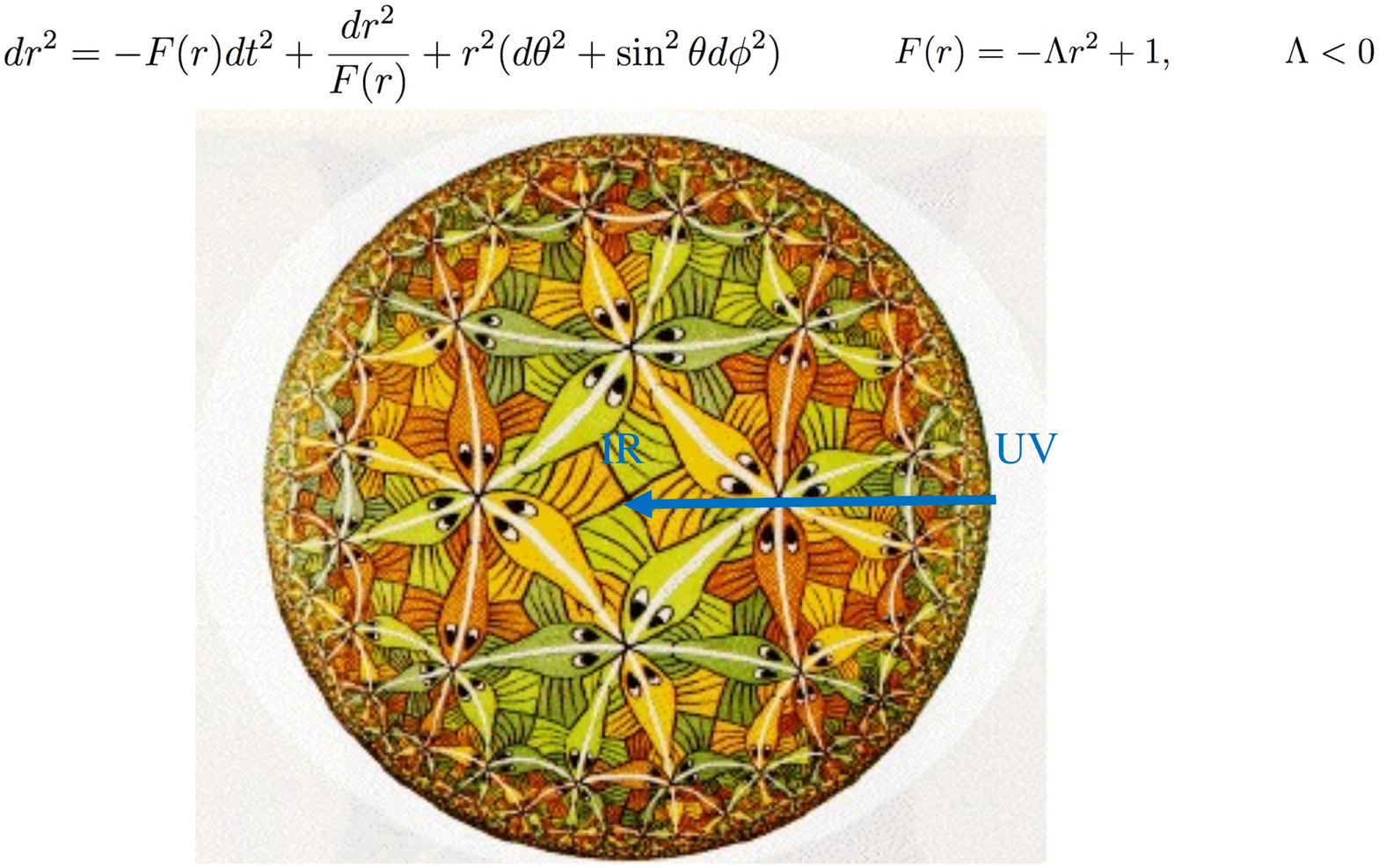}
\caption{Esher's graphical representation of the Poincare disc, a two dimensional cut through the hyperbolic Anti-de-Sitter space. The radial direction (blue arrow) is the extra dimension encoding for the scaling direction in the quantum field theory that lives on the boundary of AdS.}
\label{fig:poincaredisc}
\end{figure}

But this gravitational space-time is of a special kind: it has an "Anti-de-Sitter" (AdS) geometry. This is a maximally symmetric manifold characterized by a constant {\em negative} scalar curvature. It is a solution of the vacuum Einstein equations for a negative cosmological constant. It is a hyperbolic space with a curvature that can be represented by the "Poincare disc" representing the extra "radial dimension" (relative to the boundary) and in addition one of the space dimensions that its shares with the field theory. This was in turn famously visualized by the artist Esher with the "fishes", Fig. (\ref{fig:poincaredisc}). The radial direction is the radial coordinate of this disc (blue arrow).  Strangely, the drawing gives away that this is an infinitely large space having still a boundary: appealingly, the field theory lives on this boundary. 

Rooted in the Lorentzian signature one also encounters a typical GR "causal weirdness" motive which is actually crucial for the correspondence to work. Although the AdS boundary lies infinitely far away from the deep interior as measured along the radial dimension, a signal emitted in the deep interior arrives at the boundary after a {\em finite time lapse}. This is actually a necessary condition for the correspondence to work: the long wavelength physics in the boundary is determined by the bulk geometry in the deep interior and a causal connection is required to make this work.  

The figure also reveals a sense of fractality, scale invariance: the pattern of fishes repeats itself but the fishes shrink in size going from the middle ("deep interior") to the boundary. But this is not symmetry in the usual sense: it is "regularity" associated with the curvature of the manifold which is called "isometry" by the geometers. As an example, consider the surface of a perfect ball: in GR language, this corresponds with a two-dimensional de Sitter space in Euclidean signature. This is the other space characterized by a maximal isometry: the curvature around any point on this surface is identical. Esher just captures the unique feature of Anti-de-Sitter geometry: besides the maximal isometry that is shares with the ball this curved geometry is also {\em invariant under scale transformations}. 

This is very easy to understand. The quantity that is independent of choice of coordinates in Riemannian geometry is the {\em metric} $ds^2$. Choosing a radial coordinate $r$ that varies from $0$ in the deep interior to $\infty$ at the boundary and normalizing by defining the AdS radius to be $L=1$, the AdS metric can be written as,

\begin{equation}
ds^2_{\mathrm{AdS}} =\frac{1}{r^2} \left( \eta_{\mu \nu}   dx^{\mu} dx^{\nu}  + dr^2 \right)
\label{AdSmetric}
\end{equation}

It is very easy to check that under the scale transformation  $x^{\mu} \rightarrow \Lambda x^{\mu}, r \rightarrow \Lambda r$ the metric $ds^2_{\mathrm{AdS}}  \rightarrow ds^2_{\mathrm{AdS}}$. The metric is thus scale invariant -- you will find out soon that AdS is unique in the regard of this scale invariance of its curved geometry. 

Shortly after Maldacena's discovery, Witten played a key role in recognizing the key generality behind the correspondence. Mathematical geometrists had already realized a deep connection: the {\em isometry} of a curved space in d+1 dimension is linked to the {\em symmetry} encountered by e.g. field theories in d dimensions. The scale invariance of the AdS metric that I just highlighted imposes conformal invariance on the boundary field theory. For this mathematical precision reason,  "Anti-de-Sitter (AdS) has to correspond with Conformal Field Theory (CFT)". 

This turns into a no-nonsense computational device by the  "dictionary", prescribing in a precise mathematical fashion how to relate quantities in the boundary to those in the bulk, metaphorically like the Fourier transformation formula behind the particle-wave duality. This is derived from the "G(ubser)-K(lebanov)-P(olyakov)-W(itten)" rule that was discovered soon after Maldacena's demonstration. This dictionary is derived from a  concise "master equation", 

\begin{equation}
\langle e^{\int d^{d+1} x J(x) O(x)} \rangle_{CFT} = \int {\cal D} \phi e^{i S_{\mathrm{bulk}} ( \phi (x, r) )|_{\phi (x, r= \infty) = J(x)}} 
\label{GKPWrule}
\end{equation}

The left hand side refers to the boundary. This is the generating functional  yielding the information to compute propagators like $\langle O O \rangle$  by taking functional derivatives $\delta / \delta O(x)$. The right hand side represents the full quantum partition sum of the dual {\em quantum} gravity problem in the bulk. But the specialty is that the bulk field $\phi$ that is dual to $O$ in the boundary is constrained at the boundary of AdS to be coincident with the source $J$ of the boundary theory. Upon taking the large $N$ and 't Hooft coupling limit the r.h.s. turns into the classical saddle point. This is just the usual business of deriving the Einstein equations by extremizing $S$, in the plain-vanilla correspondence the Hilbert-Einstein action defining GR. 

A minimal example of how this works can be looked up in the introduction of Ref. \cite{holodualbook} and let me sketch here the outcomes. Depart from a scalar field $\phi$ with mass $m$ in the bulk and solve its EOM's in an AdS background.  You will find that upon approaching the AdS boundary (radial coordinate $r \rightarrow \infty$) this field falls off universally according to $f_k (r) = r^{-d -1 + \Delta} A(k) + r^{-\Delta} B (k)$ where $k$ refers to the energy-momentum in the boundary while $A(k)$ and $B(k)$ are the coefficients of the leading- and subleading components of $\phi$. 

As it turns out, the GPKW rule implies that the two point propagator of the operator $O$ in the boundary is given by $\langle O(-k) O(k) \rangle = B(k) / A(k)$,  the ratio of the subleading $B (k)$ ("response") to leading $A(k)$ ("source") boundary asymptotes of $\phi$ in the bulk. It follows that $\langle O(-k) O(k) \rangle \sim k^{2 \Delta - d -1}$.  But this is just a branch-cut, as you learned to appreciate in Section (\ref{qucritical})! $\Delta$ is here the scaling dimension of the operator $O$, according to the conventions of conformal field theory. This is determined by the classical EOM governing $\phi$: $\Delta = \frac{d+1}{2} + \sqrt{ \frac{d+1)^2}{4} + m^2 L^2}$ where $L$ is the AdS radius (setting the overall scale) and $m$ the mass of the bulk field. 

The take home message is that the {\em mass} of the field in the bulk is dual to the {\em anomalous dimension} of the operator $O$ in the boundary! It is in fact bounded by $m^2=0$ where one finds the "engineering" dimension of the free critical theory, Eq. (\ref{freecrtitfield}). Such massless bulk fields are typically associated with conserved currents in the boundary, like the  total momentum in the homogeneous space, that are subjected to "hydrodynamical protection".   
   
This illustrates the rather counterintuitive "magical mechanism" implied by the GKPW rule. Upon further exploring it, one finds that pure AdS encodes for all the generic traits of a Lorentz-invariant, zero density strongly interacting quantum critical state of the kind as you met in Section (\ref{qucritical}). However, the structure of the duality permits the anomalous dimensions to become anything: it is just pending the mass of the bulk field. These are fixed departing from "top-down" precision set ups. But what matters here is that the zero density point of departure -- the pure  CFT -- is of the strongly  interacting "quantum-supreme" critical kind as testified by the anomalous dimensions as well as universality revealed by the simple scaling forms ruling in the boundary. 

The {\em structure} of the renormalization group flow with its ensuing ramifications for the boundary (the scaling theory) is hard wired in the geometry of the bulk and only the numbers (anomalous dimensions) are pending the specifics of the UV theory (large $N$, etcetera). The take home message of the remainder will be that this pertains to the much more intricate circumstances one meets at finite temperature and especially the "signs infested" finite density circumstances. 

The above example is the most elementary GR exercise one meets in the holographic duality agenda. Upon rolling out the amazingly rich full AdS/CMT portfolio the basic moral of this example does however repeat itself systematically. Some intricate property of the boundary theory under "real life" conditions translates typically into a gravitational problem to solve in the bulk that you would not guess at all beforehand. In no time one runs into fancy, state of the art GR computation and I will keep this largely implicit in the remainder, at best I will sketch it in words. To get into it seriously, study the books and do the exercises. 

The richness of physical reality "at the boundary" has its reflection in the bulk in the form of a plethora of fields. Using the full string theoretical machinery it is possible to derive a zoo of gravitational set ups that are in a mathematically precise sense dual to "physical" boundary field theories, where all the effective potentials etcetera are quantitatively determined: the "top-down set ups". However, eventually the {\em structure} of these bulk theories is determined by the greater principle of "effective field theory". This is just nothing else than Landau's strategy to derive coarse grained free energies/actions based on symmetry principle. One can leave all the numbers as parameters and this defines the "bottom-up" agenda. Matters like the mass of the bulk fields/scaling dimensions in the boundary are just kept as free parameters while the general nature of the boundary physics is deduced in terms of {\em phenomenological} theory: the "bottom up" agenda. 

Ironically, it has happened repeatedly that particular effective field theory constructions were discovered in top-downs to subsequently get "lifted up" to the bottom up generality. Without this string theoretical guidance one just not gets the idea to look for such particular constructions. To find out how this works, have a look at  the "rasta black hole hairs" discussed in Section  (\ref{Intertwined}). 

A somewhat annoying aspect for condensed matter applications is in the fact that holographic boundary matter is by default {\em ultrarelativistic}, characterized by vanishing rest mass. It is just not known how to hardwire finite rest mass gravitationally. To get some idea regarding the difficulties, you just learned that mass in the bulk translates into the scaling dimensions of massless "stuff" in the boundary. This may give rise to pathological features when addressing physics in the boundary "descending" from a large rest mass UV physics, see e.g. Section (\ref{Homosecsector}). 

Delving deeper in the correspondence, there is yet another twist in the symmetry logic connecting bulk and boundary, all rooted in the "isometry-to-symmetry" relation. Global symmetry in the boundary is dual to {\em gauge} symmetry in the bulk. Another way of comprehending this statement is that gauge fields are just a mathematical convenience to capture {\em curvature}. The physical property in Maxwell theory is (electric-, magnetic) field strength being gauge the curvature. In GR the curvature is captured by the Riemann tensor, while the gauge freedom is associated with the choice of coordinate systems  (general covariance, diffeomorphism invariance). The action in the bulk is all about curvature (isometry) that then translates in the global symmetries controlling the boundary.  

A first example is associated with the  most basic symmetry of the boundary: its space-time itself.  Assuming a homogeneous and isotropic flat space-time, invariance under {\em global}  Lorentz transformations is then ruling. The associated Noether charges are collected in the energy-stress tensor describing the boundary matter. How to identify the gravitational dual associated with energy-stress? The key is yet again in the isometry-symmetry relation: it dualizes in the {\em metric} fluctuations of the bulk geometry. For instance, you will soon encounter the response to an applied shear stress. Shear is spatial spin 2 and the associated bulk fields correspond literally with {\em gravitational waves} -- "gravitons" -- propagating from the boundary to the deep interior of the bulk. 

But there is more physics to worry about in the boundary. In the next section we will focus on finite density, referring to stuff characterized by a conserved total charge associated with   an internal global $U(1)$ symmetry. The correspondence is rooted in string theory. It is axiomatic in string theory that such internal symmetries are descending from geometry through {\em Kaluza-Klein compactification}. This refers to a 1920's discovery:  departing from GR in 5 space-time dimensions, upon rolling up one of the dimensions into a small "$U(1)$" circle at large scales an effective theory  arises corresponding with  Maxwell electrodynamics in combination with GR describing  a four dimensional extended space-time. 

In the description at the beginning of this section we referred to the gravitational bulk of Maldacena's correspondence as AdS$_5 \times$S$_5$. It descends from a ten dimensional manifold, while  5 dimensions are compactified in a "ball" giving rise to a zoo of conserved "gauge charges". As geometrical curvature is the invariant under coordinate transformations (general co-variance), through the Kaluza-Klein construction {\em gauge curvature} -- electrical- and magnetic field strength in Maxwell theory -- turns into the invariant associated with  the compactified dimensions. The take home message is that the {\em globally} conserved internal charges in the boundary are dual to Yang-Mills {\em gauge fields} in the boundary. Density propagators are computed by {\em electro-magnetic waves} --  "photons" --  propagating in the bulk. The big deal in  Section (\ref{holoSM}) is that {\em finite density} in the boundary is associated with electrical fields emanating from the deep interior, piercing the boundary. These require an electrical monopole charge located in the deep interior -- this is the central principle behind the finite density strange metal holography. 

But there is yet one other entertaining dictionary entry that is rather easy to understand conceptually and of grave consequence: how to deal with finite temperature? The holographic magic that I just pointed out turns out to work amazingly well. Let me conclude this very short introduction in AdS/CFT with a sketch how this works.       

\subsection{Plain vanilla AdS/CFT and finite temperature.}
\label{FiniteTMaldacena}

Early in the development it was realized (again) by Witten how to encode the boundary finite temperature in the bulk. The answer: insert a Schwarzschild  black hole in the deep interior (the "middle") of AdS! This excited the string theorists because it linked AdS/CFT directly to the portfolio of black hole quantum physics: Hawking radiation, Bekenstein-Hawking black hole entropy and so forth. 

I already alluded to the holographic principle inspired by the Bekenstein-Hawking entropy of the "thermal" black hole. This set of ideas is elementary and let me summarize the essence in a bit more detail.  Take a classical, Schwarschild black hole geometry and implement free quantum fields in this space time. Given the  "accelerating" geometry near its causal horizon (Unruh effect), using the standard rules of quantum mechanics an external observer will perceive the black hole as a black body radiator with a temperature set by the {\em area} of the horizon. One can deduce the entropy of such a thermal object and this is the Bekenstein-Hawking entropy. Divide the horizon in cells with a dimension set by the Planck length, to insert one bit of information in each Planck-cell: the entropy of a macroscopic black hole turns out to be actually very large. But what matters most is that this entropy scales  with the {\em area} of the horizon and not with the {\em volume} that the black hole occupies. Once again, in quantum field theories entropy always scales with volume and it is as if a dimension is missing in the gravitational theory -- the root of the idea behind the "holographic" principle. 

This holographic entropy affair gets bigger than life in the correspondence. Mathematically it is a surprisingly simple and elegant affair. It is a matter of holographic principle that {\em equilibrium} in the boundary is dual to a {\em stationary} geometry in the bulk. Given this circumstance one can Wick rotate  both in the boundary and the bulk to Euclidean signature. GR is a-priori not relying the signature of time, although there is no causality structure in Euclidean signature. But there is no causality either in the boundary when it is in strict equilibrium. Insert now a Schwarzschild black hole in the deep interior.

\begin{figure}[t]
\includegraphics[width=0.7\columnwidth]{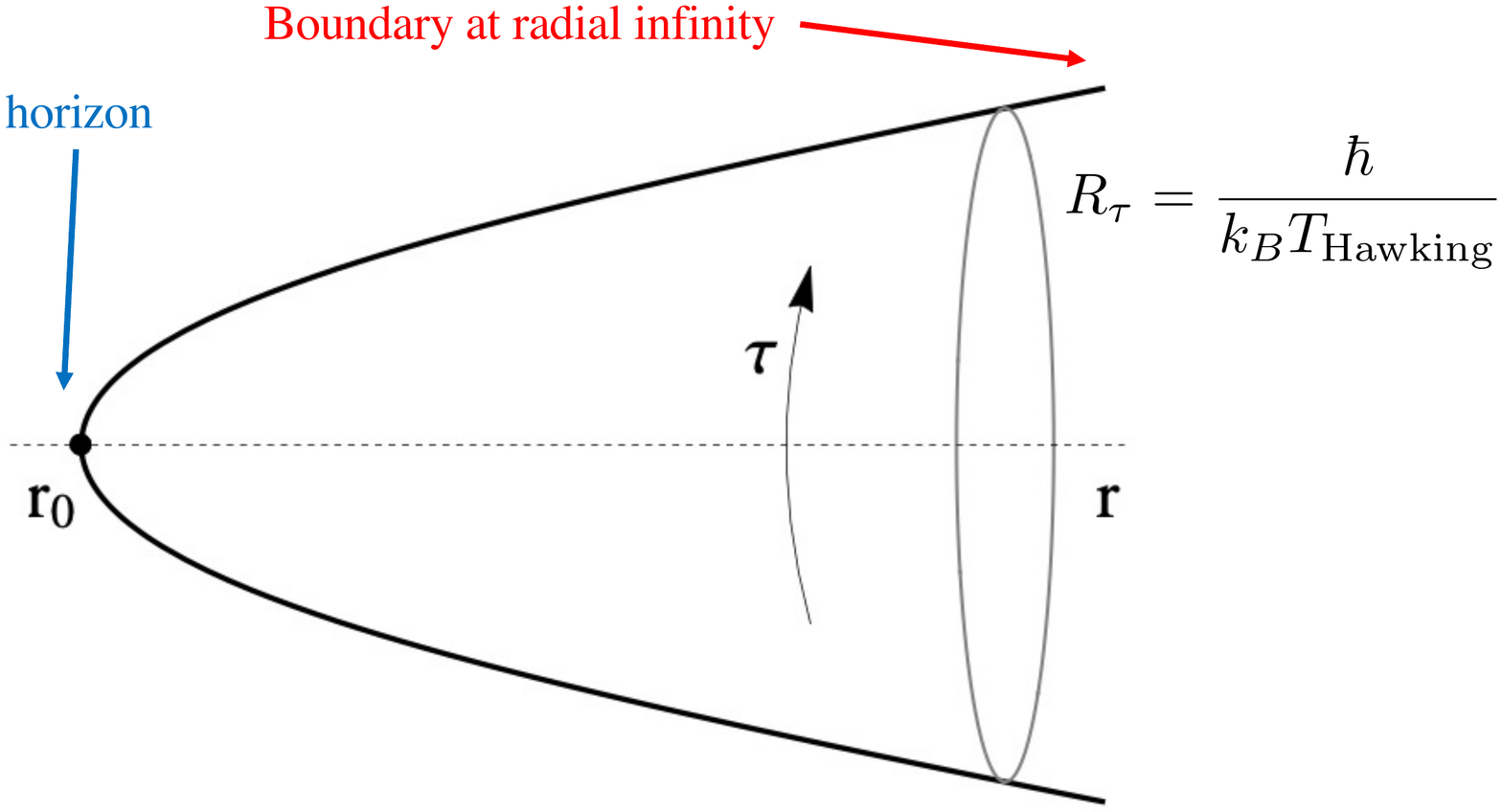}
\caption{ 
The Schwarzschild black hole metric in Euclidean signature. Euclidean time ($\tau$) forms a circle starting at the horizon $r_0$ with zero radius, growing as function of the radial coordinate ($r$) to asymptote at radial infinity where the boundary resides with a radius set by the Hawking temperature ($R_{\tau} = \hbar / (k_B T_{\mathrm{Hawking}})$. Accordingly, this determines  the temperature of the field theory in the boundary.}  
\label{fig:Hawkingeucl}
\end{figure}

One can evaluate the Schwarschild metric in Euclidean signature by analytical continuation, with the outcome illustrated in Fig. (\ref{fig:Hawkingeucl}). Focussing on the radial direction -  imaginary time dimensions one finds that the latter rolls up in a circle that shrinks to zero radius at the horizon. This circle opens up moving away from the horizon to acquire a radius set by the Hawking temperature at infinity. But the AdS boundary is at infinity: the boundary field theory lives in a geometry where the imaginary time forms a circle with radius $\tau_{\hbar} = \hbar/k_B T$. This of course rings a bell, Section (\ref{Planckiandiss})! Yet again the "isometry = symmetry" logic is at work: the "finite size  scaling"  effect of temperature in the boundary is encoded entirely in the near horizon geometry of the bulk black hole. 

The bottom line is that all known principles of the boundary thermodynamics are impeccably reproduced by the bulk thermal black holes. The entropy counts in the correct way in the elementary CFT setting as it should, according to $S \sim T^d$. But we will find out that dealing with  the rich "hairy" black holes of AdS/CMT the thermodynamic principles associated with phase transitions are impeccably reconstructed as well. E.g. the thermodynamics of holographic superconductors near the thermal phase transition is precisely captured by the Ginzburg-Landau free energy functional (Section \ref{HoloSC}).

Perhaps the most impressive result is that this success also extends to the dissipative {\em dynamics} of the macroscopic fluid formed at finite temperature. For any finite temperature liquid at macroscopic times living in a homogeneous and isotropic background we have a universal theory that dates back to the 19-th century: Navier-Stokes hydrodynamics. This describes the responses of the fluid under non-equilibrium conditions. The Lorentzian signature now matters: one has to evaluate gradient-expansion "space-time quakes" in the near horizon geometry encoding for the "deep IR" (macroscopic times and length) in the boundary.  Upon pulling this via the dictionary to the boundary the outcome is -- it impeccably reproduces the mathematical {\em structure} of Navier-Stokes! Hydro is therefore a special corner of GR. This actually poses a significant problem to the mathematicians. There is a Clay institute "milennium prize" on Navier-Stokes but not on GR ...

However, the parameters of the hydrodynamical theory are not standard departing from AdS-Schw geometry (Schwarzschild Black-hole in AdS).  The crucial part is the dissipation, encapsulated by the {\em viscosity}. Departing from a conformally invariant UV the bulk viscosity has to vanish for symmetry reasons (be aware of such "UV dependence") and one is only dealing with {\em shear} viscosity. 

This can be easily computed in linear response and the outcome is yet another simple marvel. Shear viscosity is sourced by shear stress and I already announced  that the dictionary insists this to be encoded by the gravitons propagating in the bulk. Anything dissipative in the boundary turns out be dual to bulk stuff falling through the horizon. The viscosity is therefore dual to the {\em absorption cross section of zero frequency gravitons by the black hole}. This is obviously proportional to the {\em area} of the horizon. But the entropy density $s$ is also proportional to this area and one finds that these are related by a geometric $1/ (4 \pi)$ factor. The result is the "minimal viscosity",

\begin{equation}
\frac{\eta}{s} = \frac{1}{4 \pi} \frac{\hbar}{k_B}
\label{minviscosity}
\end{equation}

Can it be simpler? Viscosity is just set by $\hbar$ and the entropy. As it turns out, this ratio is extremely small as compared to what is established in typical normal, "molecular" fluids like water and so forth. But the  crux is of course that this is a classical  fluid that is formed from strongly interacting quantum critical matter. We learned that this is subjected to the intrinsic extremely rapid thermalization in Section  (\ref{Planckiandiss}). In fact, the minimal viscosity is just Planckian dissipation in disguise. 

This follows from elementary dimensional analysis. The dimension of viscosity is set by the free energy density $f (T)$ and the characteristic momentum relaxation time associated with the presence of finite spatial gradients, $\tau_P$: $\eta (T) = f(T) \tau_P$. Dealing with a quantum critical state there is no internal energy and the free energy density is entirely entropic $f(T) = s T$. Stick in $\tau_P \simeq \hbar / (k_B T)$ and it follows that $\eta \simeq (\hbar/k_B) s$! One needs the classical  black holes  (large $N$ limit) only to determine the parametric factor $1/(4 \pi)$. 

Notice an aspect that may be at first sight confusing. Planckian dissipation means "the best heater that can be realized in principle". But this leads to an extremely small viscosity, which is in turn governing how the moving fluid converts its kinetic energy into heat. The secret is in the fact that viscosity is associated with momentum dissipation and it is proportional to the very short Planckian time and not the Planckian rate. For instance, as I already discussed in Section (\ref{FermiliquidfinT}) the viscosity in the $^3$He becomes very large in the low temperature Fermi-liquid regime. This is rooted in the extremely "gaseous" nature of this fluid,  with the momentum dissipation time set by the very long collision time  $\tau_c \sim E_F \hbar /(k_B T)^2$.  

The discovery  of the minimal viscosity by Kovtun, Son and Starinets in 2004 signals the start of the development that turned into the AdS/CMT portfolio. It caused a big splash in 2005 when evidences appeared that it is actually at work in nature. The context is the quark gluon plasma. This refers to an affair that has been heroically pursued for a long time in high energy physics. Collide heavy nuclei at a very high speed in high energy accelerators and the ensuing fireball may get so hot and dense that one gets in the deconfining regime of QCD. One anticipates that a plasma may be  realized of free-ish quarks and gluons. When the explosion time permits this may eventually behave as a hydrodynamical entity. 

Around 2005 firm evidences started to appear at the RIHC facility in Brookhaven that such a plasma is indeed realized, behaving in hydrodynamical ways. Remarkably, the experimentalists managed to measure the $\eta/s$ ratio. Based on QCD perturbative theory (morally, kinetic gas theory) one anticipated a value for this ratio that is order of magnitudes larger than what was measured. Instead, it turned out to be very close to the minimal viscosity prediction of holography. It is still a bit shady why QCD in this regime should give in to "conformal principle" -- QCD is about running- and not marginal couplings but one may envisage that at the temperatures reached in the heavy ion collision both the "asymptotic freedom" (kinetic gas physics)  and "infrared slavery" (confinement) are "balancing" in an effectively "quantum supreme" physics. 

Since 2005 this "fluid-gravity correspondence" turned into a subfield of its own. Among others it was used to sort out higher gradient hydrodynamics. Proceeding phenomenologically this turns into a complicated affair but using the gravity dual it becomes rather mechanical, solving the gravity equations   systematically. In this way, it was used to detect for instance subtle flaws in the derivations of the Landau hydrodynamics school, but also shedding light on tricky themes like how to handle entropy currents in relativistic regimes. It also works the other way around. The small viscosity implies that turbulence is around the corner: this is dual to "turbulent" near horizon geometry in the bulk. This is technically quite hard to deal with but there are reasons for such turbulent horizon to be perhaps of relevance to black hole mergers. 

All of this is described in detail in the textbooks \cite{Erdmengerbook,holodualbook,lucasbook}.  The greatest intrigue is perhaps associated with very recent  developments \cite{SonnerLiuRev} that unfolded after the books were written. Using special "out of time" correlation function one can lay its hand on the short time "quantum chaos". One can identify a "quantum Lyaponov ("scrambling) time" that in essence measures how long it takes for an observable to get completely lost in the exponentially large Hilbert space. It can be shown that this is also set by $\tau_{\hbar}$ in the holographic fluids. 

Amazingly, together with the "butterfly velocity" expressing how fast this chaos spreads in space a holographic mathematical relation ("pole skipping") shows that the macroscopic viscosity is completely set by these short time chaos quantities! This is perhaps best understood as reflecting the very rapid eigenstate thermalization in these maximally many body entangled CFT's. In essence, the hydro that requires local equilibrium already sets in at the short scales of the chaos and nothing new happens until one reaches the macroscopic scale. It is actually a bit ironic that the belief is relatively widespread in the community working on fluid gravity that this may tell stories about hydro in general. I can assure that nothing of the kind is going on in water -- it is all about quantum critical systems and the hydro associated with pure CFT's has not been identified in the laboratories.

\subsection{Large $N$ versus the UV independence of the structure of the zero density scaling theory.}

To conclude this section, the take home message is that the success of this  thermal agenda  raises the confidence in the "holographic oracle". Despite the large $N$ etcetera special conditions formally required to rely on classical gravity, the correspondence  automatically encodes for the correct {\em structure} of the phenomenological theories associated with the boundary. The {\em numbers} are in hindsight also reasonable -- only parametric factors like the $1/ (4 \pi)$ in the minimal viscosity are pending manifestly on the infinite $N$ limit. 

I never got a clear answer why classical bulk gravity works so well in reproducing the universal aspects of this zero density affair. Anything that is going on in the physical universe is far from this large $N$ limit, with its classical gravity bulk. As I emphasized, for "physical" small values of $N$ the bulk is supposed to be governed by a quantum geometry that is still in the dark, but this does not seem to matter  for the {\em structure} of the theory predicting physical observables -- the branch cut propagators at zero temperature, the Planckian dissipation at finite temperature, the thermodynamics and so forth. 

So much is clear that it is all controlled by {\em symmetry}, and zero density holography is just an impeccable symmetry processing apparatus somehow revealing the most general way this controls observables. The conformal invariance constrains the form of the thermodynamics, while it imposes the branch cut form on the $T=0$ propagators leaving the anomalous dimensions  as free phenomenological parameters. Even less obviously, it works as well for the finite temperature physics as implied by the finite size scaling of the Euclidean time circle, eventually turning into Navier-Stokes due to impeccable encoding of conserved Noether currents and the dissipation mechanism. 

We will next dig into the fermion sign invested finite density holographic portfolio. In this area we cannot check it against facts we know directly from the QFT side since nothing is known for sure. Is the same magic at work as at zero density, perhaps indicating that in densely entangled quantum supreme matter symmetry principle is governing everything that can be measured to an even greater degree that we are used to in conventional semiclassics? As we will see, the nature of this "holographic" symmetry principle does change drastically at finite density translating in a different type of "quantum critical" phenomenology. But is it reliable?  This is the big question that is begging for an answer.

\section{The revelation: the holographic strange metals.}
\label{holoSM}

We are done with the preliminaries. As announced in the introduction, at stake is whether universal emergent physical principle may be at work dealing with non-stoquastic, quantum supreme finite density matter.  Once again, this cannot be emphasized enough: because of the  signs we used to be mathematically blind.

This section should be viewed as the core of the narrative that I attempt to capture in these notes. It deals with {\em  finite density holography}. The point of departure is the zero density "Maldacena" state that I highlighted in the previous section. As I explained, this is a zero rest mass Lorentz invariant system  characterized by stoquastic quantum supremacy as testified by e.g. the anomalous dimensions. The dictionary spells out what is required in the bulk in order to raise the chemical potential in the boundary to obtain a {\em finite} density in the boundary: one has to accommodate an electrical monopole charge in the deep interior. This is like raising the chemical potential in a strongly interacting critical incarnation of graphene physics. The outcome is a state that is "infested by fermion signs" and since {\em  the zero density matter is already densely many body entangled one is not surprised that the finite density metal is also quantum supreme}. 

The outcome is that we get a glimpse of what is going on behind the fermion-sign brick wall although it is tied to specific UV circumstances that are different from what is encountered in the electron systems of condensed matter physics. However, on closer consideration one can discern various meta-principles rooted in strong emergence that may well be general to the degree that these also are at work in the laboratory systems. This is what I wish to explain in the remainder.  

This "AdS/CMT" affair developed rapidly in the string theory community in the period 2007-2013. The striking part that energized it to quite a degree is that with the eyes half closed the gross physics of these strange metals is quite similar as to that of Fermi-liquids. Initially it was a rather subconscious affair: by just experimenting around with finite density AdS/CFT, the "black holes" produced signals that were somehow familiar. 

In fact, this development accelerated by a development I was myself directly involved: the holographic "photoemission" showing the presence of Fermi-surfaces \cite{Sciencefermions09}. Although it became clear later that this is one of the occasions ruled by large $N$ UV sensitivity, it stressed the similarity with the familiar condensed matter fermiology.  Importantly, there is no doubt that the meaning of finite density in holography involves {\em finite density of fermions}, see Section (\ref{Holofermions}).   

It took myself quite a while before I became fully aware of the origin of this intuition. The revelation is that the similarity of the strange metal- and Fermi-liquid phenomenology originates in a particular type of renormalization group structure. This "covariant" RG is a rather recent insight, that was realized first by Gouteraux \cite{Gouterauxcov}. I am still in the process of fully digesting its scope -- I may be in this regard a bit ahead of the holographic mainstream where it is presently not yet widely disseminated.This will be in the limelight in the most important passage of the lecture notes: Section (\ref{EMDscaling}). Amazingly, resting on gravitational universality the claim is that the structure of this "covariant renormalization group" can be completely classified for the holographic metals. 

The weakly interacting Fermi-liquid fits this scaling logic as a special case, having a similar status as the free (Landau) critical theory in the quantum phase transition context. But this generalizes dealing with the quantum supreme metals of holography. Universality in the same guise as encountered in the strongly interacting critical state is in effect, leading to {\em simple} scaling behaviour that is yet again governed by {\em anomalous} scaling dimensions. These are however of a completely different kind than those that are at work in the Widom-Kadanoff scaling theory of Section (\ref{thermaluniversality}). 

To highlight the notion that holographic strange metals should be viewed as quantum supreme {\em generalizations of the Fermi-liquid} I will discuss the properties of these strange metals, by going down a similar list as for the Fermi-liquid in Section  (\ref{Fermiliquid}). There is so much going on with transport at finite temperature that this deserves a section of its own (Section \ref{holotransport}). The similarity with the Fermi-liquid also extends to the way strange metals become unstable in a BCS-like fashion (Section \ref{HoloSC}). But this generalizes beyond the Fermi-liquid canon in the form of baroque black hole hairs dual to the orchestrated "intertwined order" in the boundary, also worthy a separate Section (\ref{Intertwined}). 

To prepare the mind for what is coming, let me start out with an appetizer revolving around basic motives that one encounters dealing with fermionic finite density matter. It reflects my present understanding of what is going and it is surely conjectural. It is about the reconciliation of the degeneracy scale that cannot be avoided dealing with fermions and the absence of scale that appears to be a prerequisite for dense entanglement (Section \ref{fermiondeg}). To further raise the appetite I will then tell the story of how it started, in the form of the pathological yet revealing "Reissner-Nordstrom" (RN) strange metal (Section \ref{RNearly}).  

\subsection{Reconciling scaling and the fermionic degeneracy scale.}  
\label{fermiondeg}

The ingredient that appears to be unavoidable in finite density fermion systems of any kind is the notion of a {\em fermionic degeneracy scale}. In the discussion of non-stoquastic physics  in Section (\ref{fermionsigns}) it may already have occurred to the reader that  fermion signs should have  as universal consequence that the ground state is characterized by a large zero point motion energy. In the case of the Fermi-liquid this is of course obvious: given the simple Pauli principle for free fermions it follows immediately that the Fermi-energy represents this zero-point motion energy.  The first quantized "winding number" path integral offered a complimentary view. We found that the "negative probabilities" have the effect that as compared to the stoquastic case the ground state energy is pushed upward -- the alternating-in-signs sum over winding numbers encoding for the Fermi energy. 

This should however also be the case dealing with quantum supreme ground states. The many body ground state wave function will exhibit many nodes in position representation associated with sign changes. This will imply that the kinetic energy is pushed upwards resulting in a large degeneracy scale. {\em Fermionic quantum supreme states of matter should be invariably characterized by a fermionic degeneracy scale}. But we directly infer a tension. We learned that "conventional" scales like gaps, or either the rigidity scale associated with order, are instrumental in suppressing the dense many body entanglement.

But there is a loop hole.  As I emphasized in Section (\ref{Fermiliquid}), {\em physical observables of the Fermi-liquid are typically power laws}. This is rooted in a continuous RG flow towards the infrared which is not interrupted by scales (gaps). But this flow is distinct from the RG encountered in the CFT's which is controlled by {\em invariance} under conformal transformations.  This is most notable in the fact that the Fermi-energy can be discerned from macroscopic measurements. E.g. the Sommerfeld specific heat tells immediately the value of the (renormalized) Fermi energy -- in a CFT the specific heat only reveals IR data.  

In Section ({\ref{Fermiliquid}) I presented the Fermi-liquid in the conventional text book style, except then for some quantum information inspired second thoughts. The essence of this affair is: count out the number  of electron-hole excitations spanning up the Lindhard continuum and employ these to compute quasiparticle life-times using perturbation theory. There is a long tradition approaching the Fermi-liquid in an RG language but as I stressed this is yet again tied to the specific weak coupling perturbation theory shedding light on the local stability of the free Fermi gas IR fixed point. All along it has been obvious that these RG flows are of a different kind than the ones encountered at phase transitions. But how to capture the difference in a general language? 

At this point holography interferes. I already highlighted the "RG = GR" principle: the RG of the boundary theory acquires a {\em geometrical} image in the bulk. A first take home message will be that the geometries encoding for finite density matter invariably wire in the equivalent of the Fermi-energy -- this is just called the chemical potential $\mu$. the other message is that  these translate in scaling behaviours in the boundary that follow {\em closely the template of the Fermi-liquid RG flow} of the previous paragraph. 

The revelation is that the structure of this RG is captured by a very simple property of the deep interior bulk geometry. As the {\em invariance} of the bulk metric under scale transformation hardwires the scaling properties of the zero density CFT, the finite density states are encoded universally in holography by {\em covariance} of the bulk metric under scale transformation (Section \ref {EMDscaling}). This means that under scale transformation the metric is not identical to itself (as in AdS), but instead it is {\em proportional} to itself. 

This turns out to have the consequence that the knowledge of the degeneracy scale ($\mu$, $E_F$) is remembered in the RG flow, and the associated scaling behaviour. Taking this "covariant RG" as central principle one can identify a set of "primary" scaling dimensions of a different kind than the ones that are at work in CFT's.  Macroscopic properties are claimed to be governed by the so-called {\em dynamical critical exponent} $z$, the {\em  hyperscaling violation dimension}  $\theta$ and the {\em charge exponent} $\zeta$. The first one has an existence also in the stoquastic case taking however at finite density values that are unheard off in the statistical physics portfolio. The last two are unique to finite density. These can be identified in the Fermi-liquid where they acquire "engineering" dimensions. The finite density covariant geometries are claimed to be classifiable, setting bounds to the values these can take. The outcome that these scaling dimensions may become highly anomalous.

But this "covariant RG"  scaling is  a fragile affair in the Fermi-liquid because of the perturbative corrections -- as for the free critical fixed point a manifold of irrelevant operators switches on upon ascending from the IR fixed point. Dealing with the holographic strange metals universality sets in: absence of perturbative complications, resulting in simple scaling behaviour as for the strongly interacting quantum critical states. I will illustrate this dealing with the BCS-like superconducting instability in Section (\ref{HoloSC}. I argued that the simplicity underlying "stoquastic universality" is rooted in the perfect averaging of the observables (VEV's) leveraged by the exponential complexity of the states in the strongly interacting critical state. I conjecture that the key to decoding the "whispers of the holographic oracle" is in the realization that the same perfect averaging is also the "first law of the RG" governing the finite density quantum supreme states of matter.   
 
This "covariant RG" is like the egg of Columbus. It is so simple and natural in the way that it captures the marriage between "critical" like behaviour (physical properties being power laws) and the Fermi-degeneracy scale that I am tempted to view it as grand general principle. Let me now put some meat on these bones.

\subsection{The first steps: the Reissner-Nordstrom strange metal.}
\label{RNearly}

As I announced, I will not elucidate the detailed workings of the holographic duality -- the gross "landscape" as sketched in the previous section should suffice to give you a sense of orientation and I will just state the various dictionary entries that are required, and sketch what is going on in the gravitational dual. You can look up in the textbooks \cite{holodualbook,lucasbook} how it really works. Once again, it is a tight mathematical framework: solve the Einstein equations, unleash the dictionary and there is no room left to tamper. This is not like the "scenario" theories that are a mainstream in this branch of condensed matter  physics.

The plain-vanilla correspondence that started with Maldacena is at the end of the day about the CFT's, as associated with "bosonic" quantum phase transitions that occur at zero density. What has the dictionary to say about {\em finite} density? The answer is very simple: an electromagnetic monopole electrical charge has to be accommodated in the deep interior of AdS. The dictionary spells out that the electrical field lines emanating from this charge when they pierce through the boundary are associated with the charge density in the latter. The chemical potential of the boundary becomes finite. 

But what kind of stuff is sourced by this chemical potential? This is a tricky affair, and I have perceived it  myself all along as one of these items where one has to be prepared for "UV sensitivity". It will become clear soon that this finite density holography is revealing strong emergence: the whole is so different from the parts that the nature of the parts may no longer matter at all. It better be so because the UV stuff of holography (large $N$ Yang-Mills, etcetera) has no relation whatsoever with the Schr\"odinger equation "chemistry" of the electrons on the lattice scale.  Once again, the quote "UV sensitivity" refers to a breach in this central principle. Especially symmetry related affairs hard wired in the UV may turn the IR to be special, and thereby no longer generic. At several instances we will meet this in the remainder.

Surely, at zero density there are no particles to count: it is the "unparticle" CFT quantum soup. How to think about quantum statistics? In the minimal set up departing from Maldacena's large N CFT we do know that the boundary is controlled by maximal supersymmetry where the bosonic and fermionic fields are in a perfect balance, with the consequence that no fine tuning of coupling constants is required for conformal invariance.

 In addition, it is also understood that this is pertaining to fields belonging to the adjoint of the YM theory: this refers to the force carrying "gluons" that because of the supersymmetry appear as bosonic- and fermionic incarnations that are in a perfect balans.  The quarks are "in the fundamental": these can be incorporated in holography as well but this involves more fanciful "brain intersection" holographic constructions \cite{Erdmengerbook}. I will largely ignore these: it involves constructions of a higher mathematical sophistication describing a more complex physics, while up to now I have not quite seen outcomes that shed more light on the relations with condensed matter experiment. This is of course not a really good reason but it is just the state of the field. 

Supersymmetry is very fragile and the most brutal way to unbalance the fermions and bosons is by going to finite density.  But what is "pulled in" by the chemical potential? I never got a clear answer to this question. By looking at the outcomes it just appears that "there are a lot of fermions". Perhaps the best evidence is in the form of proof of principle for a Fermi-liquid to appear as an instability of the strange metal -- I will come back to this still incompletely understood electron star affair near the end of this Section. But what happened with all the bosonic fields that are present at zero density? It may be that this "doping of supersymmetric stuff" may go hand in hand with an undesired UV sensitivity. I perceive this as one of the reasons to never trust completely the finite density holographic results. 

So we have to accommodate an electrical charge in the deep interior. The simplest possible way to accomplish this in gravity is by departing from an Einstein-Maxwell bulk action. This is the familiar GR textbook affair of considering a universe characterized by gravity (GR, "Einstein theory") and the presence of electromagnetism (Maxwell).  There is then one unique outcome that is stationary: the "classic" charged black hole, already solved in the 1920's by Reissner and Nordstrom (RN). 

\begin{figure}[t]
\includegraphics[width=0.7\columnwidth]{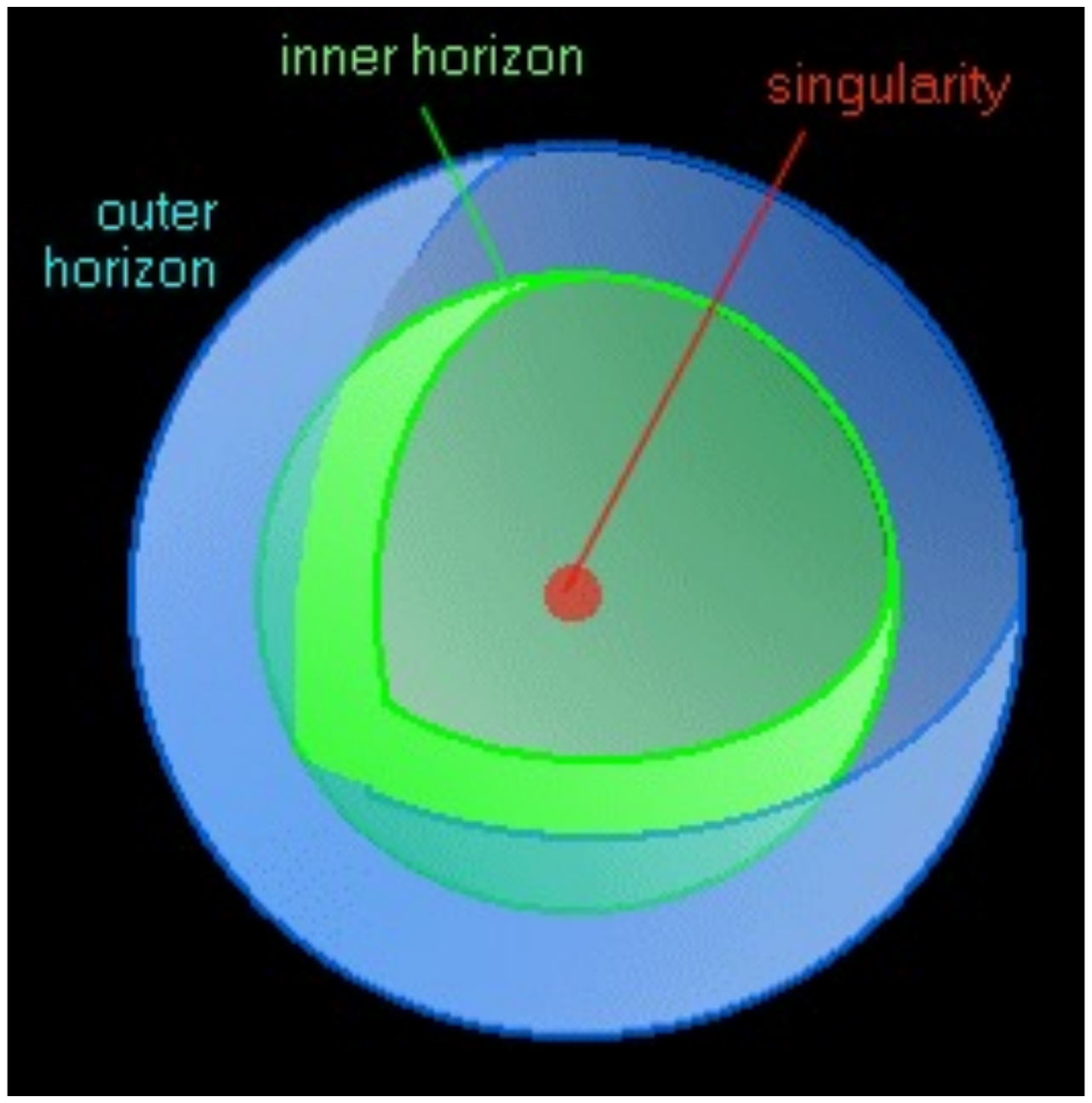}
\caption{The "finite temperature" electrically charged Reissner-Nordstrom black hole with its double horizon structure.}
\label{fig:RNblackhole}
\end{figure}

This is part of the basic GR repertoire as discussed in textbooks. While it is easy to derive the solutions,  such RN black holes have quite a bit more structure than the elementary Schwarzschild version (Fig. \ref{fig:RNblackhole}). These have {\em two} horizons: an inner- and outer one. Upon passing the outer horizon the geometry switches from time-like to space-like, flipping back to a time-like inner region, implying that the singularity can be avoided -- it even repels finite rest mass observers. But there are more puzzling features that appear to still confuse even the specialists. The inner horizon turns out to be a Cauchy horizon which means that it is not defined how to extend a geodesic through this horizon. In addition there is a famous conundrum called "mass inflation" associated with the inner-horizon: a debate has been raging since forever whether such black holes can be stable at all in the presence of in-falling energy. But these difficulties are avoided in AdS/CFT -- the boundary has no knowledge regarding what happens behind the horizon. 

This story continues by showing that a limit can be reached where the energy of the black hole coincides with the electromagnetic energy: the "extremal" RN black hole. The two horizons merge in a double coordinate singularity and the surface gravity is vanishing. A principle of black hole radiation is that the Hawking temperature is set by this surface gravity and it follows immediately that the extremal BH encodes for a {\em zero} temperature, finite density state of the boundary matter. 

But the horizon area is still finite, and although the Hawking temperature vanishes there is therefore still entropy set by the area of the horizon according to the Bekenstein-Hawking entropy governing the boundary field theory. The RN strange metal is characterized by {\em zero temperature} entropy and this is a pathological feature. In chemistry the idea that zero temperature entropy is not admissible in physical systems is  sometimes called the "third law  of thermodynamics". But this is overstated: it is a pragmatic affair, any small influence that will be overlooked will have the effect of lifting the ground state degeneracy. 

However, in the big portfolio of top-down holographic set ups that are derived from "first principle" string theoretical constructions such RN solutions do not occur. There are always other fields present beyond Einstein-Maxwell and these take care that extremal RN black holes with their zero temperature entropy are avoided -- see underneath. The RN strange metals are therefore by default pathological. But they are simple and should be looked at as toy models that do grab some of the generic features of the big family of holographic strange metals.

The early results on the RN metal did reveal already the big picture of the grand principle behind the holographic strange metals. Moreover, the outcomes directly caught my full attention since these are already quite suggestive regarding a major puzzle that seemed to be revealed already in the 1990's in experiments : the "local quantum criticality", as I will explain in a moment. Hong Liu and coworkers from MIT deserve the full credit for finding out how this can be deduced from the bulk, resting on the  notion of geometrizing the renormalization group in terms of the isometry of the bulk geometry \cite{HongScience,HongTechnical}. 

I already advertised that charged black holes associated with finite density in the boundary have a geometrical structure that is very different from the Schwarzschild black holes encoding for finite temperature at zero density.  These imprint on AdS in the following way. Near the boundary it is business as usual but upon descending towards the interior the geometry completely reorganizes upon crossing the radial coordinate associated with the chemical potential $\mu$ in the boundary. This turns out to be the way that the "Fermi degeneracy" is encoded in the bulk: $\mu$ has a role similar to $E_F$ in the boundary as we will see. 

Upon descending further along the radial direction the geometry crosses over rapidly to the {\em near-horizon} geometry of the charged black hole. For the RN black hole the metric is textbook material, revealing that the near horizon geometry is actually quite remarkable. It is called "$\mathrm{AdS}_2 \times \mathrm{R}^d$".    $\mathrm{R}^d$ refers to the space dimensions shared with the boundary, and it just refers to the fact that it is a flat space. However, the remaining time- and radial direction morph in a two dimensional Anti-de-Sitter: the $\mathrm{AdS}_2$. Focussing in on the zero temperature extremal case, the geometry is completely different from the pristine near boundary $AdS$ geometry encoding for the zero density stoquastic CFT.  This is the bulk encoding of the strong emergence hard wired in finite density holography: the scaling properties of the finite density state at scales small compared to $\mu$ are entirely unrelated to those of the zero density stoquastic type CFT. 

Given the "GR = RG" principle, "AdS" means that in the boundary one will find a CFT like scaling behaviour but in the RN $\mathrm{AdS}_2$ "throat"
(referring to the geometry close to the horizon) this only acts out in the plane spanned by time and the radial direction. Since the radial direction is the scaling direction, this implies a purely {\em temporal} scale invariance in the boundary. Let us recall the definition of the dynamical critical exponent: $t \sim l^z$, where $t$ is time and $l$ is length. We are now dealing with a system where  the correlation time $\tau_{\mathrm{cor}} \rightarrow \infty$, it is temporally critical. But the spatial correlation length is somehow finite (see Section \ref{infinitezpseudopot}). This implies that the dynamical critical exponent $z \rightarrow \infty$!

As I will discuss in Section (\ref{EELSlocqucrit}), such a kind of a behaviour was inferred from experiments on the cuprate strange metals already in the late 1980's. It is a cornerstone of the early "marginal Fermi-liquid" phenomenology \cite{MFL}, while it was observed in dynamical experiments more recently. It got the name "local quantum criticality" and this semantic was directly used in holography as well. 

This was all along a great mystery in the condensed matter context. In Section (\ref{qucritical}) I explained that one may encounter a $z$ of two or three at least in the absence of strong quenched disorder but an infinite $z$ appears to be excluded from this agenda. There is a claim that it can happen in certain dissipative settings \cite{dissXY} but this is quite controversial. When I learned about this infinite $z$ arising in this natural but unfamiliar holographic fashion I got immediately greatly intrigued, embarking definitively on the risky pursuit of concentrating my research on holography.

But in these  early days there was also quite a bit of confusion. There was initially a perception that the RN strange metal was the unique finite density metallic state predicted by holography. There was the uneasy aspect of the zero temperature entropy -- at a point there were even false claims originating in the experimentalist's community that such zero T entropy was observed. Until the present day you may find outsiders who are claiming that holography is all wrong since it predicts such entropy. This is nonsense --  we understand this much better now: the dilatons change the rules as you will see now.  

\subsection{Covariant RG:  dilatons, hyperscaling violation  and the fermionic degeneracy scale.}
\label{EMDscaling}

To make this "symmetry processing power" of holography to work one has to make sure that the bulk gravitational theory is sufficiently general to represent the universal features of the boundary phenomenology. In this regard the minimal Einstein-Maxwell gravity that predicts the Reissner-Nordstrom black hole to be the unique gravitational dual of the boundary metal is falling short. 

Once again, the string theorists have mighty mathematical machinery in the offering in the form of the "top-down" holographic set ups. As the original Maldacena construction is to physicists standards mathematically proven to express a precise duality, they constructed subsequently a plethora of such "exact" dualities involving richer physical circumstances. It is all large $N$ limit and these translate therefore to classical gravitational physics in the bulk, involving fields that may be rather unfamiliar. But one continues to discern "Einstein theory universality".  These extra fields hard wire yet other universal traits in the bulk that translate into universal properties of the boundary phenomenology. One can let again the numbers to be pending the specific UV while the structure of the deep IR theory may have  a much greater applicability.

\subsubsection{Kaluza-Klein compactification and the dilatons.}

Turning to finite density, the top-down "set ups" are univocal: even for the relatively humble affair of determining the structure of the thermodynamics in the boundary one {\em has} to take care of an additional field, the {\em dilaton}. Dilatons may be unfamiliar but they are bread and butter in string theory.  

Kaluza-Klein compactification is fundamental to all of string theory including the correspondence.  This goes back to the demonstration in the 1920's by Kaluza and Klein that when one departs from pure GR in 5 space time dimensions, upon rolling up one of the space dimensions in a small circle one obtains Einstein-Maxwell theory in 4 dimensions at scales large compared to this circle radius. Fundamental string theory can only be formulated in 10 overall dimensions, while the specific limit taken in holography typically involves classical supergravity in 11 dimensions. One has to roll up, say, 7 of these dimensions to obtain the 4 dimensional bulk gravity required to encode the boundary theory in 2+1 dimensions. In these high dimensional circumstances there are many ways to accomplish this rolling up in compact dimensions (the Calabi-Yau manifolds) and pending what one picks one arrives at different lower dimensional theories. 

However, a common denominator is that regardless how one compactifies a new scalar field drops out the Kaluza-Klein procedure that has the typical effect in Einstein theory to affect the volume of space: it is therefore called the "dilaton field".  To give an impression of how the action of such a "Einstein-Maxwell-Dilaton" theory looks like, 

\begin{equation}
S = \frac{1}{16 \pi G} \int d^{d+2} x \left[ ( {\cal R} - 2 (\nabla \Phi )^2 - \frac{ V (\Phi)} { L^2}  ) - \frac{Z (\Phi) }{4 e^2} F_{\mu \nu} F^{\mu \nu} \right]
\label{EMDaction}
\end{equation}

where $G$, ${\cal R}$ and $L$ are Newton's constant, the scalar curvature and the AdS radius respectively. $F_{\mu \nu}$ is the Maxwell field strength and $e$ the charge. The novelty is in the scalar field $\Phi$: the dilaton. In the deep interior $\Phi$ becomes typically large and the potentials acquire odd-looking forms like $ Z (\Phi) = Z_0 \mathrm{exp} (\alpha \Phi),   V (\Phi) = -V _0 \mathrm{exp} (\beta \Phi)$. 

The structure of these potentials are generic and these give rise to various general circumstances in the boundary. One possibility is that the deep interior geometry just "disappears" and these (soft-, hard-) "walls" describe {\em confining} states in the boundary: these flourished in the context of the "AdS/QCD" pursuit \cite{Erdmengerbook,holodualbook}. However, given the form of the potentials as I just quoted, as function of the free parameters like $\alpha, \beta$ these describe a vast family of "near horizon scaling geometries" that dualize in a {\em family of scaling theories describing finite density matter.}

\subsubsection{The near horizon scaling geometries of EMD gravity.}

Gouteraux, Kiritsis and coworkers \cite{EMDclass} demonstrated in their seminal work that EMD gravity can be used to classify.  A seemingly universal scaling theory for the finite density matter can be extracted, having a similar status as Kadanoff's scaling theory for the conventional critical state of Section (\ref{thermaluniversality}). This hinges on the GR = RG principle: derive the geometry in the deep interior resting on universal characteristics of the gravitational theory, and when this turns out to be a "scaling geometry" (as is the case for EMD gravity) it will dualize in a scaling prescription in the boundary. 

Following this strategy they show that the deep interior metric acquires the general form, 

\begin{equation}
ds^2_{\mathrm{EMD}} = \frac{1}{r^2} \left( - \frac{dt^2}{ r^{2d(z-1)( d - \theta )} } + r^{2\theta/ ( d - \theta )} dr^2 + dx^2_i \right)
\label{EMDmetric}
\end{equation}

where $t, r, x_i$ are (Lorentzian) time, radial coordinate and space coordinates, respectively.  The metric is responsible for thermodynamics and it has an isometry that translates into scaling laws in the boundary in terms of two free parameters $\theta$ and $z$ ($d$ is the number of space dimensions, as usual). In the top-down settings these are determined by the specifics of the dilaton potentials.  One is not surprised that in the boundary dual $z$ corresponds with the now familiar dynamical critical exponent. But pending the dilaton potentials this can now take any value $ 1 \le z < \infty$. The Lorentz invariant (zero density) and local quantum critical scaling behaviours are just the extremal cases. The news is in $\theta$, called the "hyperscaling violation exponent". It is called like this because the free energy is no longer scaling with the volume of the system $\sim L^d$ but instead with a "reduced" volume $L^{ d - \theta }$. It can take values $ d < \theta < - \infty$ without running into no go theorems -- the precise conditions are $(d-\theta) / z \ge 0, (d - \theta)(dz - d - \theta) \ge 0$ and $(z-1) (d + z - \theta ) \ge 0$

The simplest consequence for the boundary theory is that the entropy (or equivalently the specific heat $C \sim T (\partial S / \partial T)$) exhibits the following scaling behaviour,

\begin{equation}
S \sim T^{( d - \theta ) / z}
\label{hyperscalingent}
\end{equation}

Dealing with the zero density state the scale {\em invariance} of the free energy imposes that $\theta = 0$. For a Lorentz invariant theory ($z =1$) one recovers the "Debye" entropy $ S \sim T^d$ that is indeed generic for all conformal field theories. Dealing with the $z \neq 1$ zero density theories it is well known that the entropy $S \sim T^{d/z}$; the factor $z$ is there to correct for the "number of time dimensions" when the Euclidean time direction is rolled up in the thermal circle. 

However, the hyperscaling dimension $\theta$ is unfamiliar in this zero density repertoire. You may not have realized it yet but it is overly familiar in a different context: it is the "central scaling gear" controlling the algebraic properties of the Fermi-liquid! The dimension $\theta$ controls how the number of {\em thermodynamically relevant} degrees of freedom scales with the volume of the system. In the Fermi liquid these are controlled by the Fermi-surface having a dimensionality $\theta = d -1$. At every point on the Fermi-surface excitations are anchored with a linear dispersion: $z=1$. Fill these values into Eq. (\ref{hyperscalingent}) and you find $S \sim T$ regardless the number of space dimensions $d$. But this is the overly familiar Sommerfeld entropy of the Fermi liquid which scales in the same simple way regardless the number of space dimensions!

\subsubsection{Finite density and the covariant renormalization flows.}
\label{covariantRG}

What is going on here? As I already emphasized in Section (\ref{Maldacenaholo}) the deep mathematical underpinning of holography is in the relation between the {\em isometry} of the geometry in the bulk and the {\em symmetry} controlling the boundary field theory.  From a CM perspective the interest is in {\em emergent} symmetries like the conformal invariance that may be realized at a stoquastic quantum critical point. Right now we are staring at a different type of symmetry at work in the boundary system that is about a particular way the system is changing continuously under scale transformation. It is about the nature of the renormalization group, that we recognize to be already at work in the Fermi-liquid. 

To understand in what regard the isometries of finite density holography are different from those of the zero density CFT's let us zoom in on how exactly the bulk geometry depends on {\em scale transformations}. We already exercised this "RG = GR" notion for the zero density case in Section (\ref{Maldacenaholo}). There we found that the isometry of AdS is unique in the regard that the metric is invariant under scale transformation:  $ds^2_{\mathrm{AdS}} \rightarrow ds^2_{\mathrm{AdS}}$ when  $x^{\mu} \rightarrow \Lambda x^{\mu}$. We learned that the dictionary than implies that the boundary is governed by the conformally invariant field theory. 

Let us consider the simplest of all geometries instead: Minkowski flat space-time. The invariant is the metric, in Cartesian coordinates $d s^2 = \eta_{\mu \nu } dx^{\mu} dx^{\nu}$, where $\mu, \nu$ refer to spacetime directions ($\eta_{tt} = -1, \eta_{ a a} = 1$, $a$ referring to space dimensions).  Your intuition may give you the impression that flat space-time  should be independent of scale. It does transform in a smooth, continuous way under scale transformations but it is actually {\em not} invariant under scale transformation.  Perform a scale transformation on the coordinates: $x^{\mu} \rightarrow \Lambda x^{\mu}$ and it follows immediately that $ds^2 \rightarrow \Lambda^2 ds^2$. The metric is {\em proportional} to itself under scale transformation but it does not continue to be the {\em same} metric. The geometer refers to this as {\em the metric transforms  covariantly under scale transformation}. 

Let us now inspect how the finite density scaling geometry family   Eq. (\ref{EMDmetric}) behaves under scale transformation. As in the boundary, we depart from a redefinition of the spatial scale, $x_i \rightarrow \Lambda x_i$. You are now used to the meaning of the dynamical critical exponents, $t \rightarrow \Lambda^z t$. You then discover that the radial coordinate will transform according to $r \rightarrow \Lambda^{( d - \theta )/d} r$. For $z =1, \theta =0$ one recovers the scale invariance -- Eq.  (\ref{EMDmetric}) reduces to pure AdS,  Eq.  (\ref{AdSmetric}). Insert these rescalings in  Eq. (\ref{EMDmetric}) and you will find when $\theta \neq 0$,

\begin{equation} 
ds^2_{\mathrm{EMD}} \rightarrow \Lambda^{2\theta/d} ds^2_{\mathrm{EMD}}
\label{CovEMD}
\end{equation}

We have uncovered the essence of the EMD scaling geometries. When $\theta \neq 0$ the structure of the RG flows is governed by {\em covariance} under scale transformation instead of the invariance underpinning the scaling relations of CFT's. Blaise Gouteraux appears to be the first who recognized this very simple but crucial insight  (see e.g. Ref. \cite{Gouterauxcov}).  

This has various consequences for the boundary RG. A first consequence is that in a covariant flow the UV scale where it starts is remembered in the deep IR. In a scale invariant flow the information regarding  the UV scales completely disappears. This is the usual affair of UV regularization -- the IR theory is strictly independent of the short distance physics. But in the covariant flow the metric is changing via the proportionality factor associated with the scale change. This dualizes in the boundary to the phenomenon that the IR observables still know about the UV point of departure.  

This is of course wiring in  the degeneracy scale associated with fermion matter that I argued in Section (\ref{fermiondeg}). The UV scale that is remembered is the chemical potential $\mu$, the same thing as the Fermi energy of the Fermi liquid.  This is linked to the fact that conformal invariance can only arise by infinite fine tuning to a quantum critical point (modulo maximal SUSY), while for covariant scaling behaviour there is plenty of room for {\em phases} of matter that do not require anything of the kind. The Fermi-liquid is case in point.

A next big difference is that to be only {\em covariant} instead of invariant under scale transformation means that this symmetry is much less powerful in constraining the behaviour of observables. It will become clear when I further unfold the rich portfolio of physical phenomena of the finite density holographic systems that much more is going on than in the zero density CFT's. This goes  way beyond the rather simple extension of the thermodynamics   embodied by the entropy, Eq. (\ref{hyperscalingent}). Whether this implies that the IR physics becomes more susceptible for "UV dependence pollution" is a next issue -- there are reasons for concern, see  the remainder.  

Although the geometry Eq. (\ref{EMDmetric}) was discovered by solving concrete, top-down inspired holographic set ups one may actually turn it around. Dealing with the CFT's one could depart from the conformal symmetry of the boundary theory, to then discover that the only form of geometrized RG flow is in the form of the pure AdS geometry. Dealing with finite density, one can depart from the scaling structure of (say) the Fermi-liquid, in addition to be enlightened regarding the possibility of a $z \neq 1$, to then pose the question what kind of isometry can reconstruct this behaviour? Relying on the simple scaling exercise in the above, one will then concludes that the metric Eq. (\ref{EMDmetric}) is the unique outcome in the same guise that pure AdS is the unique geometrization of the CFT RG flow. 

Should it be for this reason that Eq. (\ref{EMDmetric}) represents the truly universal "library" of RG flows? This appears to be a quite defendable conjecture dealing with {\em homogeneous} geometries. However, as I will discuss at length in the next two sections one has to cope with a very different side of gravity in case that the space is not homogeneous and/or isotropic. Einstein theory shows here its real face in the form of a system of non-linear partial differential equations and it becomes very hard if not impossible to write down transparent, closed solutions like Eq.  (\ref{EMDmetric}). Electrons in solids fall in this "inhomogeneous" category and it may well be that yet different scaling geometries arise that may be more complex given that symmetry exerts less control. Presently, nobody has an answer to this question and it is a main challenge of our numerical holography effort in Leiden, see Section (\ref{holotransport}). 

\subsection{The holographic strange metals as generalized Fermi-liquids.} 
\label{genFL}

I just announced that the phenomenology controlled by the "covariant" scaling is quite a bit richer than for the CFT's. In fact, it is a vast landscape that is not at all completely explored. One can yet identify a number of general features. The overarching message is  that the gross "organization" of such metallic states rests on a template formed by the Fermi-liquid, which is then extended by turning the scaling dimensions to become anomalous relative to the "engineering" dimensions of the Fermi-liquid.  Let me put some meat on the bare bones, focussing on particular aspects of this phenomenology. A lot more is known and I refer to the textbooks for a more exhaustive discussion. 

\subsubsection{The thermodynamics: embarrassing the marginal Fermi-liquid.} 

If anything stands a chance to be truly universal it is the simple but very powerful formula Eq. (\ref{hyperscalingent}) for the entropy. I already unveiled that it captures the scaling of the thermodynamics of the Fermi-liquid in an extremely efficient fashion. There is no need to hassle with the integrals involving Fermi-Dirac distributions as is done in the elementary textbooks. 

In all its simplicity, this formula can be used to great effect in the empirical arena of the cuprate strange metals. This is really the subject of the last section but this is so elementary that it deserves to be included already at this point, if not only because it illustrates the amazing powers of the covariant RG. 

One immediately infers that by assuming $d - \theta$ to be finite it follows immediately from Eq. (\ref{hyperscalingent}) that $S \sim T^0$ in a local  ($ z \rightarrow \infty$) quantum critical system: the temperature independence implies that there is zero temperature entropy. This is precisely the origin of the $T=0$ entropy pathology of the Reissner-Nordstrom strange metal that of course also submits to the scaling logic. 

But as I will discuss in detail in Section (\ref{EELSlocqucrit}) there is direct experimental evidence supplied by electron loss spectroscopy for local quantum criticality in the cuprate strange metals. This was actually already the central pillar of the very early "marginal Fermi-liquid" theory \cite{MFL}. This got later a more precise identity within the Hertz-Millis scheme of Section (\ref{hertzmillis}): the Fermi-liquid quasiparticles decay in the continuum of local quantum critical fluctuations associated with a QCP by a Yukawa coupling \cite{Varmaloops}. By involving in essence second order perturbation theory, one then finds that the entropy coming from the quasiparticles diverges logarithmically through their effective mass. 

However, in this mindset it is completely worked under the rug that this critical continuum represents also thermodynamically relevant degrees of freedom. It follows immediately that since $z \rightarrow \infty$ the marginal Fermi liquid entropy should be temperature independent! The electronic specific heat  in the strange metal regime was measured a long time ago and it appears to be Sommerfeld $S \sim T$ -- initially it was without further thought assumed to reflect a Fermi liquid. 

This actually amounts to a no-go theorem  for this marginal Fermi-liquid affair, in fact largely overlooked until the present day. But there appears to be still a problem of principle in the cuprate metal context: how to reconcile $z \rightarrow \infty$ with a Sommerfeld entropy?  Top-down holography offers here a remarkable solution, that is also quite elegant viewed from the string theoretical side. It was  introduced in the AdS/CMT context by Steve Gubser and Fabio Rocha \cite{Gubser}. It is therefore called the Gubser-Rocha model (also "conformal to AdS$_2$"). This is the same Gubser as the G of the GKPW dictionary -- he unfortunately deceased at a young age through a mountaineering accident. 

This top down involves a RN extremal black hole in 11 dimensional supergravity that after compactification just "touches" the space plane of the 4D holographic bulk with the outcome that the horizon shields the singularity in such a way that the area of the horizon becomes zero at zero temperature. This has therefore devoid of zero temperature temperature. However, one finds that besides $z \rightarrow \infty$ also $\theta \rightarrow - \infty$ such that the ratio $- \theta / z \rightarrow 1$: according to Eq. (\ref{hyperscalingent}) this results in a "Sommerfeld" entropy $S \sim T$, regardless of space dimensionality! 

Although I have no clue what $\theta \rightarrow - \infty$ means, the fact that this GR metal is top-down consistent means that it can arise in a physical theory. So much is clear that the cuprate strange metal cannot possibly be controlled by the $\theta = d-1$ Fermi-surface hyperscaling violation. Although the mainstream in the condensed matter community has been taken it for granted that Fermi-surfaces control the strange metals, on closer consideration thermodynamics excludes directly this possibility.  

\subsubsection{The generalized Lindhard excitation spectrum of strange metals.} 
\label{holochargeexc}

Following the template of the Fermi-liquid the next property to look at is the spectrum of charge excitations. It can be measured by electron-loss spectroscopy (Section \ref{EELSlocqucrit}) and it figures prominently in the textbook treatment of the Fermi-liquid, Section (\ref{RGFermiliquidzeroT}). At stake is the charge density - charge density propagator or dynamical charge susceptibility, $\chi_{\rho} (q, \omega)$  in a large range of momenta $q$ and energies $\omega$.

\begin{figure}[t]
\includegraphics[width=0.7\columnwidth]{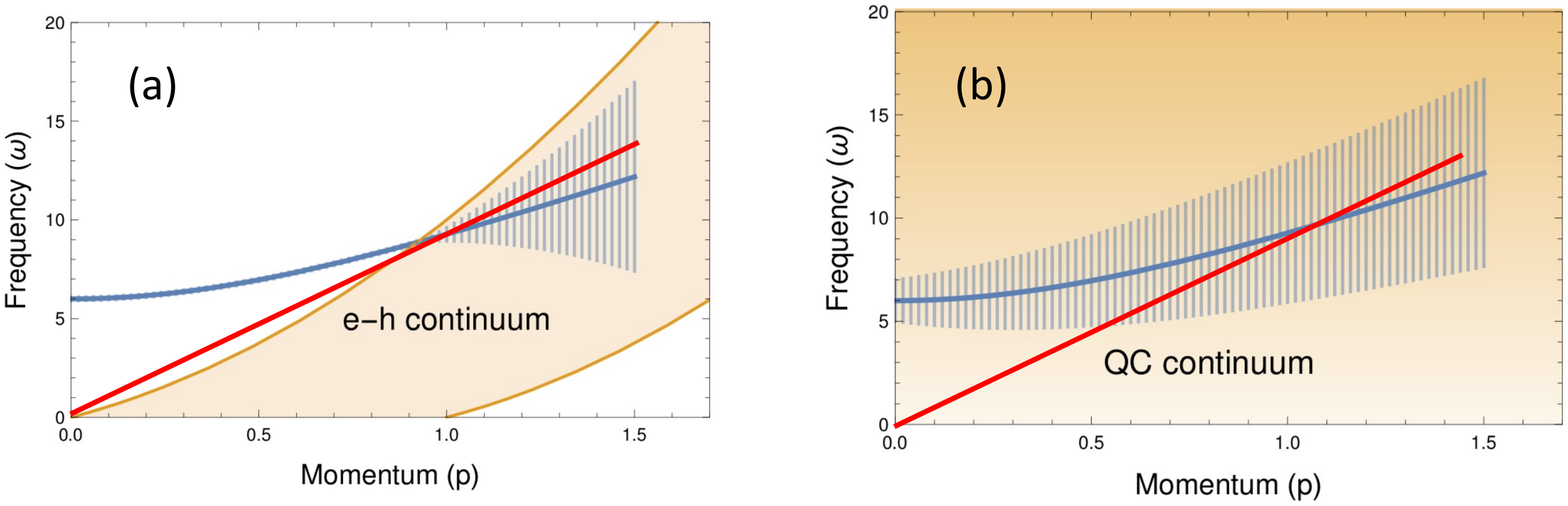}
\caption{The charge response. In (a) the Lindhard spectrum formed by the electron-hole excitations around the Fermi-surface is indicated. In addition, the sepctrum is characterized byzero sound (red line) that turns into the plasmon in the charged Fermi-liquid \cite{Holoplasmons}. The response in a $z \rightarrow \infty$ holographic strange metal is characterized also by the zero sound/plasmon but the Lindhard continuum is supplanted by a continuum characterized by anomalous scaling dimensions.}
\label{fig:lindhardfig}
\end{figure}

In the discussion of the Fermi-liquid  we learned that this revolves in first instance around the charge susceptibility of the Fermi gas  $\chi_0 ({\bf q}, \omega)$. In the Galilean continuum this is captured by the Lindhard function that counts the number of electron-hole pairs that can be created with a center of mass momentum $q$ and energy $\omega$, see the left panel of Fig.  (\ref{fig:lindhardfig}) for the imaginary part. As I emphasized in the discussion, originating in the "classicalness" of the fixed point physics (FL phenomenology) of the interacting Fermi-liquid one only has to incorporate time dependent mean field though the RPA formula  $\chi  ({\bf q}, \omega) = \chi_0 ({\bf q}, \omega)/ \left( 1 - V_q \chi_0 ({\bf q}, \omega) \right)$. For a repulsive short range interaction $V_q$  we found that this reveals the propagating zero-sound mode at $T=0$ (red line), promoted to the plasmon (blue line) dealing with the Coulomb interaction.  According to the Linhard function, the frequency dependence is an algebraic affair at larger momenta: $\mathrm{Im} \chi_0  ({\bf q}, \omega) \sim \omega$ in $d = 3$.

Highlighting the interpretation of the holographic strange metals as generalized Fermi-liquids, the take home message is that their charge excitation spectra has a gross organization that {\em coincides} with that of the Fermi liquid! 

Invariably one finds the zero sound mode, automatically dropping out of the dictionary. There is no need to add interactions since holographic strange metals are by default in more than one sense strongly interacting. But this is rooted in symmetry that is as always impeccably processed by the correspondence. It is just hydrodynamical principle: in a homogeneous space total momentum is conserved while also total charge is conserved and this implies the existence of a sound mode, now in the zero temperature fluid. Using RPA means in various holographic guises, this can be promoted to the plasmon: see e.g.  Ref. \cite{Holoplasmons}. 

The existence of sound is generic and therefore not remarkable. The remarkable part is that  the governance of covariant RG implies that a spectrum of excitations arises at finite momenta and frequencies being similar to the Lindhard spectrum! As the latter, these are  incoherent excitations characterized by {\em power-law} behaviour.  But the scaling dimensions are now set by the anomalous dimensions in a "hyperscaling" fashion -- set by thermodynamic exponents -- in analogy with the Kadanoff scaling.  

More precisely, dealing with charge related  responses there is room for one more anomalous dimension in the EMD bulk next to the thermodynamic $\theta$ and $z$: the "charge exponent" $\zeta$ \cite{EMDclass}. This is expressing that the electrical charge is running under RG. This is a bit of a delicate affair. The electric charge quantum is {\em locally} conserved, "protected by Gauge invariance", and is not supposed to run. This has the ramification  that the electrical charge quantum  can be deduced from macroscopic measurements; e.g. there is a factor $2 e$ in the Ginzburg-Landau free energy that becomes observable by measuring the quantized magnetic flux associated with a superconducting vortex.  To see $\zeta$ at work one has to lift the charge conservation. This has then intriguing consequences, as for instance for the Aharonov-Bohm effect as analyzed by Phillips and coworkers \cite{PhilipABzeta}, next to modifying the scaling behaviour of the charge responses. To keep things as simple as possible I will ignore here $\zeta$ -- its ramifications can be easily retrieved from the literature. 

From the covariant RG at work in the EMD scaling geometries of the previous section one can immediately deduce the zero temperature scaling from of this incoherent excitation spectrum. It actually submits to the energy-temperature scaling in precisely the sense you learned in the context of quantum criticality. This is hardwired in the way that the horizon of the thermal black hole evolves in the deep interior scaling geometry.  The result is for the regime for $\omega << \mu$ and $\omega > T$ that the spectral function scales according to, 

\begin{equation}
\mathrm{Im} \chi_{\mathrm{inc}}  ({\bf q}, \omega) = \omega^{( d - 2 - \theta +z)/z} \mathrm{F}  \left( \frac{\omega}{|q|^z} \right)
\label{EMDchargescaling}
\end{equation}

Remarkably, this does also apply to the Fermi-liquid Lindhard function. As before $\theta = d -1$ (Fermi surface) and $z=1$,  while the scaling function relating frequency and momentum $F (\omega / q^1) = \omega / q$: the formula recovers the  well stablished scaling behavior $\chi" \sim \omega/q$  of the Lindhard continuum at energies small compared to $E_F$! This is quite remarkable: the scaling of the charge excitations is determined by the thermodynamics ($\theta$, $z$) in a way that is entirely unrelated to the way correlation function exponents are tied to thermodynamic scaling dimensions at stoquastic quantum critical points. 

The scaling function $ \mathrm{F} (x)$ is obviously important but we know little about it -- it is just not computed yet for EMD backgrounds. A particularly interesting case is the limit embodied by local quantum criticality, $z \rightarrow \infty$. This has only been thoroughly explored for the RN strange metal. I will come back to this at greater length in Section (\ref{infinitezpseudopot}).

 A first take home message is that at higher frequencies these "generalized Lindhard excitations" at least for $z \rightarrow \infty$ fill up the whole $\omega, q$ plane, as indicated in  the right panel of Fig. (\ref{fig:lindhardfig}). This has the consequence that even at $q=0$ the plasmon living at its high energy plasmon frequency can decay by "falling apart" in the "generalized Lindhard spectrum", the analogue of Landau damping. The plasmon of the Fermi-liquid is infinitely long lived at $q=0$  in the homogeneous background because of the kinematical mismatch with the Lindhard spectrum. 

However, considering the low frequency limit odd things are happening as further explained in Section (\ref{infinitezpseudopot}). The issue is that for $z \rightarrow \infty$ a spatial scale can still be discerned, translating in a characteristic momentum $q_{\mu} \simeq \mu$ as highlighted by Hong Liu and coworkers \cite{honglocal}. For $q < q_{\mu}$ the low frequency/temperature response is completely dominated by the hydrodynamical response. As will be discussed in the next section this is in turn at finite temperature governed by Planckian dissipation and minimal viscosity, just as the zero density case.   However, a rather abrupt cross-over takes place to a regime $q > q_{\mu}$, characterized by a scaling determined by a {\em momentum dependent} frequency/temperature exponent.  For instance, in two space dimensions  $\mathrm{Im} \chi  ( q, \omega)  \sim \omega^{2 \nu_q - 1} $ where $\nu_q = (1/2) \sqrt{ 5 + 4 q^2 - 4 \sqrt{ 1 + 2q^2} }$.

\subsubsection{ Strange metals turning into superconductors: "quantum critical BCS".}
\label{HoloSC}

The first result signalling that AdS/CFT could have dealings with electrons in solids was the discovery of holographic superconductivity, independently by Hartnoll, Herzog, Horowitz ("H$^3$") and (again) Gubser in 2008, see the books. This actually triggered the AdS/CMT frenzy that followed in the string theory community. Given the iconic status of BCS in the Fermi-liquid portfolio, it is perhaps the best stage to highlight its deep similarities with the holographic mechanism. I will hit this home towards the end of this passage. 

Gravitationally it is a remarkable affair. The reader may be aware of the "no-hair theorem". It is mathematically proven that stationary black holes cannot carry detailed traits: they are like elementary particles, completely characterized by overall quantities including their mass, charge, linear- and angular momentum. But this theorem depends critically on flat asymptotics: there is no no-hair theorem dealing with AdS asymptotics!

A general way to capture spontaneous symmetry breaking is implied by linear response theory: when there is a response without a source one is dealing with an ordered state. In Section (\ref{AdSCFTgen}) I sketched the way how fields dualize via the dictionary to the bulk,  finding out that the source and response are associated with the leading- and subleading components of the bulk fields asymptoting  near the boundary.  

One then discovers that the "VEV without a source" can only happen when the bulk field acquires a finite amplitude in the bulk, in essence describing that the black hole acquires an atmosphere formed from this stuff. It is a rather unusual atmosphere as compared to what one encounters in the cosmos: it extends all the way to infinity along the radial coordinate, with a density falling off with a power law. 

According to the dictionary the "breaking of  $U(1)$" associated with superconductivity  dualizes in a complex scalar field in the bulk. Now it comes: the (positive) mass-squared of such a bulk field will in first instance code for the scaling dimension in the boundary. However, when one inserts such a field in the strongly curved background near the horizon, one may run into the "BF bound": this is the gravitational phenomenon that the curvature may translate into the mass term in the action turning negative into an equivalent flat manifold. This is the usual Mexican hat affair, but now evolving in the bulk near the horizon. The outcome is that upon lowering the temperature in the boundary the near-horizon curvature is growing until the BF bound is violated and the black hole acquires a spontaneous finite amplitude "scalar atmosphere" that turns out to asymptote near the boundary precisely in the desired fashion! 

This is a miracle. In the bulk  gravitational machinery is at work that was not realized in the canonical agenda of GR before -- the BF bound violation triggering the hair. This is then processed by the dictionary to a boundary physics that is precisely right. This is just like a mathematical clockwork where all the gears work together to produce miraculously the correct physical outcome.   

Given that the correspondence is mercilessly precise with anything that is controlled by symmetry, this holographic superfluid/superconductor (one can effectively gauge the boundary) is impeccably  reproducing the phenomenology: one finds perfect conductors, the fluxoids/quantized vortices are reproduced, etcetera. This was even used  to shed light on superfluid turbulence in two dimensions \cite{HongSFturb}! In fact, it became clear later that this is only the tip of the iceberg. In Section (\ref{Intertwined}) I will summarize the state of the art of holographic symmetry breaking, finding that the most natural and general bulk dualizes into a boundary order that is strikingly similar as the "intertwined" orders observed in underdoped cuprates.  

But the Bardeen-Cooper-Schrieffer theory of conventional superconductors goes a step further than just the symmetry controlled part of the phenomenology. As I reviewed in Section (\ref{BCSbasics}), it predicts that at $T_c$ a gap opens in the spectrum of Fermi-liquid quasiparticles. This gap is proportional to the amplitude of the order parameter being {\em exponentially small} in the UV parameters: $ \Delta \sim \exp{ (- E_F/V )}$, where  V is the attractive interaction.  But this instability will occur always, regardless the smallness of $\Delta$: the principle that the Fermi-liquid as state of matter is only unstable in an absolute sense in the presence of an attractive interaction. In addition, it introduces the large coherence length that is unique for fermionic superconductors. 

The take home message is that also in this regard holographic superconductivity is a close cousin of the BCS/Fermi liquid version. In fact, the observable physics is so close to BCS  that it can be easily misidentified as a BCS superconductor! To observe the sharp difference unusual experiments have to be implemented which turn out to be very difficult to realize in the lab. Let me explain.  

In Section  (\ref{BCSbasics}) I discussed on purpose the BCS mechanism in scaling language. As a reminder, the focus is on the propagator $\chi_p ( \omega )= \langle b^{\dagger} b \rangle_{q=0, \omega}$ associated with the pair operator $b^{\dagger} = \sum_k c^{\dagger}_{k, \uparrow}c^{\dagger}_{-k, \downarrow}$. For the Fermi gas this quantity is characterized by a marginal scaling dimension $\alpha_p =0$. This implies that its zero frequency real part is logarithmically divergent in the IR -- the "Cooper Logarithm" -- and by plugging this in the RPA formula one immediately deduces the exponential gap function.  

The way that  for temperatures above $T_c$  the RPA pair susceptibility behaves is shown in Fig. (\ref{fig:HoloSCsusc}), panel A for the imaginary part of $\chi"_p$ of   $\chi_p ( \omega, T )$ as function of energy and temperature. The way this works is that a resonance starts to develop for decreasing temperature, becoming discernible at a temperature $\sim 5 T_c$. Upon further lowering temperature it moves down in energy while its width is decreasing. Right at $T_c$ it turns into a delta function at zero frequency, that would turn into a tachyon at lower temperatures signalling linear instability. This can all be captured in terms of time dependent Ginzburg Landau, involving a damping associated with the Landau damping in the pair channel. The crucial assumption is that the phase transition is governed by Landau mean field, which makes sense since the large coherence length suppresses the critical fluctuations. 

\begin{figure}[t]
\includegraphics[angle=-90,width=0.7\columnwidth]{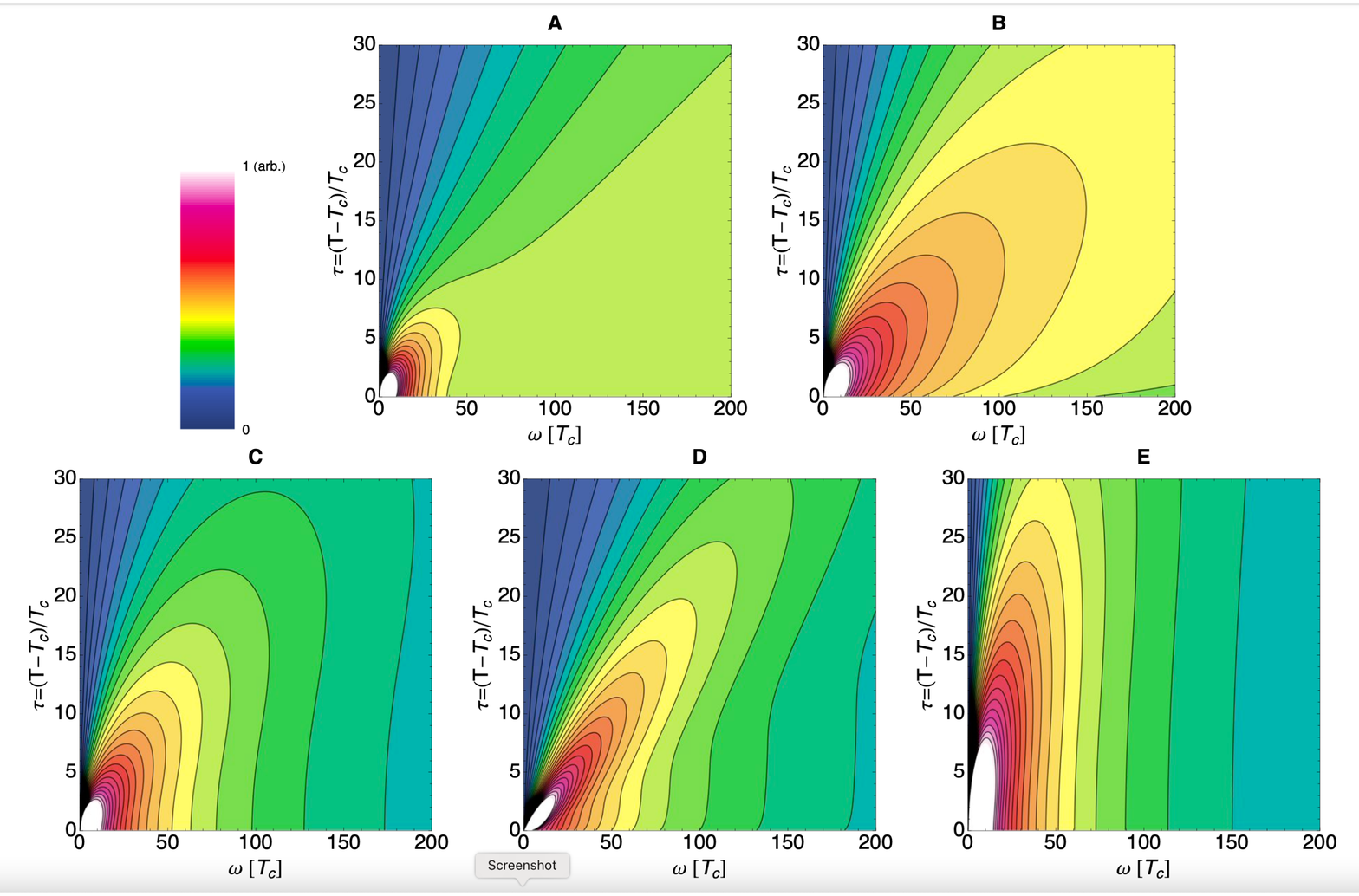}
\caption{The imaginary part of the pair susceptibility (false colors) as function of energy $\omega$ and (reduced) temperature $(T-T_c)/T_c$  in the metallic state approaching the transition to the superconducting state \cite{holoBCS}. Several cases are compared: "plain vanilla" BCS (A), Strongly coupled "Hertz-Millis" (B), quantum critical BCS (C), "large charge" (D) and "small charge" (E) holographic cases. See the text for further explanations.  }
\label{fig:HoloSCsusc}
\end{figure}

Let us now turn to the typical dynamical response of holographic superconductors in this regime. We addressed this in Leiden in 2011 in an effort involving both string theorists and condensed matter specialists. This resulted in a paper which is also the source of Fig.'s (\ref{fig:HoloSCsusc},\ref{fig:HoloSCscaling}) that was at first instance aiming at the experimental community \cite{holoBCS}. We pointed out that in order to look directly for the origin of superconductivity at a high temperature a particular type of "spectrometer" should be constructed. This was picked up by several key players, but for practical reasons it did not materialize, see underneath. I perceive it in hindsight as an optimal stage to illustrate the workings of the finite density scaling universality according to holography. 

According to the dictionary, the pair propagator in the boundary is associated with the scalar field in the bulk responsible for the instability. This is  a non-conserved order parameter that is behaving like the simple example of a scalar field as explained in Section (\ref{Maldacenaholo}). The mass of the bulk field sets the scaling dimension of the corresponding (pair) operator in the boundary. This scaling dimension is in turn pending on the particular set up -- in principle one needs a top down but you will see that it reveals a general message. 

A generic set up is to depart from a finite density RN metal, coupled to the scalar "pair" field. We actually mobilized a particular top-down set up: see the paper \cite{holoBCS}. The result for $\mathrm{Im} \chi_p (\omega, T) = \chi"$ is shown in Fig. (\ref{fig:HoloSCsusc}), panel D. You notice that it looks similar as the BCS result with the "relaxational" order parameter peak becoming intense and sharp upon approaching $T_c$ (the "white blob" near the origin). We also included the so-called "small charge" incarnation of the holographic superconductor, that relies on "alternate quantization". This is a rather fanciful holographic affair, but the effect in the boundary is simple. Using the $\chi_p$ of the "plain vanilla" holographic superconductor as $\chi_0$ in the RPA formula one can adjust the outcome with an effective interaction "$V$" that acts like a repulsive interaction in the BCS setting, reducing the holographic $T_c$: panel E in  Fig. (\ref{fig:HoloSCsusc}). 

\begin{figure}[t]
\includegraphics[angle=-90,width=0.7\columnwidth]{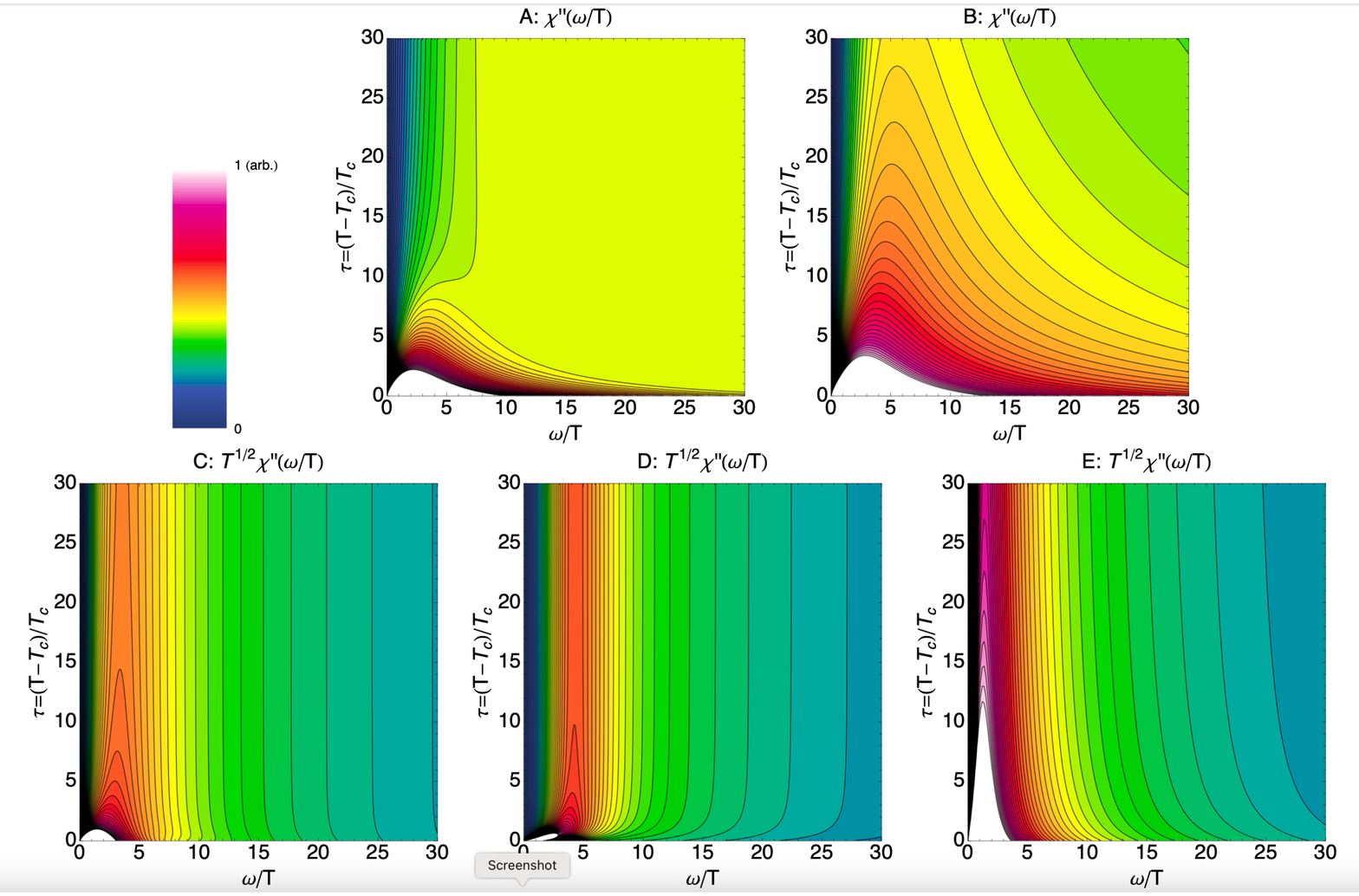}
\caption{As in Fig. (\ref{fig:HoloSCsusc}) but now subjected to $E/T$ scaling  \cite{holoBCS}. The take home message is that all cases except B obey simple "universal" energy-temperature scaling, the difference being in the anomalous dimension of the pair operator. The outlier is case B: this is the outcome of a "maximally scale invariant", strongly coupled implementation of the Hertz-Millis quantum critical affair, illustrating how serious perturbative corrections wreck the energy-temperature scaling. See the main text.}
\label{fig:HoloSCscaling}
\end{figure}

Comparing BCS (panel A) with the holographic outcomes (panels D,E), where are the differences? These get sharply in focus by applying an energy-temperature scaling collapse. Take the ratio $\omega/T$ as the $x$ axis and $T^{\alpha_p} \chi"_p (\omega, T)$ for the $y$ axis and adjust $\alpha_p$ to arrive at a scaling collapse. The outcome is shown in Fig. (\ref{fig:HoloSCscaling}). This reveals that in all three cases energy-temperature scaling is in effect in the normal state at energies and/or temperatures large compared to $T_c$. This scaling is only interrupted by the lingering instability towards superconductivity. This is revealed by the "contour" lines in the false colors becoming precisely parallel in the upper right quadrant of these panels. 

What is the difference between BCS and the holographic cases? In BCS (panel A) the scaling dimension is of course  marginal, $\alpha_p =0$ (the uniform light green area), while for the holographic cases the normal state/strange metal scaling dimension is {\em relevant} peaking towards low frequency with scaling dimension $\alpha_p = 1/2$ in both cases (panels D,E)!
 
This can be captured in a very simple construction.  In the same year that holographic superconductivity was discovered, we pointed out  \cite{qucritBCS}  that one can easily generalize BCS relying on its scaling logic, calling it "quantum critical BCS". This just departs from the wisdom that one can keep the mean field machinery explained in the above, modifying it by considering a generalization of  the pair susceptibility of the normal state, $\chi_p^0 (\omega)$. We asserted that it could be a scaling function but now one with an unknown scaling dimension of the pair operator precisely in the guise of the covariant RG,

\begin{equation}
\chi_p^0 (\omega) \sim \omega^{2 \Delta_p}
\label{qucritBCS}
\end{equation}

When $\Delta_p = 0$ this recovers BCS. For the "irrelevant" $\Delta_p > 0$ one needs a quite finite interaction for the instability to occur (e.g., in a semimetal like zero density graphene $2\Delta_p  = 1$). However, $\Delta_p$ can in principle be negative: the "pair operator is relevant in the deep IR". When one now solves the gap equation one finds out an {\em algebraic} expression for $T_c$, instead of the logarithm. Instead of this marginal, fine tuned borderline affair that is at work in the Fermi gas the normal state is now {\em genuinely unstable} towards superconductivity. 

One may not be surprised that this implies that a quite  humble "pairing glue" may now gives rise to a  "high" T$_c$. This was actually the main take home message of the quantum critical BCS paper. To explain the shear magnitude of T$_c$ in the cuprates has been a holy grail in this community. But invariably it was assumed by everybody that the pair susceptibility of the metallic state should be BCS like, which in turn implies that one has to invent a very large attractive interaction, the "superglue" theme that gave rise to much guess work. But we know that the normal state is not a Fermi-liquid. Why should the pair operator in the metal be marginal?  

The quantum critical BCS result is shown  in panel C of both figures, using $\Delta_p = - 1/2$ and a 2D CFT model capturing the $E/T$ scaling,. Besides some small details associated with the evolution of the relaxational peak one infers directly from especially the collapsed data (Fig. \ref{fig:HoloSCscaling}) that this precisely captures the response of especially the "plain-vanilla" holographic case (panel D). 

The take home message up to this point is that holographic superconductivity is rooted in the same impeccable simple scaling logic as weak coupling BCS, the difference being that the scaling dimension of the pair operator can take a-priori any value. Obviously, when it is irrelevant in the IR the superconducting instability is suppressed -- the BF bound is not reached dealing with a too massive bulk scalar. The marginal BCS case requires infinite fine tuning. The natural situation with regard to the "physics of T$_c$" is therefore according to holography associated with a relevant pair operator.  

We are not done yet. I emphasized in Section (\ref{FLdiagrams}) that the neat and simple scaling of the Fermi-liquid is special to the weak coupling limit. When interactions become substantial a plethora of irrelevant operators kicks in that can be addressed with perturbation theory. This is the analogue of the perturbative interactions destroying universality above the upper critical dimension in the stoquastic critical case having as ramification that simple scaling will fail. 

This is addressed quite professionally  in this paper \cite{holoBCS}; compared to the other cases this was in technical regards the tour de force. I referred in section (\ref{FLdiagrams}) to the Migdal-Elaishberg theory relying on the smallness of the Migdal parameter: when the characteristic energy scale associated with the "glue" is small compared to $E_F$ one ends up with a self-consistent perturbation theory that can be solved in closed form. In Section (\ref{hertzmillis}) I discussed the  main stream Hertz-Millis idea: couple the Fermi-gas to the quantum critical fluctuations associated with a stoquastic quantum phase transition. Chubukov and coworkers argued that these fluctuations should be bounded from the above by a small energy scale, introducing thereby a small Migdal parameter.  In turn, the pairing glue originates in the scale invariant fluctuations and one can now assert that the "glue function" to be used in the Migdal-Eliashberg theory is itself scale invariant, $\lambda ( i \omega) \sim 1/ | \omega |)^{\gamma}$. 

This creates the circumstances for a  "UV" that is as scale invariant as possible within the Hertz-Millis context. But the trouble is that the Fermi-gas itself is scale full. The result for the pair susceptibility is shown in panel B in Fig.'s (\ref{fig:HoloSCsusc},{\ref{fig:HoloSCscaling}). The unscaled version does not look all that different from the other cases but the big difference is revealed by the scaled outcome: one sees directly from panel C in Fig. (\ref{fig:HoloSCscaling}) that there is no longer energy-temperature scaling,  no parallel contour line to be discerned in the upper right quadrant! This illustrates in a lively fashion the detrimental effects of the perturbative corrections on scaling universality, now in the context of the finite density systems.  

The take home message is that a measurement of the dynamical pair susceptibility in the normal state, in the approach to the transition would be an ideal way to get an experimental handle on the key questions: find out whether there is the type of scaling collapse highlighted in Fig. (\ref{fig:HoloSCscaling}) and if so determine the scaling exponent. As a first example of a rather generic "syndrom", the type of experiments that would reveal the "smoking gun" evidence for quantum supreme matter cannot be easily implemented in the laboratory, you will meet more of it in Section (\ref{highTcxep}). In principle a device can be manufactured that in principle can mine this pair susceptibility information in the form of the "Ferrell-Scalapino-Goldman" device revolving around the second order Josephson effect, see the paper  \cite{holoBCS}. Several experimental groups tried, but up to now it failed for the annoying reason that for chemistry reasons it has proven impossible to fabricate the required thin insulating barriers. 

There was actually {\em computational} evidence presented for the "relevant" pair susceptibility. I referred to cluster DMFT in Section (\ref{Compmethods}). In fact, under the supervision of the father of the cluster version --  Mark Jarrell who also deceased at a young age -- direct evidence for relevant quantum critical  BCS was reported in 2011 \cite{JarrellBCS}. DMFT is in the hands for a large community where the obligatory stuffs get cited thousands of times. This paper is only cited $\sim 50$ times! It is easy. Mankind is genetically a religious species, and BCS gap functions have acquired an untouchable status in the minds of the specialists, comparable to the status of Jesus in the heads of christians. 

\subsubsection{ The holographic Leiden-MIT fermions and the UV dependence.}
\label{Holofermions}

The pair susceptibility illustrates a general circumstance: for various practical reasons it is difficult, if not impossible to measure any type of {\em dynamical} response functions over a large kinematical range in the laboratories. We also encountered the "primitive" charge susceptibility. Away from $q=0$ where it is readily available by optics only very recently machinery became available to measure it with a sufficient energy resolution (Section \ref{EELSlocqucrit}). The dynamical magnetic responses can be probed by inelastic neutron scattering but this is impeded by a signal to noise ratio that is so bad that  the signal cannot be detected in the cuprate strange metal.

In fact, the only dynamical information that is readily available with very high resolution pertains to the {\em fermion} operators: the momentum space angular resolved photoemisson (ARPES) and its real space counterpart Scanning Tunneling Spectroscopy (STS) (see Section \ref{Cupratefermions}). The next big splash following immediately after the holographic superconductivity  was  the discovery of the "Leiden-MIT holographic fermions". I was myself directly involved. This started  in 2007 when I got on speaking terms with Koenraad Schalm and I fired at him the question "any chance that you know how to compute fermion propagators holographically". Koenraad knew how to do this -- it got delayed significantly by a bug in the code but eventually we nailed it with help of the graduate student Mihailo Cubrovic. Subsequently we managed to land the paper in Science \cite{Sciencefermions09}: the first ever string theory work that got published in this prestigious journal. 

However, in the mean time greatly competent competitors had gotten the same idea: Hong Liu, John McGreevy and their brilliant students at MIT. They managed to take apart the numerical results to actually explain how it all worked in terms of bulk language. I already alluded to their penetrating insights in Section (\ref{RNearly}) identifying the $z \rightarrow \infty$ local quantum criticality as encoded in the throat geometry of the RN black hole. They explained also how this landed in the structure of the fermion two point functions in the boundary. 

 Although this pulled me into the holographic pursuit, shortly thereafter it was clarified that these fermions are an instance where the IR phenomenology is actually {\em critically dependent on the large $N$ UV.}  This was quite deceptive: initially it looked like that the holographic outcomes were remarkable look-alikes of the experimental results. But it became clear later that this was a deception (see also Section \ref{Cupratefermions}).  
 
 ARPES is a close proxy to measuring  the probability to remove an electron from the system at a particular energy and single electron momentum. STS measures a proxy to the probability to add and remove an electron, but now locally in space ($  {\bf r}  =  {\bf r'} $). Both techniques are not ideal measurements, one has to be aware of various caveats but by and large the experimentalists have a fair understanding what these are. These therefore yield direct information on the central pillar of the classic diagrammatic perturbation theory of "particle physics" applied to electron systems: the one electron Green's function, 

\begin{equation} 
i G_{\alpha, \beta} ( {\bf r} t, {\bf r'} t' ) = \langle  \Psi_0^N | \hat{T} \left[ \psi_{\alpha} ( {\bf r} t) \psi^{\dagger}_{\beta} ( {\bf r'} t') \right] | \Psi_0^N \rangle
\label{elgreensfunc}
\end{equation}

specializing to zero temperature ($ | \Psi_0^N \rangle$ is the ground state)  while  $\hat{T}$ is the time ordering operator and  $\psi^{\dagger}_{\alpha} ( {\bf r} t)$ creates an electron at ${\bf r} t $ with spin $\alpha$. STS yields the local in space spectral function while ARPES delivers the momentum space spectral function for electron removal, associated with Eq. (\ref{elgreensfunc}) in the frequency domain. 

Long before these experimental techniques were developed, the quantum field theorists  who developed the diagrammatic theory in the 1950's appointed this object as organizing principle: the "double lines" in the standard diagrams. But this rests on converging perturbation theory around the free fixed point: it only works departing from a SRE product "classical vacuum". 

I presume the reader is familiar -- it works the same way in the high energy context  as in condensed matter, the main difference being that in the latter case one aims at the finite density Fermi gas. One departs from the non-interacting system, with the single particle electron states given by the quantum mechanical band structure $H_0 = \sum_{{\bf k}, n, \sigma} \varepsilon_{{\bf k}, n} c^{\dagger}_{ {\bf k}, n, \sigma} c_{ {\bf k}, n, \sigma}$ where $n$ is a band label that I will suppress in the remainder to save on writing indices. One then switches on the interactions $\sim H_1$. Under the condition that  the perturbation theory is {\em converging}  the effects of the interactions can be lumped together in the {\em self-energy} $\Sigma$,

\begin{equation}
G ( {\bf k}, \omega ) = \frac{1} { \omega - \varepsilon_{\bf k} + \mu- \Sigma ({ \bf k}, \omega ) }
\label{selfenergyG}
\end{equation}

The imaginary part of $G$ is the spectral function that is measured experimentally.  The existence of the quasiparticles is signalled by the presence of "poles".  The self-energy is a complex function,  $\Sigma = \Sigma' + i \Sigma"$ and $\Sigma"$ encodes for the inverse life time of the quasiparticle -- the Kramers-Kronig consistent $\Sigma'$ will shift around the energy of the quasiparticle encoding for the generic mass enhancement of the quasiparticle relative to the bare particle. In the spectral function one will find a Lorentzian peak at this QP energy with a width $\sim \Sigma"$, the  "particle pole".  

When $\Sigma"$ decreases sufficiently fast upon approaching zero energy the quasiparticle is underdamped ("it exists") becoming infinitely long lived at the Fermi-surface. Departing from this fixed point one can get away with second order perturbation theory at low energy.  The excited electron/hole decays in the Lindhard continuum thereby picking up the power law behaviour. Accordingly, one is dealing with a near ideal gas like situation where the life time of the electron is set by the collision rate with other electrons, which is easily computed to be $\Sigma" (\omega, T) \sim g^2 \left( (\hbar \omega)^2 +  (2 \pi k_B T )^2 \right)$.  The dispersion $\varepsilon_k \sim v_f k$ and it follows self-consistently that the interacting electron system renormalizes in the free quasiparticle gas. 

Anticipating on the discussion regarding the experimental landscape in cuprates (Section \ref{highTcxep}), the evidence coming from STS and ARPES for the existence of such quasiparticles {\em deep in the superconducting state} is overwhelming. These appear to closely follow the expected behaviour for the Bogoliubov fermions of the BCS theory: $\gamma_k^{\dagger} \sim u_k c_k^{\dagger} + v_k c_{-k}$. Intriguingly, these spring into existence when the superconducting order develops. However, the question is whether these also exists in whatever incarnation in the strange metal above $T_c$. 

Also in the normal state, pending the direction of momentum one still finds peaks moving as function of momentum (the "nodal fermions") and the habit is widespread to jump to the conclusion that there are still quasiparticles, Fermi surfaces and all of that. But this is a tricky affair. For instance, the CFT Fermions at zero density (Fig. \ref{fig:fermionbranchcut}) do exhibit peaks in the spectral function that disperse as function of momentum. The "unparticleness" is however encoded in the energy dependence of the line shape, branch cut propagators have an analytical structure that cannot be written in the perturbative self energy form Eq. (\ref{selfenergyG}). Such lineshapes are for practical reasons difficult to capture in experiment and I will discuss very recent progress demonstrating the unparticle nature of the strange metal in Section (\ref{Cupratefermions}).   

The Leiden-MIT fermions seemed to get close initially but it became later clear that this is actually a large $N$ pathology.  What is going on? This early work departed from the RN strange metal. According to the dictionary fermions are inherently quantum mechanical also in the bulk and one finds out that the two point fermion propagators in the boundary dualize in Dirac {\em quantum mechanical waves} "falling" to the black hole in the bulk. 

The MIT group reconstructed how this works in the bulk in terms of the "matching construction". For the Dirac fermions propagating in the bulk one can capture the effects of the change in geometry upon entering the RN throat in terms of an effective flat space Schr\"odinger equation: the "geometrical domain wall" signalling this geometry change at radial coordinate $\sim \mu$ then translates in a large potential barrier felt by the fermion. This acts like a box "clamping" the Dirac waves at the boundary and the outer side of the barrier producing standing waves, corresponding with a tower of fermion states in the boundary. These can however tunnel through the barrier, landing in the near horizon regime where these fall through the horizon: the rate by which this happens  translates into the inverse life-time of the quasiparticles in the boundary. 

The bottom line is that the holographic fermion propagators reproduce the self-energy form Eq. (\ref{selfenergyG}). The $\varepsilon_{\bf k}$ are associated with the radial "standing Dirac waves" while the self energy is set by the tunneling to the near-horizon geometry. The action is at the holographic equivalent of the (large) Fermi-momentum and this probes the geometry at length smaller than the local length I discussed towards the end of Section ( \ref{holochargeexc}). The deep infrared fermion propagator has here a similar odd form, depending on the fermion wavevector: $\Sigma (k, \omega) \sim \omega^{2 \nu_k -1}$ where $\nu_k \sim \sqrt { (1/\xi )^2 + k^2 }$ with $\xi \sim 1/ \mu$ being the local length. 

The picture that emerges is that quasiparticles do exist but by second order perturbation theory (the tunneling in the bulk)  they do encounter a "quantum critical heat bath" (the near horizon strange metal degrees of freedom) in which they eventually decay. This is in tune with the old marginal Fermi-liquid phenomenology, the difference being that the anomalous dimension governing the self-energy is now a free parameter, set by $k \simeq k_F$. But for $\omega \rightarrow 0$ $\Sigma" \rightarrow 0$ and one can identify a Fermi surface -- the big deal of our Leiden numerical work. 

The tricky part  is that the exponent $\alpha = 2 \nu_{k_F} -1$ can vary all the way from 2 (the Fermi-liquid value) to smaller than 1. The case $\alpha =1$ is the marginal case: by measuring the width of the peaks in momentum space as function of energy one sees such a damping in the ARPES (nodal fermions) of optimally doped cuprates, the stronghold of the MFL view. However, $\alpha$ can be less than one and then the self energy is more IR relevant than the dispersion: in the scaling limit $G \rightarrow 1/\Sigma$ and one is dealing with a "naive" emergent branchcut in the deep IR. For the insiders, this is coincident with the central wheel  in the highly fashionable "SYK model" affair. Frankly, being aware that this is precisely the instance where {\em real} holography (to my impression, the SYK claims in this regard  are over-interpretations based on mathematical coincidences) drowns in large N pathology, I am most hesitant to take any of the SYK developments serious. Here is the reason that you don't find any reference to SYK in these lecture notes.  

Because of the similarity with the MFL-style analysis of ARPES  this looked early on quite promising. But I remember well sitting in on a talk by the string theory celebrity Joe Polchinski, explaining a paper he had written with Faulkner \cite{Semiholo}. Rarely ever did I encounter such a disappointment dealing with theories of physics.

Joe explained that the above phenomena can actually be understood directly in the boundary field theory. But the big deal is that these are {\em critically dependent} on the fact that one is dealing with a supersymmetric Yang-Mills theory {\em in the large $N$ limit}! This is just UV sensitive special effect of the large $N$ limit and from the way it springs in existence one can directly conclude that this has nothing to say about the way that quantum supreme metals formed from electrons in solids behave. 

This large $N$ affair is as follows. Given that one is dealing with a non-Abelian gauge theory one may not be surprised that even in the large $N$ versions one is still encountering the generic physics of such theories: the vacuum may be {\em confining or deconfining}. All the "quantum supreme" powerlaw stuff is associated with the deconfining regime. It is well understood how confinement works in  holography: in essence, the geometry of the deep interior just vanishes being replaced by "walls", geometrical structure  reflecting  the waves emanating from the boundary. This is the bulk "box" I referred to in the above and the various standing waves that form along the radial direction dualize in towers of states in the boundary corresponding with the gauge singlet "mesons" of the confining phase. 

In fact, in this particular setting the fermions are the supersymmetric partners of the gauge degrees of freedom in the adjoint: in top-down set ups these can be for instance be identified with the SUSY partners of gravitons. The big deal is that in the large $N$ limit the interactions between the gauge singlets ("pion exchange") is completely suppressed and the (s)mesons live infinitely long. This is the origin of the free limit in holography. 

But the real bad news is still to come. As Joe explained, the Leiden-MIT fermions are nothing else than "mesonic resonances" that are developing in the deconfining state upon approaching the confinement transition tuned by varying the zero density  fermion anomalous dimension towards the unitary (free CFT) limit. But the nature of these resonances is determined by a large N pathology. One can demonstrate that all vertex corrections are $1/N$ suppressed and only under this condition the simple affair of quasiparticles decaying in a heat bath by second order perturbation theory makes sense! This is actually the same short cut taken in MFL -- we know that this is a-priori unreasonable, it is intuitive guess work. It is controlled in holography, unfortunately, by a small parameter that has no meaning in condensed matter ($1/N$). 

This story continuous with the construction of the dual of the Fermi-liquid that followed soon thereafter: the "electron star" (see Chapter 11 in \cite{holodualbook}). It may be already obvious that deep in the confining regime it is very simple to construct a real Fermi-liquid directly in the boundary. One is dealing with the towers of non-interacting fermionic mesons. One can subsequently just fill up these states using the Pauli principle to find an impeccable Fermi-gas. The subtle question is, how is this finite density Fermi-gas encoded in the bulk?  The fermions are $1/N$ suppressed and therefore individual fermions do not contribute to thermodynamics, let alone that collective (density etc) responses have anything to do with fermion loops: this is special effect of the Fermi-liquid. But the key is that a {\em macroscopic} assembly of fermions in the bulk can take over the ground state. 

In the special, unphysical limit that the fermion charge becomes infinitesimal the bulk becomes tractable: upon lowering the fermion mass in the bulk, at a critical point the RN black hole "uncollapses" in the "electron star". This is nothing else than the Tolman-Oppenheimer-Volkoff solution for the neutron star, resting on the Thomas-Fermi equation of state for the Fermi gas, but now extended by incorporating the electrical charge. This describes indeed the free Fermi-gas in the boundary. In fact, in this limit the electron star takes over from the black hole when the zero density fermion scaling dimensions become such that the Leiden-MIT probe limit indicates that Fermi surfaces start to form.  

At the least this provides proof of principle that the holographic strange metals know about the fermion signs since these can be the birth place of Fermi-liquids. But even within the confines of the large $N$ theory the precise workings of these holographic Fermi-liquids is far from settled. Upon increasing the fermion charge one finds out that one has to deal with a {\em quantized} electron star. It turns out to be very difficult to corner this because of a variety of gravitational difficulties one encounters dealing which such highly quantal matter sources. This is presently still an open problem. 

The take home message is that one should be at all occasions acutely aware that the strong emergence physics suggested by holography may be polluted by "UV dependence".  Seemingly general traits of the IR theory may still depend in a critical way on the oddities of the UV theories of the string theorists. The good news is that the motives that spoil the fermions are tied to $1/N$ suppressed features. None of these seem to play a role in the leading order (in $N$) collective sector that delivers the charge excitations, transport, pair susceptibilites and so forth. Still, holography may be a worthwhile guide book telling us how to navigate our minds out of the quasiparticle tunnel vision. But this navigation tool can not be trusted, at unexpected instances it may be quite unreliable.              
 
This terminates the first part of the discussion of the physics of quantum supreme states of matter according to holography. What remains to be done is the presentation of holographic transport theory, the most highly developed part of this agenda. But to get this in proper perspective the last technical hurdle has to be overcome: spoiling the homogeneity and isotropy of the space manifold. This is presently still a frontier where much is yet to be sorted out.

\section{ Breaking the translational symmetry and holographic transport theory.}  
\label{holotransport}

The study of macroscopic transport properties has been dominating in the field of applied holography. This has caused a bit of a tunnel vision among the string theorists -- as a condensed matter "customer" I wish to know especially what holography has to say about  the spectroscopic responses highlighted in the previous Section. But compared to the large body of holographic literature on macroscopic transport these aspects are lagging behind. In part this bias is due to an unfamiliarity with experimental techniques in this community. But it also reflects culture: the string theory tradition has been propelled by mathematics and a lot more can be done in the form of analytical results in the bulk dualizing in the long wavelength -- long time regime of the boundary. This is a greatly intriguing affair by itself. 

Where are the reservations regarding transport coming from, viewed from the phenomenological arena? It is an old wisdom that "the conductivity is the first thing that is measured but the last thing that is explained." The origin of the trouble is that transport revolves around  "nearly" {\em conserved quantities}:  charge, total linear- and angular momentum. The consequence is that systems that are microscopically very different may still exhibit transport properties that are qualitatively very similar. The crucial data containing the differences are "filtered" by the conservation laws to a degree that only quantitative information can be mined. These numbers are in turn highly sensitive to details: these reflect the  break down of the conservation laws, and this is a non-universal affair.

I will start the discussion with a pedestrian introduction to the generalities of transport theory. This relies at the end of the day on wisdoms of 19-th century physics.        But it may be beneficial for especially the condensed matter reader to be reminded of the basics. The Fermi-liquid based transport theory has been of course a phenomenal success story of twentieth century physics, explaining in great detail how it works in normal metals. However, a folklore evolved where the difference between transport generality and the specifics of the, in a  way, pathological Fermi-liquids got snowed under. In Section (\ref{Drudetransport}) I will spell out how "nearly" conserved total momentum together with charge conservation imposes the universal Drude transport. Charge conjugation symmetry is the other crucial ingredient (Section \ref{CCSinc}) -- I will emphasize that one has to keep eyes wide open in this regard dealing with the ultra-relativistic nature of holographic liquids in Section (\ref{Homosecsector}).

I will then zoom in on the highlights of holographic transport (Section \ref{Holohydro}). A most striking, extreme  contrast is predicted between the gaseous transport physics of Fermi-liquids versus the holographic predictions for finite temperature quantum supreme metals.  The latter are in the grip of the extremely fast quantum thermalization -- the Planckian dissipation as highlighted in Section (\ref{FiniteTMaldacena}) -- with the ramification that even in the presence of substantial disorder these behave like {\em hydrodynamical} liquids, that are expected to be characterized by the minimal viscosity. Given this extreme contrast, this is the arena of choice to look for the experimental smoking guns, but yet again the complicated chemistry is in the way of fabricating the required nano-transport devices. 

In the final Section (\ref{holoUmklapp}) I will focus in on a frontier of the AdS/CMT pursuit: the way that periodic potentials influence holographic transport, the "Umklapp" motive. Here the mathematical hell  breaks loose in the form of the bulk Einstein theory turning into a complicated affair of non-linear partial differential equations. This is presently largely uncharted and poorly understood. I will illustrate this with some very new results for the simplest case, the Reissner-Nordstrom metal on a square lattice which is already littered with interesting surprises.

\subsection{The basics of transport theory: conservation laws.}
\label{Drudetransport} 

The detailed explanation of the way that the transport of charge and heat works in normal metals is among the great success stories of solid state physics. The Boltzmann kinetic theory departing from the Fermi-liquid was cooked to perfection in the 1950's turning into a semi-quantitative framework explaining in detail how various transport behave pending the specifics of the electron system. Generation after generation got trained in this art and it seems that in the course of time  a certain blindness developed with regard to distinguishing the generalities that govern any form of transport and the specifics of the gaseous physics in its most extreme form underlying Fermi-liquid transport. 

Invariably, the macroscopic transport in metals is of the "Drude" kind. A student taking a bachelor course in solid state physics may have the impression that Drude transport is uniquely associated with the weakly interacting Fermi-gas, the Sommerfeld model.  It does happen that I hear condensed matter physics professors claim that a Drude transport implies that one is dealing with a Fermi-liquid!  It is actually the case  that Paul Drude formulated his theory in 1900, before Planck identified his constant. He departed from a dilute classical plasma, resting on the fact that with kinetic gas theory one can keep track of all the details. But the outcome is actually completely generic: it is controlled by {\em conservation laws}, protecting the "hydrodynamical soft modes". The theory is generic, it applies to literally {\em anything}. Only the  numbers are pending the specifics of the microscopic nature of the electron system. 

This is very simple  but the intervention of holography was in this regard for me personally quite beneficial in helping me to think outside the "quasiparticle box". Transport in holographic systems is entirely detached from the usual weakly interacting quasiparticles but it was at least psychologically an eye opener to see that it eventually boils down to the same Drude response as for the textbook Fermi-gas. It also alerted me regarding widespread habits that had developed in the CM community in the course of time, such as the rather cavalier way to completely ignore the "transport vertices" in computing the current response functions entirely in terms of "bare" fermion-loops, referring to "local models". This is for instance the community standard in e.g. the use of dynamical mean field theory. Frankly, I am unconvinced whether any of it relates to transport in the laboratory. 

Let me present here a very elementary overview of the principles controlling transport before I turn to the specifics of the holographic metals. 

\subsubsection{Drude transport: generic at finite density.} 

Transport means that stuff is moving from A to B in a finite amount of time. One meets immediately the first Noether charge identified by mankind: when space is {\em homogeneous}, it looks everywhere the same, the total momentum of the stuff is {\em conserved}. It is Galilean invariance: your velocity is not changing when you fly through empty outer space. 

The current is actually about the transport of mass, assuming for simplicity that the stuff has a finite rest mass. The most natural force to accelerate such stuff is gravity. But electrons are very light and the gravitational force is vanishingly small as compared to the electromagnetic forces that can be exerted on the charged electrons. In X-ray tubes, synchrotrons and so forth one lets electrons accelerate in the vacuum in an electrical field and everybody knows the outcome. Assume a system of electrons at a density $n$ and one can relate the total charge current $\vec{J}$  to total momentum $\vec{P}$ by, 

\begin{equation}
\vec{J} = n e \vec{v} = \frac{n e}{m_e} \vec{P} 
\label{currentdef}
\end{equation}

However, when electrons move in solids with thermal energies they experience a space where the translational symmetry is weakly broken -- how this works requires microscopy. This implies that their total momentum acquires a finite lifetime, $\tau_P$. This is all one needs to know to write the EOM for the total momentum,

\begin{equation}
\frac{d \vec{P}}{dt} + \frac{1}{\tau_P} \vec{P} = e \vec{E}
 \label{DrudeEOM}
\end{equation}        

where $\vec{E}$ is the applied electrical field. Assume this source to oscillate in time like $\vec{E}  (t ) = \vec{E} (\omega) e^{i \omega t}$ and look for a response $\vec{P} ( t) = \vec{P} (\omega) (t) e^{i \omega t}$ and it follows immediately that $ ( i \omega  - 1/ \tau_P)   \vec{P}(\omega) = e E (\omega)$.  The AC (optical) conductivity $\sigma (\omega)$ becomes ,

\begin{eqnarray}
\vec{J} (\omega) & =  & \frac{n e}{m_e} \vec{P}  = \sigma (\omega) \vec{E} (\omega) \nonumber \\
\sigma (\omega) & = & \frac{ {\cal D}_D}{ \frac{1}{\tau_P} - i \omega}
\label{Drudelong}
\end{eqnarray}

Where ${\cal D}_D$ is the general quantity called the "Drude weight". In this specific example it is coincident with the plasma frequency squared of the simple electron plasma ${\cal D}_D = \omega^2_p = n e^2 /m$ but in general this "screening power" of the electron system may involve  microscopic specifics. The optical conductivity is a complex function $\sigma (\omega) = \sigma_1 (\omega) + i \sigma_2 (\omega)$, where the real part  ($\sigma_1$) represents the dissipative part of the response. This is the "half Lorentzian" centred at zero frequency with a width $ 1 /\tau_P$: $\sigma_1 (\omega) =   {\cal D}_D \frac{1/\tau_P}{ (1/ \tau_P)^2 + \omega^2}$. 

Taking now the DC limit  ($\omega =0$) it follows immediately for the DC resistivity that $\rho = 1/\sigma_1 (\omega =0) = 1 /{\cal D}_D \times 1/\tau_P$, the familiar text book expression (${\cal D}_D$ and $\tau_P$ are typically expressed in the specific dimensions of the Fermi gas). It is a habit by condensed matter transport specialists to assume without further thought that one is dealing with Drude transport. But the only way to find out whether this is indeed the case is by measuring the optical conductivity. The outcome Eq. (\ref{Drudelong}) represents an analytical function that is uniquely related  via Eq. (\ref{DrudeEOM}) to the existence of a momentum that is sufficiently long lived to conclude that  it actually controls the transport. 

This simple ploy can be further formalized using the Mori-Zwanzig memory-function formalism. One can look this up, see e.g. the book \cite{lucasbook} having a strong focus on transport. The take home message is that it revolves entirely around modes that are to zero-th order protected by conservation laws, bringing into account  the "weak" violation of the conservation law using perturbation theory. It highlights the  difference with other forms of perturbation theory by promoting the hydrodynamical soft modes to the drivers seat. 

One important generalization is that the relaxation rate may dependent on time itself. In the above we assumed that the origin of the momentum relaxation is static: $1/\tau_P$ is assigned to "elastic scattering". Using the memory matrix formalism it is easy to demonstrate that the momentum relaxation can be captured by an "optical self-energy" $\hat{M}$,

\begin{equation}
\sigma (\omega)  =  i \frac{ {\cal D}_D}{ \hat{M} (\omega) +  \omega}
 \label{genDrude}
\end{equation}   

One recovers the "simple Drude" by asserting that $\hat{M} (\omega) = i /\tau_P$. However, this KK consistent quantity is in general frequency and temperature dependent. For instance I will explain soon that due to the presence of a periodic lattice (Umklapp scattering) $\hat{M}$ becomes in a Fermi-liquid  $\mathrm{Im} ( \hat{M} ) \sim (\hbar \omega)^2 + ( k_B T )^2$, it behaves as the quasiparticle collision rate. 

A final general wisdom is associated with the fact that the optical conductivity and the charge susceptibility/dielectric function are tied together by the continuity equation. According to the Kubo formalism the conductivity and charge susceptibility are  associated with the current-current  and density-density correlation functions, respectively.  But currents and densities are tied together by the continuity equation that is expressing charge (number) conservation, $\partial_t \rho + \vec{\nabla} \cdot \vec{\rho} = 0$. Accordingly, one can relate the response functions by, 

\begin{eqnarray}
\sigma (\omega, {\bf  q}) & = &  i \frac{\omega}{ q^2} \Pi   (\omega, {\bf  q}) \nonumber \\
\chi_{\rho}  (\omega, {\bf  q}) & = & \frac{\Pi   (\omega, {\bf  q}) }{ 1 - V_q \Pi   (\omega, {\bf  q})} \nonumber \\
\varepsilon   (\omega, {\bf  q}) & = & 1 - V_q \Pi   (\omega, {\bf  q})
\label{polprop}
\end{eqnarray}

Where $V_q \sim 1/q^2$ brings into account that the zero sound that resides in $\Pi$ gets promoted to the plasmon. The infinitely long lived sound at $q \rightarrow 0$ in the Galilean continuum translates in the delta function peak at $\omega =0$ that one finds in his limit in $\sigma_1$.  
Different from $\chi_{\rho}$ that can be measured over a large momentum range by EELS, because of the fact that the light velocity is much 
larger than typical material velocities one can with optical means only measure $\sigma (\omega, q =0)$. In fact the (longitudinal) conductivity shows the zero sound mode as a resonance at finite momentum as well as the imprint of the (generalized) Lindhard continuum.

In summary,  the take home message is that this Drude response is actually completely generic for any system at a finite density characterized by a "weak" breaking of translational symmetry. On this level it has no relation whatsoever to the issue whether one has the extremely gaseous excitations of the Fermi-liquid or either with the densely entangled unparticle soup of the holographic metals. The transport in the latter is at energies compared to the chemical potential as much dominated by a Drude peak as it is in a simple metal like copper. 

\subsubsection{Charge conjugation symmetry and the "incoherent" conductivity.} 
\label{CCSinc}

Is there any way for a physical system to evade a Drude response while the conductivity is finite? There is just one symmetry circumstance allowing this: {\em charge conjugation symmetry}. This is again very simple to understand. Imagine a system composed of a reservoir of positive- and negative charges that can move freely, while these reservoirs carry precisely the same net charge. Apply an electrical field: the $+$ and $-$ charges move in precisely opposite direction and an electrical current is running.  However,  since these reservoirs are otherwise identical the {\em center of mass of the combined system is not moving}. Hence, the electrical current $\vec{J}$ {\em decouples completely  from total momentum} $\vec{P}$.   

What to expect for the optical conductivity under such circumstances? Let us consider the case of massless, non interacting Dirac fermions. An example in condensed matter physics is graphene. Graphene is of course famous for having such "Dirac cones" in the band structure. But there is an issue with the interactions. Although in much of the gigantic graphene engineering effort it is taken for granted  that interactions should be ignred, departing from the (poorly screened) Coulomb interaction one can show that the interactions are marginally irrelevant \cite{juricicgraph}. The IR fixed point is free but interactions switch on rapidly upon raising temperature or energy. But this will not matter for the electrical conductivity as you will see in a moment. 

\begin{figure}[t]
\includegraphics[width=0.7\columnwidth]{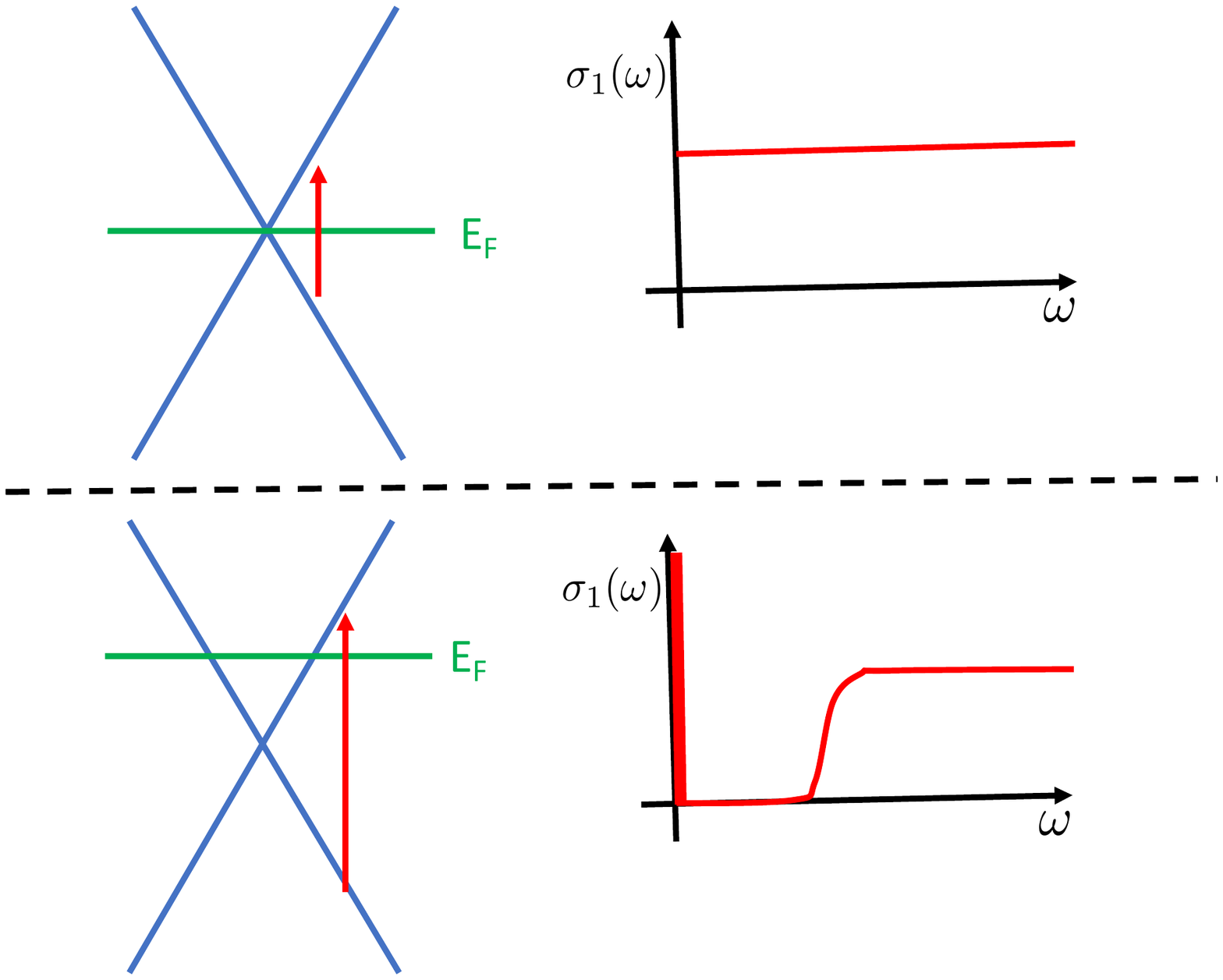}
\caption{The optical conductivity dealing with free Dirac fermions. At zero density (upper panel) the optical conductivity is due to "interband transitions" (red arrow) giving rise to an energy independent conductivity in two space dimensions. At a finite density (lower panel) these are exponentially suppressed at low energy due to Pauli blocking and this missing spectral weight lands in the Drude peak at zero frequency.}
\label{fig:diracfermopt}
\end{figure}

The conductivity is now determined by the particle-hole excitations that can be created at a given energy, see Fig. (\ref{fig:diracfermopt}). In a condensed matter language these correspond with interband transitions that are organized in a particular fashion. It is easy to count this out with the result that at zero temperature $\sigma (\omega) \sim \omega^ {(d - 2 )/z}$. For Dirac fermions $z=1$ but this scaling expression also applied  for instance to the case that two parabolic bands touch at the charge neutrality point as in bilayer graphene  where $z=2$. It follows immediately that in two space dimensions (as in Graphene) the optical conductivity is independent of frequency (Fig. \ref{fig:diracfermopt}), a well known fact that has been confirmed experimentally.

But imagine now that we turn into it into a strongly interacting CFT such that the free Dirac fermions will turn into the branch cuts, what happens with the electrical conductivity? Yet again symmetry exerts total control: next to energy conservation we know that the electrical charge is locally conserved, "protected by gauge-invariance". This implies that the current operators cannot acquire anomalous dimensions and the implication is that the engineering scaling of the free limit is actually universal. 

Assuming universality, a ramification is that the optical conductivity should obey the following scaling relation \cite{PhilipCFTcond} ,
 
\begin{equation}
\sigma_{\mu=0} (\omega, T) = \frac{Q^2}{\hbar} T^{(d-2)/z} \; \Sigma ( \frac{\hbar \omega}{k_B T})
\label{CFToptcond}
\end{equation}

where $Q$ is the electrical charge. The energy/temperature cross over function $\Sigma (x)$ has to be of such a form that for $x <<1$ $\sigma \sim T^{(d-2)/z}$ while for large x $\sigma \sim \omega^{(d-2)/z}$: only the prefactors can be different in both limits .  The take home message is that the conductivity acquires in $d=2$ this universal  frequency independent form in the presence of charge conjugation symmetry tied to zero density. This is surely confirmed by holography; in the plain vanilla "Maldacena CFT" one finds that the prefactors I just mentioned are the same due to a bulk electromagnetic duality \cite{Subirelmagdu}. 

\subsubsection{Thermal transport and the Wiedemann-Franz law.}

Charge conjugation has striking consequences dealing with {\em thermal} transport. The electrical field is special in the regard that it pulls the positively- and negatively charged reservoirs in opposite directions. But imagine that we could source these with e.g. gravity: both reservoirs would move in the same direction resulting in mass transport implying that total momentum takes control: this would again result in a Drude type transport. This can be achieved by applying thermal gradients. The positive and negative "particles" carry the same amount of entropy and a heat current will start to flow "overlapping" (in the memory function language) with total momentum. 
 
 Given a Drude conductor, the way that thermal- and electrical conduction hangs together is canonical. This is generically governed in linear response by the following  tensor of transport coefficients. In the presence of an electrical field $E$ and temperature gradient $\nabla T$ an electrical ($J$) and heat ($Q$) current will start to flow according to, 
 
 \begin{eqnarray}
 \left( \begin{array}{c}
 J \\
 Q \\
 \end{array} \right)
&  =
 \left( \begin{array}{cc}
 \sigma & \alpha T \\
 \alpha T &\bar{\kappa} T \\
 \end{array} \right)
 \left( \begin{array}{c}
 E \\
 - \nabla T/T
 \end{array} \right) 
 \end{eqnarray}
 
where $\bar{\kappa}$ and $\alpha$ are the thermal conductivity and thermo-electric conductivity that is behind the thermopower. To illustrate the workings of Drude transport a bit further, in a relativistic system one obtains the following expressions for the various optical conductivities \cite{holoDctrans}. Define the energy density an pressure as $\mathcal{E} + P$ and the charge density and chemical potential as $Q$ and $\mu$, the entropy density $s$ while the momentum relaxation  rate is $\Gamma$, 

\begin{eqnarray}
\sigma (\omega ) & = & \frac{Q^2} {\mathcal{E} + P} \frac{1}{\Gamma - i \omega} + \sigma_Q \nonumber \\
\alpha (\omega ) & = &\frac{Qs} {\mathcal{E} + P} \frac{1}{\Gamma - i \omega}  - \frac{\mu}{T} \sigma_Q \nonumber \\
\bar{\kappa} (\omega) & = & \frac{s^2 T} {\mathcal{E} + P} \frac{1}{\Gamma - i \omega} + \frac{\mu^2 }{T} \sigma_Q
\label{thermoeltransport}
\end{eqnarray}

One discerns the Drude form for $\sigma$: substitute $Q^2 \rightarrow (ne)^2$ while for a system characterized by a large rest mass $\mathcal{E} + P \rightarrow m_e n$ and the Drude weight becomes the familiar $\omega^2_p = ne^2 /m_e$. The $\sigma_Q$ contributions are special for zero rest mass systems -- I will come back to this soon.

One discerns that all these coefficient share the same momentum-relaxation factor $1/ (\Gamma - i \omega)$, the differences are entirely in the weights. One discerns that when all these quantities are available one can actually find out from only DC transport data whether one is dealing with a Drude transport. All one needs in addition are the thermodynamic quantities charge density and entropy that can be determined independently. 

This includes the Wiedemann-Franz(WF) ratio $L = \kappa / (\sigma T) = (s/Q)^2$, this just counts the entropy carried by the current per unit of charge. Given a Sommerfeld entropy and a temperature independent carrier density this becomes just a number --  in the Fermi gas this corresponds with the Lorentz number $L = \pi^2 k_B^2 /3e^2$. 

But let us now turn to zero density, as  controlled by charge conjugation symmetry. The charge density $Q = 0$ while the entropy density is finite at finite temperature. The Drude part in $\sigma$ is vanishing: what actually remains is the zero density "incoherent" component, Eq. (\ref  {CFToptcond}): this is referred to in Eq. (\ref{thermoeltransport}) as $\sigma_Q$ where "$Q$" refers rather awkwardly to "quantum critical".  This is inspired by the branch cut form of Eq. (\ref  {CFToptcond}) but it only relies on the fact that the Drude is completely suppressed. There is of course scale invariance wired in even in the free limit by the absence of a mass scale in the free dispersion. This naming echoes desires early in the development to somehow connect holography to zero density QCP's.  

The bottom line is that at zero density the Drude thermal conductivity and the "branch cut" electrical conductivity completely loose the "Drude" synchrony, exhibited by a strongly temperature dependent WF ratio. This was observed to be indeed the case in graphene at charge neutrality \cite{Kimhydro}.
 
\subsubsection{The holographic "quantum critical second sector" at finite density.}
\label{Homosecsector}

I already alerted the reader that one should be always on the outlook for "UV dependence" in holography, the fact that the large $N$ CFT at zero density may cause special effects also in the strong-emergence IR of the holographic finite density metals. I already explained the disastrous effect of large $N$ in obscuring the physics of the fermions in Section (\ref{Holofermions}). However, in this transport context there is yet another symmetry condition hard wired in the UV, being  less threatening but yet again  giving rise to a pathology when one wants to apply holographic wisdoms to e.g. cuprate electrons. Electrons have a rest mass $\sim 1$ MeV, but holography can only be made to work for ultrarelativistic matter with vanishing rest mass: the zero density SUSY CFT's. 

As I stressed over and over again, AdS/CMT is the story of what may happen when one departs from such a zero density, zero mass densely entangled stuff that is subsequently "doped" to finite density. The closest CM proxy would be graphene: when it would form a strongly interacting zero density critical state it could become a literal incarnation  of a holographic strange metal at finite density. This does not seem to be the case: for good reasons finite density graphene appears to be a rather perfect Fermi-liquid. 

Let us first find out what to expect for the optical conductivity when the {\em free} zero density Dirac electron system is put at a finite density.  Shift up the chemical potential/Fermi energy in the Dirac cone (Fig. \ref{fig:diracfermopt}). Because of the Pauli blocking the "vertical" transitions giving rise to the frequency independent optical conductivity at zero density are completely suppressed. An absolute "hole" appears in the spectral weight, being exponentially suppressed at finite temperature for $\omega <  E_F$, recovering above $E_F$. Given the f-sum rule one can immediately conclude that this weight will accumulate in the Drude peak forming a delta function peak at zero frequency in the homogeneous background. 

This is {\em not} the way it works with the fundamental QED response of real electrons. In order to discover the positrons one has to exceed the pair production threshold: at lower energy the density of electron-positron pairs  is suppressed exponentially. But this is a special trait of the non-interacting theory. What has holography to say about it? 

The outcome is remarkable and a bit mysterious. Instead of the hard, exponential pair production gap $\sim E_F$ the pair production spectrum "smears" in a powerlaw affair! The result is that in the deep IR as probed by DC transport two conducting fluids exist in {\em parallel} ("anti-Mathiessen"): on the one hand a Drude sector, living however side-to-side with an incoherent sector that is invariably irrelevant towards the IR. This is the "$\sigma_Q$" in Eq.  (\ref{thermoeltransport}): $\sigma_Q (\omega =0, T) \sim T^{\alpha}$, a branch cut submitting to $\omega/T$ scaling. As shown by Ref. \cite{Richardsourcing}, by taking particular combinations that can be computed of electric fields and temperature gradients one can separately  source both fluids, thereby rendering their existence to be observable. 

This is completely relying on holography -- it is still a bit of a mystery of how to interpret the separate existence of these two fluids in the field theory language. The Drude part is easy, but what distinguishes the "incoherent sector" so that it has an existence by itself? So much is clear that it is somehow controlled by an  {\em emergent} charge conjugation symmetry, a necessary condition to avoid the Drude logic.       

This "two-sector" affair caused at least in my environment initially quite some confusing: is the incoherent pair-production affair a generic property of the densely entangled "generalized Fermi liquid" (e.g., Ref. \cite{Holoplasmons})? A simple argument demonstrates that this cannot possibly be the case: it is a  UV sensitive affair, and the existence of the second sector is critically dependent on the fact that the UV degrees of freedom (zero density CFT) are ultrarelativistic, characterized by the absence of rest mass. 

The argument is relying yet again on symmetry. The two sectors emerge in  homogeneous space: the Drude part is here the delta function at zero and the pair production sector appears in the optical conductivity as a spectrum of excitations that grows algebraically with energy. Consider now a system of UV degrees of freedom having a rest mass that is 9 orders of magnitude or so larger than the energy in the experiment while at the scale of this rest mass one is dealing with particle physics: the electrons in solids. Under these circumstances the momentum- and electrical currents have to overlap for the full 100\%: at least at any finite temperature where one can use hydrodynamics {\em all} the spectral weight has to be in the Drude part.  The fundamental electrons just behave like the band structure graphene electrons: pair production is exponentially suppressed, the density of positrons in an electron system at a temperature  of a couple of Kelvins is of course completely vanishing.  

 Hence, the second sector is at least in the case of a homogeneous spatial background a special effect associated with the ultrarelativistic nature of the holographic UV. This causes more shortcomings when applied to the condensed matter electrons. Yet another ramification of the large rest mass of the electron is the decoupling of charge and spin. At most one has to account for the small spin-orbit coupling. But when the rest mass is vanishing one encounters the spin-orbital locking of the helical Dirac states and the spins disappear as  separate decrease of freedom: in holography there is no room for the physics of spins, including (anti) ferromagnetism. 
 
 The technical difficulty is that it is just not known how to incorporate finite rest mass in the zero density holographic set ups, a necessary condition to get a view on the effects of a  finite mass UV. For the time being, all one can do is to just not trust anything that is rooted in the presence of the "second sector".  As you will see, upon breaking the translational symmetry of the spatial manifold there are other ways that incoherent excitations may arise in the AC conductivity but also in these cases one  should be aware of the possible "pollution" due to the massless UV. 
 
 When discussing holographic superconductivity in Section (\ref{HoloSC}) I worked another troubling aspect rooted in this ultrarelativistic affair under the rug.  Deep in the fully gravitationally backreacted, low temperature holographic superconductor one finds an {\em emergent} deep interior scaling geometry. At zero temperature the "scalar hair" takes over completely from the black hole -- this applies for instance to the Einstein-Maxwell-Scalar theory where one would find the RN extremal black hole when the scalar is switched-off. In essence a "halo scalar hair star" is formed, characterized by a  "hole" forming in the deep interior surrounded by the scalar hair.  This deep interior in turn dualizes in incoherent low energy excitations in the boundary. There is a claim that these can be classified \cite{blaiseSFuni} and pending the specifics of the set up it may even happen that this "rediscovers" the  pristine zero density $AdS$ geometry. 
 
This is one of the mysteries of holography. As I explained in the beginning (Section (\ref{SREorder}) spontaneous symmetry breaking is exquisitely associated with the "classical" SRE product state vacua. Is it so that holography is telling us that part of the densely entangled strange metal is "grabbed" by this classicalness, leaving however behind a reconstructed densely entangled affair being responsible for the incoherent low energy excitations deep in the superconductor? Or does it have dealings with "UV rubbish", associated with the baroque large N etcetera CFT? The answer is at the time of writing still in the dark. 

\subsection{Breaking translations in holography.} 
\label{Holohydro}

In so far holography was involved, up to this point we were dealing invariably with the way it works in the translationally invariant homogeneous space. Given the universality of Drude transport, anything characterized by broken charge conjugation symmetry has to behave like a perfect conductor given the conservation of total momentum implied by the homogeneity of the spatial manifold.  
 
But electrons in solids do not live in such a space: these invariably encounter the lattice formed by the ions. This may be in the form of a perfectly periodic crystal structure, that may also be subjected to imperfections: the quenched disorder.  We have learned to appreciate holography as a symmetry processing machinery: as it turns out, the (in)homogeneity of space is in this regard a {\em critical} symmetry condition.   

This is deeply rooted in the way that the Einstein theory in the bulk is stitched together. The first confession one encounters in any GR textbook is that the results that are highlighted are all exclusively tied to {\em highly symmetric} circumstances. This is presented with a great degree of mathematical discipline. The isometries of the problem translate in Killing vectors and when there are sufficiently many of them the mathematical abyss of the Einstein equations as a system of highly non-linear partial differential equations ("PDE's") can be reduced to an ordinary differential equation that can be solved on the backside of an envelope: e.g., the Schwarzschild solution, the FLRW cosmology and so forth. But if one breaks the homogeneity and isotropy of the spatial manifold hell breaks loose: one has somehow to "tame the PDE's".    

The same applies to non-stationary conditions in general. Dynamical GR is not at all charted. A case in point one encounters in the fluid-gravity duality that I discussed in Section (\ref{FiniteTMaldacena}). The non-stationary near horizon gravity is identified to be in precise dual relationship with the Navier-Stokes hydrodynamics in the boundary. But the best known of all dimensionless parameters is found in hydro: the Reynolds number $R$ that is inversely proportional to the viscosity. Given the minimal viscosity $\eta /s \sim \hbar$ the viscosity can become quite small at low temperature and it is easy to get into a flow regime governed by large Reynolds number: but this implies that the fluid flow is in a {\em turbulent} regime. But given fluid-gravity duality this implies that the near horizon gravity is also exhibiting turbulent complexity! 

The fluid-gravity specialists are still struggling getting this under control, although there is progress. I heard claims that during black hole mergers the conditions may be met for such turbulent horizons to play a role. Ironically, this is  challenging the good taste of the mathematical community. There is a Millennium prize by the Clay institute on Navier-Stokes,  revolving around the turbulent regime. But we learn that Navier-Stokes is actually a tiny, very special part of GR. There is no Millennium prize for GR! 

Yet again, in the present time the big GR effort is in first instance focussed on black hole mergers given the arrival of the gravitational wave detectors. Numerical GR solutions are an absolute requirement to interpret the signals, and also in this regard this heroic effort has been much helped  by a good fate. It took roughly half a century of intense effort by a small expert community to get the numerical GR codes working, being just in time to interpret  the first merger signals! In fact, these only work involving specific simplifying circumstances and a worldwide effort is presently evolving  to further generalize this numerical RG. 

As it turns out, as a spin-off of this effort efficient algorithms appeared that do not quite work for the dynamical problems while these are good enough for the stationary problems with low spatial symmetry associated with equilibrium holography. A few dispersed shots were launched some ten years ago to demonstrate proof of principle that this holographic numerical GR does work but not much happened since them. So much is clear that the breaking of translations interferes critically with holography. Given all the extra work to be done in the bulk it is not surprising that new holographic principle may arise being critical for the application to the electron systems in solids. Metaphorically it is like trying to explain the physics associated with the band structures of real solids having only insights into the quantum mechanics of particles in a box and the Pauli principle. 

The reason that this did not trigger a big effort appears to be rooted in cultural factors. The string theory tradition has all along been propelled by mathematics and there is no computational tradition. The research culture is quite different. Computational physics is in a way much closer to experimental physics than to the mathematical culture of string theory. As in experiment one needs a small army to get the machines -- the codes -- working. When the codes start working these are like the experimental machines: in essence black boxes that produce data. One has then to systematically collect the data to then  try to find out whether one can discern a phenomenological framework revealing the deeper meaning  behind the data. This is quite different from the mathematical tradition where by solving the equations one obtains directly an overview of anything that can happen. 

The bottom line is that when it became clear that a further exploration of AdS/CMT had to rely on numerics the string theory community at large shied away and went elsewhere. We felt responsible in Leiden and we have been investing in a professional computational holographic effort in the form of a program package written in efficient  programming language (like C++, the industry standard among string theorist is mathematica!) running on supercomputers. The first results are obtained making above all clear that we have seen only the tip of the iceberg. I will report on some of it at the end of this chapter 

Transport properties are at the centre of this "inhomogeneous numerical GR".  But before we get there let us first step back to recollect the effects  of the breaking of translations  in the familiar Fermi-liquids in order to get the differences in sharp perspective.  

\subsubsection{Momentum relaxation in Fermi-liquids.}
\label{FLUmklapp}

According to the universal Drude logic, in order to have a finite resistance the macroscopic current should have a relaxing total momentum which in turn requires a space lacking translational invariance. This pertains to literally anything -- one could attach electro-statically charged cat furs to elephants that one then throws out of a space ship in outer space. This flow of charged elephants represents perfect conduction. The issue is that the {\em numbers} in the Drude conductor do know that low temperature Fermi liquids do behave quite differently from charged elephants. 

The big deal is the periodic ion lattice. I will largely ignore the impurities -- quenched disrder appears to be a bit of an afterthought in e.g. the cuprate metals and the "Umklapp" is just more involved and interesting than weak disorder. The Fermi-liquid is in this regard different from any  classical ("molecular")  fluid in the regard that due to quantum mechanics the breaking of translations by a periodic lattice is {\em irrelevant at the IR fixed point.} At zero temperature a macroscopic Fermi-liquid living in a perfectly periodic potential forgets entirely the existence of such a potential turning into a perfect metal: the residual resistivity disappears in the absence of impurities.

This is rooted in the fact that the states near  $k_F$ that do the work are living in momentum space: these are delocalized in real space and this averages out the periodic potential. The thermally excited quasiparticles can only dissipate the total momentum at a quasiparticle collision where the lattice can absorb part of their centre of mass momentum -- of course two particle momentum exchange does not affect the total momentum otherwise. Consider a potential $V = V_G \cos ( \vec{G} \cdot \vec{R} )$ where $\vec{G}$ is the Umklapp wave vector. The operator that dissipates momentum is of the form,

\begin{equation}
\hat{O} = \int (\Pi_{i=1}^4 d^d p_i) c^{\dagger}_{p_1}  c^{\dagger}_{p_2}  c_{p_3} c_{p_4} \delta ( p_1 + p_2 - p_3 - p_4 - G )
\label{FLUmklapp}
\end{equation}

But to relax the total momentum  this action at the large Umklapp vector should get coupled to the macroscopic current. This was elaborated by Lawrence and Wilkins in the 1970's \cite{UmklappFL}   in terms of "Umklapp efficiency". The single particle continuum states near $k_F$ acquire an admixture of the Umklapp copies $ | k \rangle \rightarrow  | k \rangle + \delta_{V_0} | k \pm G  \rangle$ in the guise of weak potential scattering. In case that e.g. $G < 2k_F$ one finds this to be associated with the "efficiency" $\Delta_G \simeq \frac{\pi}{4} \frac{G}{k_F} \frac{V_G}{E_F}$. The rate by which momentum is absorbed is set by the two-quasiparticle collision rate, $1/\tau_c \simeq ( k_B T )^2 / (\hbar E_F)$. The momentum relaxation rate becomes then $\Gamma_G  = \Delta_G / \tau_c \sim T^2$. This is the story behind the $T^2$ resistivity, dominating at low temperatures in conventional metals before the phonons take over the momentum relaxation. 

One infers that  $\Gamma_G  \rightarrow 0$ because $\tau_c \rightarrow \infty$ when $T \rightarrow 0$: the IR irrelevancy of Umklapp and the effective Galilean invariance in the deep IR is therefore an emergent global symmetry. There is profundity in this simple story: in a classical fluid this cannot happen. 

But we are not done yet. Dealing with $^3$He one immerses the fluid in a truly translationally invariant background and I highlighted in Section (\ref{FiniteTMaldacena}) that this behaves at a finite temperature as a Navier-Stokes fluid, characterized however by a very large viscosity $\eta_{FL} \simeq nE_F \times \tau_c$. The momentum exchange between the quasiparticles that is responsible for the viscosity takes a time $\tau_c$,  becoming very long at low temperatures. The associated collision length $l_c \sim v_F \tau_c$ is easily of order of many microns at low temperatures. But it is extremely difficult to render crystals to be so perfect that they are devoid of any disorder on such large scales. Generically the {\em elastic} mean free path $l_{mf}$ is much {\em smaller} than the collision length. At the "collision" with the impurity the individual quasiparticle dumps its momentum in the lattice,  long before the system finds local equilibrium, the condition for the collective hydrodynamical flow behaviour. The fate of total momentum is in this regime controlled by the loss of {\em individual} quasiparticle momentum.  

This is the underpinning of the standard kinetic transport theory that is entirely revolving around how individual quasiparticles loose their single particle momentum. Only very recently extremely clean conductors became available where it is claimed that $l_c < l_{mf}$ so that one can look for the signatures of hydrodynamical electron transport in mesoscopic transport devices, graphene being the prime example \cite{Graphenehydro}. But one should be acutely aware that this extremely "gaseous" behaviors where quasiparticles can fly forever is extremely special for the Fermi-liquid. Despite the folklores imprinted by 80 years of Fermi-liquid success, why should any of this extreme behaviour survive in anything that is not a Fermi-liquid?     

\subsubsection{Minimal viscosity and the shear drag resistance in holographic fluids.}
\label{sheardrag}

In our daily human world we encounter all the time the resistance due to fluids, be it that it limits the speed of cars or either that we need pumps to get water through a pipe. At least in the regime of low Reynolds number with its smooth flows this is due to {\em shear drag}. The breaking of space translations due to e.g. the walls of the pipe cause gradients in the flow and when the viscosity is finite this dissipates the fluid kinetic energy into heat. 

To estimate this resistance one can rely on very simple dimensional analysis. The viscosity can be converted into the kinematic viscosity (or diffusivity) through $\nu = \frac{\eta}{\mathcal{E} + P} = \frac{\eta}{mn}$ in the relativistic and non-relativistic fluid, respectively. The kinematic viscosity has the dimension of diffusion, $\left[ \nu \right] = \mathrm{m}^2/\mathrm {s}$. Introduce now the length associated with the distance where the breaking of translations becomes manifest,  $l_{\eta}$: this can be e.g. the radius of the pipe. The momentum relaxation rate can now be estimated to be of order, 

\begin{equation}
\Gamma_{\eta} \simeq \frac{\nu}{l^2_{\eta}} = \frac{\eta}{ m n l^2_{\eta}}
\label{viscousGamma}
\end{equation}

Consider now any holographic fluid living in a system where translational invariance is broken by a low density of impurities such that the typical wave vectors $\sim G$ are quite small as compared to e.g. the chemical potential. I already explained in Section (\ref{FiniteTMaldacena}) the triumphant minimal viscosity predicted for the finite temperature CFT. The argument revolved around a bulk universality: the viscosity is set by the absorption cross section of zero frequency gravitons by the black hole and this scales with the area of the horizon as does the entropy. Therefore $\eta / s = A_{\eta} \hbar /k_B$ where $A_{\eta} = 1 / (4 \pi)$. But this is completely universal, it also applies to the finite temperature (and even the extremal zero temperature) black holes that are dual to the finite density strange metals!  

Plugging minimal viscosity in Eq. (\ref{viscousGamma}) and using relativistic units it follows for the momentum relaxation rate setting the Drude conduction,

\begin{equation}
\Gamma_{holo} \simeq \frac{A_{\eta} s} {(\mathcal{E} +P ) l^2_{\eta}}
\label{holoviscGamma}
\end{equation}

The momentum relaxation rate and thereby the temperature dependence of the resistivity is according to holography in this regime just proportional to the entropy! It is very simple but it should be a bit of a shock for practitioners of conventional metal transport theory.  

This is quite suggestive in the context of cuprates, see Section (\ref{linresemp}). In the strange metal regime the entropy has been measured to be Sommerfeld, linear in $T$. We know experimentally that the Drude weight is temperature independent: it follows from Eq. (\ref{holoviscGamma}) that therefore the resistivity should also be linear in temperature, the  holy grail of the cuprate strange metal. In fact, we appear to know all the numbers except $l_{\eta}$. To get that $\Gamma \simeq k_B T / \hbar$ $l_{\eta}$ has to be of order of a reasonable couple of nanometers \cite{JZsciPost19}.  

As will become further substantiated underneath this "shear drag" appears to be a rather universal result in holography at least dealing with small wavevector components of the symmetry breaking potential. Here one can actually discern a long wavelength universality rooted in the "hydro", where one get quite some insight. Momentum relaxation is then associated with a {\em shear} stress exerted by the spatial inhomogeneities on the fluid flow. Imagine water flowing through a channel littered with a periodic array of obstacles: velocity gradients transversal to the flow will develop dissipating the flow given that the viscosity is finite. 

The image of this in the gravitational dual can be tracked to quite a degree with analytical means yielding further insights.  You learned from e.g. the minimal viscosity mechanism that the spatial shear components of the boundary energy stress tensor are dual to {\em gravitons} in the bulk. In a homogeneous space gravitons are massless and zero mass in the bulk encodes for the conservation of total momentum. Now we encounter a simple wisdom that appears to have been overlooked completely in the history of GR. How does an Einstein universe looks like when the Energy stress on the r.h.s. of the Einstein equation is associated with matter that breaks translational invariance? In other words, what happens when the universe would be filled with a "cosmological" crystal? 

The answer is \cite{crystalgravity}: this corresponds with the {\em Higgs phase} of gravity where the background geometry takes the role of the gauge fields and the crystal that of the matter. In a way it is the most intuitive way to understand Higgsing in general: the crystal forms a "preferred frame", a coordinate system that is imprinted in the geometry by gravitational backreaction. As in the usual Yang-Mills setting, the geometrical curvature can now only be accommodated in the form of "gravitational fluxoids" as the Abrikosov quantized magnetic fluxes in a superconductor. 

At long wavelength this spatial frame fixing has as ramification that the graviton acquires a {\em Higgs mass}. Massive gravity has a history in cosmology. This was however plagued by inconsistencies related to the (implicit) assertion  that also time translations (unitarity) should be broken which is an unphysical affair. But  exclusively fixing the spatial frame is healthy --  it is just "crystal gravity"\cite{crystalgravity}. One now discerns the universality in the workings of the dictionary. In the boundary one applies a periodic potential with a  small wavevector $G$. This dualizes in the bulk to a "fixed spatial frame" Einstein geometry having the universal long wavelength (small $G$) ramification that the graviton acquires a mass. The mass of the graviton in turn dualizes into a finite life time for the macroscopic total momentum: the dictionary entry for the $\Gamma$ in the Drude response Eq. (\ref{thermoeltransport}) is identified.

Using an effective field theory ploy the graviton mass can be easily incorporated in the bulk -- it just parametrizes the  $l_{\eta}$ of Eq. (\ref{holoviscGamma}). Although we understood this less well at the time, in Ref. \cite{Davisonhydro} we worked out an explicit holographic set up illustrating precisely the above mechanism. We departed from the Gubser-Rocha strange metal, that I already advertised in Section  (\ref{EMDscaling}) as the only loop hole that I know to reconcile $z \rightarrow \infty$ with a Sommerfeld entropy, $s \sim T$. We added the  massive gravity term to the bulk action, to find a {\em perfectly linear resistivity} all the way up $k_B T \simeq \mu$. It is just explicit holographic proof of principle demonstrating that indeed the  very simple Eq. (\ref{holoviscGamma}) is all one needs to know. 

The big picture is that perhaps nowhere else one discerns such a sharp contrast in the physics of conventional Fermi liquid metals and the predictions of holography for the densely entangled "quantum supreme" metals. In a Fermi-liquid one approaches at low temperature the extreme gaseous fluid behaviour more closely than anywhere else: the quasiparticles behave completely independently up to the collision length scale $l_c$ becoming very large at low temperature. The ramification is that the momentum relaxation is entirely due to single particles scattering against the impurities. But the rapid equilibration of the holographic fluid as rooted in the Eigenstate thermalization of the densely entangle "unparticle soup" has the consequence that at microscopic scales the fluid behaves cooperatively as described by hydrodynamics. This is then in a way  a simplifying  circumstance since such a fluid behaves like a macroscopic "molecular" fluid like water flowing through a space with obstacles.  

The difficulty is to distinguish these very different transport regimes in experiment. At distances $>> l_{\eta}$ in both cases one gets the simple diffusional  Ohm's law behaviour. The culprit is the quenched disorder that cannot be avoided. Consider e.g. the flow of water through a fast flowing shallow river with lots of boulders. The net transport will be eventually Ohmic and indistinguishable in this regard from the flow of gas in comparable circumstances since total momentum will be completely destroyed on large scales. One can estimate the characteristic dimensions where this cross-over takes place in cuprates  \cite{JZsciPost19}, finding that (optimistically) hydro flows cannot be discerned at scales larger than a micron.  

One therefore has to employ submicron transport devices to look for the signatures of hydrodynamical flows. For instance, when the ploy in the above would be realized the implication would be that the viscosity itself would become very small at low temperature since we just know from experiment that the Sommerfeld type entropy becomes very small as in a Fermi gas, $s \sim T/\mu$. It is then relatively easy to get into a flow regime characterized by large Reynolds numbers by e.g, injecting a current through a small constriction. Although the precise ramifications are not worked out one would expect to see signatures of this in the nano-transport machinery. But this is very hard to realize because of the difficulty to reliably "nano-engineer" chemically complicated substances like the copper oxides. 

Yet another simple ramification is in the fact that the minimal viscosity implies that the viscosity is vanishing when the entropy is vanishing: asserting the absence of zero temperature entropy this implies that the viscosity is vanishing in the $T \rightarrow 0$ limit. The holographic strange metals turn into {\em perfect fluids} at zero temperature and perfect fluids are perfect conductors! In sharp contrast with Fermi-liquids there should not be a residual ($T=0$) resistivity. This may relate to strange anomalies in cuprate transport that I will discuss when I turn to experiment. 

Finally, all along there is the warning of UV sensitivity: is the minimal viscosity in the finite density systems indeed universal or somehow tied to e.g. the large $N$ limit? There is a reason for concern \cite{JZsciPost19}. I gave the simple dimensional analysis argument for the minimal viscosity in the zero density CFT in Section (\ref{FiniteTMaldacena}). The dimension of viscosity is set by the free energy density times a characteristic momentum relaxation time; the former is purely entropic in a CFT $\sim sT$ while the time is just $\tau_{\hbar} \simeq \hbar / (k_B T)$ and it follows that  $\eta / s \sim \hbar$. But in a finite density system the free energy at low temperature is dominated by the density, $f \sim \mu n$, and combining this with the Planckian time one gets $\eta \sim 1/T$, growing not as rapidly as in the Fermi-liquid with decreasing temperature but still divergent in the zero temperature limit. Especially Hartnoll has been stressing this, arguing that the minimal viscosity may be an artefact of the large $N$ limit. But this is controversial -- yet again, an experimental verdict would be most useful in shedding light on what actually amounts to a deep question in quantum gravity.     

\subsection{The Umklapp-mechanisms according to inhomogeneous holography.}
\label{holoUmklapp}

How to proceed from the long wavelength but otherwise unstructured momentum dissipation encoded by massive gravity? Other holographic ploys became fashionable that do somehow incorporate more structure while these are still resting on the simple spatially homogeneous bulk. These "linear axions" and "Q lattices" do break translations and dissipate momentum but in such a way that one can still compute matters with simple ordinary differential equations. I will ignore these here -- it is motivated by the ease of computation while it is completely in the dark what kind of physics is described in the boundary. Whatever it is, it is surely not something that is in any obvious related to the conditions met in the electron systems.     

As in the Fermi-liquids the interest is in the first place in the mechanism leading to momentum dissipation by the periodic lattice, the "Umklapp". This is a condition invariably encountered in any solid. Specifically for the copper oxides, these crystals are  far from perfect but the degree of disorder appears to vary quite a bit between the different subfamilies. However, it appears that the typical strange metal properties are rather independent of the degree of quenched disorder. The basic Umklapp departing from the perfectly periodic lattice appears to be the big deal. But to get a handle on how this works one {\em has} to sacrifice the homogeneity and resort to the numerical RG. 

As  I already stressed, this is hard work and very little is known. The effort in Leiden just started to get on steam and we are nearly finished with a first such study of the most basic holographic system of the kind. We considered the RN strange metal in two space dimensions on a background formed by a square lattice single harmonic electrostatic potential,

\begin{equation}
V = V_0 \left( 1 + A  ( \cos ( G_x x) + \cos ( G_y y) )  \right)
\label{squarelatpot}
\end{equation}
 
 with a strength $A$ and an Umklapp wavevector of magnitude $G$.  As I explained in Section  (\ref{holoSM}) its special deep interior scaling geometry is pathological and it is presently nearly completely in the dark how the Umklapp works in other strange metals characterized by arbitrary $z$'s and $\theta$'s. The RN affair is however already interesting and surprising, showing at the least that momentum dissipation by Umklapp is governed by an entirely different physics compared to the Fermi-liquid "$T^2$" case that I outlined in Section (\ref{FLUmklapp}).

\subsubsection{The $z \rightarrow \infty$ holographic "pseudopotential".}
\label{infinitezpseudopot}

Yet again the big deal is in the scaling properties of the strange metal. The RN metal is characterized by $z \rightarrow \infty$, a condition that is of interest since there appears to be direct evidence for such scaling in the cuprate  strange metals. But what does this mean for the way that such a metal reacts to an external potential both in terms of screening properties and Umklapp momentum relaxation? The text books only tell us how this works in Fermi-liquids: I already alluded to the "Umklapp efficiency" while the screening of the external periodic potential is also a canonical affair. 

The latter is in essence governed by Fermi-pressure. The linear response of the charge density to a potential with wavevector $G$ is given by the real part of the charge susceptibility at zero frequency and it follows immediately from the Lindhard function that $\mathrm{Re} \chi_{\rho} (\omega =0, G) = \partial n /\partial \mu$, the compressibility of the Fermi gas modulo possible "$2k_F$" enhancements due to nesting coincidences. It is in essence spatial scale independent and it takes care that the charge density is smoothened out compared to a non-degenerate system. 
 
 However, the Fermi liquid is characterized by the "covariant" scaling dimensions $\theta = d -1, z=1$. What to expect for $z \rightarrow \infty$? This refers to the local quantum criticality affair. By definition $\tau_{\mathrm{cor}} =  l_{\mathrm{cor}}^z$: although $\tau_{\mathrm{cor}}$ is diverging for $z \rightarrow \infty$ it is not at all clear what is going on with the {\em spatial} organization of the system. The usual correlation length looses its meaning in a local quantum critical system. 
 
 Yet again, Hong Liu and coworkers \cite{honglocal} showed their muscles by discovering a quite surprising structure hard wired in the bulk geometry of the RN system. Perhaps the greatest wonder of GR is that causality structure, the way that cause leads to effect, is hardwired in the Lorentzian signature geometry itself. The grand master of this affair is Roger Penrose who devised the Penrose causality diagrams but also got rewarded a Nobel prize in 2020 for his "trapped surfaces", causal objects that are at the heart of the singularity theorems. The take home message is that causality structure may be used to extract universal principle even when explicit solutions are beyond reach. In the greater context of whether one can trust holography, when a particular boundary physics is rooted in such causality structure in the bulk one better takes it very serious as an affair having the best chance to reveal correct general principle  ruling in the boundary.
 
 Pristine AdS is already exhibiting strange causal attitudes: it is a GR classic that although the radial direction extends to infinity it takes only a finite time to travel from the boundary to the deep interior, a motive that is crucial for the correspondence. Turning to the extremal RN-AdS geometry, Hong Liu and coworkers discovered the following causal structure hard wired in this geometry. Take two points in the near-horizon geometry with a spatial separation $l$. Launch light rays at these two points and follow their nul-geodesics to the boundary. One discovers that when these points are farther apart than a distance $\xi  = \pi / (\sqrt 2 \mu)$ (in $d =2$) the ensuing light cones arriving at the boundary will not overlap! This implies that one is dealing with patches of size $\xi$ in the boundary theory that do not correlate with each other at distances larger than their size. In conventional units, $\xi \simeq 1/ k_F$ and the universal prediction is that the system will react quite rigidly to a potential with $G > 1/ \xi$ while it becomes very soft at larger length scales!
 
This causality structure turns out to be universal for {\em all scaling geometries characterized by $z \rightarrow \infty$} \cite{Leidennum21}. The hyperscaling violation does not affect this causal structure at all -- more reason to take it serious. 

\begin{figure}[t]
\includegraphics[width=0.7\columnwidth]{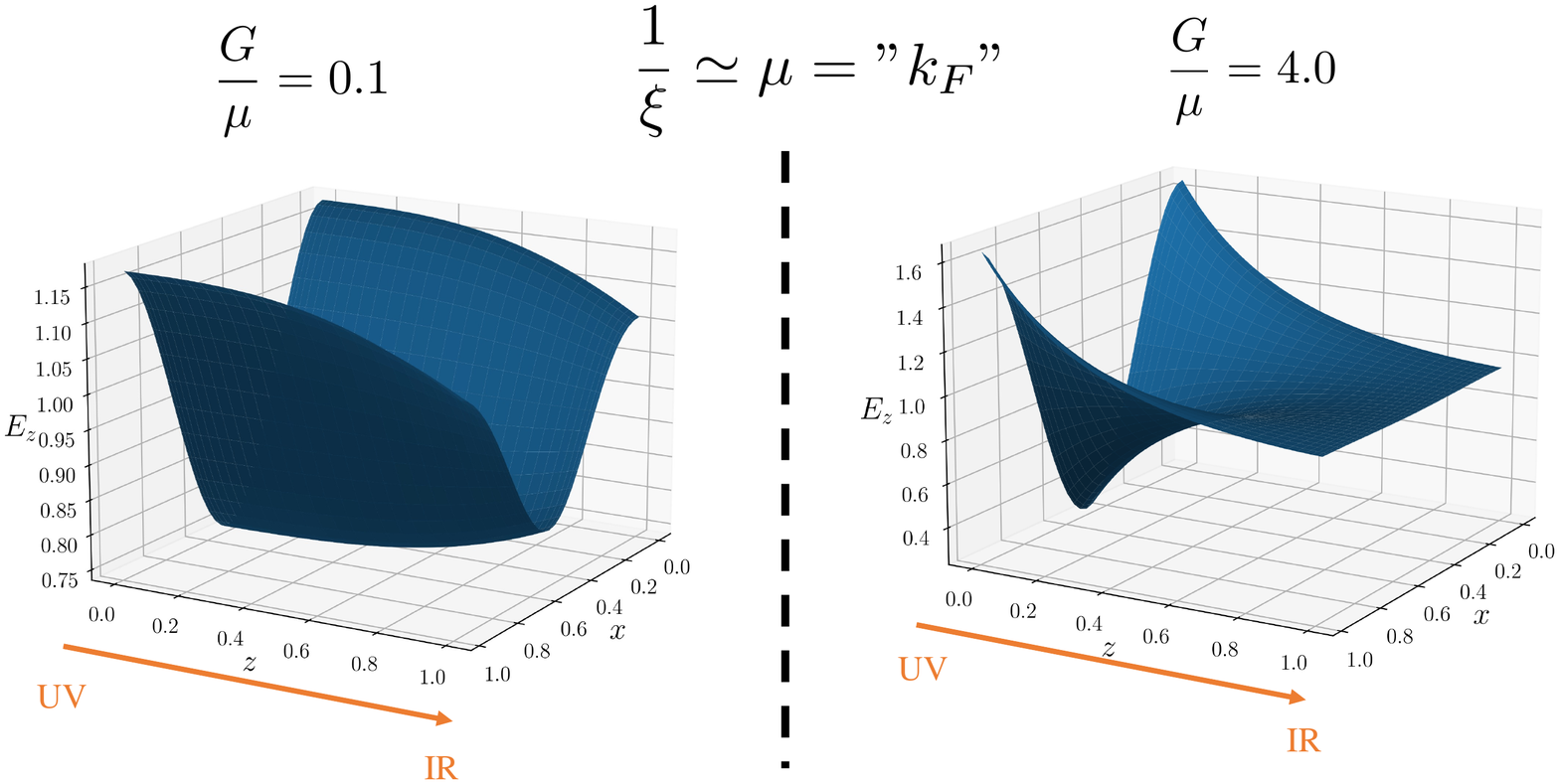}
\caption{The electrical field in the bulk as function of radial direction represents the scale dependent effective charge density in the boundary -- the geometrization of the RG flow  \cite{Leidennum21}. For $z \rightarrow \infty$ one finds that for wave vectors  being larger- and smaller than the inverse of the local length the charge modulation in the deep IR is strongly reduced and barely affected, respectively.}
\label{fig:chargevsG}
\end{figure}

The ramifications for the boundary can be illustrated in a spectacular manner by reference to the geometrization of the RG flow  \cite{Leidennum21}. The electrical field in the bulk is dual to the charge density in the boundary, and by plotting the way that the spatial structure of this electrical field varies along the radial direction one gets an image of the RG flow of the density in the boundary theory. A typical result is shown in Fig.(\ref{fig:chargevsG}). In the right panel the situation is indicated for a harmonic potential with a $G > 1/ \xi$: a charge modulation is present at the boundary but when one dives to the horizon one sees that it becomes very small. At low energy there is barely any response to the potential and the system is nearly homogeneous. However, dealing with a $G < 1/ \xi$ (left panel) the UV  response (density in the boundary) is comparable, but upon descending along the radial direction the modulation stays very large all the way to the horizon. The periodic potential hits hard also in the deep IR where the transport resides!  

This kind of $z \rightarrow \infty$ screening behaviour is of course entirely different from the Fermi-liquid wisdoms at work in normal metals. One may contemplate possible consequences. The first one coming to mind is the screening of impurity potentials: short wavelength components of the impurity should be strongly suppressed in the IR while long wavelength components survive. This may offer a surprising clue to a long standing puzzle in the cuprates. Potential disorder should be detrimental for d-wave superconducting order because of pair breaking. However, it appears that it is remarkably insensitive to the degree of this disorder. The issue is however that the pair breaking is due to large momentum exchanges tied to the short wavelength components of the impurity potential: when these are suppressed only the forward scattering remains which is harmless for the pairing. 

\subsubsection{The Umklapp momentum relaxation in RN strange metals.}

The next issue is, how does such inhomogeneity affect the life time of total momentum and thereby the Drude transport? As in the Fermi-liquid one expects a "memory function logic" insisting that the momentum relaxation rate is coming from an effective coupling between the macroscopic current and a momentum sink living at the Umklapp wavevector $G$, 

 \begin{equation}
\Gamma = g_G \mathrm{Lim}_{\omega \rightarrow 0} \frac{\mathrm{Im} G^R_{T T} (\omega, G)}{\omega}
\label{memfieUmklapp}
\end{equation}

$g_G$ is an effective coupling while the remainder is the slope of spectral function associated with the momentum absorber -- this is universally associated with the spatial shear component of energy-stress tensor $T_{xy}$. 

The outcome for the thermoelectric  DC transport can be numerically computed  \cite{Leidennum21}. Using all thermo-electric transport coefficients one can use Eq. (\ref{thermoeltransport}) to separate the incoherent ($\sigma_Q$) and Drude contributions: one finds that for $T \le 0.1 \mu$  it is completely dominated by the Drude part and since both the charge- and entropy density $Q, s$ can be computed independently from  the thermodynamics one can isolate the momentum relaxation rate $\Gamma$ -- $Q$ and $s$ are approximately temperature independent in this low temperature regime. 
 
 \begin{figure}[t]
\includegraphics[width=0.7\columnwidth]{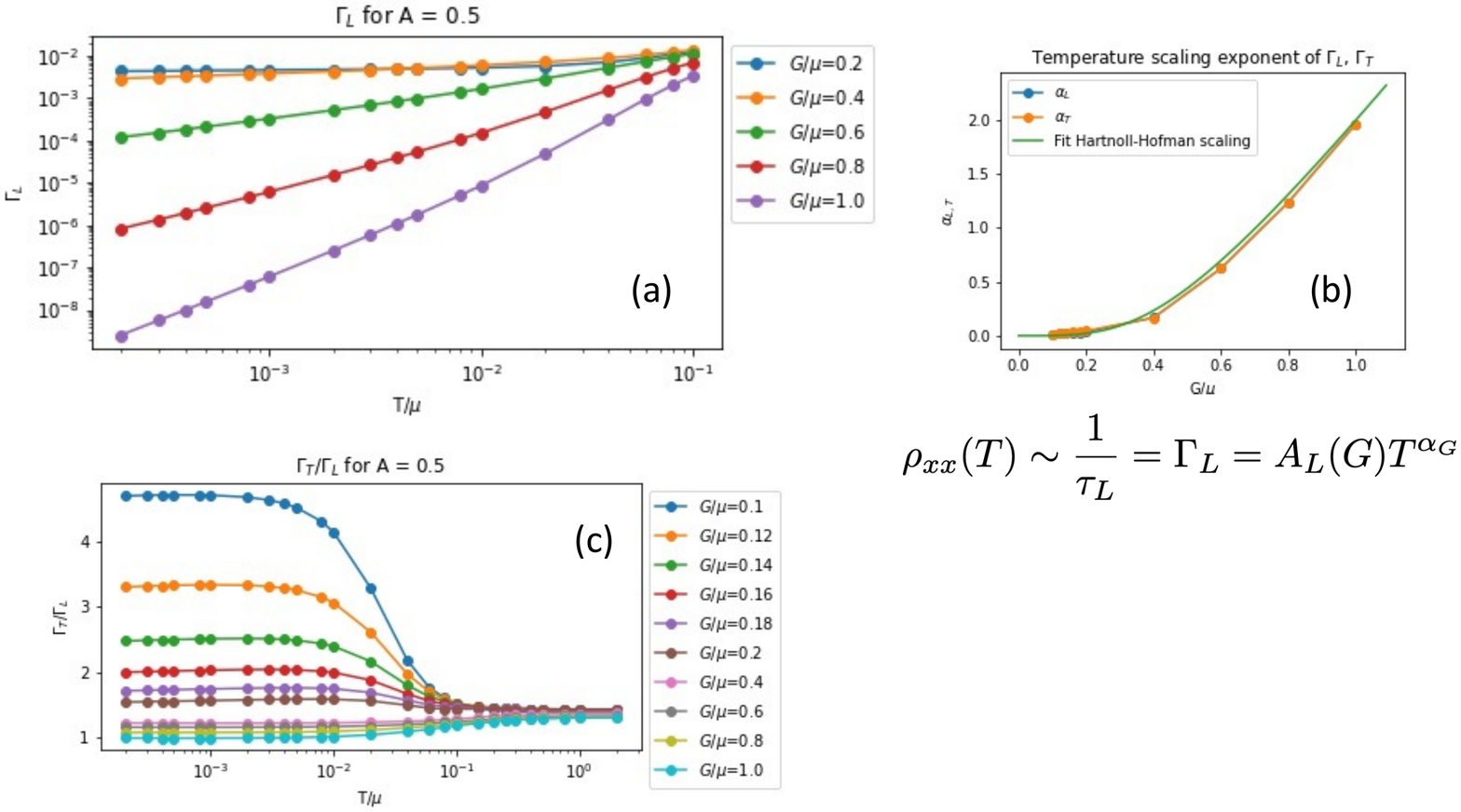}
\caption{Tentative results  \cite{Leidennum21} for the momentum relaxation rates associated with the RN strange metal in the square lattice background. (a) The longitudinal momentum relaxation rate $\Gamma_L$ rate as function of temperature for varying periodicity $G$. (b) The scaling with temperature follows the predictions for the "generalized Lindhard" deep IR for large G. (c)   When the deep IR "crystal lattice" becomes strong for small $G$'s the transversal ("Hall") momentum relaxation rate $\Gamma_T$ becomes large compared to $\Gamma_L$ which is associated with the resistivity.}
\label{fig:Umklapptransport}
\end{figure}
     
Compared to e.g. the way that Umklapp dissipates momentum in a Fermi-liquid the outcome is greatly surprising. The dependence on the strength of the potential $A$  (Eq. \ref{squarelatpot}) is an unremarkable $A^2$, suggesting that $g_G$ is governed by second order perturbation theory. However, the dependence on $G$ is spectacular, see Fig. (\ref{fig:Umklapptransport}a) noticing that it is plotted on a log-log scale: the low temperature $\Gamma$ varies by 6 orders of magnitude!  This revolves around a change of behaviour occurring at the wavevector associated with the local length scale $\sim 1 /\xi \simeq \mu$. 

As suggested early on by Hartnoll and Hofman \cite{Hartnollhofman}, in the regime that $G \ge 1/\xi$ the momentum sink ($G^R$ in Eq. \ref{memfieUmklapp}) is associated with the "generalized Lindhard excitations", exhibiting scaling behaviour associated with the scaling geometry realized in the RN near horizon geometry. You already encountered that in the context of the Leiden-MIT fermions in Section (\ref{Holofermions}). We now need the EOM's of classical fields propagating in this background and the outcome is that for $\omega =0$ using the $\omega/T$ scaling,

\begin{eqnarray}
\Gamma & \sim &T^{2\nu_G -1} \nonumber \\
\nu_k & = & \frac{1}{2} \sqrt{ 5 + 4k^2 - 4\sqrt{ 1 + 2k^2} }, \;  k \ge 1/ \xi 
\label{GammaRN}
\end{eqnarray}

as for the fermions, one observes that this exponent is in this regime only dependent on the Umklapp momentum and not on $\theta, z, \zeta$. This is special for the RN throat. Up to $T \simeq 0.1 \mu$  the momentum relaxation of the Drude conductivity closely approaches a perfect power law $\Gamma = A_{\Gamma} T^{\alpha_G}$. The result is that the computed  $\alpha_G$ closely tracks the prediction Eq. (\ref{GammaRN}), see Fig. (\ref{fig:Umklapptransport} b). Since $\alpha_G > 0$ in this regime  the Umklapp is irrelevant and as in the Fermi-liquid "Galilean invariance is an emergent symmetry at the IR fixed point".

However, when $G$ becomes small compared to $1/\xi$ one finds a completely different behaviour: according to Fig.(\ref{fig:Umklapptransport}a) the resistivity becomes approximately temperature independent in the Drude regime! This is largely responsible for the big change at low temperature, at $T \simeq 0.1 \mu$ the $\Gamma$'s are varying by no more than an order of magnitude as function of $G$. 

Yet again, we know how this works for sufficiently small $G$: this is the shear drag regime explained in Section (\ref{sheardrag}). This is the regime governed by the hydrodynamics and the bulk universality insists that one is dealing with massive gravitons describing the shear drag in the boundary. We can rely on Eq. (\ref{holoviscGamma}) observing that $\eta \sim s$ (minimal viscosity) and as I repeatedly emphasized RN is suffering from the zero temperature pathology: $s$ is temperature independent and therefore $\Gamma$  is roughly temperature independent. There is an issue with the length $l_{\eta}$: in general the effective Umklapp in the deep IR may run under renormalization adding temperature dependence to this quantity. But one may then argue that for $z \rightarrow \infty$ purely spatial quantities such as $l_{\eta}$ should be marginal, temperature independent. 

The numerics is just demonstrating that the crossover from "generalized Lindhard" momentum dissipation to the shear drag occurs rather suddenly at a critical Umklapp periodicity set by the local length $\xi$: at length scales smaller than $\xi$ one is dealing with Eq. (\ref{GammaRN}) switching rather abruptly to the shear drag regime for larger lengths. 

The take home message is that the influence of Umklapp potentials on momentum relaxation in holographic strange metals is entirely different from the usual Fermi-liquid "particle physics" affair. At the same time we also realize that specifically the RN metal is highly pathological. There is a big need to understand how this works for the arbitrary EMD scaling geometries  that I highlighted in Section (\ref{holoSM}). Obviously, one has to rely on numerical GR but this at present completely uncharted: the exploration of this vast landscape is the central challenge for the computational AdS/CMT effort in Leiden. 

For completeness, let me shortly summarize the few other transport wisdoms that have been looked at in this "inhomogeneous" context.  This is all restricted to exploratory work limited to RN set ups  \cite{Leidennum21} . 

\subsubsection{The Hall momentum relaxation rate.}
\label{holomagnetotransport}

 At stake is in first instance DC transport in magnetic fields in the presence of a lattice. The big deal is that the Lorentz force induces a momentum that is transversal to the electrical field -- in fact, it sources angular momentum.  This communicates with the {\em anisotropy} of the spatial manifold associated with the background potential. A crucial aspect is that the relationship between spatial translations and rotations is {\em semi-direct}: finite translations cannot be distinguished from rotations. But this implies that when the effective (in the IR) Umklapp potential becomes weak the isotropy of space effectively restores. Henceforth, such {\em angular} momentum dissipation is inherently non-linear and it can only be addressed with the numerical GR. 

It is an easy exercise to show that for a Drude conductor in an anisotropic spatial manifold  this just implies that there is a separate "transversal" or "Hall" relaxation rate $\Gamma_T$ that combines with the cyclotron frequency in the transport coefficients, next to the usual "longitudinal" relaxation rate $\Gamma_L$  highlighted in the above.  This is acknowledged in Fermi-liquid kinetic theory  where one invokes "angular" collision integrals to capture the effects. However, the breaking of rotations enters through the anisotropy of the Fermi-surface and the natural outcome is that $\Gamma_T$ is typically quite similar to $\Gamma_L$. This became a big issue in high Tc in the early 1990's when it was discovered that $\Gamma_T \sim T^2$, contrasting with the "Planckian" $\Gamma_L \sim T$ in the optimally doped strange metals. 

Prelimenary results  \cite{Leidennum21} indicate that when the IR Umklapp becomes large for small G the $\Gamma_T$'s deduced from the DC magneto-transport   computed for the RN metal on the square lattice become significantly larger than the $\Gamma_L$'s : see Figure (\ref{fig:Umklapptransport}c). The transversal rate $\Gamma_T$ is deduced by fitting the Hall angle to the Drude expression for this quantity. Upon decreasing the ordering wavevector $G$ the imprint of the lattice in the deep IR is rapidly growing (e.g., Fig. \ref{fig:chargevsG}). The outcome is that for $G = 0.1 \mu$ the transversal rate $\Gamma_T$ becomes nearly an order of magnitude larger than the longitudinal rate.  Different from a Fermi-liquid this unparticle matter reacts strongly to the breaking of the isotropy of the spatial manifold.  

 But there is yet another universal aspect associated with the "shear-drag" hydrodynamical regime. In hydrodynamics vorticity/circulation is dissipated by the same shear viscosity as the shear in the flow itself. The consequence is that both rates are governed by the same $\eta$ in Eq. (\ref{holoviscGamma}). This is clearly reflected in the results shown in Fig. (\ref{fig:Umklapptransport}c): in the  "throat regime" ($T << \mu$) the $\Gamma_T/\Gamma_L$ ratio becomes temperature independent. 

The behaviour of the Hall angle in near optimally doped cuprate strange metals caused quite some stir in the early 1990's  by the observation that $\Gamma_T \sim T^2$, contrasting with the "Planckian" $\Gamma_L \sim T$. Insisting that the transport is of a hydrodynamical nature, given the observation that both angular- and linear motion is governed by the same viscosity, the only way that  a  different temperature dependence can arise is by invoking a running in the RG sense of the effective Umklapp potential in the deep IR. The relaxation rates are determined in addition by the length $l_{\eta}$. Since the transversal momentum relaxation is strongly non-linearly realized it is natural that it will show a different RG flow than the longitudinal one when the effective potential is itself temperature dependent. 

It remains to be seen whether such a behaviour can be realized in a $z \rightarrow \infty$ holographic setting. In fact, magnetotransport measurements are playing a key role in the most recent developments: see Section (\ref{qucritphase}). On the one hand these supply the best available evidence for gross behaviours that are supporting the quantum supreme matter idea: the strange metal as a {\em phase} of matter. But at the same time, zooming in on the details of the magneto-transport surprising  behaviours are found which appear to be also beyond the explanatory power of holographic transport theory in its present state -- stay tuned. 

\subsubsection{The optical conductivity in the presence of a lattice.}  
\label{holotransinh} 

I already emphasized that the optical conductivity  of any system formed from large rest mass UV degrees of freedom such as electrons will show the perfect fluid response where all the spectral weight is in the Drude "delta function at zero". In order to find anything else translational symmetry has to be broken. In the above we focussed on the low energy Drude response but Umklapp has also another consequence in conventional metals: bands are formed and one will find always the "interband transitions" at higher energy. These of course are also responsible for the optical response of conventional semiconductors and insulators: it is associated with "bound charges"  forming localized dipoles that absorb the radiation. 

As I will discuss in the experimental Section (\ref{highTcxep}), in the cuprates the response associated with the conduction electrons is exhibiting a quite perfect Drude peak at low frequencies. However, getting above 50 meV or so, this is taken over by such a "bound response" which is actually the best branchcut characterized by an anomalous scaling dimension that has ever been observed in metals! From the argument in the previous paragraph it follows as matter of principle that this is somehow originating in the Umklapp: metaphorically, it is like interband transitions that by the magic of quantum supremacy have turned into a scaling affair. 

Does holography shed any light on this affair? Upon ascending in energy the effective strength of the periodic potential is increasing (the irrelevancy towards the IR) and the non-linearities in the bulk GR will grow: one has to rely on the numerical GR. So much is clear from the little work that has been done that interesting things are going on related to this "unparticle interband transitions" question but it is rather poorly understood.

Actually, the seminal work showing that the numerical GR could be made to work addressed the (longitudinal) optical conductivity in the RN metal with only an unidirectional periodic potential \cite{Horowitzoptics}. These authors picked large $G$'s where the potential is effectively strongly suppressed, but they observed that besides the Drude peak there is a non-Drude "tail" at energies that are still compared to $\mu$: this has nothing to do with the pair creation continuum ($\sigma_Q$). They claimed that this was a conformal-like tail but this turned out to be not quite the case. 

The only other study \cite{DonosOptics} also departed from the same set up but looked at what happens for $G < 1/ \xi$. Here the effective Umklapp is stronger and accordingly there is more weight in this Umklapp induced incoherent part. Sharp resonances develop at $q = 0$, that were interpreted as Umklapp copies of the zero sound mode by these authors. This turns out to be incorrect  \cite{Leidennum21}: we have been studying the optical conductivity as function of {\em momentum} and this shows that these resonances have nothing to do with the sound mode. At the moment of writing it is completely in the dark what these are. 

The take home message of this section is that the understanding of holographic transport in this non-linear regime of strong IR periodic potentials is in a rudimentary stage. At the same time, this is in experiment the most prominent source of information providing evidence for quantum supreme matter physics, see Section (\ref{highTcxep}). So much is clear that "homogeneous" holography is falling short in providing explanations for the experimental observations, while so little is understood presently regarding inhomogeneous holography that it is impossible to arrive at any definitive conclusion. 

\section{Intertwined order and  black holes with Rasta hair.} 
\label{Intertwined}

I already discussed the remarkable mechanism of holographic spontaneous symmetry breaking. I focussed on the first example that was identified: the holographic superconductor which is dual to a black hole with scalar hair (Section \ref{HoloSC}). But the superconducting order is only the tip of the iceberg. Yet again inspired on top-down set ups, Donos and Gauntlett \cite{Donoslattice}  found out a natural mechanism that leads to the {\em spontaneous breaking of translations} in holography specifically in two space dimensions. As normal matter is prone to form crystals at low temperature, so do the holographic strange metals. 

One anticipates that this again involves  technical hardship: also when the cause is spontaneous symmetry breaking these "holographic crystals" require numerical GR in the bulk given the inhomogeneous nature of the spatial manifold. Accordingly, the study of such states did not get beyond demonstration of principle and there is surely no systematic understanding presently available.    

Yet again, the outcomes are remarkably suggestive in the context of the cuprates. In the underdoped cuprates one can identify the "pseudogap regime" taking over from the strange metal when temperature is lowered (Fig. \ref{fig:cupratephasedia}). During the last quarter century much of the experimental effort went into characterizing electronic ordering phenomena other than the superconductivity that are present here. It developed into a holy grail, the "pseudogap as answer to everything." 

I am myself of the opinion that it is to a degree a distraction. The fact that it got so much on the foreground is the outcome of a process fuelled by the "economy of science". Meaningful employment is the scarce good in this trade and much of the existent laboratory equipment is geared to detect order. The ensuing employment program gave it more gravitas than it deserves. 

I do have a right to be opinionated: arguably the theme was born by work I completed as a young postdoc. It was actually quite a harsh experience, starting with the publication  \cite{ZaGun} delayed by 1.5 years by an absurd fight with Physical Review  Letters referee's, while since then I got repeatedly exposed to attempts to rob me from the IP rights. But eventually it landed me a Spinoza prize, the highest scientific distinction in the Netherlands.  For the young people, this is just how human society works, even in physics. 

But it is  too much honour for it to be highlighted in the final section that summarizes the experimental situation with regard to "quantum supreme" signals in the data. Let me therefore shortly summarize here the experimental situation. 

It was a process of discovery: the type of orders that are encountered are in part of an exotic kind that were not known before the late 1980's. In addition there are multiple forms of order that are in a particular fashion synchronized. Metaphorically it is like a "symphony of order parameters", designated as "intertwined order." As I just emphasized it started out in 1987 with my discovery of what is now called the "spin stripes" \cite{ZaGun}. According to semiclassical mean field (Hartree-Fock)  insulating "stripe" domains are formed  in a doped Mott-insulator showing the antiferromagnetic spin order of the insulator, separated  by magnetic domain walls where the carriers get localized. This is a first example of "intertwinement', the spin and charge order are in an unconventional way in synchrony. 

This was later followed by claims that are in part still controversial. At the pseudogap temperature $T^*$ (Fig. \ref{fig:cupratephasedia})  the onset of {\em parity} breaking has been claimed, as well as  the occurrence of a type of order involving spontaneous diamagnetic "loop currents" that break time reversal (see underneath). The latest addition is the {\em pair density wave} (PDW): superconducting order at finite wave vector, appearing in synchrony with the charge order. 

Despite numerous attempts, it is difficult (if not even impossible) to explain it departing from standard "SRE" semiclassical mean field theory. For instance, there is basically a no go theorem for PDW's both in the weak- and strong coupling limits. Remarkably, already the handful of first  attempts to address holographic crystallization reveal that  with the exception of spin order (prohibited by the vanishing rest mass) not only the portfolio of exotic orders but also the intertwinement arise naturally as instability of the RN strange metal! It is to a degree even difficult to avoid it. Is this shear coincidence or is so that the quite baroque ("Rasta") black hole hair associated with this order in the bulk is signalling a deep message?  

After many years of trying to understand the "intertwinement" in the established canon, what strikes me most is that in the holographic incarnation it arises as a "weak coupling" instability from a rather featureless metallic state in holography, in the guise of the BCS-like behaviour of the superconductivity (Section \ref{HoloSC}). This suggests that the order-wise featureless quantum supreme metal carries already the seeds of the complexity of the ensuing order -- unimaginable in the case of a Fermi-liquid. This is I will emphasize discussing the experimental situation.   Whatever, let me present here a short overview of this remarkably rich holographic physics of order. 

\subsection{Intertwinement by the  bulk $\theta$ term.}
\label{holoPDWs}

The bottom-up holographic portfolio rests on the principles of effective field theory. As for the Landau free-energy one relies on locality (gradient expansion) and one can add any term based on whatever fields one deems to be of potential relevance that leave the action invariant under symmetry operations. Next to the usual suspects (Hilbert-Einstein, Maxwell, scalar fields, etc.) one also encounters {\em topological} terms that qualify. In uneven space-time dimensions these are the Chern-Simon terms, famous for being the effective field theory behind the fractional quantum Hall effects in 2+1D. But in even  dimensions including our 3+1D universe these take the form of "theta terms", famous for the role they play in QCD. In this dimension these are of the form  $\sim \theta \varepsilon_{\mu \nu \lambda \sigma} F^{\mu \nu} F^{\lambda \sigma}$. In non-relativistic notation these are of  the $\vec{E} \cdot \vec{B}$ form that may be familiar from the unusual electrodynamics in topological band insulators. 

The "axions" have a long history in QCD. This amounts to the idea that the topological angle is determined by a {\em dynamical} field $\chi$, $\theta \sim {\mathcal \theta} (\chi)$. When this complex scalar axion field  $\chi = |\chi | \exp{i \phi_{\chi}}$ condenses such that $| \chi | \neq 0$ the topological term switches on. By inspecting particular top-down set ups, Gauntlett {\em et al.} \cite{Donoslattice} found that such terms naturally arise in the 4D holographic bulk, including specific forms of the potentials. As it turns out, these have as the first effect that the BF "black hole hair" instability shifts to {\em finite momentum} corresponding with {\em crystallization} in the boundary!   

But the profundity is still to come. The theta term in the bulk couples three fields: next to the axion $\chi$, it involves also the bulk electrical- and magnetic field strengths. The electrical field is dual to the charge density in the boundary but the magnetic fields encode for the occurrence of spontaneous {\em orbital currents}. The condensation of the axion itself is dual to the breaking of {\em parity} in the boundary. Here is the origin of the intertwinement that I will illustrate in a moment with a striking visual impression. 

\begin{figure}[t]
\includegraphics[width=0.9\columnwidth]{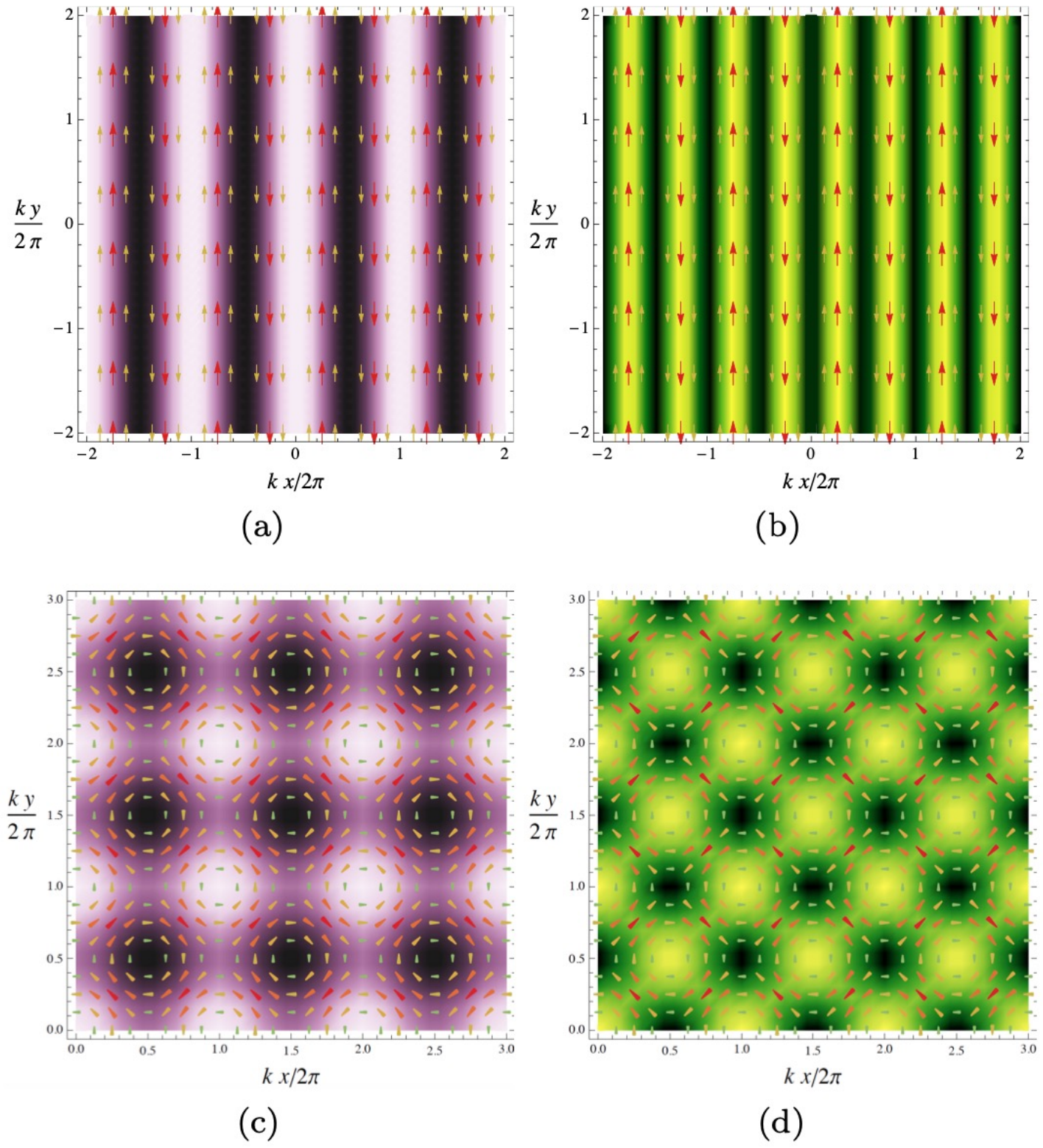}
\caption{The holographic intertwined order for a unidirectional- (a,b) and square lattice (c,d) spontaneous crystallization \cite{HoloPDW} . In panels (b,d) the charge density is indicated by the fals colors  In addition the spontaneous orbital currents are indicated by the arrows: the current patterns have a double periodicity as compared to the charge density: for the square lattice this is the same symmetry pattern as for the "d-density waves" in the condensed matter tradition. In (a,c)  the amplitude (false colours) and phase (arrows) of the superconducting order parameter is indicated that also doubles the unit cell, being precisely out of phase with the diamagnetic currents.}
\label{fig:PDWcai}
\end{figure}

But there is yet more room in the bulk: in the axion part of the action one has the freedom to {\em gauge} the  axion field \cite{HoloPDW}  by including a St\"uckelberg term in the action $( \partial_{\mu} \phi_{\chi} - A_\mu )^2$ -- in CM this is known as the Josephson action, it turns out that the Swiss theorist St\"uckelberg figured this out already in 1938. This has the effect that at the moment the axion condenses holographic superconductivity sets in, albeit at a finite momentum as well. This is the pair density wave!

It is not only the case that all these orders set in simultaneously at $T_c$ -- the parity breaking being the primary order parameter -- but these are also intertwined in the cuprate sense of the word. This is illustrated in Fig. (\ref{fig:PDWcai}). One can as well address a undirectional-  ("stripy") or a square lattice translational symmetry breaking. The green false colours in the right panels indicate the charge density  modulation, the "crystal" in the literal sense. As most easily inferred from the unidirectional case, the arrows in the left panels indicate the spontaneous currents that have twice the periodicity of the charge modulation. The purple false colours indicate the {\em pair density wave} having the same periodicity as the currents, being however precisely out of phase. This is just like the "spin stripes", with the currents and the PDW taking the role of the antiferromagnet. The same logic governs the square lattice where now the currents form closed loops, ordering in a similar way as envisioned in the loop current claims. 

Surely, this has an image in the bulk in the form of a highly textured "corrugated" black hole hair in the bulk: this "black hole with Rasta hair" is the most complex fanciful stationary black hole solution known to mankind! Is it shear coincidence that it gets so close to this similarly unprecedented "order circus" encountered in the cuprates? This is so intriguing that I allow myself to be a bit religious just for the pleasure of it -- it is the heavens that speak to us in a piece of rusted copper ... 

\subsection{The holographic Mott insulators.}
\label{holoMott}

It is not yet the end of the story:  when it comes to black hole complexity one can shift to even a higher gear. I discussed the crucial role that the Mott insulator plays in the high Tc problem in Section (\ref{Mottness}). The industry standard is to depart from Hubbard type models like Eq. (\ref{Hubbardmod}). But these are quite specific, in fact quite oversimplified models that were introduced as toy models being taken in the mean time more literal than intended when these were formulated. The Mott insulator is actually resting on a much more general principle that is independent of microscopic modelling. In full generality, the electron Mott insulator is a {\em crystal formed from electrons with a crystal structure that is commensurate with the underlying ionic lattice}. 

It just examplifies the general idea of {\em commensurate pinning} that is also occurring in meat-and-potatoes classical systems.  It is very simple. Form a "balls and springs" solid with a particular lattice structure and lattice constants. In the spatial continuum this will be characterized by its Goldstone bosons -- the massless acoustic phonons. But embed it now in a background containing a potential that precisely fits the spontaneous crystal lattice, and this will shift in  such away that its "atoms" are lying precisely in the external potential minima. The effect will be that the phonons will acquire a {\em commensuration gap}, associated with the energy required to rip the crystal from the background. High energy physicist know this general phenomenon as the "pseudo-Goldstone bosons", referring to the general behaviour of Goldstone bosons in an explicit symmetry breaking field. 

The ensuing state is clearly insulating, characterized by the commensuration gap which is a better name than the "Mott gap". Moreover, symmetry wise it is a featureless state because the external potential has already explicitly  broken the translational symmetry. The only additional ingredient  one meets dealing with  electrons are the spins at low energy.    

This suggests a window of opportunity to construct Mott-insulators in holography. We just learned that holographic matter can be tailored to crystallize. All one needs to do is to switch the external lattice potentials of the previous section and see what happens. This was accomplished in Leiden, with Sasha Krikun in the drivers seat  \cite{holoMott}. For technical reasons the focus was on a unidirectional lattice -- in the mean time we also got it to work for a square lattice pushing the bulk numerical GR. The outcome is that it works precisely as anticipated. In essence it is like the intertwined order (without PDW's) where the commensurate background potential  acts to further enhance the charge modulation. 

A difference with the Hubbard model variety is that because of its ultrarelativistic nature holographic matter does not support spins that can go their own way. The role of the spins is taken instead by the orbital loops and we identified the analogue of superexchange finding that it takes a much smaller energy to "flip" the orbital loops than the charge gap. 

But there is more complexity around the corner. The "spin stripes" I advertised in the above should be viewed in full generality as "higher order commensurate Mott-insulators". This is referring to well established principle in a subject that was big in the 1970's: the study of crystals having a periodicity that does not quite match the background potential. Imagine a mismatch of e.g. a quarter of a lattice constant. One can formulate ball-and-spring models in such a background ("Frenkel-Kontorova models"). These are intrinsically non-linear and one finds typically the formation of soliton like textures. These are called {\em discommensurations}: the crystal stays locally commensurate, forming commensurate domains, while the mismatch is stored in narrow areas (the discommensurations) that form a regular discommensuration lattice. This is precisely describing the Hartree-Fock spin stripes, called like this in the first Zaanen-Gunnarsson paper \cite{ZaGun}, referring in fact to classical "striped" discommensuration patterns. 

Especially dealing with a one dimensional modulation, the "Devil's staircase" was identified in such systems. Vary the commensurate mismatch continuously. At various rational ratio's (like the 1/4 of the example) a higher order commensuration point is realized and a discommensuration lattice forms. But now change the ratio a bit: what happens is that a discommensuration lattice forms inside the commensuration lattice. And so forth, the outcome is a fractal hierarchy of such discommensuration lattices, the Devil's staircase. Amazingly, such a Devil's staircase can be realized departing from the holographic matter \cite{holoMott}!

\begin{figure}[t]
\includegraphics[width=0.9\columnwidth]{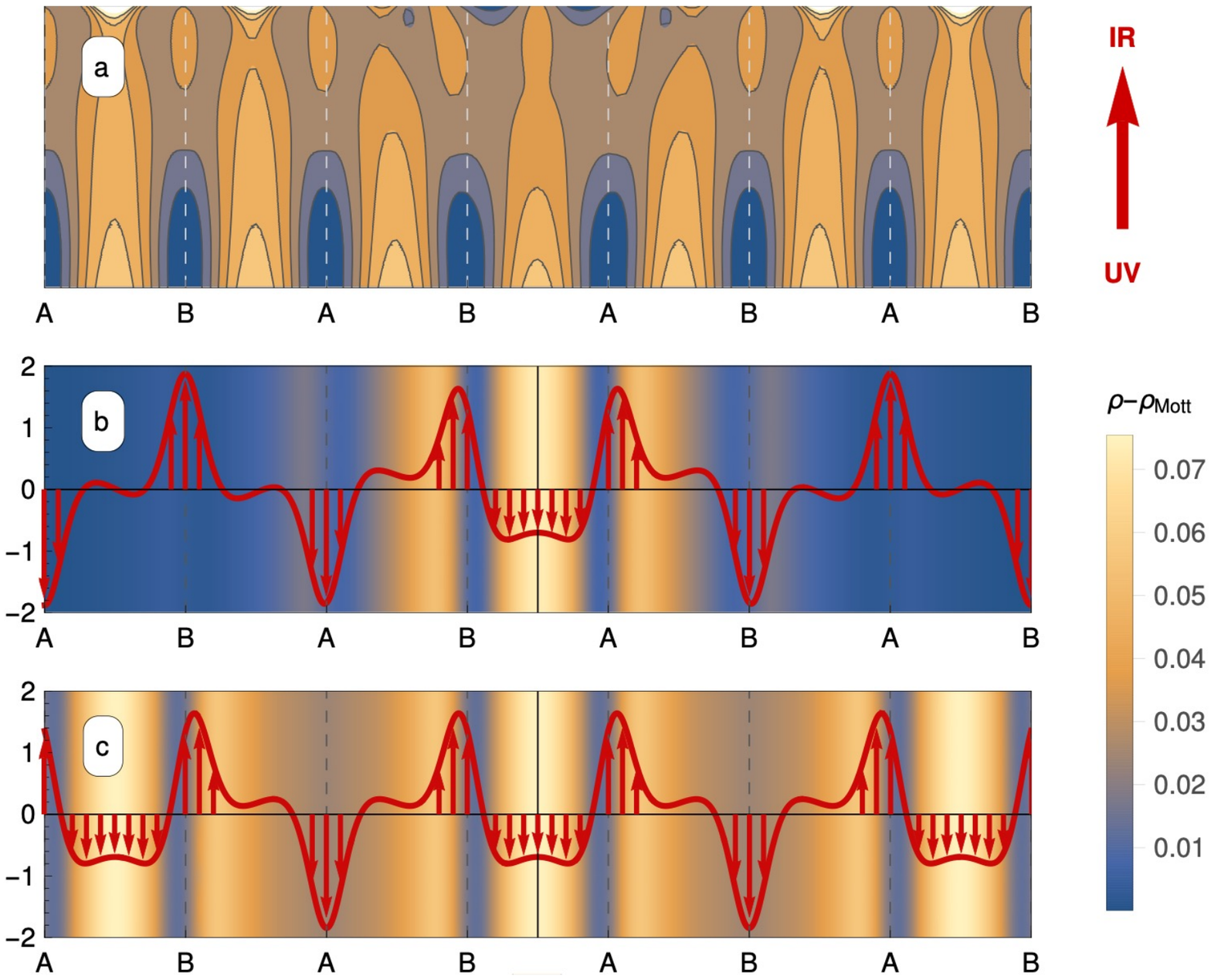}
\caption{The "stripes" in the holographic Mott insulator \cite{holoMott}. When the external- and spontaneous translational symmetry breaking are incommensurate discommensurations form. In (a,b) we show the structure of one such discommensuration associated with a unidirectional symmetry breaking. (a) The electrical field in the bulk as function of radial direction yielding the RG flow associated with pinning and the formation of the discommensuration in the middle. (b) The difference of the charge density compared to the commensurate Mott insulator (false colors) and the current order (red line, arrows) revealing that the discommensuration is a domain wall in the staggered current order.  (c) These "stripes" form a regular pattern, shown here for a three lattice constant discommensuration lattice.}
\label{fig:holostripe}
\end{figure}

For the Hubbard-Hartree-Fock stripes the charge discommensurations are at the same time domain walls in the antiferromagnet formed by the spin system. Amazingly, the same is happening in the "holographic stripes"  but now involving the orbital currents. These form an antiferromagnetic pattern in case of the spontaneous crystal where the direction of the current flips traversing a maximum in the charge density,  see Fig. (\ref{fig:PDWcai}). In Fig. (\ref{fig:holostripe}) the outcomes are shown for the "stripes" in the holographic Mott insulator. We depart from a unidirectional translational symmetry breaking as in Fig. (\ref{fig:PDWcai}b) adding a similar unidirectional potential which is however incommensurate. In panels (a,b) we focus in on a single discommensuration. In panel (a) the electrical field in the bulk is shown as in Fig. (\ref{fig:chargevsG}) providing an image of the RG flow but now related to the pinning physics. The primary modulation wavevector is $\sim \mu$ and the external potential would by itself die-off in the IR. But the spontaneous symmetry breaking is relevant to the IR and one sees from the figure that these get clamped together with each other midway the radial direction: the RG view on commensurate pinning!

 But this clamping is interrupted midway the space direction: the discommensuration. Focussing on the charge density (false colors in panels b,c) the charge does indeed accummulate at the discommensuration, Focussing now on the currents (arrows) one discerns that the discommensuration corresponds at the same time with a {\em domain wall} in the staggered current order. This is precisely what is happening in the cuprate spin stripes, after identifying the staggered antiferromagnetism with the staggered current order. Surely, these stripes form together an orderly striped structure, Fig. (\ref{fig:holostripe}). 
 
As far as I am aware, looking in the bulk this "stripy" black hole sets the complexity record for a  stationary geometry in all of gravitational physics ...
  
 What happens with the excitation spectrum in the holographic Mott insulator? The optical conductivity shows that the Drude peak associated with the sliding mode in the continuum shifts up to a finite energy: this is just the expected pinning "Mott" gap. I already discussed at length the generic "incoherent" second sector in Section (\ref{Homosecsector}), with a big warning sign given that it is a specialty of ultrarelativistic matter. However, we are now dealing with a deep IR/near horizon geometry that is grossly reconstructed because of the highly non-linear impact in the bulk of the translational  symmetry breaking. The outcome is that although the Drude conduction is completely gapped an incoherent sector is left behind at low energy showing the typical power law conductivity of such incoherent stuff, with a resistivity that is power law divergent towards zero temperature. 

As I emphasized in Section (\ref{CCSinc}), {\em emergent charge conjugation symmetry} is a necessary condition for having finite conductivity exempted from a Drude transport. The most direct, simple observable is that the {\em Hall effect will vanish}. This is very easy to see. Consider two equally large basins with opposite charge and apply an electrical field and a magnetic field in a perpendicular direction. Due to the Lorentz force as may positive- as negative charges will rain down on the boundary and no Hall voltage will arise. To check this a real two dimensional lattice is required. In the mean time we got this working and we find that indeed the Hall effect vanishes upon entering the holographic Mott insulator. 

This is directly communicating with highly serendipitous recent experimental results by Popovic and coworkers on the cuprate spin stripes \cite{Popovic}. These show typically a remnant superconductivity that can however be completely removed in magnetic fields $\sim 30$ Tesla which are available at the high field labs. They find that in this "cleaned up"  spin stripe phase a slowly diverging resistivity in tune with the expectations for the incoherent stuff. But more strikingly, {\em the Hall effect is abruptly disappearing} upon entering this phase. This may  be the first and best example of emergent quantum supreme matter that persists down to the lowest temperatures. This begs for a large experimental effort, measuring as many physical properties as possible in these difficult circumstances of very large magnetic fields and ultra low temperatures. 

\section{The fog of war: the experimental situation.}
\label{highTcxep}

The specific appeal of this whole affair is to find out whether quantum supreme matter is realized in nature. The high Tc superconducting cuprates take in this regard the role of fruit fly. Above all these are the "ground zero" of the intellectual crisis that developed during the last thirty years, where it became increasingly clear that the established paradigm of "classical matter" (the ESR products) fails in explaining a large body of experimental findings. In a parallel development unfolding in the computational community we became increasingly aware of the fundamental brick wall in the form of the fermion sign problem as highlighted in Section  (\ref{fermionsigns}). We better accept the fact that the mathematical tool box required to decode the way in which nature works in this realm used to be completely empty. 

This has now changed by the arrival of AdS/CMT. But to my opinion it is not quite reliable as a navigational device.  I like to put forward the following metaphor. Imagine it would have been extremely hard to measure the electrical- and magnetic properties of aluminum metal at sub-Kelvin temperatures,  while it would have been easy to measure up the stuff in the core of a big neutron star. It is not certain what is happening out there but with a little luck these experimentalists would have discovered that the quarks form at these high densities a rather weakly coupled Fermi-liquid that upon further cooling would turn into a flavor-colour locked superconductor: another invention by Wilczek, it is a BCS superconductor driven by the attractive nature of gluon exchange, accommodating the zoo of quantum numbers of the  standard model in an unconventional (in the BCS sense) order parameter. 

But it is then obviously a long way to go all the way to aluminium, finding out that  electrons have not much of internal symmetry as compared to the zoo of flavors and colors of QCD. Instead, there is  band structure, the fact that electrons exchange phonons giving rise for strange reasons (from the QCD perspective) to attractions that can even overwhelm the huge Coulomb repulsions. But both cases are governed by the same overarching emergence 'meta' principles. There are Fermi-surfaces and quasiparticles in the normal state, the condensates are governed by the principles of spontaneous symmetry breaking, etcetera. In this analogy,  AdS/CFT is like the neutron star matter while the high Tc-related phenomena are like aluminium.

The limitations of the AdS/CFT navigator are obvious. Instead of dealing with the zero density CFT "parts" that at finite density turn into the holographic strange metal "wholeness', one departs from the condensed matter electrons that by being subjected to the strong lattice potential (the "Mottness") may be forced into the quantum supremacy realms. I already highlighted various manifestations of "UV sensitivity" that cause matters to become different by principle. But it may well be deeper. The central principle revealed by AdS/CMT is the covariant scaling. I am a believer in the universality of $\theta$ and $z$ anomalous dimensions (see underneath) but "covariant RG" is much less constraining than the full fledge invariance under scale transformation which is at the heart of the notion of the stoquastic critical state universality. 

The first question to ask is, is it really sure that the cuprate electrons are not submitting to the established canon of "classical" matter? For most of its history it was taken for granted that some contrived version of the meat-and-potato diagrammatics would eventually explain everything. Although large parts of the high Tc community is still living in this tunnel, the data that have been streaming in very recently make this proposition in the mean time untenable, as I will discuss underneath. These new results are suggestive of the "meta" emergence principles suggested by AdS/CMT to be at work in the cuprates: (1) The strange metals appear to be {\em phases of matter} in the same general sense as Fermi-liquids are (Section \ref{qucritphase}), (2) These appear to be governed by the covariant scaling principle revealing extremely anomalous dimensions (Section \ref{zinftyEELs}) while the finite T transport reveals the Planckian dissipation,  (3) Although photoemission has been historically the experimental source for the (non-quantum supreme) presence of particles, very new results indicate that this may be misleading (Section \ref{Cupratefermions}). 

But ironically, these gross properties are revealed through experimental signals that to quite a degree are {\em not at all explained by AdS/CMT} in its present state, as I will emphasize in the discussion.  It may be that these reside in the "inhomogeneous", strongly non-linear bulk regime as discussed in Sections (\ref{holotransport}, \ref{Intertwined}) that is barely explored. But I would not be surprised when eventually it will turn out that the cuprates are in a quite different "covariant universality class" where phenomena occur that are unheard off in the "QCD-like" version. 

\subsection{Cuprate strange metals: quantum critical phase versus quantum critical point.}
\label{qucritphase}

Already since the 1990's the idea that the "strangeness" of strange metals originates in the presence of a quantum critical point somehow related to the Hertz-Millis ploy explained in Section (\ref{hertzmillis}) has been influential. There are quite a number of examples especially in the heavy fermion family where the basic idea is very credible \cite{LohneysenRMP}. One does find (mostly) magnetic order parameters that do undergo a zero temperature transition. A "quantum critical wedge" showing anomalous properties is anchored at the quantum critical point. The associated cross-over lines in the coupling constant-temperature plane are clearly discernible, even showing the tell tale signs of a (heavy) Fermi-liquid re-emerging in the renormalized classical regime. As I already emphasized, this does not imply that these are fully understood  but there is no doubt that the "strangeness" is eventually rooted in the QCP. 

Until rather recently it was a community consensus that the cuprates were also in the grip of a quantum phase transition of the kind at optimal doping.  The phase diagram published as part of the " community consensus review" that appeared in Nature in 2015 \cite{Naturecons15}  is reflecting this, see Fig. (\ref{fig:cupratephasedia}). The purple "strange metal" area  is indicated as the wedge. On the underdoped side the $T^*$ line decreases roughly linearly with doping, suggestive that it hits zero temperature at the "critical doping" $p_c$. The reflex has been to associate the pseudogap temperature  $T^*$ that reveals itself in a form of incomplete gapping affecting nearly all physical properties with the onset of order at a classical phase transition. But this continues to be a confusing issue -- it is far from settled whether this directly involves the intertwined order package that I shortly discussed in Section (\ref{Intertwined}). 

A difficulty is that as far as one can say all these orders have disappeared at dopings that are significantly lower than $p_c$.  It is still impossible to completely surpress the superconductivity near optimal doping because of the extremely high magnetic fields that are required to get above the (mean field) $H_{c2}$. One   therefore has to look for the signatures of the vanishing order deep in the superconducting state but despite intense effort nothing was ever detected. One idea is that the order parameter is just very hard to detect with standard experimental machinery: the "hidden order" -- the loop currents presented in Section (\ref{holoPDWs}) are an example. 

All along there was however an ideological bias. Until the arrival of the holographic strange metals all that could be imagined was the QCP. At least in my mind there was all along an uneasiness. In so far such metallic QCP's are tractable (in essence, implicitly departing from a Fermi-gas  plus order parameter) at the end of the day the quantum critical behaviour is an infrared affair. Upon rising energy or temperature a regime which is Fermi-liquid like should re-emerge: this is typically observed in the heavy fermion QCP systems. But not so in the cuprates: for instance, the linear "Planckian" electrical resistivity extends all the way to the temperatures where the crystals melt. 

Another uneasy affair was the lacking of any experimental signature of the crossover line in the overdoped regime for $p > p_c$. The crossover from the strange metal (purple) to a presumably Fermi-liquid like overdoped regime (white) is just a product of imagination: it is not supported by literally {\em any} empirical support.  The experimental characterization of the metals in the overdoped regime  has been on the move since 2015. In a remarkable pace, evidences accumulated supporting a quite different basic view than the traditional "QCP". 

 Apparently,  at dopings $p < p_c$ a first type of strange metal phase is realized that is prone to become "BCS-like" unstable to the intertwined order. It is BCS-like in the following sense. Intrinsic to quantum phase transitions is the general notion of "fluctuating order". Upon zooming in on the dynamical susceptibility associated with the order parameter one should find strong enhancements, the $E/T$ scaling and so forth as discussed in Section (\ref{qucritical}) in the metallic state up to the temperatures where the metal is "strange". However, in a weakly coupled BCS  superconductor (or equivalently, a "nesting" type density wave instability) such correlations disappear rapidly at temperatures above the thermal transition. Only upon closely approaching this transition the RPA ("paramagnon") style relaxational peak develops, compare with the discussion of "quantum critical BCS" in Section (\ref{HoloSC}).  
 
 With the caveat that the experimental probes are far from perfect, nobody has managed to pick up signals of the quantum critical fluctuating order kind over the large temperature range where  the "linear in $T$ resistivity" strange metal is realized above $T^*$.  This is indicating a "weak coupling like" development of the order.  The correlations now pertaining to the intertwined order are then expected to rapidly peter out above the transition temperature in the strange metal. This is also what is found dealing with the holographic intertwined order and even the Mott-insulators as discussed in Section (\ref{Intertwined}) at temperatures above their $T_c$'s. One learns that this weak coupling-like behaviour supersedes the Fermi gas. But the take home message of the  "Rasta hair" of Section (\ref{Intertwined}) is that very different from the Fermi gas such weak coupling like instabilities may support complex, multi-order parameter instabilities. 
 
\begin{figure}[t]
\includegraphics[width=0.9\columnwidth]{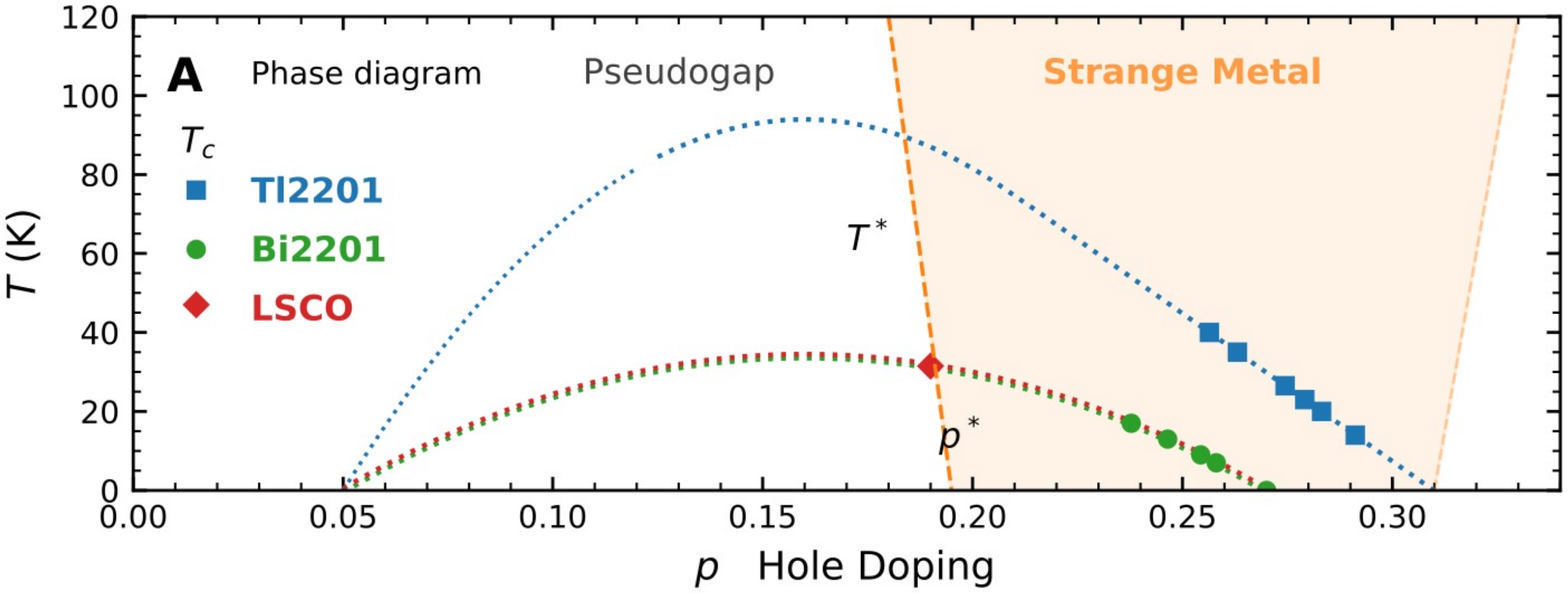}
\caption{The high Tc phasediagram updated on basis of magnetotransport measurements \cite{HusseyMR}  emphasizing that the metal in the overdoped regime is a {\em phase} of matter, different from the metallic phase below optimal doping ($p_c$) that is prone to "pseudogap instability" at low temperatures.}
\label{fig:husseyphasedia}
\end{figure} 
 
 The news that broke in 2019 is  the experimental discovery \cite{Sudijump} of a {\em discontinuous} first order like zero temperature phase transition that  appears to take place at $p_c$. The discoveries that followed leave no doubt that a strange metal phase of a different kind takes over in the overdoped regime. Eventually, at very high dopings beyond the end point of the superconducting dome there are signs that this merges into a Fermi-liquid like affair, see Fig. (\ref{fig:husseyphasedia}).

This transition is by itself most unusual: even holography has presently nothing to say about it. It was first seen in ARPES \cite{Sudijump} in the form of a dramatic change occurring at rather high energies in the electron spectral functions. Below $p_c$ in the normal state the spectral functions associated with the "antinodal" momentum directions are completely incoherent "unparticle" affairs. Above $p_c$ this suddenly changes, showing quasiparticle signatures -- whether these are real quasiparticles is a different issue, see underneath. This was quickly followed by specific heat evidence for the sudden collapse of the remnant pseudogap upon crossing into the overdoped state (also supported by ARPES). Instead of the "pseudogap" order to be the culprit, the moral appears to be that the underdoped metal is prone to a BCS-style intertwined order instability while the overdoped metal is only unstable in the superconducting pair channel. 

The highly unusual aspect of this "first order like" change is that many macroscopic properties such as the superconducting $T_c$, the resistivity and so forth are varying in a perfectly smooth manner around $p_c$: these are expected to jump upon crossing a zero temperature first order transition. However, very recently it became clear that such signatures are revealed in {\em magneto} electrical transport properties. A discontinuous change has been observed in the Hall relaxation time at $p_c$ \cite{SebastianHall}, supported by magneto-resistance measurements. The latter \cite{HusseyMR}  supply direct evidence for the strangeness of the overdoped metal, in the form of the "Planckian quadrature": the magneto-resistance appears to scale with $\Delta \rho \sim k_B T \sqrt{ 1 + ( \mu_B B /k_B T )^2}$ ($B$ is the magnetic field strength), a poorly understood "Planckian" scaling behaviour also observed in a QCP system. Yet, in the overdoped metal it occurs in the whole doping range: it is a property of this metallic {\em phase}. Unpublished results by Hussey's group show that this Planckian quadrature magnetoresistance  suddenly disappears below $p_c$ being supplemented by a Drude type "modified Kohler's rule", i.e. governed by the  $\Gamma_{L,T}$ discussed in Section (\ref{holomagnetotransport}).

I perceive this as the best empirical evidence for the cuprate metals to be forms of quantum supreme matter. Ironically, it resolves the ideological divide that has haunted the subject since the very beginning. On the one hand, on the gross level the physics appears to be similar to the Fermi-liquid affair and a large part of the community  found in this comfort to run variations on fermiology to explain the observations. However, a better educated (mostly theoretical) company struggled with all kind of inconsistencies. Upon getting informed by holography regarding the existence of the generalized Fermi-liquids hidden "behind the fermion sign brick wall" this ambiguity evaporates.

But this does not mean that one can read off all the answers from the holographic handbook. Holography has at this moment in time {\em nothing} to say about the "strange" first order transition at $p_c$. Neither does it shed any light on the difference between the underdoped- and overdoped strange metals. Although holography is revealing with regard to the "Hall angle" question (Section \ref{holomagnetotransport}) it does not tell anything about the "quadrature" magnetoresistance.   Yet again there is irony: holography does not shed any light on the magneto-transport and ARPES data that reveal the separate existence of two types of strange metals.   

\subsection{The collective dynamical responses: currents and charge density.}
\label{zinftyEELs}

More is needed to find out whether these strange metals have anything to do with the quantum supreme matter of holography. As I emphasized repeatedly, the main signal coming from the "black holes" of holography is that {\em phases} of matter exist that are revealing {\em simple scaling behaviours} of the same kind as the ones encountered at strongly interacting stoquastic quantum phase transitions. Realizing that this emerging simplicity originates in this case in the  exponential complexity of the states erasing all details in the observables, observing  such scaling behaviour in the strange metals should trigger the alarms. 

According to holography the gross organization of these "quantum critical phases" follows the pattern of Fermi-liquids, but these are now characterized by anomalous dimensions. At the QPT's these scaling dimensions are unique numbers, part of the universality class set by symmetry and dimensionality. According to holography,  these generically should vary {\em continuously} dealing with a  quantum critical {\em  phase} pending the changing microscopic conditions associated for instance with the doping level. 

In fact, the universality realized at the thermal phase transitions was discovered experimentally. The thermodynamic exponents are easy to measure and the 1970's critical theory explained it after the fact. But dealing with the quantum incarnations the life in the laboratory is less easy. You should now appreciate that this information is rather hidden in simple measurements like DC transport, where it is "shrouded" by conservations laws, Section (\ref{holotransport}). 

Thermodynamical information is "primus inter pares", but it is actually quite difficult to isolate the electronic entropy at the high temperatures of the strange metal regime. The old tour de force determination of the high temperature specific heat by Loram and Tallon appeared to show such an impeccable scaling, but deceptively in the form of a simple Sommerfeld entropy that was by reflex interpreted as evidence for a Fermi-liquid. But I argued forcefully in Section (\ref{genFL}) that this has to be a deception being aware that $z \rightarrow \infty$ relying on the holographic scaling form Eq. (\ref{hyperscalingent}) for the entropy. Being aware, one can now turn it around to realize that the simplicity of this Sommerfeld entropy is {\em supporting} this generalized notion of universality. 

But to get a full view on the presence (or not) of simple scaling it is intrinsic to the quantum incarnation that one has to have access to dynamical linear response functions over a large range of frequency, temperature and momentum, avoiding conditions controlled by conservation laws. This refers to hardship in the laboratories -- little information of the kind is available for the reason that it is very difficult to measure. In Section (\ref{HoloSC}) I illustrated the powers of this affair focussing on the iconic case of superconductivity. When the required "pair susceptibility spectrometer" would have been routinely available in the labs, the observation of $E/T$ scaling revealing anomalous scaling dimensions  would have had a similar impact as the observation of anomalous dimensions at thermal transitions. 

In fact, to obtain an "unfiltered" view on scaling properties one would like to measure properties that are {\em not} protected by conservation laws. The pair susceptibility I just referred to is a point in case. Similarly, the spin fluctuations associated with  (non-conserved) antiferromagnetism qualify. These are quite literally in the driver's seat in Sachdev's book on quantum phase transitions \cite{QPTSachdev}.  These can be measured by inelastic neutron scattering, which yields a perfect realization of a linear response measurement. However, the technique is infamous for its bad to signal to noise ratio. Although it was the technique of choice in the early intertwined order exploration where the spin correlation become structured and strong (spin stripes), up to now nobody has been able to resolve the signal in the strange metal regime. In fact, the very recent  EELS experiments measuring the charge dynamical susceptibility that I will highlight underneath are at the moment stand alone in this regard. 

\subsubsection{The linear resistivity and the Drude behaviour. }
\label{linresemp}

I have already repeatedly referred to the oldest and in a way still most obvious indication that something highly unusual is going on: the DC resistivity being linear in temperature all the way from $T_c$ to temperatures where the crystals melt (e.g., Section \ref{sheardrag}). The difficulty is to explain why it is so {\em simple}. In a Fermi-liquid the temperature dependence of the resistivity is governed by the way in which individual quasiparticles loose their momentum. By principle this is a strongly temperature dependent affair: the "Umklapp $T^2$" at the lowest temperatures, taken over by the phonons that can cause a variety of temperature dependencies to eventually saturate when the inelastic mean free path becomes of order of the lattice constant. Especially the violation of the latter Mott-Ioffe-Regel (MIR) limit is striking. Around 400K or so this is reached in the cuprates but there is no sign of it in the data -- the resistivity stays perfectly linear. 

To find out what is going on {\em dynamical} information -- optical conductivity -- is crucial also in this case.   From DC transport it is just impossible to pin point the general nature of the transport physics and one encounters frequently misinterpretations. The optical conductivity measurements leave no doubt: at low frequencies these reveal a textbook Drude lineshape, Eq. (\ref{Drudelong}). This was already noticed when the first measurements became available in the late 1980's. Pushing the limits of the effective resolution in a recent study \cite{Heumenopt}, it was concluded that at least 90\% of the low energy spectral weight is at least at temperatures below the MIR captured by the Drude conduction, see Fig.(\ref{fig:Heumenopt}). 

\begin{figure}[t]
\includegraphics[width=0.9\columnwidth]{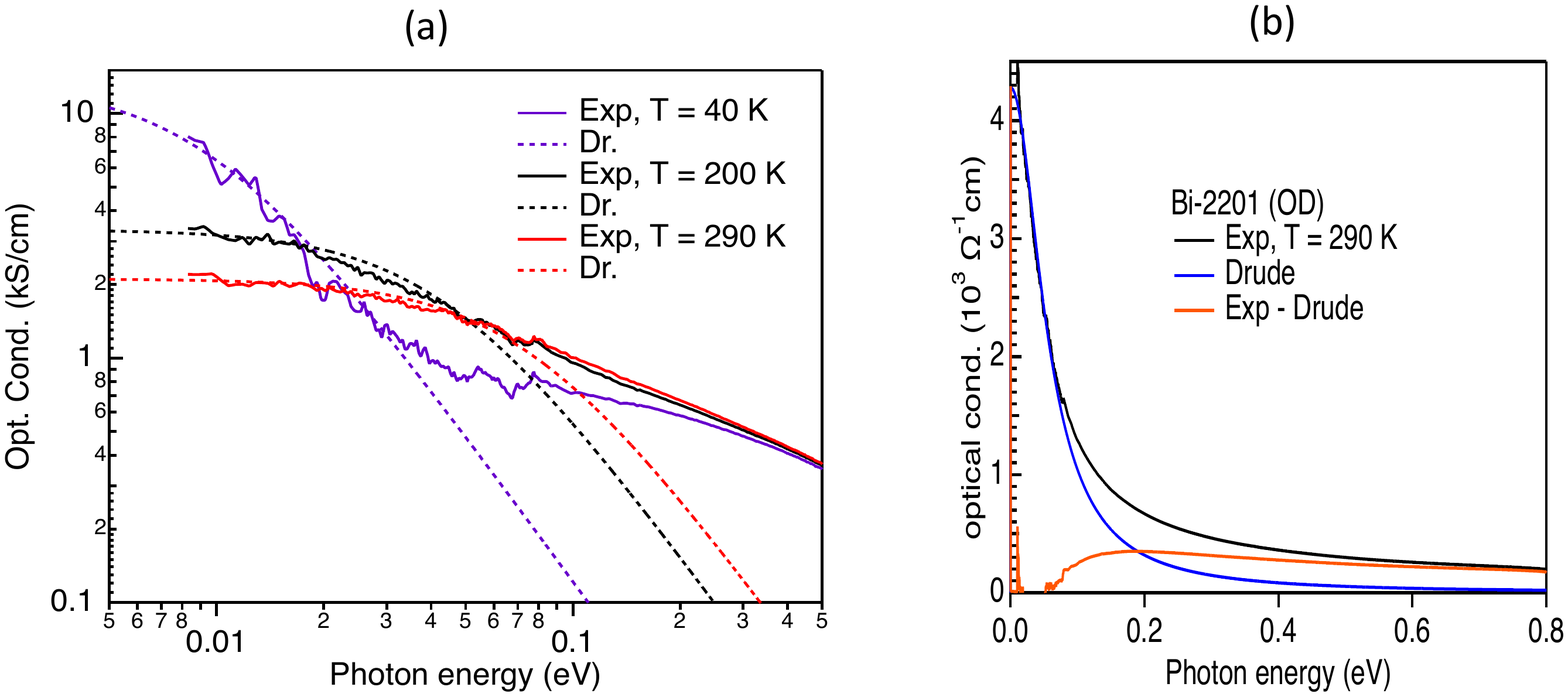}
\caption{The real part of the optical conductivity of cuprate metals \cite{Heumenopt}. (a) The low frequency part is perfectly fitted with a simple Drude form over a large range of temperatures and doping. (b) After subtracting this Drude part (blue) of the overall signal (black) a contribution is left over that is a perfect branch cut (red) above a gap $\sim 50$ meV.}
\label{fig:Heumenopt}
\end{figure}

The DC conductivity is then the product of the Drude weight and the momentum relaxation rate $\sigma (T, \omega=0) = {\cal D} \tau_P$. The only way to separate these two factors is again by direct measurement of the frequency dependence -- studies that only use DC information guessing the Drude weight based on Fermi-gas relations are just not to be trusted. It follows from this optical study \cite{Heumenopt} that ${\cal D}$ is nearly temperature independent while it increases in a linear fashion over the doping range all the way from mildly underdoped up to the end of the overdoped regime, varying smoothly at $p_c$: claims that there would be sudden change in carrier density based on Hall effect measurements are flawed -- these changes are actually happening in the Hall relaxation rate $\Gamma_T$ \cite{SebastianHall}.

The linear resistivity is therefore entirely due to the momentum relaxation rate, and this can be shown to be very close to the Planckian dissipation time $\tau_p \simeq \hbar / (k_B T)$ \cite{Heumenopt}.  I highlighted this as the natural time associated with the "conversion of work in heat" in densely entangled systems. The difficulty is however that it is now governing the life time of {\em total momentum}: by hydrodynamical principle this is only finite when the "deep IR" translational symmetry is broken. By default it cannot be truly universal, it is somehow tied to the way that the translational symmetry breaking is "processed" by the strange metal: this is just implied by the fact that it is a Drude transport. 

The only available internally consistent explanation I am aware of is the minimal viscosity "shear drag" suggested by holography, Section  (\ref{sheardrag}). It is far from proven fact -- the first feature that one would like to observe in the laboratory is that the electron system behaves like a hydrodynamical fluid but this is a tall marching order for the experimentalists. Another difficulty is related to the phonons: upon raising temperature these become available as extra momentum sink next to the Umklapp scattering, and why do these not imprint on the temperature dependence? Even more pressing, above the MIR temperature it is quite questionable whether one is dealing with a Drude conductor \cite {Delacretazbad} and yet again the issue is why nothing is seen in the data crossing over from simple shear drag at low temperature to something else in the "bad metal" high temperature regime. 

\subsubsection{Optical conductivity and the branch cut.} 
\label{optbranchcutcupr}

But the optical conductivity  reveals more: in fact, the most perfect branch cut response I have ever encountered in these realms, signalling that scaling takes over from the Drude response in the energy range $\sim 50$ meV all the way to a very high energy of roughly 1 eV \cite{vdMarelopt03} . This is illustrated in Fig. (\ref{fig:Heumenopt} ) \cite{Heumenopt}: the low energy part can be highly accurately fitted with a Drude peak  with a frequency independent $\Gamma$. Upon subtracting this from the measured $\sigma_1$ a higher energy part is left over that can be fitted with a  branch cut characterized by a smeared low energy gap,

\begin{eqnarray}
\sigma_{\mathrm{exp}} (\omega) & = & \sigma_D (\omega) + \sigma_{\mathrm{incoh}} \nonumber \\
\sigma_D  (\omega) & = & \frac{{\cal D}_D}{ \Gamma - i \omega} \nonumber \\
\sigma_{\mathrm{incoh}} & = & {\cal D}_{\mathrm{incoh}}  \left( \sqrt{\Delta_{\mathrm{incoh}}^2 - \omega^2} - i \Gamma_{\mathrm{incoh}} \right)^{-\alpha_{\mathrm{incoh}}}
\label{optcondbranchcut}
\end{eqnarray}

Here $\Delta_{\mathrm{incoh}} \simeq 0.05$ meV  and together with $\Gamma_{\mathrm{incoh}}$ this  parametrizes the smeared gap at the low energy end. Although the modelling of the cross over from the Drude to the branch cut part is ambiguous it is crucial that Eq. (\ref{optcondbranchcut}) fits both the real- and imaginary part of the measured conductivity. At energies well above $\Delta$ the branch cut takes over and this signals itself through a frequency independent phase angle that lights up in the data : $1/ (i \omega)^{\alpha} = \exp{ (- i \alpha \pi/2) } /  | \omega |^{\alpha}$. 

The scaling dimension $\alpha_{\mathrm{incoh}}$ shows the signs of qualifying as a genuine anomalous dimension. At optimally doping it is approximately $2/3$. However, in the recent study \cite{Heumenopt} the optical conductivity is systematically investigated in a large doping regime. Further evidence for the overdoped metal above $p_c \simeq 0.19$ continuing to be strange is in the finding that the "conformal tail" extends all the way to the highest doping while its relative spectral weight is barely changing. But these data indicate that  $\alpha_{\mathrm{incoh}}$ is continuously  varying as function of doping up, varying in the single layer BISCO from $\simeq 0.4$ at very low doping $p \simeq 0.05$  to $\simeq 0.8$ at the highest dopings where superconductivity has disappeared. 

Such branch cuts with varying exponents are  suggestive that we are indeed dealing with a  quantum supreme metal in the spirit of the holographic strange metals. But there is still quite some distance to go. As I emphasized in Section (\ref{holotransport}) the presence of the ionic lattice is a {\em necessary} condition to find this optical spectral weight. I also explained in Section (\ref{holotransinh} ) that presently it is still to be found out whether general principle can be extracted for such spectral weight from holography, since "inhomogeneous" holography is still poorly understood. 

Especially with regard to this "optical branch cut" I would not be surprised when eventually it turns out that this is beyond the capacity of holography. After all, the Umklapp that is at the origin of the branch cut is in the electron systems the key ingredient behind the "Mottness" that is the prime suspect for the "release" of the fermion sign driven quantum complexity as explained in Section (\ref{Mottness}). A strikingly mysterious fact is that the branch cut appears to extent up to energies as high as 1 eV where one gets close to the UV physics of the electrons in Cu-O unit cells.  This invokes UV sensitivity -- I would not be surprised when it will  eventually become clear that although the general scaling principles revealed by holography are applying we are dealing in the condensed matter systems with a different universality class of the covariant scaling agenda than the ones revealed by the large $N$ CFT's.

\subsubsection{EELS and the local quantum critical charge response.}
\label{EELSlocqucrit}

This brings us to the recent  pioneering effort to measure the dynamical charge susceptibility using electron energy loss spectroscopy  (EELS) \cite{AbbamonteEELS1,AbbamonteEELS1}. Obviously, the first object one would like to know as a theorist is this very basic property. EELS is the way to go, since this spectroscopic technique has the potential to measure the response over  basically the whole parameter range of frequency, temperature and especially also momentum as of relevance to the condensed matter electrons.  One can wonder why this was not accomplished already  25 years ago in the high Tc context. This is  because of  politics. EELS is a rather old technique and it was hard to get funding for the necessary upgrades. After the success of the Abbamonte group this seems now to be quite different -- with a little luck there will be soon much more data of a higher quality. 

One should also keep an eye on the Resonant Inelastic X-ray Scattering (RIXS). This technique is characterized by a million fold or so larger carbon footprint but it is at the instrument builders frontier involving gigantic beamlines at last generation synchrotrons. The difficulty is that it reveals much more "dirty" information -- it is not a simple linear response affair and it is still littered with interpretational difficulties. 

\begin{figure}[t]
\includegraphics[width=0.9\columnwidth]{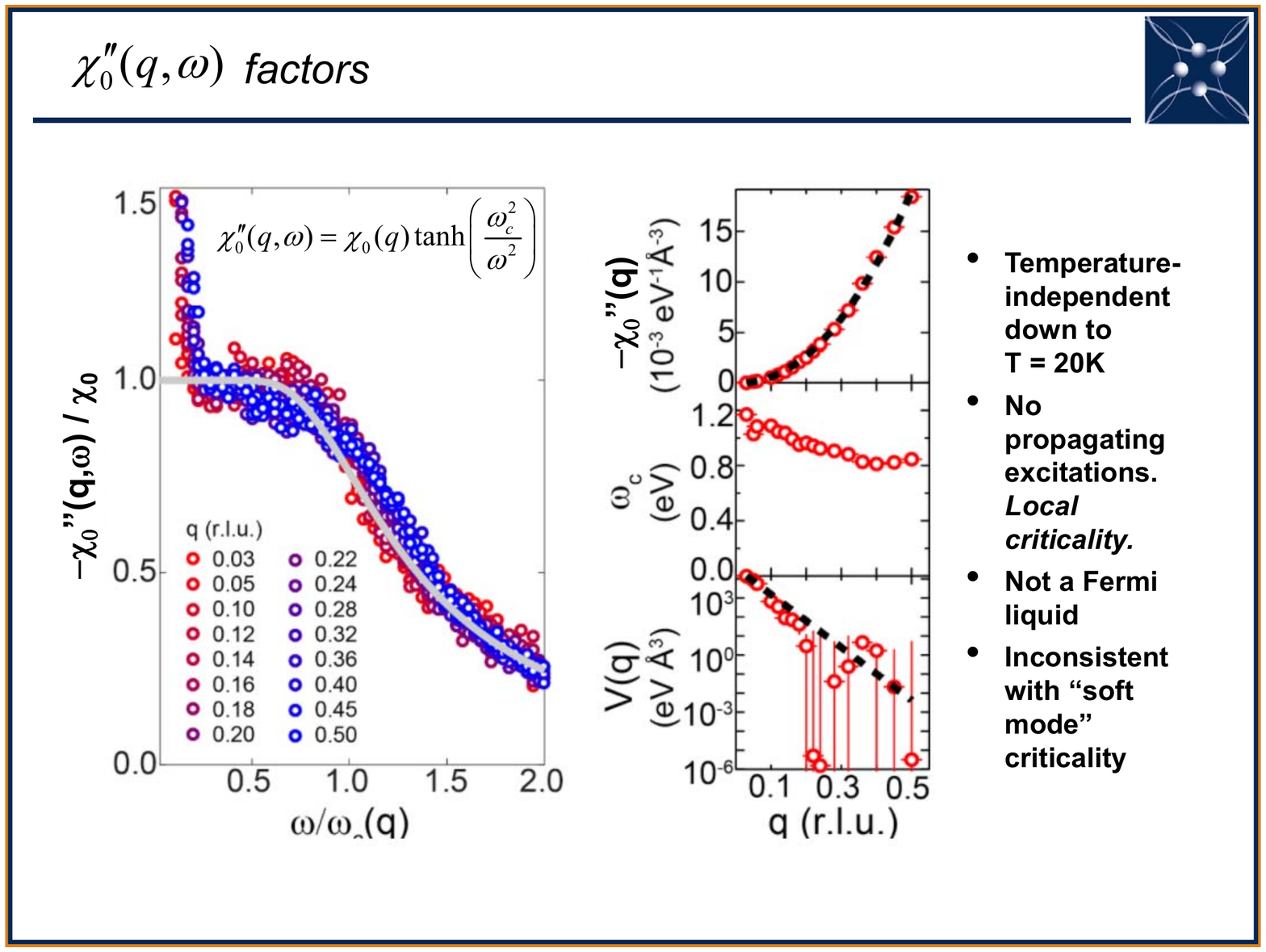}
\caption{The electron loss function in an optimally doped cuprate, in a large range of frequencies and momenta \cite{AbbamonteEELS1}. 
The data show an impeccable scaling collapse for all measured momenta, demonstrating the local quantum criticality governing the charge density response.The frequency dependence is is marginal -- frequency independent -- up to a UV cut off $\omega_c \simeq 1$ eV.}
\label{fig:AbbamonteEELS}
\end{figure}

The main outcome of the pioneering EELS experiment is shown in Fig. (\ref{fig:AbbamonteEELS}) obtained in an optimally doped cuprate \cite{AbbamonteEELS1}. This should be directly compared with the (longitudinal) prediction for the spectral function associated with dynamical charge susceptibility of the Fermi-liquid with the plasmon and especially the Lindhard continuum. It looks entirely different: instead of the Lindhard continuum with its strongly dispersing upper bound as function of momentum and linear frequency dependence (Section \ref{holochargeexc}, 
Fig. \ref{fig:lindhardfig})  one finds a continuum that is (a) {\em independent} of frequency up to a cut off $\sim 1$ eV (the "marginal" frequency scaling dimension) and especially (b) it appears to be entirely momentum independent up to momenta halfway the Brillouin zone as highlighted by the scaling collapse in the figure. 

I stressed the universality of the plasmon rooted in total momentum conservation and the long ranged Coulomb interaction. The claim of Abbamonte et al.  is that the plasmon, known  from optics to have  already  an unusual short life time at zero momentum, gets completely overdamped very rapidly upon increasing momentum. This is strikingly different from the behaviour in Fermi-liquids where the Landau damping sets in at rather large momenta.  

These observations constitute the best, direct experimental confirmation for the claim that the cuprate strange metal is local quantum critical, the $ z \rightarrow \infty$ scaling acting prominently in Section (\ref{holoSM}). I perceive it actually  as the most direct evidence for the cuprate strange metals to be of a holography-type  quantum supreme nature.  Such scaling behaviour is surely beyond the realms of semiclassical "particle physics" while it is also beyond what is possible in stoquastic quantum critical states. I remind the reader of the general scaling arguments explained in Section (\ref{EMDscaling}) insisting that the hyperscaling violation exponent $\theta$ has to come to rescue to prevent the zero temperature entropy to take over. The diverging dynamical critical exponent is just the anomalous dimension of choice to once and for all disqualify quasiparticles. 

Yet there are still quite some issues that are far from settled. How does it relate to the branchcut found in the same energy range in the optical conductivity? This would translate via the continuity equation into a (transversal) polarization propagator scaling like $1 / (i\omega)^{1 + \alpha}$ at $q =0$, quite different from the marginal scaling suggested by the  EELS data. The dogma is that at $q=0$ one moves the electron fluid uniformly and  it is impossible to discern a different transversal- and longitudinal response. The only loop hole in this apparent truism I am aware off involves the goldstone bosons: when there is structure in the unit cell the acoustic phonons are subjected to Umklapp and optical phonons appear at zero wave-vector. It is of course well known that the transversal- and longitudinal optical phonons are different at zero momentum. I emphasized in Section (\ref{holoMott}) that the Mott insulator is fundamentally   an electron crystal where the acoustic phonons are gapped by commensurate pinning. Could it be that the observation of a very different transversal- and longitudinal response is somehow anchored in the optical phonons of the electron crystal?  In other words, is this a hitherto unrecognized ramification of the Mottness of Section (\ref{Mottness}) persisting in the metallic phase? 

Finally, the first EELS results on the doping dependence are also surprising \cite{AbbamonteEELS2} , emphasizing further the differences between the transversal and longitudinal  response. The claim is that at high temperatures it all looks the same as for optimal doping. But upon lowering temperature changes are seen as function of doping. The momentum independence is within the resolution of the experiment not affected, but in the underdoped regime the authors find that at lower temperatures the charge susceptibility undergoes a reorganization on a large energy scale. The lower energy part changes drastically, showing  signs of becoming relevant in the same sense as discussed for the holographic superconductivity (Section \ref{HoloSC}), bending upwards as function of decreasing energy. But this makes sense, just indicating that an instability towards charge order starts to develop. The intriguing part is that this seems to happen over a large momentum range:  that the "stripy" charge order wavevectors are picked up is apparently a last minute affair. 

The temperature evolution in the overdoped regime is the real surprise: upon lowering temperature a "hole" appears in the low energy spectral weight. This adds further mystery to the overdoped strange metal: one would expect to see this scale-full behaviour to imprint on the deep IR but nothing is seen in the DC data at the characteristic temperature indicated by the EELS data. Altogether the big deal is that the $E/T$ scaling is grossly violated: these changes happen on an energy scale $\sim 0.5$ eV which are very large as compared to the temperatures $\sim 100$ K where it sets in both on the under- and overdoped side. This is reminiscent  of the spectral weight transfer effects that are tied to Mottness.       

I expect further advances in a near future on this frontier. The results of the Abbamonte group are obtained using a low energy reflection mode EELS machinery. This has its disadvantages as compared to the very high energy transmission EELS. The latter is an old technique, that used to have a bad energy resolution of order of 0.5 eV. But with modern electronic optics this can be much improved and presently several groups are building such new rigs inspired by the success of Abbamonte's machine.        

\subsection{The fermions: ARPES and STS.}   
\label{Cupratefermions} 

Considering what experimentalists can do in the lab, the only spectroscopic techniques that deliver the required dynamical information with astonishing momentum, energy and temperature resolution are the "fermionic" photoemission (ARPES) and scanning tunneling spectroscopy (STS) technique. I already discussed these in Section  (\ref{Holofermions}) arriving at the conclusion that because of the large $N$ UV sensitivity holography is in this regard not a useful navigational aid. 

ARPES is just the most direct way to find out whether one is dealing with a Fermi-liquid. The quasiparticles manifest themselves as poles, Lorentzian peaks that disperse around as function of momentum mapping out the band structure. Their widths reflect the imaginary part of the self energy, typically varying like $\omega^2$. In fact, when ARPES started to work in the 1970's it had a great impact directly demonstrating that band structure  does exist in a quite literal fashion in simple electron systems. 

Because of practical circumstances such as the two dimensional nature of the cuprate electron systems ARPES works particularly well in copper oxides. When in the mid 1990's high resolution data became available these seemed to show a lot of dispersing peaks suggestive of band structure, including "Fermi-surfaces" of the kind expected from the DFT band structure computations. This had a big impact -- one heard people say that without ARPES nobody would have gotten the idea to explain "high Tc" in terms of Fermi-liquid style quasiparticles. After the fact the quasiparticles  were again occupying the main stage -- the fermiology main stream I already alluded to.  

Ironically, very recently a case is developing indicating that these "moving peaks" may be quite deceitful in this regard. As the example of the CFT fermions (Fig. \ref{fig:fermionbranchcut}) illustrates, dispersing peaks do not prove anything regarding quasiparticles. Whether one is dealing with quasi-particles or either "un-particles" is encoded in the analytical form of the {\em line shapes}. Especially the energy dependence of the ARPES line shapes is for reasons associated with the imperfections of the technique not easy to nail down empirically. 

To recognize the best established "unparticle signatures" in the lineshapes one has to focus in on the underdoped regime, while it is pending the direction of the momentum. One distinguishes in the Brillouin zone the directions corresponding with the lattice directions in real space -- the "anti-nodal" direction -- from the zone diagonal "nodal direction". The d-wave superconducting gap is at maximum at the anti-nodes while it is vanishing at the nodes, explaining the terminology. Independent of doping,  rather sharp peaks are found to disperse near the nodes. 

However, at the anti-nodes there is a lot of action. This is the point of departure for the recent  detection of a "first order like" transition by ARPES that I announced in the above \cite{Sudijump}.   Above the critical doping one discerns at the anti-nodes also "quasiparticle" peaks for a fixed momentum as function of energy ("energy distribution curves", EDC's). But upon crossing $p_c$ into the underdoped regime these abruptly vanish being replaced by featureless "unparticle" continua -- the signature of the "first-order like" transition  (Section  \ref{qucritphase}). 

Remarkably, when temperature is decreasing through the superconducting transition sharp peaks appear also at the antinodes, growing from the incoherent spectral weight. Remarkably, the spectral weight of these peaks appears to scale with the {\em superfluid density} \cite{ZXShenSFdesnity}. This is completely beyond anything that one can discern from our understanding of conventional BCS superconductivity.  There is no doubt that deep in the superconducting  state  {\em genuine} quasiparticles of the Bogoliubov kind are present. The sharp peaks spanning up a whole "Fermi-surface"  seen in  ARPES are consistent with  the phase sensitive "quasiparticle interferences"
measured in STS demonstrating directly their quantum mechanical wave nature. But this is in the greater context not surprising: the BCS-like superconducting state is a SRE product and as a vacuum it has to support quasiparticles. These appear to reconstruct in the superconducting state from a normal state where they are absent at least in the underdoped regime.  

In turn, these have bearings on another set of observations playing a key role in the belief system of a group who is in the grip of trying to prove that the Hertz-Millis style QCP is the culprit. At very low temperatures and high magnetic fields quantum oscillations are observed suggestive of the existence of a Fermi surface associated with a genuine Fermi-liquid. One deduces from these measurements the kind of mass enhancements expected for such a QCP. The trouble is however that it is established beyond any doubt that these systems are still deep in the "vortex fluid" regime given the magnetic fields that can be achieved. This means that these systems are still fully-fledged superconductors with a relatively low density of free magnetic fluxoids that take care of the dissipative properties of the fluid. Hence, the  "Bogoliubons" stabilized by this superconductivity are persisting and although it is still a bit in the dark how these manage to form a Fermi-surface this is obviously not telling anything about what is going on  in the normal state at room temperature. 

This is the instance to introduce the stunningly strange nature of the actual line shapes in the metallic state. Although noticed early on \cite{ARPESdichearly} this is so strange that it is completely overlooked by large parts of the community. Let us focus on the "unparticle" antinodal continua in the underdoped strange metal. Instead of measuring the line shape as function of energy one can also fix energy and vary momentum ("momentum distribution curves", MDC's). Also at the anti-nodes the MDC's correspond with rather sharp Lorentzian peaks, with a width comparable to the nodal fermions, despite the fact that energy wise these are smooth branchcut like continua! This is just plainly mysterious. 

Measuring the MDC linewidths is for technical reasons easier than to study the EDC's. High precision measurements were carried out of the energy dependence of these momentum space widths $\Gamma_k (\omega)$, especially in the nodal regime. At optimal doping   $\Gamma_k (\omega) \sim \omega$, the traditional main stay of the marginal Fermi-liquid school of thought: this is obtained by assuming free fermions decaying by second order perturbation theory in the "marginal" continuum of Section  (\ref{EELSlocqucrit}). This story then continuous claiming that transport and so forth is just like in the Fermi-liquid except that now the single particle life time is set by this   inverse life time $\sim \omega$ or $T$, as usual. One can then continue to blame all the strangeness to be caused by this anomalous decay of the quasiparticles: this is the central assertion of the old marginal Fermi-liquid idea \cite{MFL}. 

\begin{figure}[t]
\includegraphics[width=0.9\columnwidth]{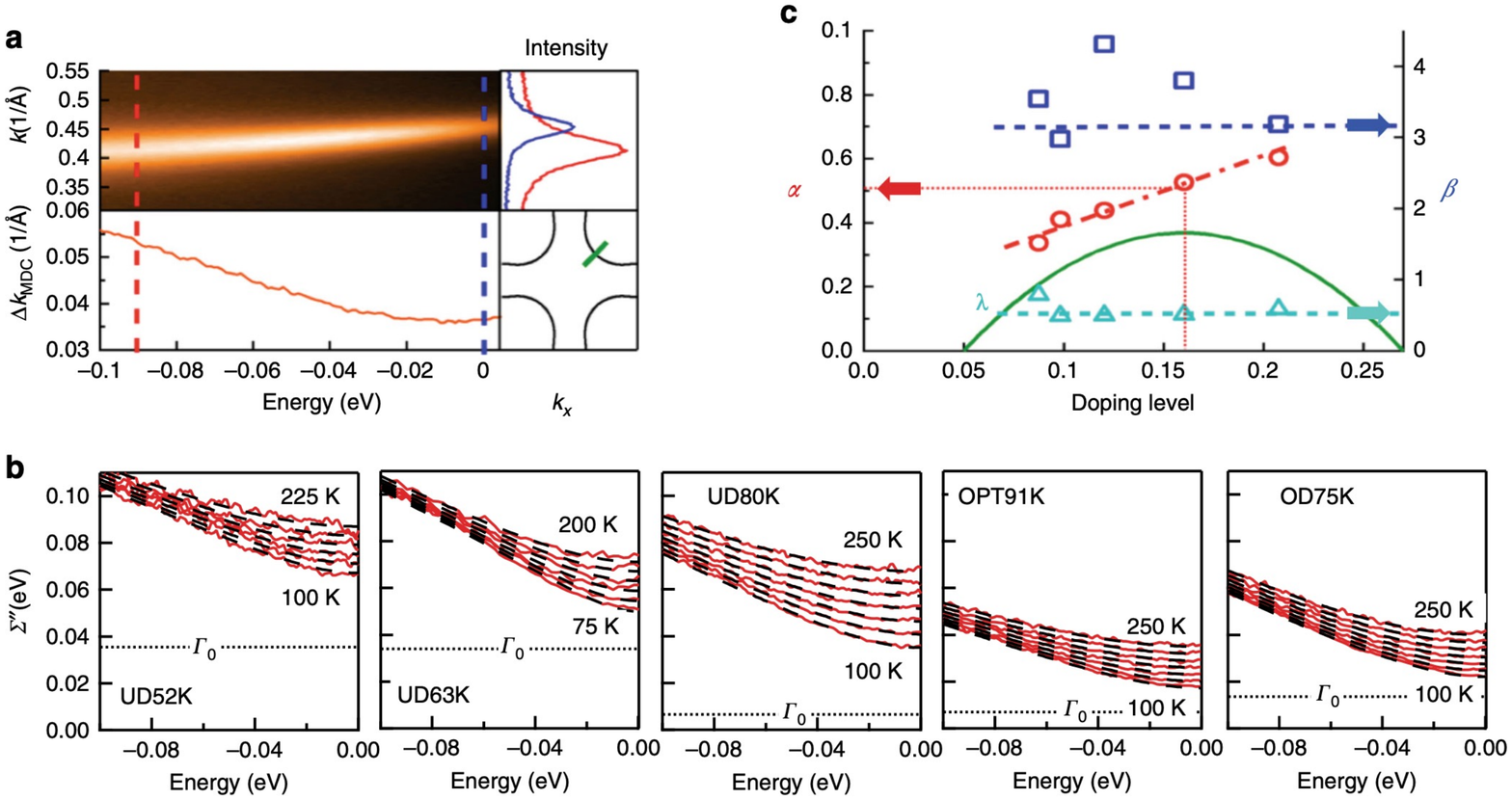}
\caption{The measurement of the momentum-width of the "quasiparticle" peaks by angular resolved photoemission  \cite{Dessaunodal}. (a) False colour imagine of the spectral function near the nodes as function of momentum and energy, illustrating the extraction of the momentum width $\Gamma_k$, see main text. (b) The momentum width as function of doping, showing (c) that only the exponent $\alpha_p$ (red line) is strongly doping dependent.}
\label{fig:Dessaunodliq}
\end{figure}

However,  recently it was found that the scaling of this MDC width is actually rather strongly doping dependent (the "nodal liquid") \cite{Dessaunodal}. The results are shown in Fig. (\ref{fig:Dessaunodliq}). Defining $\Gamma_k (\omega) = \Gamma_0 + \lambda \left( (\hbar \omega)^2 + (\beta k_B T)^2 \right)^{\alpha_p}$ one obtains high quality fits. $\Gamma_0$ is a small "elastic" broadening while the remainder shows the $E/T$ scaling. The parameter $\beta$ relating temperature and energy is $\sim \pi$ and rather doping independent, as is the "coupling" $\lambda$. However, the exponent $\alpha_p$ is strongly doping dependent, ranging all the way from $0.3$ to $0.7$ in the indicated doping range --  the "marginal" value $0.5$ at optimal doping appears just as a coincidence. 

Yet again, this signals smoothly varying anomalous dimensions indicating that one is dealing with a quantum critical {\em phase}. It is in fact of the same kind as found for the  RN strange metal holographic fermions \cite{HongScience}  of Section (\ref{Holofermions}). One is tempted to interpret this involving the "second order perturbation theory" decay of quasiparticles in some kind of quantum critical heat bath. But there is perhaps no better way to understand the deep theoretical inconsistency invoked in such an intuitive assignment as for the reasons discussed in Section (\ref{Holofermions}), identifying it as a large $N$ pathology. 

But let's get back to the EDC's. I already highlighted the strange disconnect between momentum- and energy line shape so manifest for the underdoped antinodal response. But what happens along the nodes? In a recent high precision EDC line shape study near the nodes it was revealed that a similar line shape anomaly is at work here \cite{HeumenARPES}. As it turns out the nodal EDC's measured with very high quality ARPES are of the following form, 

\begin{eqnarray}
A(k, \omega) & =  &  \frac{1}{\pi} \frac{ W (k, \omega) \Gamma_k (\omega)} { (\omega - v_F k)^2  + \Gamma^2_k (\omega)} \nonumber \\
W(k, \omega) & = & A_W + B_W \omega^{\alpha_W}
 \label{Wdefinition}
\end{eqnarray}

Near the nodes these are characterized by a linear dispersion $\varepsilon_k = v_F k$ while $\Gamma_k (\omega)$ refers to the MDC width that I just discussed. This is just business as usual  except for the factor $W (k, \omega) $ in the numerator.  As it turns out, this encodes for a "fat" powerlaw tail ($\alpha_W > 0$). This appears to be at maximum around optimal doping, gradually diminishing until $W \simeq 1$ in the strongly overdoped regime. 

The take home message is that the "branchcut EDC/quasiparticle pole MDC" that is so obvious at the antinodes in the underdoped metallic regime is actually still at work even in the nodal regime up to high dopings. Given the difficulties to determine the EDC lineshapes this was just hidden from view until very recently.  Once again, this frequency dependence is reminiscent  of the "unparticle" branch cut spectra of  CFT's (e.g., Fig. \ref{fig:fermionbranchcut}). However, for a CFT the MDC's are also of the branch cut form, while here the momentum dependence is quasiparticle like. Although there is not enough data available for a mathematical proof -- for Kramers-Kronig reasons the spectral functions above $E_F$ are  required -- one can make the case that it is virtually impossible to associate this with an outcome for a perturbative self-energy controlled by the free limit. 

 It is presently completely unclear how to discern the underlying physical principle that is responsible for this momentum wise quasiparticle -- energy wise unparticle behaviour. So much is clear that given the latter one cannot possibly jump to the presently widespread community belief largely based on ARPES that the cuprate strange metals are in essence Fermi-liquids with some special effects. To be continued.

\section{Epilogue}

I have apparently managed to capture the reader who got all the way to this point in this text with the narrative. I hope I got across the excitement as I perceive it. At stake is whether mankind is in the process of discovering that forms of macroscopic matter exist that are in the grip of the essence of quantum physics: dense many body entanglement. Although still hypothetical it is convenient to give it a name.  My proposal is to call it  "quantum supreme matter". 

Once again, without equations we cannot observe these realms of reality, and in this regard the news comes from the mathematical frontier. Holography is enlightening in the regard that it demonstrates possibility.  The AdS/CMT equations reveal the existence of a set of physical principles that are entirely different from the paradigm born in the 1950's revolving around the emergence of {\em classical} matter (in the quantum information sense) from the quantum physics governing the microscopic scales. As a bonus, it is as a theoretical construct quite cool: "using black holes as quantum computer." 

Eventually, all what really counts is whether nature wants to submit to these rules, the prime suspects being found in the strongly interacting electron systems of condensed matter physics. As I emphasized especially in the final experimental section, the case is suggestive but far from conclusive. Yet again, holography is far from perfect  as a GPS to navigate this landscape given all the "UV sensitivities" that have been identified, and especially those that are still in the realms of the unknown-unknowns. At the same time, it is {\em stand alone} as a mathematical machine that is telling stories regarding the physical world behind the "quantum supremacy brick wall". 

In my perception, further progress in the foreseeable future is in first instance to be expected from the condensed matter laboratories: the experimentalists are presently in the driver seat. Although camouflaged as lecture notes for a graduate course, my real agenda has been to produce a manifest intended to catalyze  the community effort by attempting to capture the big picture in the simplest possible words. Besides the experimental community,  another group of potential stake holders is the community that is engaged in getting the quantum computer to work. My impression is that the affair highlighted here is not quite on their radar. This quantum supreme matter qualifies to be on top of their benchmarking list, as testing ground for the early generations of quantum simulators. 

I have presented it as a body of ideas that are to quite a degree settled in, at least in the various sub-communities. However, in one crucial regard this has been a bit misleading. The AdS/CMT expert will have noticed that I forwarded a strong claim that is presently not at all a community consensus. I am  convinced that I have managed to decode the essence of the message hidden in the "whispers of the holographic oracle". 

Yet again, from the very beginning my intrigue in holography was fuelled by a sense of eerie familiarity that I could all along not quite nail down, until recently. After realizing it, I perceive it as a no-brainer. As all great achievements in physics, it is very simple but it takes a while before our ape-brain gets used to it. The key is contained in the 1970's notion of universality  governing thermal phase transitions of the right kind. Universality actually means that the physics becomes extremely simple, as captured by simple scaling relations requiring a minimum of information: the story of dimensionality and symmetry. As a historical accident, this arose in an era that physicists were not aware of mathematical complexity theory.  I am not claiming a mathematical proof, but I find it obvious. {\em A maximal exponential complexity of the unitary evolution spawns minimal complexity in the read out}, using the quantum computer as metaphorical device. 

Once again, observables are expectation values of few-point correlators -- the VEV's -- and our relatively thorough understanding of stoquastic quantum critical states reveals that the delocalization in the exponentially large Hilbert space spawns a perfect averaging, wiping out all information other than the most basic properties of symmery and dimensionality. It is Einstein's "our Lord is subtle but not malicious" in overdrive. What should be completely beyond the capacity of human comprehension turns in the physical world into a highlight of Wigner's "unreasonable effectiveness of mathematics in the natural sciences".

The latter is embodied by holography. After getting used to the way it works, in no time one becomes acutely aware of its Platonic qualities. As I tried to get across in these notes, it is endowed with the aesthetics of mathematics, it is elegant and above all {\em simple}. It is the reason for finding only simple equations in this text:  aspects that really matter are captured in basic GR structures like the metric of a Schwarzschild black hole in an AdS space. 

At zero density one learns that one has to fine tune to a quantum critical point in order to rid the system of energy scales that are detrimental for the quantum supremacy. One then realizes that at finite density that there is one scale that cannot be avoided: the Fermionic degeneracy scale. Given the sign problem it is also clear this this scale will not necessary spoil the exponential complexity. How to reconcile these requirements? 

The black holes offer the answer: the structure of the renormalization group being encoded in the structure of the curved geometry in the bulk, revealing a highly mathematical disciplined unique outcome: the {\em covariant RG}. This lay hidden in the EMD classification that was discovered some ten years ago -- I am grateful to Blaise Gouteraux for alerting me regarding this overarching, deceptively simple principle. I just added to this the realization that this also captures the weakly interacting Fermi liquid, yet again getting the idea due to earlier work.   Subir Sachdev who has been my teacher in many regards in this subject matter was at a point quite busy with this affair in reversed gear: trying to make the point that there are hidden Fermi-surfaces in holographic strange metals, with the  emphasis on the similarity of the scaling properties. 

This then leads to the notion of "generalized Fermi-liquids". As  the Landau free critical theory forms a template for the strongly interacting state, so does the Fermi liquid reveal the way that the observables of the quantum supreme generalization are basically organized. As for the stoquastic case, the latter involves anomalous dimensions and so forth, and although a lot more is going on than in the case of the stoquastic incarnations  the same sense of {\em universality} (the perfect averaging of the VEV's) is in effect.  

What should the experimentalists look for? {\em The key is the "unreasonable simplicity" revealing this generalized universality notion}. In fact, this has been all along responsible for the intrigue with the high Tc data: the linear resistivity, the optical branchcut, more recently the EELS data and e.g. the "Planckian quadrature" in the magneto transport. But I did emphasize that to detect the quantum supreme physics one needs to know where to look for it. It is different from the present emphasis on detecting order which is tied to the classicalness of the SRE product states. In the first place, this stuff will show its face best in {\em dynamical} linear response information. In so far this information is available it is encouraging, but the challenge for the instrument builders is to expand this repertoire. The detection of the pair susceptibility as discussed under holographic superconductivity may be inspirational as an illustration of what is required. Similarly, a main contrast between "classical" and quantum supreme physics is perhaps in transport: the "minimal viscosity hydro" versus the extreme gas physics of conventional metals. Yet again, this involves cutting edge instrumental building, but now in the realms of quantum transport devices. 

Let me zoom out even further. In fact, all of this is no more than {\em diagnostics}: the signals to look for that may prove or disprove the existence of quantum supreme matter. But  is it good for anything beyond the satisfaction of the curiosity of a small band of physicists? Mankind should be more grateful to our forefathers than they actually are. For instance,  without the work that revealed the nature of "SRE product matter" the electronics revolution culminating in the magical little boxes called smart phones would not have happened. It spawned semiconductor physics -- equipping the engineers with the curiosity driven insights of our clan worked miracles.  But now we may be on the verge discovering states of matter obeying a completely different rule book. Will this further the control of mankind over the natural world, for its material benefit? 

I perceive this as the 64 k$\$$ question. Of course the famous quote from a 19-th century representative in the British parliament that it was not to be expected that Maxwell's equations would ever earn a penny of tax income is of course applicable also in this context. But I am not overly optimistic. The gross message seems to be  that to our present state of understanding quantum supremacy of matter just means that humble observers that can only control circumstances after the collapse are confronted with the {\em loose of control} in a most absolute sense. What else can this stuff do than convert useful work into heat in the most efficient way as permitted by the laws of physics?  

\begin{acknowledgments}

As I emphasized I reported on the outcomes of a large effort of various sub communities in physics. Accordingly, I am indebted to a large number of individuals who taught me crucial insights. There are just too many to list -- feel good when you infer praise in the text acknowledging your name. The 2021 advanced topics of theoretical physics class of the dutch Delta Institute of Theoretical Physics (DITP) deserves special mention. I used them to test drive a prototype of these lectures  and the response of the students was a delight. This was instrumental for the improvement of the presentation, as testified by a comparison between the draft (first) version of the notes and this improved version.  This project has received funding from the Netherlands Organization for Scientific Research (NWO/OCW).

\end{acknowledgments}

\end{document}